\documentclass[
aps,
prd,
twocolumn,
twoside,
a4paper,
showpacs,
superscriptaddress,
groupedaddress
eqsecnum,
reprint,
showkeys,
preprintnumbers,
nofootinbib
]{revtex4-1}

\input{Main.sty}

\input{Main.cls}

\newcommand{\nn}{\nonumber}

\begin{document}

\preprint{IFT-UAM/CSIC-20-94}
\title{The clustering dynamics of primordial black boles in $ N $-body simulations}

\author{Manuel Trashorras}
\email{manuel.trashorras@csic.es}

\author{Juan Garc\'ia-Bellido}
\email{juan.garciabellido@uam.es}

\author{Savvas Nesseris}
\email{savvas.nesseris@csic.es}

\affiliation{Instituto de F\'isica Te\'orica UAM-CSIC, Universidad Auton\'oma de Madrid, Cantoblanco, 28049 Madrid, Spain}

\date{\today}



\begin{abstract}


We explore the possibility that Dark Matter (DM) may be explained by a non-uniform background of approximately stellar-mass clusters of Primordial Black Holes (PBHs), by simulating the evolution them from recombination to the present with over 5000 realisations using a Newtonian $ N $-body code. We compute the cluster rate of evaporation, and extract the binary and merged sub-populations along with their parent and merger tree histories, lifetimes and formation rates; the dynamical and orbital parameter profiles, the degree of mass segregation and dynamical friction, and power spectrum of close encounters. Overall, we find that PBHs can constitute a viable DM candidate, and that their clustering presents a rich phenomenology throughout the history of the Universe. We show that binary systems constitute about 9.5\% of all PBHs at present, with mass ratios of $ \bar{q}_{\rm B} = 0.154 $, and total masses of $ \bar{m}_{\rm T,\,B} = 303\,M_\odot$. Merged PBHs are rare, about 0.0023\% of all PBHs at present, with mass ratios of $ \bar{q}_{\rm B}= 0.965 $ with total and chirp masses of $ \bar{m}_{\rm T,\,B}= 1670\,M_\odot$ and $ \bar{m}_{c,{\rm M}} = 642\,M_\odot $ respectively. We find that cluster puffing up and evaporation leads to bubbles of these PBHs of order 1 kpc containing at present times about 36\% of objects and mass, with hundred pc sized cores. We also find that these PBH sub-haloes are distributed in wider PBH haloes of order hundreds of kpc, containing about 63\% of objects and mass, coinciding with the sizes of galactic halos. We find at last high rates of close encounters of massive Black Holes ($ M \sim 1000\,M_\odot$), with $ \Gamma^{\mathrm{S}} = \unit [(1.2 \substack {+5.9 \\ -0.9}) \times 10^{7}] {yr^{-1} Gpc^{-3}}$ and mergers with $\Gamma^{\mathrm{M}} = \unit [1337 \pm 41] {yr^{-1} Gpc^{-3}} $.

\end{abstract}

\keywords{primordial black holes, dark matter, black hole binaries, black hole mergers, $ N $-body simulations}

\maketitle



\section{Introduction}
\label{sec: Introduction}

One of the most pressing problems in cosmology is that of the nature of Dark Matter (DM), first suggested in the thirties by Zwicky from the observation of motion anomalies in the Coma galaxy cluster in Ref.~\cite{Zwicky:1933gu}, and by Rubin in the seventies from the observation of an excess velocity in the tails of galaxy rotation curves in Ref.~\cite{Rubin:1970zza}. Evidence for DM is now abundant, from structure formation, galaxy dynamics, lensing, and Cosmic Microwave Background (CMB) observations (see Refs.~\cite{Massey:2007wb,Ade:2015rim,Aghanim:2019ame,Aghanim:2018eyx}), largely in agreement with a DM density of $ \Omega_c h^2 = 0.120\pm 0.001 $ (see Ref.~\cite{Aghanim:2018eyx}).

For four decades now there have been a myriad attempts to explain the nature of DM, with possible candidates spanning a huge difference in order of magnitude, with masses from $ \unit [\order {10^{-22}}] {eV} $ to $ \unit [\order {100}] {\msun} $. From new weakly interacting particles (WIMPs) such as ultra-light axions (see Ref.~\cite{Hui:2016ltb}) emerging in some extensions of the Standard Model of particle physics, to hypothesised Massive Halo Compact Objects (MACHOs) such as massive Black Holes (BHs) (see Ref.~\cite{Bertone:2010zza}) populating galactic halos.

However, and in spite of the very numerous efforts, no viable DM particle candidate has been found in experiments like the Large Hadron Collider (LHC), nor a convincing signal has been detected either in ground or underground facilities, directly or indirectly, with cryogenic detectors such as the Cryogenic Dark Matter Search (CDMS), or liquid Xenon detectors like LUX, nor any annihilation probes like Fermi have found an excess signal in the sky that may be attributed to it. So overall, no experiment has yet observed a statistically significant direct detection of a DM particle candidate \cite{Lin:2019uvt,Schumann:2019eaa}. However, the XENON1T detector recently reported an excess over known backgrounds consistent with the solar axion model, albeit only at a low statistical significance of $\sim3.5\sigma$ \cite{Aprile:2020tmw}.

Yet, ever since the first LIGO observation of Gravitational Waves (GWs) (see Refs.~\cite{Abbott:2016blz,TheLIGOScientific:2016htt,Belczynski:2016obo}) from the merger of $ \unit [\order {10}] {\msun} $ BHs and subsequent detections, and additional merges observed in LIGO runs O1 and O2, another hypothesis, that is, that DM may be partially or entirely made of Primordial Black Holes (PBHs), is gaining ground, and a number of models been proposed (see Refs.~\cite{Clesse:2015wea,Bird:2016dcv,Clesse:2016ajp,Sasaki:2016jop,Carr:2016drx,Garcia-Bellido:2017fdg}).

Experiments like LIGO-VIRGO herald a new era of GW astronomy in which new observations previously unavailable in the electromagnetic channel become now visible in the gravitational one, thus opening a powerful and far reaching window to the Universe. In particular, it is expected that, should DM be made of PBHs, then this new observational probe would be open to explore DM in novel ways~\cite{Clesse:2020ghq}.

Primordial Gravitational Waves (PGWs) arising from the interaction of these PBHs may also prove to be a useful tool to constrain the period of inflation in the early Universe, through their abundance, clustering, slingshot and merger event rates and dynamical profiles. Moreover, a Stochastic Gravitational Wave Background (SGWB) arising from the PBH-PBH interaction history from the Early to the late Universe may be detected in the future (see Refs.~\cite{Clesse:2016ajp,Clesse:2018ogk}) by spaced based interferometers, Fast Radio Bursts (FRB) (Refs.~\cite{Connor:2016rhf,Munoz:2016tmg}), Pulsar Timing Arrays (PTA) (Refs.~\cite{Kramer:2010tm,Schutz:2016khr}) and $ \unit [21] {cm} $ experiments (Refs.~\cite{Madau:1996cs,Tashiro:2012qe,Gong:2017sie,Hasinger:2020ptw}).

Observational evidence of a DM component of the Universe is abundant: from purely dynamical probes such as galaxy rotation curves and star velocity dispersions to weak and strong gravitational lensing of distant galaxies, cosmic microwave background spectrum of temperature anisotropies, structure formation history, type Ia supernovae distance measurements and galaxy, cluster and quasars surveys probes, such as the baryon acoustic oscillation scale, redshift space distortions and Lyman- $ \alpha $ spectroscopy. All of these probes allow for the possibility that all of DM may consist of BHs in the stellar mass-range, particularly if BH clustering and BH mass spectra are considered (see Refs.~\cite{Belotsky:2018wph,Desjacques:2018wuu,Calcino:2018mwh,Garcia-Bellido:2017imq}).

If so, these PBHs may offer a rich phenomenology not only able to explain the nature of DM, but also to explain some questions regarding early structure formation~\cite{Inman:2019wvr}, the radial profiles of galactic cores, ultra-faint dwarf galaxy dynamics, correlations between the X-ray and Cosmic Infrared Background, as well as BH merger rates, mass spectrum and low spins (see Ref.~\cite{Clesse:2017bsw} for a review of these issues).

Our paper is organised as follows: In Section~\ref{sec: Theoretical background} we present the theoretical background of PBH formation, while in Section~\ref{sec: Methodology} we describe how our simulations were designed and performed, their properties, the choice of the Initial conditions (ICs) and the Evolution Snapshots (ESs) as well as their limitations. We also, for better intuition of our results, provide animations of our simulations in the public repository \href{https://www.dropbox.com/sh/lrgxwcgx27mv097/AAA49x_wNSadknlFPVYaKYema?dl=0}{here}. Once performed, the analysis of the simulations can then be grouped in two broad categories, one regarding the separate populations that arise in the simulations, and another regarding the dynamics.

Concerning the former, Section~\ref{sec: Population statistics} is devoted to the identification of differentiated populations arising in the simulations, which are mainly those of clustered and ejected PBHs, bounding and mergers of PBHs, evaporation rates, parent trees and merger trees, while Section~\ref{sec: Close encounters} deals with the close encounter dynamics where two PBHs passing sufficiently near to each other produce a merger pair or a binary pair.

Concerning the latter, Section~\ref{sec: Dynamical distributions} is dedicated to the study of the mass, position, velocity and density profiles of PBHs, segregated by their population, as well as to the degree of mass segregation and dynamical friction, while in Section~\ref{sec: Orbital distributions} we examine the orbital distributions of PBHs and its repercussions for both either transient or stable binaries. In Section~\ref{sec: Background DM implications} we present estimates of the hyperbolic encounter and merger event rates for a medium-sized galactic halo and comoving cosmological volume.

Finally, in Section~\ref{sec: Conclusions} we present our conclusions, summarizing our findings and comment on the feasibility of PBHs as a DM candidate according to our results.



\section{Theoretical background}
\label{sec: Theoretical background}


\subsection{Formation mechanisms}
\label{subsec: Formation mechanisms}

First proposed in the sixties when Zel'dovich and Novikov (see Ref.~\cite{Zeldovich:1967}) realised that BHs could very well have been formed in the early Universe, catastrophically growing by accreting the surrounding plasma during the radiation era from the extremely high gravitational forces. It was later, in the seventies, that Hawking (see Ref.~\cite{Hawking:1971ei}) and Carr (see Refs.~\cite{Carr:1974nx,Carr:1975qj}) found that a sufficiently overdense region or inhomogeneity in the early Universe could undergo gravitational collapse providing the seeds for formation of these BHs, more commonly referred to as PBHs in this context.

Several mechanisms have been proposed in the literature to form these PBHs in the early Universe (see Refs.~\cite{Carr:1996aa,GarciaBellido:1996qt}), such as domain walls, cosmic strings, and vacuum bubbles. In other models, however, they are formed by the gravitational collapse of overdense regions in the Universe either in the radiation era or, less frequently, the matter era. The existence of such overdensities is a consequence of certain inflationary models, and could be, if discovered, used to probe dynamics during inflation.

A typical scenario in which PBHs may form is as follows. During inflation high peaks in the primordial curvature power spectrum produce overdensities that grow and collapse gravitationally in the radiation era~\citep{GarciaBellido:1996qt}. These high peaks are produced naturally by a number of inflationary models, such as in hybrid inflation by having a phase transition at the symmetry-breaking field (see Ref.~\cite{Clesse:2015wea}), single-field inflation with an inflection point (see Refs.~\cite{Garcia-Bellido:2017mdw,Ezquiaga:2017fvi,Ezquiaga:2018gbw}) or by having the inflaton coupled to gauge fields (see Refs.~\cite{Garcia-Bellido:2016dkw,Garcia-Bellido:2017aan}). The clustering and mass spectrum of these PBH is highly dependent on the shape of the primordial curvature power spectrum; a narrow peak will typically produce a nearly monochromatic spectrum of PBHs, while a broader peak will produce PBHs with a wider mass range of $ \unit[\order {0.01}] {\msun} - \unit[\order {1000}] {\msun} $. This mass power spectrum could then evolve by hierarchical merging (see Ref.~\cite{Doctor:2019ruh}) right after recombination, but more importantly, from PBH masses that would greatly increase by accreting baryons from the dense plasma in the early Universe. The largest of these PBHs would then become Super Massive Black Holes (SMBHs) and InterMediate Mass Black Holes (IMBHs), provide the seeds for galaxies and accelerate cosmological structure formation at early times. These PBHs would still have a very small cross section with matter and are an ideal candidate for DM.

More generally speaking, PBHs may form with different masses greater than the Planck mass $ M_\text{P} = \unit [1.094 \times 10^{-38}] {\msun} $ (see Ref.~\cite{Carr:2005zd}) and with very low spin at different epochs as the size of the cosmological horizon increases, forming a broad spectrum, but their most frequent masses must be smaller than $ \unit [\order {10^{4}}] {\msun} $ at formation, since otherwise they would have altered significantly the early Universe dynamics by injecting energy to the medium before recombination which would be apparent in the Cosmic Microwave Background.

However, available PBHs masses in the late Universe are shifted with respect to their masses at formation due to Hawking evaporation, merger history and accretion. At present, PBHs masses may range from the largest supermassive BHs at the cores of the largest galaxies of about $ \unit [\order {10^{11}}] {\msun} $, up to the tiniest possible BHs with masses larger than $ \unit [\order {10^{–18}}] {\msun} $, which is the smallest non-evaporated BH mass possible at present (see Ref.~\cite{Page:1976df}).

Note that this mass range is much broader than that of the more familiar astrophysical BHs formed with large spins as the remnants of the most massive stars in core-collapse type II supernova, and whose masses are restricted to be heavier than about $ \unit [5] {\msun} $ but lighter than $ \unit [\order {30}] {\msun} $ (see Ref.~\cite{Rhoades:1974fn}). Therefore, an observation of very low spin BHs and BH masses out of this range would provide significant support to the PBH hypothesis. This is particularly true for the lower bound, as there is no known astrophysical mechanism that may create BHs in the sub-solar mass range, while, through merger, spinning BHs in the LIGO-VIRGO mass range are possible, even if very unlikely, although impossible without spin, specially in the high mass range.



\subsection{Observational constraints}
\label{subsec: Observational constraints}

There are many observational constraints on uniformly distributed PBHs which restrict their mass distribution. These constraints can be grouped into four different categories:
\begin{enumerate}[label=\roman*)]
	\item \textit{Evaporation constraints} arising from the fact that all PBH with a mass of less that $ \unit [\order {10^{-19}}] {\msun} $ would have evaporated by the Hawking mechanism at present-day (see Ref.~\cite{Carr:2009jm})
	\item \textit{Lensing constraints} from femtolensing of gamma-ray bursts (see Ref.~\cite{Barnacka:2012bm}) and millilensing of compact radio sources (see Ref.~\cite{Wilkinson:2001vv}), as well as microlensing constraints from the EROS collaboration (see Ref.~\cite{Tisserand:2006zx}), the OGLE collaboration (see Ref.~\cite{Wyrzykowski:2011tr}), HSC Andromeda observations (see Ref.~\cite{Niikura:2017zjd}), Kepler observations (see Ref.~\cite{Griest:2013aaa}) and caustic crossings (see Ref.~\cite{Oguri:2017ock}).
	\item \textit{Dynamical constraints} from white dwarf disruption (see Ref.~\cite{Graham:2015apa}), neutron star capture (see Ref.~\cite{Capela:2013yf}), wide halo binary disruption (see Ref.~\cite{Quinn:2009zg}), dynamical friction on PBHs (see Ref.~\cite{Carr:1997cn}) and compact stellar systems in ultra-faint dwarf galaxies and Eridanus II (see Ref.~\cite{Brandt:2016aco,Li:2016utv}).
	\item \textit{Accretion constraints} from the energy injection by PBHs in the early Universe before photon-electron decoupling by spherical accretion (see Ref.~\cite{Ali-Haimoud:2016mbv}) and accretion disks (see Ref.~\cite{Poulin:2017bwe}) on PBHs, as well as radio (see Ref.~\cite{Gaggero:2016dpq}) and X-rays (see Refs.~\cite{Gaggero:2016dpq,Inoue:2017csr}) emission from present PBHs in the late Universe.
\end{enumerate}

Moreover, while a monochromatic PBH mass power spectrum of PBHs is excluded by observations, this kind of spectrum is not very realistic since gas accretion and merger histories will both broaden and shift the mass power spectrum over time, not mentioning that PBHs can form in a broad spectrum of masses to begin with, as previously mentioned. In either case it has been found that a PBHs mass spectrum spanning a few orders of magnitude in the range of $ \order {0.01-1000} $ which for any particular mass does not reach more than 40\% of the DM energy density is still viable (see Refs.~\cite{Clesse:2015wea,Clesse:2017bsw,Garcia-Bellido:2017fdg,Carr:2019kxo,Garcia-Bellido:2019tvz}).

Also, these constraints are calculated for a monochromatic mass distribution. It has been found that a recalculation of these constraints for a wide mass distributions tends to have the effect of relaxing some of these exclusion curves (see Refs.~\cite{Garcia-Bellido:2017imq, Garcia-Bellido:2017xvr,Calcino:2018mwh}).

In addition to this, the constraints act on different epochs in the history of the Universe. For instance, most lensing and dynamical constraints such as wide halo binaries disruption act on redshifts $ z \approx 0 $, constraints from type IA supernovae act on redshifts $ z \approx 1 $, constraints from X-rays and the Cosmic Infrared Background act of redshifts $ z \approx 10 $, and accretion constraints from the Cosmic Microwave Background act on redshifts $ \order {1000-10^{4}}$, see Ref.~\citep{Garcia-Bellido:2018leu}.

Overall, this amounts to a picture where PBHs could survive these constraints and explain all of DM either if the mass spectrum evolves in redshift through merging and accretion, enough not to be excluded by these constraints from accretion at early times as well as those from lensing and dynamical constraints at late times, PBHs mean masses growing from $ \unit [\order {0.01-10}] {\msun} $ in the early Universe to $ \unit [\order {10- 1000}] {\msun} $ in the late Universe.

Moreover, clustered PBHs in the stellar mass range can comprise the totality of DM. Present constraints indicate that a monochromatic distribution of PBHs can account for 100\% of DM in the stellar mass range of $ \unit [\order {10}] {\msun} $, once one takes into account that clustering relaxes lensing constraints significantly~\citep{Garcia-Bellido:2017xvr,Calcino:2018mwh}.

More importantly, since these PBHs may form over an extended mass spectrum, it can be found that PBH may comprise the totality of DM even if a fraction of these objects have masses greater than $ \unit [\order {100}] {\msun} $, where the accretion, dynamical friction, wide binaries disruption and ultra faint dwarfs constraints start to be significant for monochromatic PBH distributions. The LIGO-VIRGO discovery of BHs in the stellar mass cannot for the moment discern if these objects are of primordial or astrophysical origin.

That however, might change in the near future as it has been proposed that bursts with millisecond durations could be explained by the GW emission from the hyperbolic PBHs encounters in dense clusters, such as the ones considered this analysis, see Ref.~\cite{Garcia-Bellido:2017qal,Garcia-Bellido:2017knh}. These encounters have a very characteristic spectrum that could allow detection by both AdvLIGO and, in the future, LISA depending on the duration and peak frequency of the emission. Studying in detail these hyperbolic encounters and simulating these dense PBHs clusters is one of the main motivations of our work and we will present our results in detail in Sections~\ref{sec: Close encounters} and \ref{sec: Background DM implications}.



\section{Methodology}
\label{sec: Methodology}

We now proceed to describe our simulations, the essential properties of which are outlined in Section~\ref{subsec: Simulation properties}, the particular choice of ICs in Section~\ref{subsec: Simulation Initial Conditions} and the information available in the time slices in Section \ref{subsec: Simulation evolution snapshots} along with comments on the advantages, disadvantages and limitations of our methodology.

In order to compute the trajectories of the gravitating bodies in our simulations we have made used of \astrograv\mbox{} (see Ref.~\cite{Calvert:2019}), a \textit{Java} numerical $ N $-body code that makes use of a Verlet integration algorithm: a time-reversible symplectic integrator method for solving Newton's equations of motion with good numerical stability. As we discuss also later, this code is purely Newtonian and does not include any General Relativity (GR) effect, such as GW emission.

In particular, we set a single configuration of $ N_{\ocap} = 1000 $ bodies, and let it evolve throughout the entire history of the Universe, with ICs determined by the bodies' masses, positions, velocities and radii. These parameters, amounting to 7 degrees of freedom, are chosen as follows.
\begin{enumerate}[label=\roman*)]
	\item For the masses: 1 degrees of freedom, taken from a random realisation of a Log-Normal (LN) distribution with $ (\mu, \sigma)_{\mass} = \unit [(2.0, 1.5)] {\msun} $.
	\item For the 3D position vectors: 3 degrees of freedom, taken from a random realisation of an isotropic, centred Multivariate-Normal (MN) distribution with $ \vec{M}_{\vel} = \unit [\vec{0}] {pc} $ and $ \vec{\Sigma}_{\vel} = \unit [1.0 \times \mathbb{1}_{3}] {pc^{2}} $.
	\item For the 3D velocity vectors: 3 degrees of freedom, taken again from a random realisation of an isotropic, centred Multi-Normal distribution, whose parameters can be determined under some assumptions by making use of the Virial Theorem.
	\item The radius correspond to the Schwarzschild radius for the object mass, which is completely determined by the object mass, and does not add any extra degrees of freedom.
\end{enumerate}

Then, we repeat the process for $ N_{\rcap} = 5000 $ realisations in the same IC configuration, in order to improve our statistics. Note that these choices and their approximate range of values agrees with some inflationary models that produce PBHs \cite{Clesse:2015wea}. Note however that, as we will see in Section~\ref{subsec: Simulation Initial Conditions}, it is more practical for both positions and velocities to use instead of isotropic Multi-Normal distributions, the equivalent Maxwell-Boltzmann (MB) distributions over an isotropic 3D vector field.

The code then computes the evolution of the gravitational system in a purely classical framework within a static background, unlike other codes such as \gevolution\mbox{} (see Ref.~\cite{Adamek:2016zes}), a full relativistic code, or \gadget\mbox{} (see Ref.~\cite{Springel:2005mi}), a cosmological hydrodynamical code which adapts classical dynamics to an expanding background. The choice of \astrograv\mbox{} was made as it has a series of advantages over the aforementioned two codes.

First, the spatial regions where these PBH clusters may form are dense enough that they quickly decouple from the expansion of the Universe and stay decoupled throughout the DM and dark energy eras, so that the relativistic effects in an expanding background would not add up to significant increase in accuracy.

Second, the code computes the system evolution much faster, since the orbital dynamics of the orbiting bodies are treated classically, the full GR treatment being computationally far more time consuming. In the face of the large number of bodies in each realisation $ N_{\ocap} = 1000 $, the large number of realisations per model $ N_{\rcap} = 5000 $, and the long evolution times of $ \unit [\order {10^{10}}] {yr} $, makes the simulation time-efficiency an item of crucial importance.

Third, even if GR effects are not properly simulated in the code, such as precession of the periastron, frame-dragging, binary spin-coupling, etc, are relevant in the strong field regime, the PBH clusters in our simulations are disperse enough to begin with and only disperse further over time, that only a very low number of these close approaches do occur in between time snapshots. This is so since the mean distance in between objects is typically $ \unit [\order {0.1}] {pc} $ and the mean minimum distance over the whole simulation time is around $ \unit [\order {0.01}] {pc} $, which for the range of masses simulated in our code, of order $ \unit [\order {10^1}] {\msun} $, leaves the magnitude of such effects completely negligible for the vast majority of trajectories.

Last, body collision and merger are, unlike in the alternatives, already implemented in the code, a feature of particular importance in the densest PBH clusters even if events of this kind are very rare, having found that the probability of a particular PBH to merge during the full time interval covered by the simulations is of $ \order {10^{-5}} $.

However, the collisions are treated classically in the code, and do not exhibit inspiralling binaries and loss of kinetic energy that leads to the emission of gravitational energy. The latter increases the likelihood for mergers as it tends to draw the bodies closer to each other, meaning the total amount of mergers in our simulations is underestimated and must be understood as a lower bound of the actual merger event rate.



\subsection{Simulation properties}
\label{subsec: Simulation properties}

We proceed now to comment on the time and space resolution of our simulations, their implications for the data extraction, and constrain the range of validity where we may trust our findings, focusing in particular on whether the evolution of PBHs is dynamically constrained to a range where the code behaves as desired.



\subsubsection{Time-resolution}
\label{subsubsec: Time-resolution}

The code has an adaptive time resolution $ \delta t $, which depends on the forces at $ i^{\textrm{th}} $ -step and is not to be confused with the time interval at which data is regularly stored, $ \Delta t $. Dependent on this $ \delta t $, that is, the time-step in between the iterations of the Verlet integrator the global error in position and velocity grows with $ (\delta t^{2}) $.

We have, however, performed tests prior to the simulations to ensure that the error during the whole evolution period does not accumulate enough to meaningfully alter the simulation output by re-running the simulations with time-steps smaller than in the final simulations by a factor of a hundredth, with no significant effect, with position and velocities typically differing by less than $ 0.01\% $ from their original values in the final output, but at the expense of both CPU time and memory.

\subsubsection{Space-resolution}
\label{subsubsec: Space-resolution}
The code is a pure $ N $-body integrator and therefore there is no theoretical minimum resolution or spatial separation as in other hydrodynamical codes. However, we can consider a minimum spatial resolution, approximately corresponding to the maximum impact parameter derived from the effective cross section of two orbiting PBHs to become gravitationally bound. This relativistic constraint is due to the power emission of the incoming body, which is absent in the codes classic framework. The threshold cross section $ \sigma $ (see Ref.~\cite{Mouri:2002mc}) for the PBH pair binding is
\begin{equation}
	\sigma = \left(\frac {85 \pi} {3} \right)^{2/7} r_{\textrm{S}}^{2} \left(\frac {v} {c} \right)^{-18 / 7} \approx \pi \Delta b^{2},
	\label{eq: Binding-resolution}
\end{equation}
where $ b $ is the limiting impact parameter and the Schwarzschild's radius is
\begin{equation}
	r_{\textrm{S}} = \frac{2 G m}{c^{2}},
	\label{eq: Schwarzschild radius}
\end{equation}
where $G$ is Newton's constant, $m$ the mass of the object and $c$ the speed of light in vacuum. Should the incoming object approach hyperbolically another object with a smaller impact parameter than the threshold, the simulated hyperbolic trajectory would clearly not track the actual bounded trajectory that one would observe. Considering the most unfavourable case in the range
\begin{align}
	0.1\unit [] {\msun} & \leq m \leq \unit [100] {\msun}, \\
	0.1\unit [] {km / s} & \leq v \leq \unit [1000] {km / s},
	\label{eq: Evolution snapshots dynamical range}
\end{align}
of masses and velocities in the simulations (see Figure~\ref{subfig: Cluster mass segregation mass weighted} and \ref{subfig: Cluster dynamical friction mass weighted}), then Eq.~\eqref{eq: Binding-resolution} renders an effective spatial resolution of $ \Delta x = \unit [100] {AU} $, less than the mean minimum distance in-between bodies, of $ \unit [\order {1000}] {AU} $.



\subsection{Simulation Initial Conditions}
\label{subsec: Simulation Initial Conditions}


\begin{table*}[t]
	\centering
	\begin{tabular}[c]{
	| c | l p{20 mm} | c
	| c | l p{20 mm} | c
	| c | l p{20 mm} |}
	\cline{1-3}\cline{5-7}\cline{9-11}
	\multicolumn{3}{| c |}{Mass ICs distribution:} &
	\quad &
	\multicolumn{3}{c |}{Position ICs distribution:} &
	\quad &
	\multicolumn{3}{c |}{Velocity ICs distribution:} \\
	[0.5ex]
	\hhline{===~===~===}
	$ \dist_{\mass} \left(m_{i} \right)_{0} $ &
	\quad $ \mu_{\mass}  $ &
	\quad $ \unit [2.0] {\msun} $ &
	\quad &
	$ \dist_{\pos} \left(x_{i} \right)_{0} $ &
	\quad $ \mu_{\pos} $ &
	\quad $ \unit [1.6] {pc} $ &
	\quad &
	$ \dist_{\vel} \left(v_{i} \right)_{0} $ &
	\quad $ \mu_{\vel} $ &
	\quad $ \unit [2.0] {km / s} $ \\
	$ \textrm{[LN]} $ &
	\quad $ \sigma_{\mass} $ &
	\quad $ \unit [1.5] {\msun} $ &
	\quad &
	$ \textrm{[MB]} $ &
	\quad $ a_{\pos} $ &
	\quad $ \unit [1.0] {pc} $ &
	\quad &
	$ \textrm{[MB]} $ &
	\quad $ a_{\vel} $ &
	\quad $ \unit [1.2] {km / s} $ \\
	[0.5ex]
	\cline{1-3}\cline{5-7}\cline{9-11}
	\end{tabular}
	\caption{
	Mass, position and velocity IC distribution parameters. Distributions plots are shown in Figures~\ref{subfig: Initial mass distribution profile}, \ref{subfig: Initial position distribution profile} and \ref{subfig: Initial velocity distribution profile}. A specific realisation of such distributions is shown in Figures~\ref{subfig: Initial mass distribution scatter}, \ref{subfig: Initial position distribution scatter} and \ref{subfig: Initial velocity distribution scatter}.
	}
	\label{tab: Simulation parameters}
\end{table*}

\begin{figure*}[t]
	\centering
	\subfloat[3D ICs positions for a random realisation.]{
	\hspace*{-0.00cm}
	\includegraphics[width = 0.50\textwidth]
	{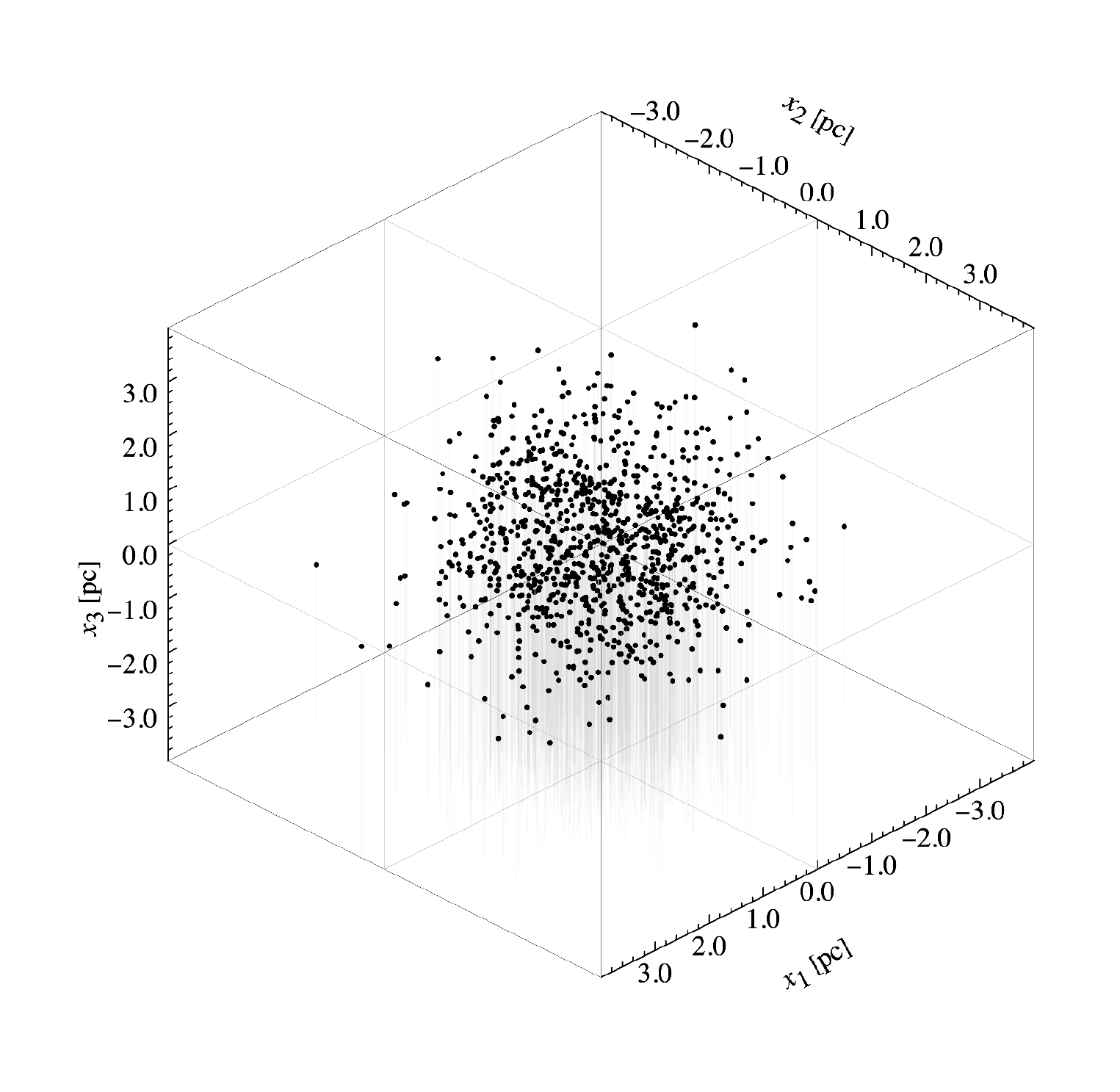}
	\label{subfig: Initial distribution positions}}
	\subfloat[3D ICs velocities for a random realisation.]{
	\hspace*{-0.00cm}
	\includegraphics[width = 0.50\textwidth]
	{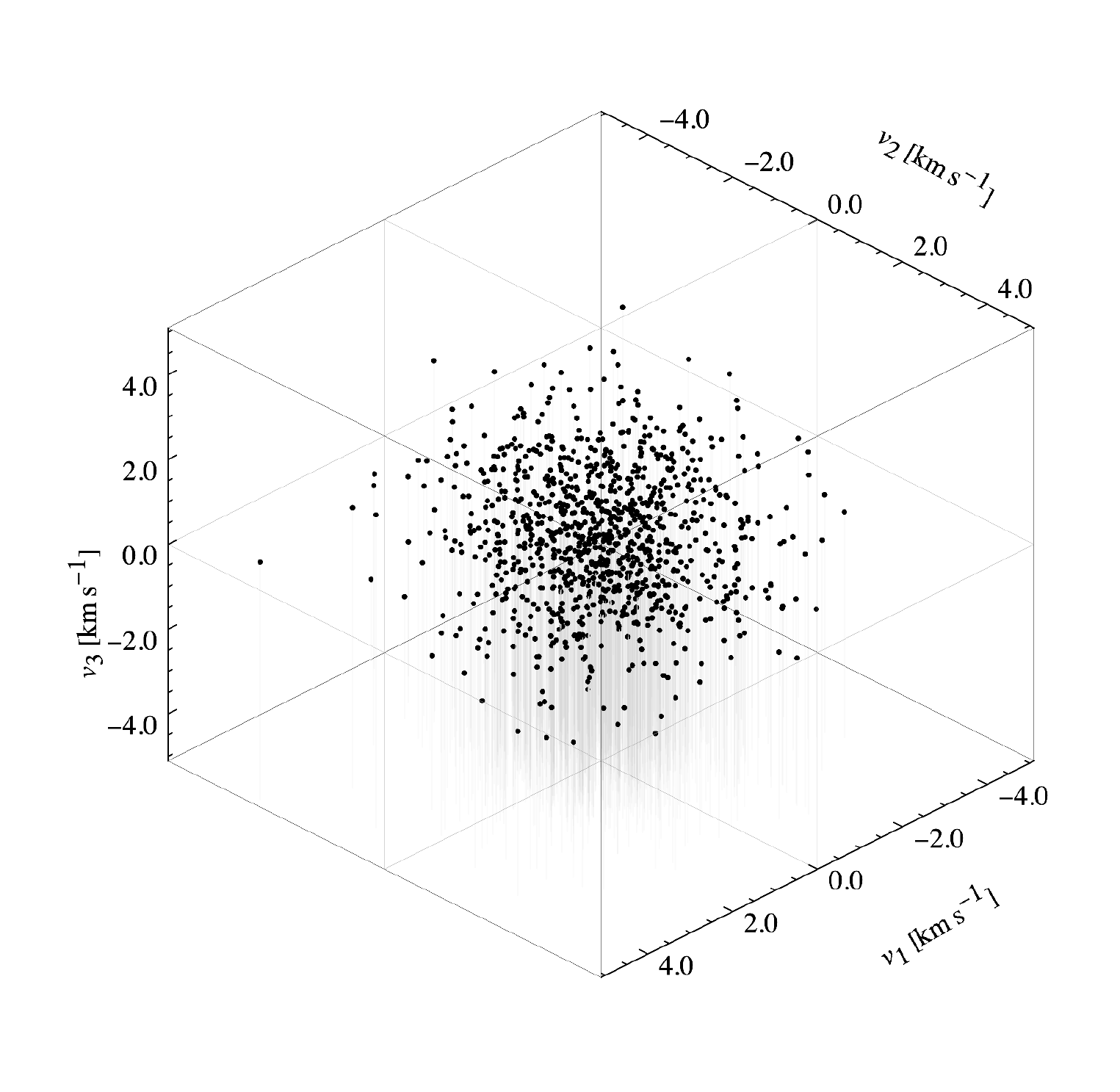}
	\label{subfig: Initial distribution velocities}}
	\caption{
	Cluster IC positions and velocities.
	Both the position and velocities correspond to a sample of an isotropic Maxwell-Boltzmann distribution, further expanded in Figure~\ref{fig: Initial position distribution}. Note that there is no correlation at the IC time between the masses and positions, so more massive PBHs are not yet favoured to cluster closer to the cluster barycentre. Also, the velocity distribution computation is illustrated in Figure~\ref{fig: Initial velocity computation} and while Figure~\ref{fig: Initial velocity distribution} shows the actual velocity profile.
	}
	\label{fig: IC distributions}
\end{figure*}

\begin{figure*}[t]
	\centering
	\subfloat[Analytical PBH cluster universal cumulative mass curve.]{
	\hspace*{-0.00cm}
	\includegraphics[width = 0.50\textwidth]
	{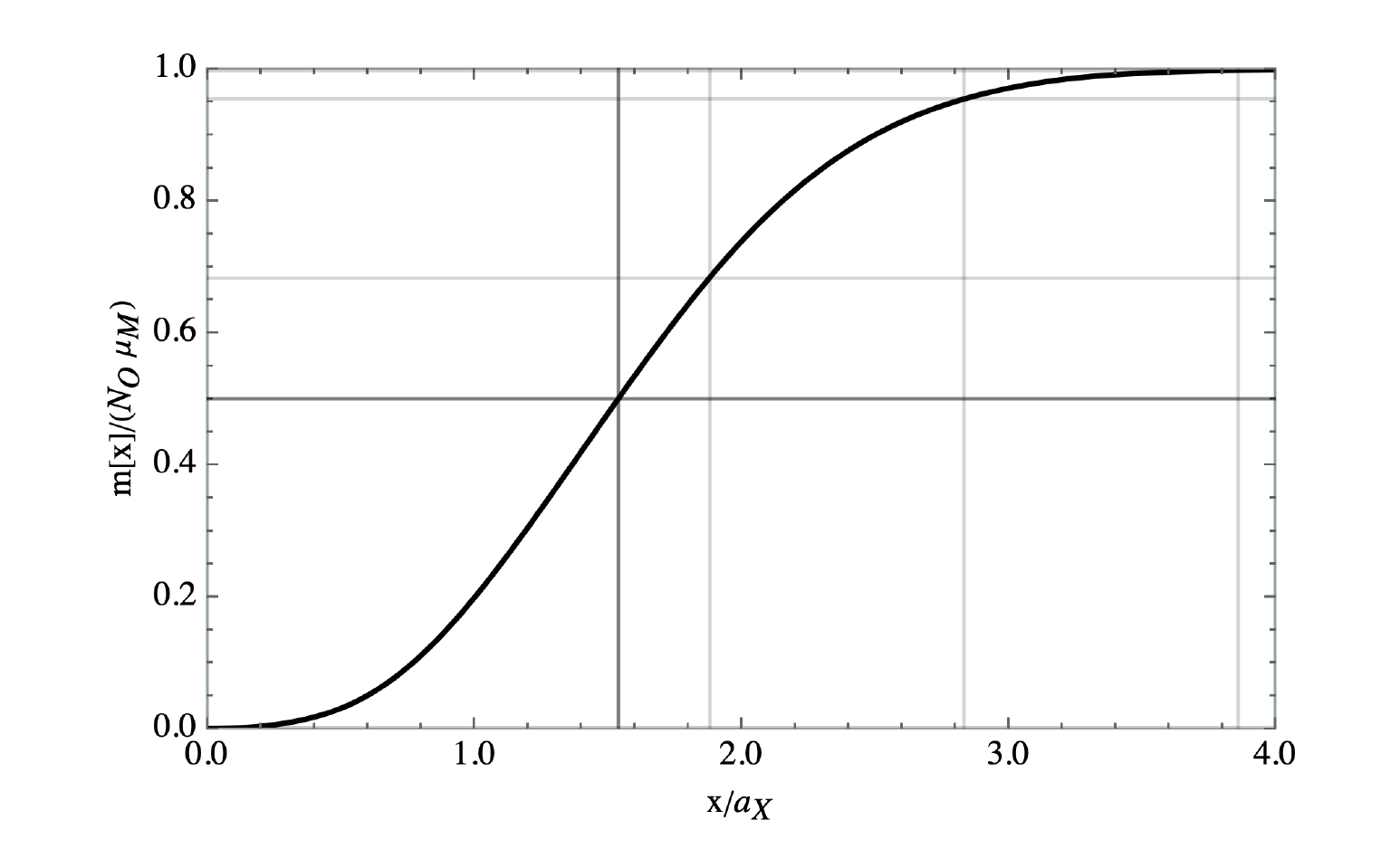}
	\label{subfig: Cluster cumulative mass curve}}
	\subfloat[Numerical PBH cluster universal rotational velocity curve.]{
	\hspace*{-0.00cm}
	\includegraphics[width = 0.50\textwidth]
	{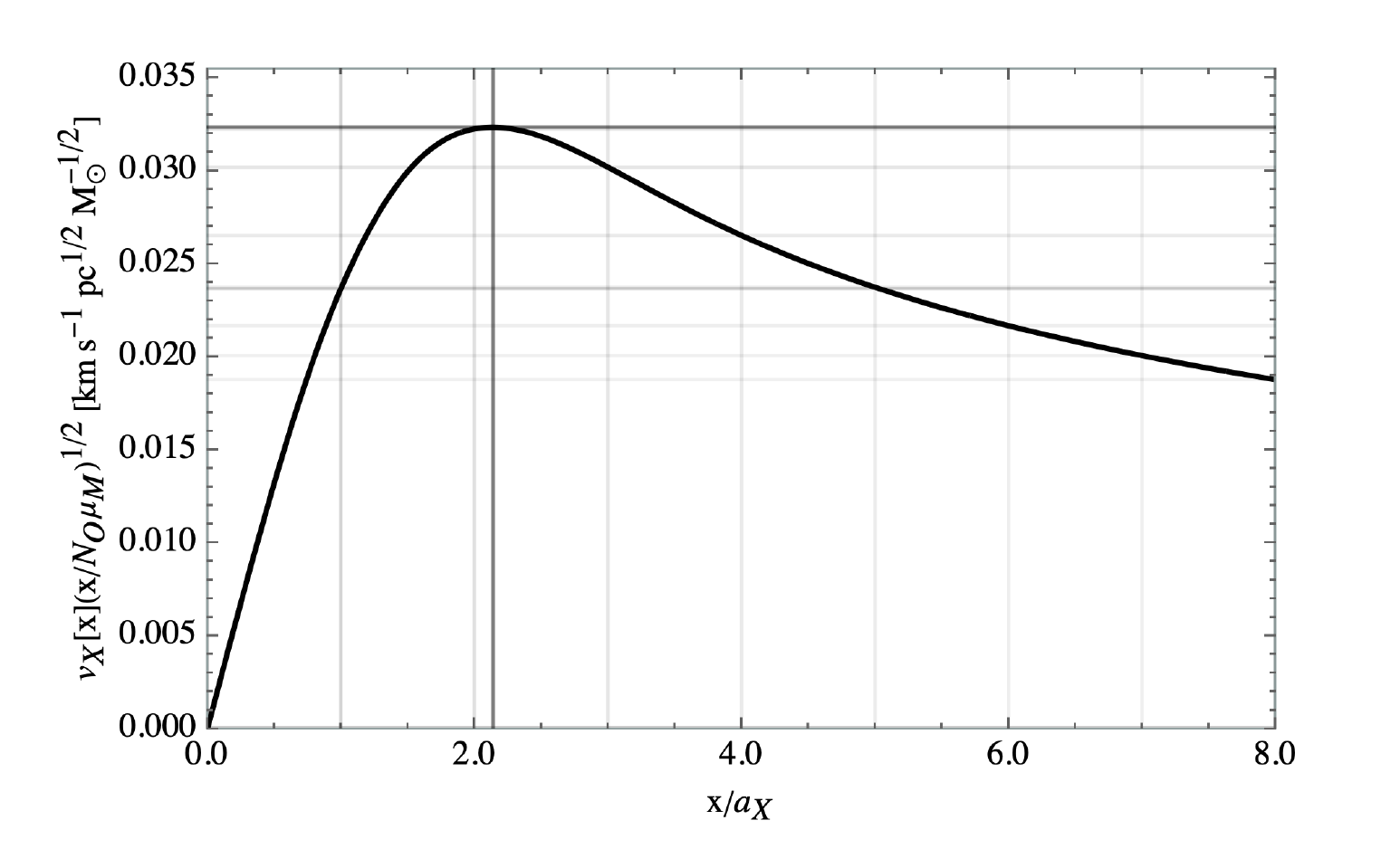}
	\label{subfig: Cluster rotational velocity curve}}
	\caption{
	Universal cluster cumulative mass and rotational velocity curves from Table~\ref{tab: Simulation parameters}, computed for a perfectly traced distribution of PBHs computed from Eq.~\eqref{eq: Inner shell mass} where $ N_{\ocap} \rightarrow \infty $. Here, $ N_{\ocap} $, the object mean logarithmic mass, $ \mu_{\mass} $, and cluster mean displacement from the barycentre, $ \sigma_{\pos} $. The dark grey line marks the curve-value at the radii containing 50\% the cumulative mass, while the light grey lines do so for 68\%, 95\% and 99\% the cumulative mass.
	}
	\label{fig: Cluster universal curves}
\end{figure*}


We proceed now to describe the generation of the simulation ICs, and the choice of distributions for all the fundamental parameters within it. The dynamics of each of the $ i $ gravitating bodies, where $ i = 1, 2, ..., N_{\ocap} $, in the simulation box is fully determined by the simulation ICs. The number of bodies in the simulation box is $ N_{\ocap} = 1000 $, and $ 4 N_{\ocap} $ free parameters are enough to completely determine the realisation IC and its subsequent evolution
\begin{enumerate}[label=\roman*)]
	\item $ N_{\ocap} $ free parameters for the mass, $ m_{i} $, one per body in the simulations, chosen from a random sample of a global Log-Normal distribution.
	\item $ 3 N_{\ocap} $ free the position coordinates, $ x_{i} $, chosen from a random sample of a global Maxwell-Boltzmann distribution for the position moduli with isotropic randomised directions.
\end{enumerate}

These parameters completely determine
\begin{enumerate}[label=\roman*)]
	\item $ N_{\ocap} $ dependent parameters for the radii of the objects, $ r_{i} $, corresponding to the Schwarzschild radius of the individual body, thus naturally following a Log-Normal distribution of mass.
	\item $ 3 N_{\ocap} $ dependent velocity coordinates, $ v_{i} $, chosen from a single random sample of a shell-specific isotropic Maxwell-Boltzmann distribution, determined by the number of bodies, $ N_{\ocap} $, the individual masses, $ m_{i} $, and the individual positions $ x_{i} $, or conversely, an isotropic shell-specific Multi-Normal distribution, where by shell-specific we mean radially dependent with respect to the PBH cluster barycentre.
\end{enumerate}

The choice of mass, position, velocity and radius parameters is then repeated for each of the $ N_{\rcap} = 5000 $ different realisations that comprise the simulations. An example of the cluster's position and velocity distribution is shown in Figure~\ref{fig: IC distributions}. This choice of IC leads to an initial density $ \rho_{0} \approx 3 N_{\ocap} \mu_{\mass} / 4 \pi a_{\pos}^{3} $ of $ \unit [\order {1000}] {\msun pc^{-3}} $ (see Figure~\ref{subfig: Quasi-static cluster density profiles}). In the remaining of this subsection, we proceed to detail the particular choice of the underlying distributions from which this parameters are extracted. A summary of the input distributions and realisations is given Table~\ref{tab: Simulation parameters}.



\subsubsection{Mass distribution}
\label{subsubsec: Mass distribution}


\begin{figure*}[t!]
	\centering
	\subfloat[IC mass Log-Normal distribution, $ \dist_{\mass} \left(m_{i} \right) \vert_{\mu_{\mass}, \sigma_{\mass}} $.]{
	\hspace*{-0.00cm}
	\includegraphics[width = 0.50\textwidth]
	{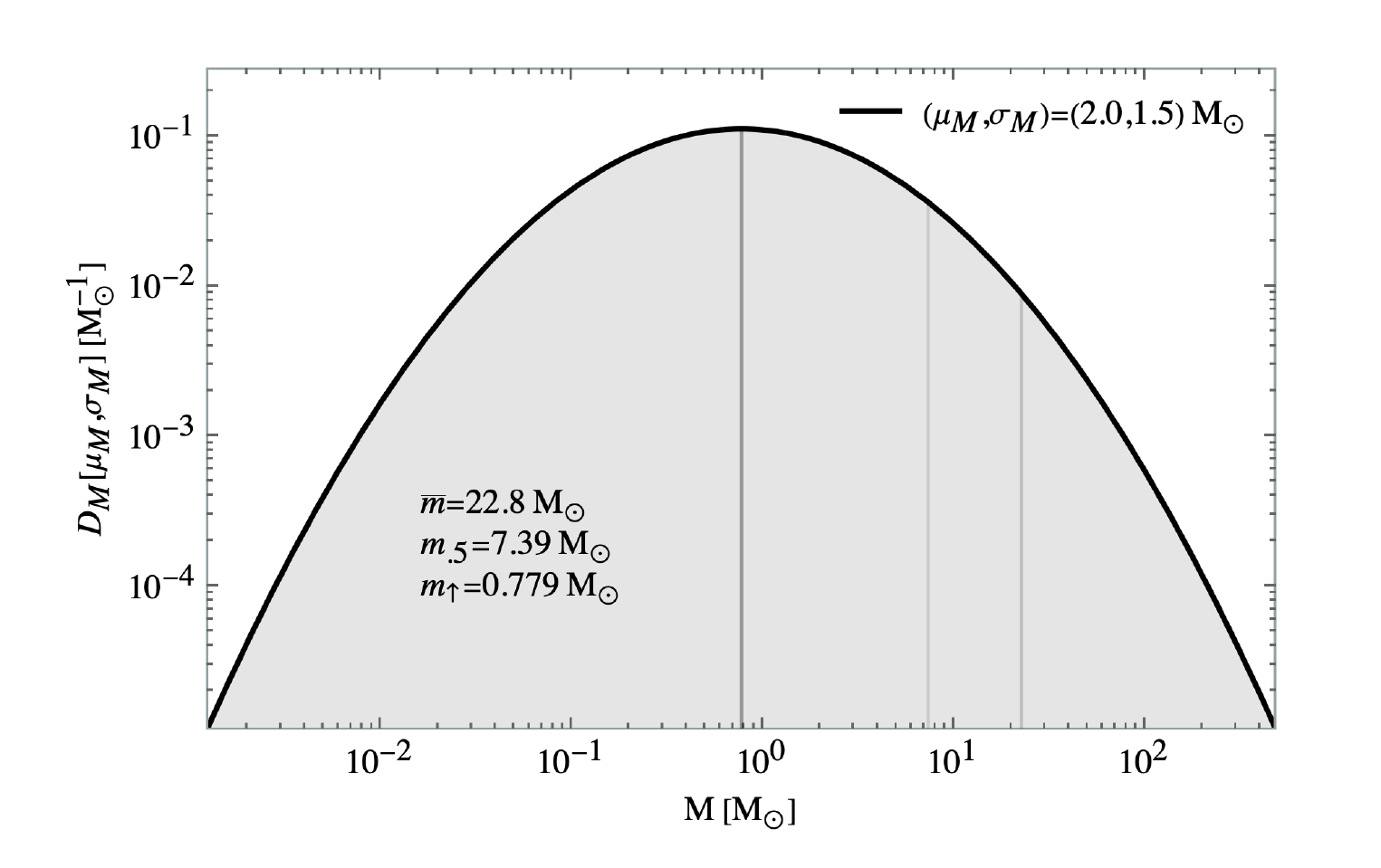}
	\label{subfig: Initial mass distribution profile}}
	\subfloat[A random realisation's IC mass profile.]{
	\hspace*{-0.00cm}
	\includegraphics[width = 0.50\textwidth]
	{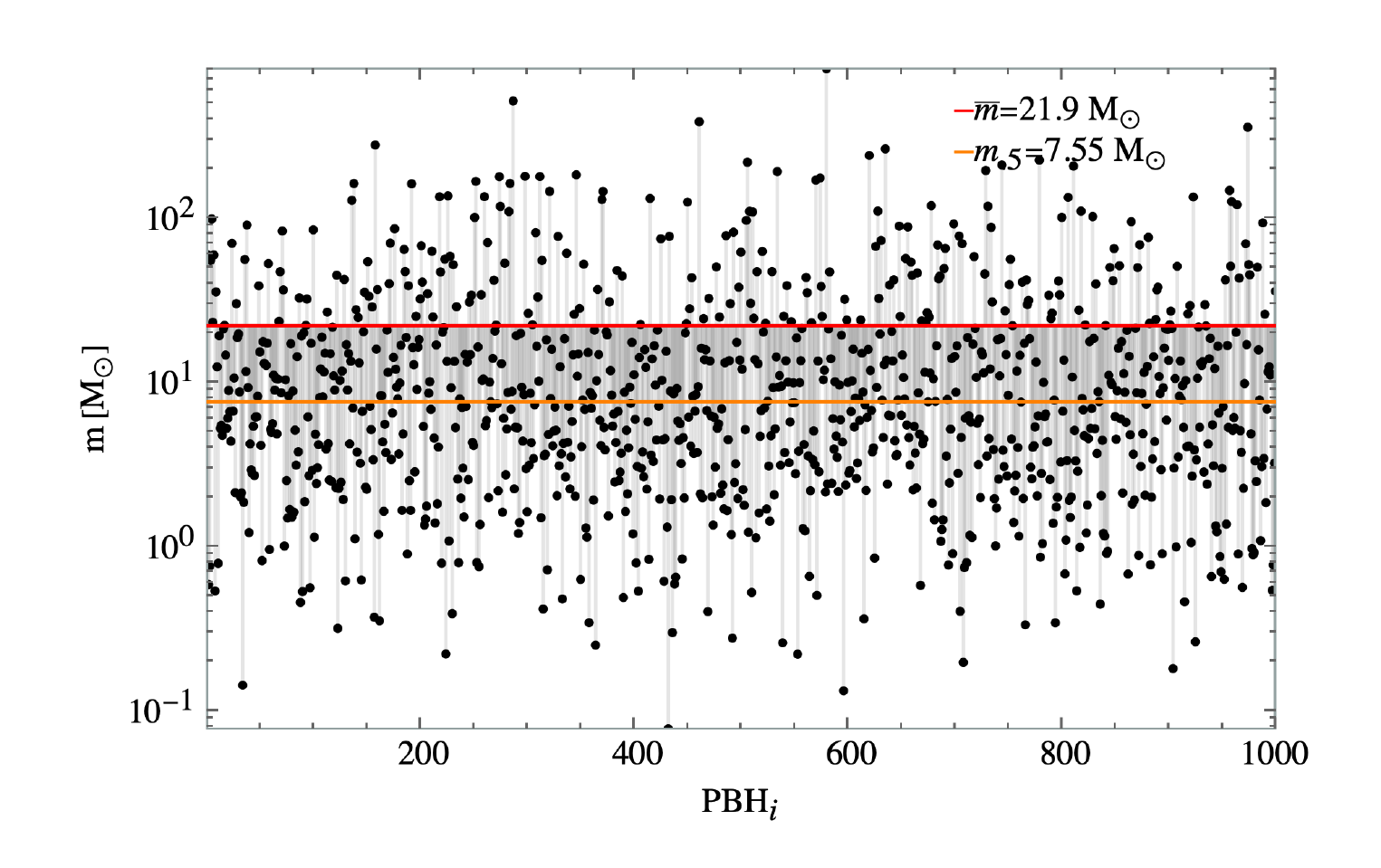}
	\label{subfig: Initial mass distribution scatter}}
	\caption{
	Mass IC profile from Table~\ref{tab: Simulation parameters} and an example realisation, from the logarithmic mean $ \mu_{\mass} = \unit [2.0] {\msun} $ and deviation $ 	\sigma_{\mass} = \unit [1.5] {\msun} $. The thick, normal and thin grey lines signal the distribution mode, median and mean respectively. The orange and red lines signal the realisation median and mean respectively. Note the large hierarchy between modal, median and mean masses due to the large $ m \rightarrow \infty $ tail of the distribution, skewing the mean to large values, which results in a large variance of the total mass per realisation. The total mass is $ m_{\tot} = \sum_{i} m_{i} = \sum_{i} (m_{i}) = \unit [2.19 \times 10^{4}] {\msun} $.
	}
	\label{fig: Initial mass distribution}
\end{figure*}


The initial mass profile $ \dist_{\mass} \left(m_{i} \right)_{0} $ of the cluster is obtained from a random realisation of a Log-Normal distribution for each realisation
\begin{equation}
	\dist_{\mass} \left(m_{i} \right) = \frac {\normalisation_\mass} {m_{i}} \exp \left(-\frac {\Delta m_{i}^{2}} {2 \sigma_{\mass}^{2}} \right),
	\label{eq: Mass distribution}
\end{equation}
where $ \Delta m_{i} = (\log m_{i}-\mu_{\mass}) $ is the PBH logarithmic mass deviation from the mean and $ \normalisation_{\mass} = (\sigma_{\mass} \sqrt{2 \pi})^{-1} $ is a normalisation factor. The total mass in each realisation is then, on average, $ M_{\mathrm{T}} = N_{\ocap} \mu_{\mass} $. In particular, for our simulations, $ \mu_{\mass} = \unit [1.5] {\msun} $ and $ \sigma_{\mass} = \unit [2.0] {\msun} $, as laid out in Table~\ref{tab: Simulation parameters} and shown in Figure~\ref{fig: Initial mass distribution}.

In the previous distribution, however, one should be reminded that $ \mu_{\mass} $ is the mean of the natural logarithm of the mass, and not the mean mass itself. Since the actual mean mass $ \bar{m} $ is be a parameter of great importance in the computation of the background number of PBHs in a typical galactic halo and it is in any case of greater physical meaning, it will be computed from the Log-Normal distribution parameters $ \mu_{\mass} $ and $ \sigma_{\mass} $ with the expression
\begin{equation}
	\bar{m} = \exp \left(\mu_{\mass}+\frac {1} {2} \sigma_{\mass}^2 \right),
	\label{eq: Log-Normal mean}
\end{equation}
while, similarly, the actual variance of the mass can be computed from the distribution parameters with
\begin{equation}
	\bar{\sigma}^{2}_{\mass} = \left(\exp \sigma_{\mass}^2 -1 \right) \exp \left(2\mu_{\mass}+\sigma_{\mass}^2 \right) ,
	\label{eq: Log-Normal variance}
\end{equation}
where in both cases the over-bar indicates that the quantity is computed in the linear (physical) scale. In the linear scale one can similarly extract the median value of the mass,
\begin{equation}
	\tilde{m} = \exp \left(\mu_{\mass}\right),
	\label{eq: Log-Normal median}
\end{equation}
as well as the the modal value of the mass,
\begin{equation}
	\hat{m} = \exp \left(\mu_{\mass}-\sigma_{\mass}^2 \right),
	\label{eq: Log-Normal mode}
\end{equation}
as a function of the Log-Normal distribution parameters. Note that, generally speaking, a Log-Normal distribution has wide tails and as such $ \bar{m} > \tilde{m} > \hat{m} $.



\subsubsection{Position distribution}
\label{subsubsec: Position distribution}


\begin{figure*}[t!]
	\centering
	\subfloat[IC position Maxwell-Boltzmann distribution, $ \dist_{\pos} \left(x_{i} \right) \vert_{\mu_{\pos} \leftrightarrow a_{\pos}} $]{
	\hspace*{-0.00cm}
	\includegraphics[width = 0.50\textwidth]
	{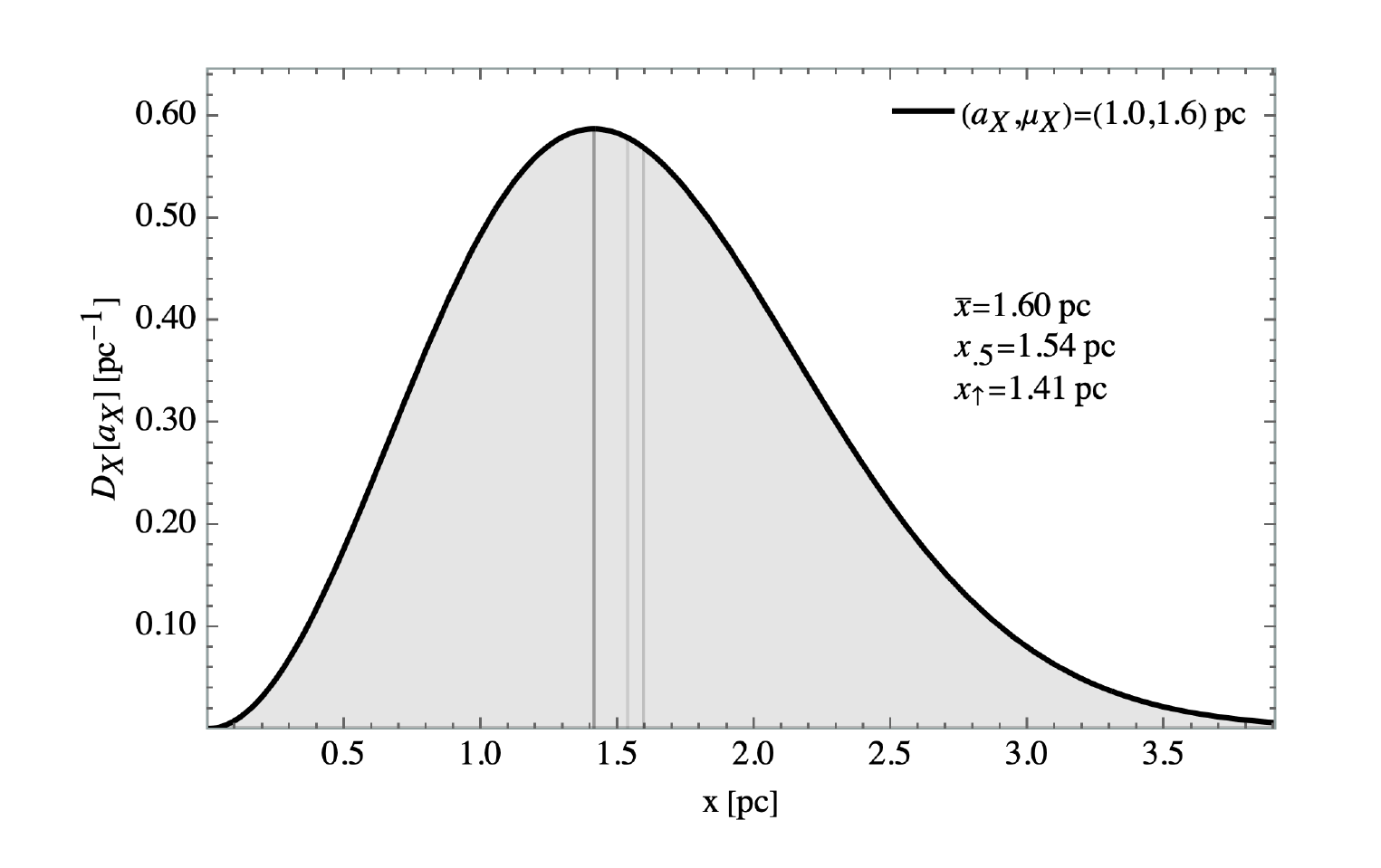}
	\label{subfig: Initial position distribution profile}}
	\subfloat[A random realisation's IC position profile.]{
	\hspace*{-0.00cm}
	\includegraphics[width = 0.50\textwidth]
	{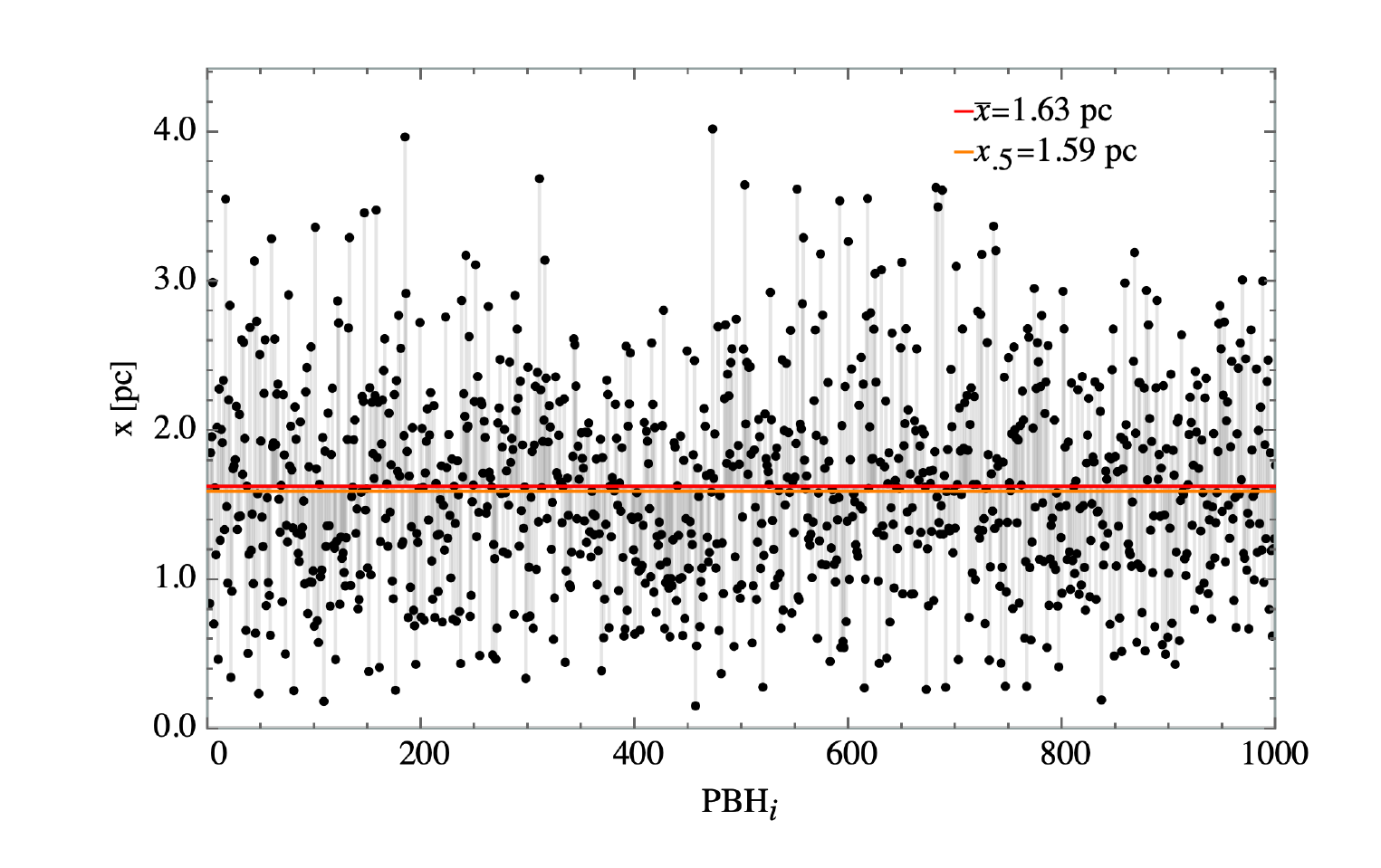}
	\label{subfig: Initial position distribution scatter}}
	\caption{
	Position IC profile from Table~\ref{tab: Simulation parameters} and an example realisation, from the mean $ \mu_{\pos} = \unit [1.6] {pc} $ and scale parameter $ a_{\pos} = \unit [1.0] {pc} $. The thick, normal and thin grey lines signal the distribution mode, median and mean respectively. The orange and red lines signal the realisation median and mean respectively. To to obtain the radial position abundance one needs only to multiply the previous distribution by the surface element $ dS = 4 \pi x^{2} $, which enhances the tail. Note the hierarchy between modal, median and mean masses is much reduced with respect to the mass case, due to the smallness of the $ x \rightarrow \infty $ tail of the distribution, which results in a small variance of the total simulated volume covered in the ICs.
	}
	\label{fig: Initial position distribution}
\end{figure*}

The initial position profile $ \dist_{\pos} \left(x_{i} \right)_{0} $ of the cluster is obtained from a random sample of a 3D Multi-Normal distribution for each realisation, with density
\begin{equation}
	\dist_{\pos} \left(\vec{x}_{i} \right) = \normalisation_{\pos} \exp \left(- \frac {\Delta \vec{x}_{i} \Sigma_{\pos}^{-1} \Delta \vec{x}_{i}^{T}} {2} \right),
	\label{eq: Multi-Normal distribution}
\end{equation}
where $ \Delta \vec{x}_{i} = (\vec{x}_{i}-\vec{\mu}_{\pos}) $ is the PBH position displacement from the cluster center of mass and $ \normalisation_{\pos} = (\sqrt{2 \pi \left| \Sigma_{\pos} \right|})^{-1} $ is the normalisation factor. In our simulations, we consider purely spherical $ \Sigma_{\pos} = \sigma_\pos^{2} \mathbb{1}_{3} $ and centrally aligned $ \vec{\mu}_{\pos} = 0 $ clusters of PBHs.

In practice, this profile is equivalent to that of isotropic vector field with the norms following a Maxwell-Boltzmann distribution, with the Maxwell-Boltzmann scale parameter $ a_{\pos} $ given by
\begin{equation}
	\mu_{\pos} = \bar{x} = 2 a_{\pos} \sqrt{\frac {2} {\pi}},
	\label{eq: Maxwell-Boltzmann mean}
\end{equation}
where $ \mu_{\pos} $ is the mean position of the distribution.

The variance of the position distribution can be directly computed from the scale parameter to which it is directly proportional with
\begin{equation}
	\sigma^{2}_{\pos} = \frac {3\pi-8} {\pi} a_{\pos}^{2},
	\label{eq: Maxwell-Boltzmann variance}
\end{equation}
while the median position of the distribution can be computed from the scale parameter with the similarly linear relation
\begin{equation}
	\tilde{x} = 1.53 a_{\pos},
	\label{eq: Maxwell-Boltzmann median}
\end{equation}
and the modal position of the distribution is then
\begin{equation}
	\hat{x} = \sqrt{2} a_{\pos} ,
	\label{eq: Maxwell-Boltzmann mode}
\end{equation}
as a function of the scale parameter. Note that a Maxwell-Boltzmann distribution has as well $ \bar{x} > \tilde{x} > \hat{x} $.

Finally, the probability distribution density $ \dist_{\pos} \left(x_{i} \right) $, as a function of the mean position $ \mu_{\pos} $, is then given by
\begin{equation}
	\dist_{\pos} \left(x_{i} \right) = \normalisation_{\pos} x_i^{2} \exp \left(-\frac {4 x_i^{2}} {\pi \mu_\pos^{2}} \right),
	\label{eq: Position distribution}
\end{equation}
where $ \mu_{\pos} $ is the mean position of the distribution and $ \normalisation_{\pos} = 32 / (\pi^{2} \mu_\pos^{3}) $ is a the normalisation factor. This equivalence is due to the fact that the $ \chi $ distribution for $ k = 3 $ degrees of freedom, i.e. the dimensionality of the simulation box, of the vector modules reduces precisely to the Maxwell-Boltzmann distribution,
\begin{equation}
	\dist_{\textrm{MN}} [\vec{0}_{3}, \sigma^{2} \mathbb{1}_{3}] (z) \sim \dist_{\chi, 3} [\sigma] (z) = \dist_{\textrm{MB}} [\sigma] (z).
	\label{eq: Distribution equivalence}
\end{equation}

For simplicity, we compute the IC with the latter Maxwell-Boltzmann distribution as it is more computationally economic given the symmetry of the problem. In particular, the simulated PBH clusters have been chosen such that, $ \mu_{\pos} = \unit [1.6] {pc} $ and $ a_{\pos} = \unit [1.0] {pc} $, (see Table~\ref{tab: Simulation parameters} and Figure~\ref{fig: Initial position distribution}).



\subsubsection{Velocity distribution}
\label{subsubsec: Velocity distribution}


\begin{figure*}[t!]
	\centering
	\subfloat[IC shell-specific Maxwell-Boltzmann velocity distribution, $ \dist_{\vel} \left(v_{i} \right) \vert_{\mu_{\vel} \leftrightarrow a_{\vel}} $]{
	\hspace*{-0.00cm}
	\includegraphics[width = 0.5\textwidth]
	{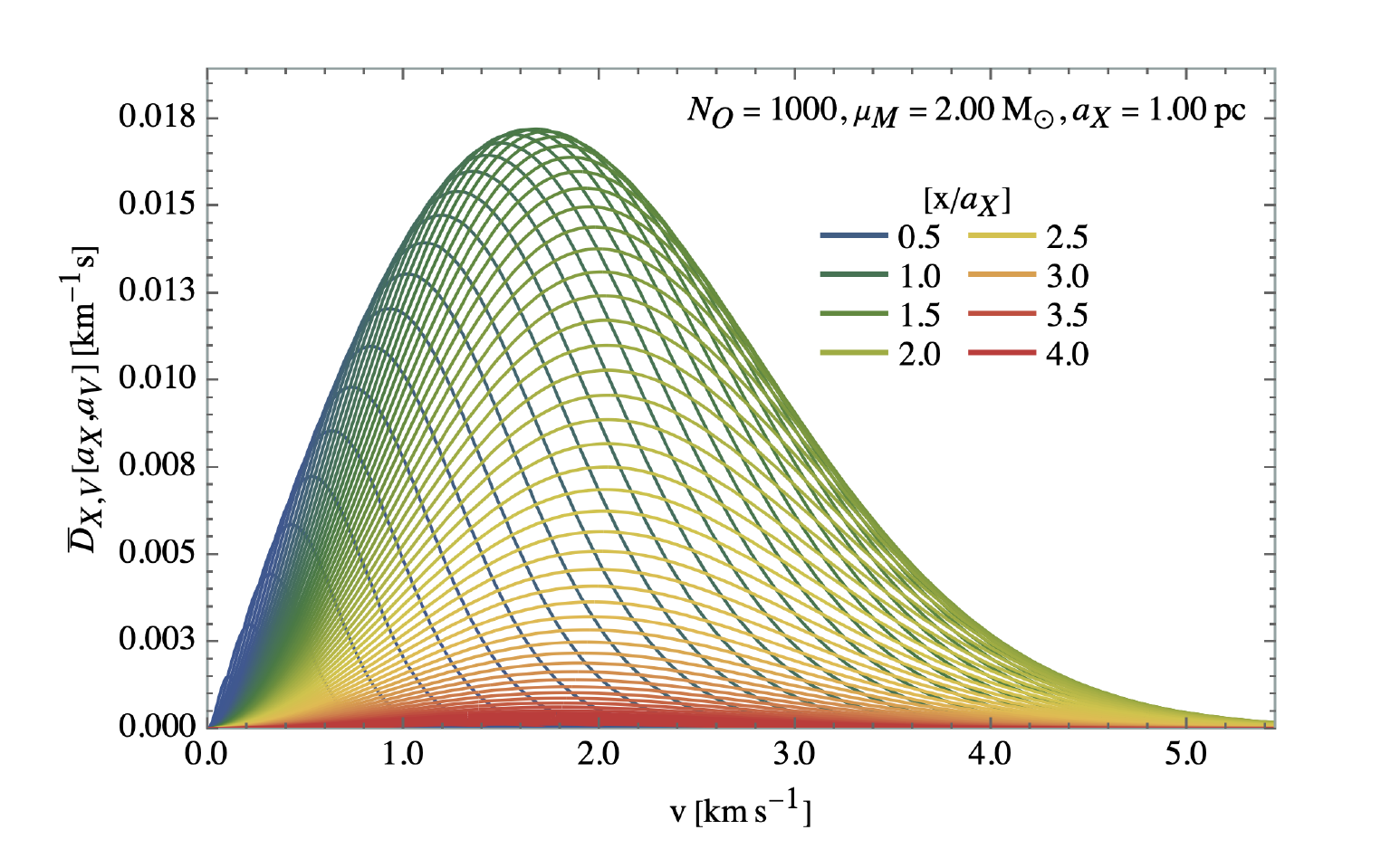}
	\label{subfig: Initial velocity distribution bins}}
	\subfloat[IC global velocity QMaxwell-Boltzmann velocity distribution]{
	\hspace*{-0.00cm}
	\includegraphics[width = 0.5\textwidth]
	{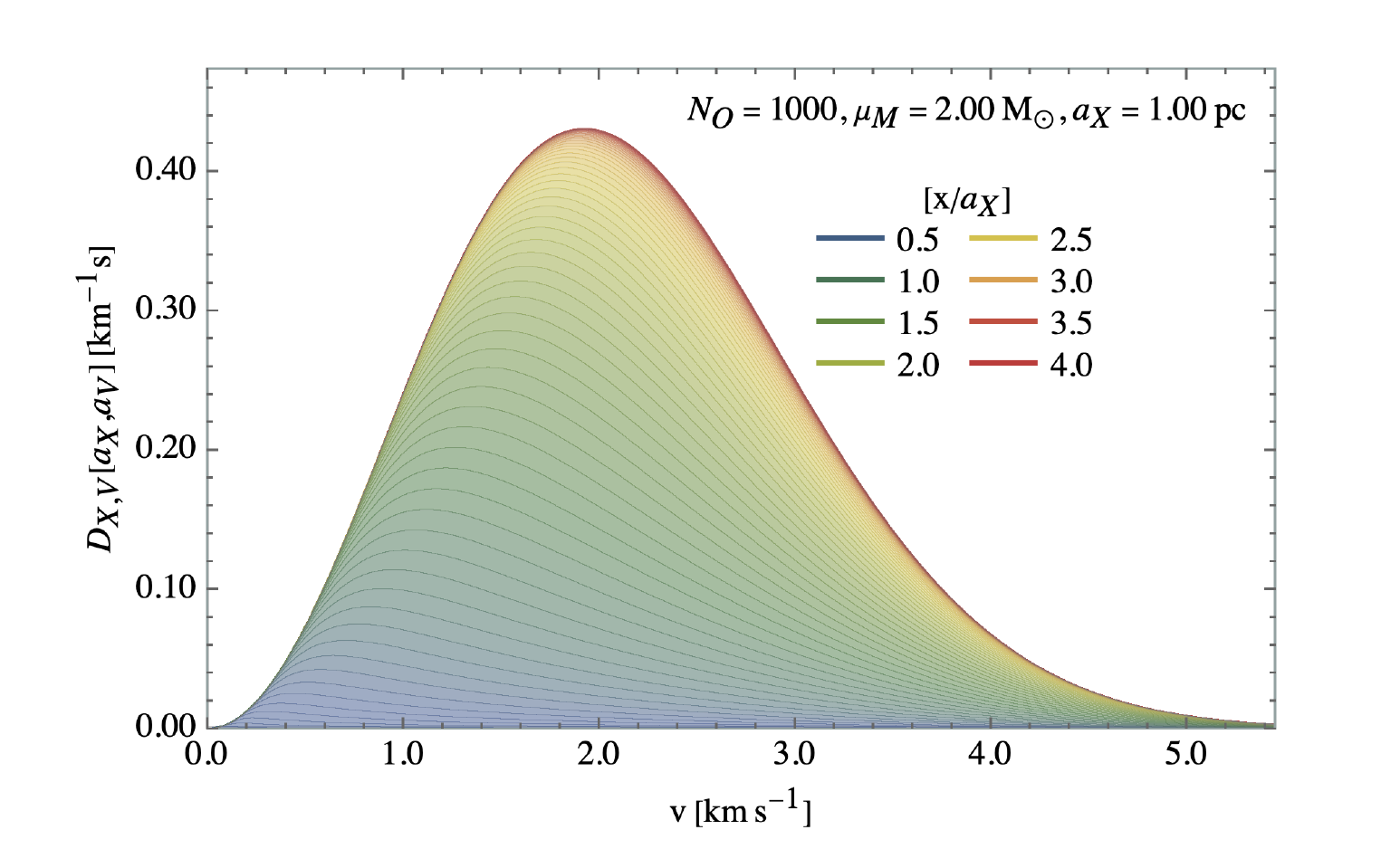}
	\label{subfig: Initial velocity distribution sums}}
	\caption{
	Velocity IC shell-by shell (left) and integrated distributions (right) computed from from Table~\ref{tab: Simulation parameters} distributions, at 64 linearly spaced radial bins in the range $ 0 \leq x \leq 4 	\sigma_{\pos} $. The latter is obtained by integration of the shell-specific Maxwell-Boltzmann velocity distribution with a density kernel corresponding to the bin surface area $ dS = 4 \pi x^{2} $, from $ x = 0 $ to $ x = 8\sigma $.
	Both the core and the outskirts of the PBH cluster roughly understood as the regions with $ 0 \leq x \leq a_{\pos} $ and $ 2 a_{\pos} \leq x \leq 4 a_{\pos} $ respectively) are not heavily populated, the core exhibits very low velocities in the range $ v \in \unit [(0.0, 1.0)] {km / s} $ while the outskirts exhibits intermediate velocities in the range $ v \in \unit [(1.0, 3.0)] {km / s} $, whereas the populated intermediate bins do show the largest velocities in the range $ v \in \unit [(1.0, 5.0)] {km / s} $.
	}
	\label{fig: Initial velocity computation}
\end{figure*}

\begin{figure*}[t!]
	\centering
	\subfloat[IC velocity QMaxwell-Boltzmann distribution, $ \dist_{\vel} \left(v_{i} \right) \vert_{\mu_{\vel} \leftrightarrow a_{\vel}} $.]{
	\hspace*{-0.00cm}
	\includegraphics[width = 0.5\textwidth]
	{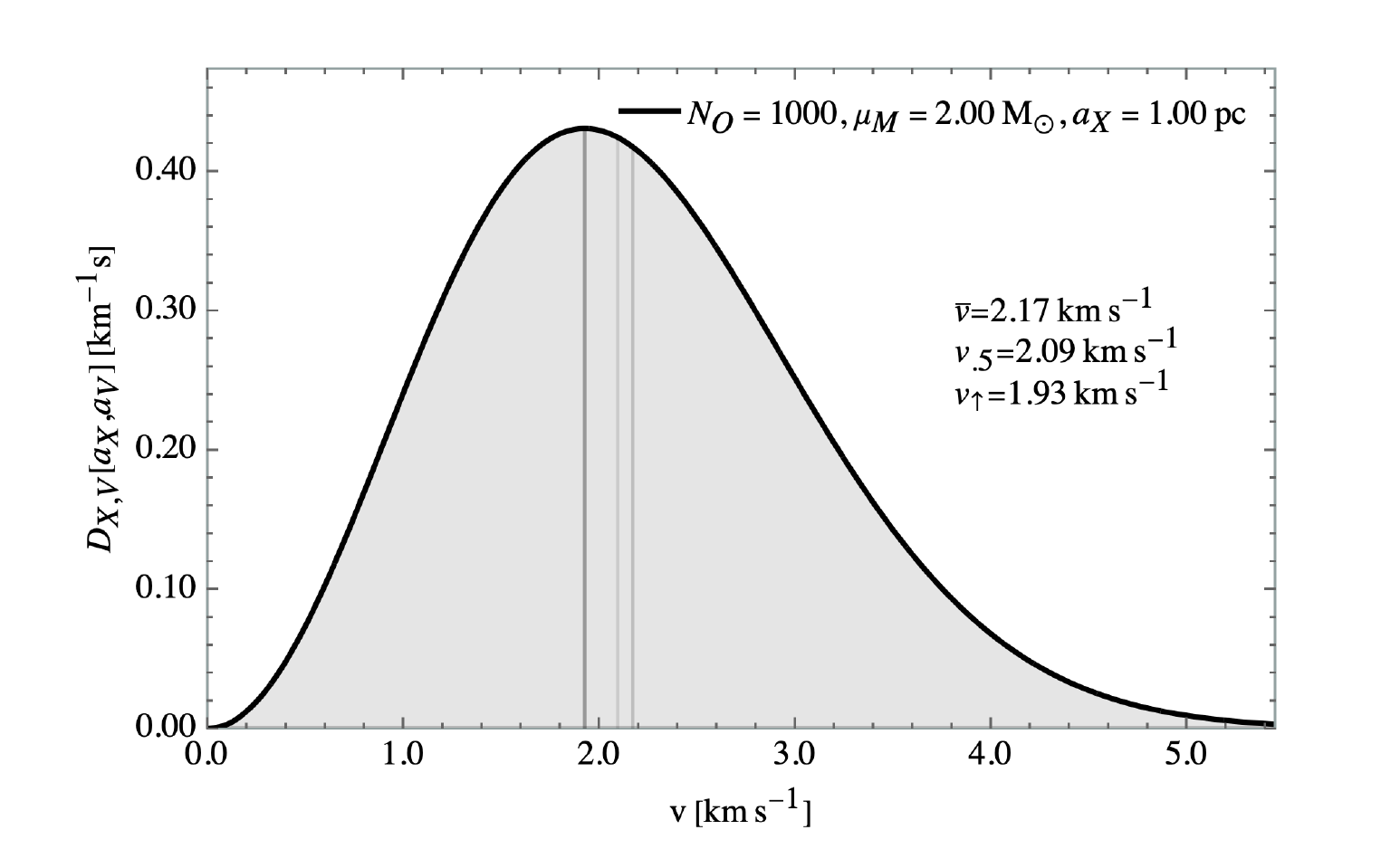}
	\label{subfig: Initial velocity distribution profile}}
	\subfloat[A random realisation's IC velocity profile.]{
	\hspace*{-0.00cm}
	\includegraphics[width = 0.5\textwidth]
	{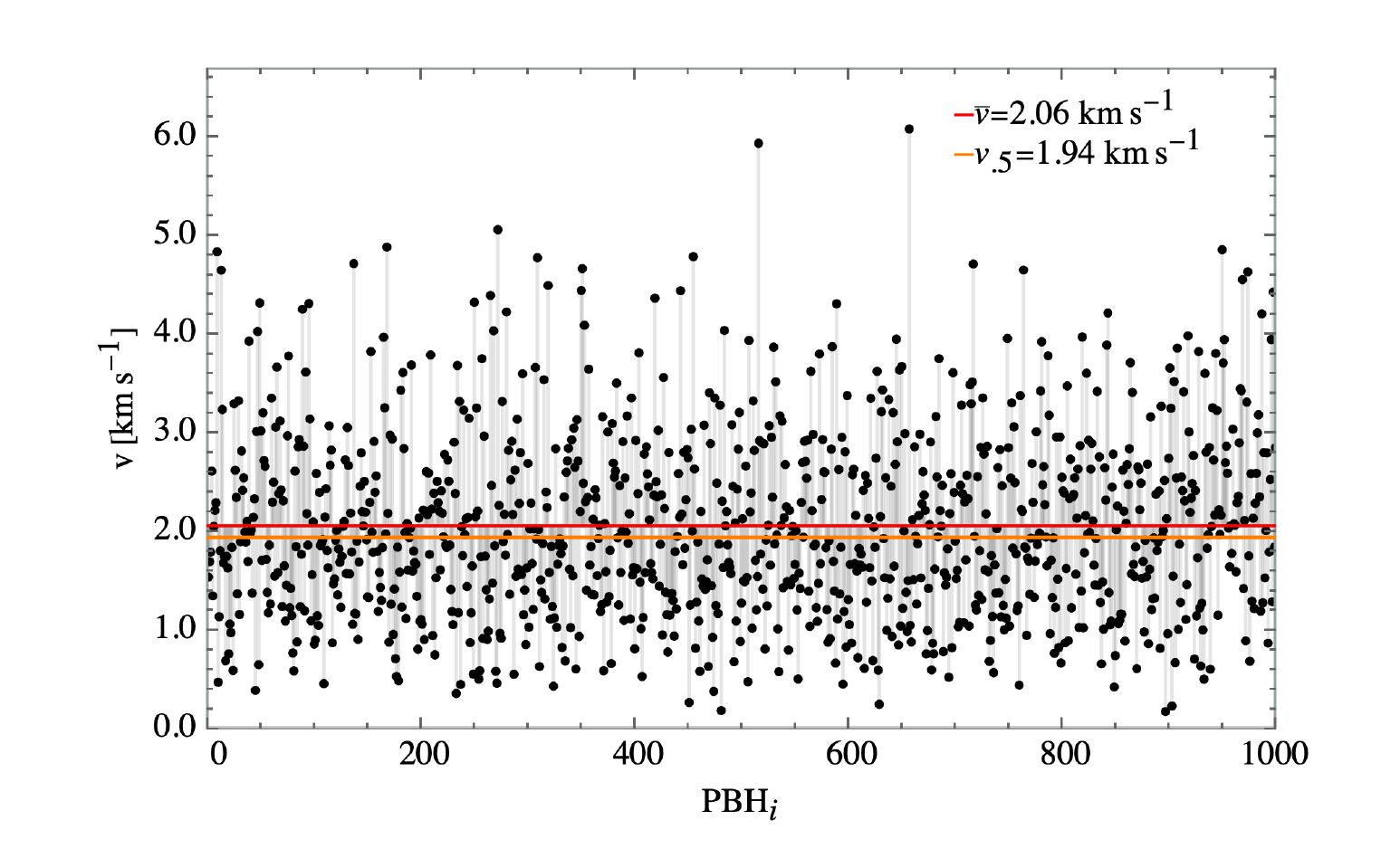}
	\label{subfig: Initial velocity distribution scatter}}
	\caption{
	Velocity IC distribution profile from computed from Table~\ref{tab: Simulation parameters} and an example realisation, from the mean $ \mu_{\vel} = \unit [2.0] {km / s} $ and scale parameter $ 	 a_{\vel} = \unit [1.2] {km / s} $. The thick, normal and thin grey lines signal the distribution mode, median and mean respectively. The orange and red lines signal the realisation's median and mean velocities respectively. Note, again, the hierarchy between modal, median and mean masses is much reduced with respect to the mass case, due to the smallness of the $ v \rightarrow \infty $ tail of the distribution, which results in a small variance of the total simulated velocities in the ICs.
	}
	\label{fig: Initial velocity distribution}
\end{figure*}


The initial velocity profile $ \dist_{\vel} \left(v_{i} \right)_{0} $ of the cluster is obtained as well from a random sample of a Maxwell-Boltzmann distribution for each realisation, which, as a function of the mean velocity $ \mu_\vel $ is given by
\begin{equation}
	\dist_{\vel} \left(v_{i} \right) = \normalisation_{\vel} v_i^{2} \exp \left[-\frac {4 v_i^{2}} {\pi \mu_\vel^{2}} \right],
	\label{eq: Velocity distribution}
\end{equation}
where $ \normalisation_{\vel} = 32 / (\pi^{2} \mu_\vel^{3}) $ is a normalisation factor. Note that both the distribution scale parameter, the velocity variance and the modal can be similarly obtained as it was the case in the position distribution in Eqs.~\eqref{eq: Maxwell-Boltzmann mean}, \eqref{eq: Maxwell-Boltzmann variance} and \eqref{eq: Maxwell-Boltzmann mode} respectively with the substitution $ \pos \rightarrow \vel $, being the same underlying distribution.

Assuming that the position distribution is a nearly homogeneous, spherically-symmetric matter distribution, that the cluster is sampled well enough that the distribution of objects traces accurately enough the prior distribution and that the Virial Theorem holds, then the mean velocity is calculated from the rotation curve of the galaxy, shown in Figure~\ref{fig: Cluster universal curves}. Such is the case of our simulations, where the masses of the individual PBHs in the cluster do not vary wildly in between them and there are enough of objects to accurately sample the distribution, with the mass evenly distributed across the cluster in a monopolar distribution.

The mean rotational velocity of the distribution, $ v_{i} $, in shells of radius $ {x_{i}} \in (0, x_{\mathrm{max}}) $ from the cluster barycentre is
\begin{equation}
	v_{i}^{2} = \frac {12 \pi G \mu_M} {5 x_{i}} \int_0^{x_{i}} m (x') dx',
	\label{eq: Mean square velocity}
\end{equation}
where the average cumulative mass, $ m_{i} $, in spheres of radius $ {x_{i}} \in (0, r_{\mathrm{max}}) $ from the cluster barycentre is then
\begin{equation}
	m_{i} (x_{i}) = \langle m_{i} \rangle N_{\ocap} \int_{0}^{x_{i}} \dist_{\pos} \left(x'_{i} \right) 4 \pi x'^{2} dx'.
	\label{eq: Inner shell mass}
\end{equation}

It is more frequent, however, to express the mean velocity in terms of the Maxwell-Boltzmann scale parameter, $ a_{\vel} $, which, from Eq.~\eqref{eq: Mean square velocity} and Eq.~\eqref{eq: Inner shell mass}, is given by
\begin{equation}
	\mu_{\vel} = 2 a_{\vel} \sqrt{\frac {2} {\pi}} = \sqrt{\frac {3 G m_{i} (x_{i})} {5 x_{i}}},
	\label{eq: Velocity scale parameter}
\end{equation}
Then, the initial, shell-dependent velocity distribution function, can be obtained by replacing the in velocity distribution of Eq.~\eqref{eq: Velocity distribution} the shell-dependent scale parameter using Eq.~\eqref{eq: Velocity scale parameter}

It can be shown, once Eq.~\eqref{eq: Velocity scale parameter} is integrated out to $ x_{i} \rightarrow \infty $ that the resulting global velocity distribution is not a Maxwell-Boltzmann distribution, but that it can be well approximated by one with $ \mu_{\vel} = \unit [2.0] {km / s} $ and $ a_{\vel} = \unit [1.2] {km / s} $, (see Table~\ref{tab: Simulation parameters}, Figure~\ref{fig: Initial velocity computation} and Figure~\ref{fig: Initial velocity distribution}), which we denote by Quasi Maxwell-Boltzmann distribution (QMB).

Additionally, it should be noted that the individual positions and velocities of each object were displaced by means of a suitable change of frame of reference such that the PBHs in the simulation box, in each realisation, have zero total linear momentum
\begin{equation}
	\sum_{i} \vec{P}_{i} = 0, \\
	\label{eq: Conservation of linear momentum}
\end{equation}
so that the barycentre of all particles remains static throughout the simulation and the PBH cluster in particular does drift minimally from its initial position as it expels PBHs on one-to-one encounters. This transformation is performed as well to solve one practical problem, and that is that it reduces the computational time and data storage needed for the analysis of the output, since tracking individually the cluster and ejected PBHs with respect to the separate cluster frame of reference is faster in our analysis pipeline if the simulation as a whole is in the rest frame. This is because, as the positions and velocities transformation from the simulation box frame of reference to the cluster one skips the intermediate step of computing the simulation box barycentre at each time slice and it is precisely operating with position and velocities what takes up most of the computational time and storage.

We do as well perform as well a suitable rotation of the frame of reference so that there is zero total and angular momentum
\begin{equation}
	\sum_{i} \vec{L}_{i} = 0,
	\label{eq: Conservation of angular momentum}
\end{equation}
for similar reasons, so that there is no bulk total rotation throughout the simulation and the cluster and ejecta PBHs develop a minimal intrinsic rotation themselves as PBHs expelled from the cluster in one-to-one encounters. Note that again, the particular choice of a frame of reference in which the total angular momentum is zero has no impact on the simulation observables or dynamics, as it should be the case.



\subsubsection{Body radii choice}
\label{subsubsec: Body radii choice}

In the code, mergers are treated classically and thus a merger occurs when the distance between the bodies centre of mass is less than the sum of their radius, set in Section~\ref{subsec: Simulation Initial Conditions} as their Schwarzschild radius. Therefore, if two bodies $ (m_{i}, r_{i}^{\mathrm{S}}) $ and $ (m_j, r_{j}^{\mathrm{S}}) $ with $ i, j \in (1, 2, ..., N_{\ocap}) $ approach each other to less than the threshold
\begin{equation}
	r^{\min} = r_{i}^{\mathrm{S}}+r_{j}^{\mathrm{S}} = 2 G \frac {m_{i}+m_{j}} {c^{2}},
	\label{eq: Schwarzschild radius threshold}
\end{equation}
below which, a merger is then registered by the $ N $-body code.



\subsection{Simulation evolution snapshots}
\label{subsec: Simulation evolution snapshots}


\begin{table*}[t!]
	\centering
	\begin{tabular}[c]{| p{5mm} p{12mm} p{5mm} | p{18mm} p{18mm} p{18mm} | p{5mm} p{12mm} p{5mm} | p{18mm} p{18mm} p{18mm} |}
	\hline
	\multicolumn{12}{| c |}{Simulation time-runs:} \\
	[0.5ex]
	\hline\hline
	$ r $ & $ (i^{-}, i^{+}) $ &
	$ N_{t} $ &
	$ \unit {t_{i^{-}}} [yr] $ &
	$ \unit {t_{i^{+}}} [yr] $ &
	$ \unit {\Delta t_{r}} [yr] $ &
	$ r $ & $ (i^{-}, i^{+}) $ &
	$ N_{t} $ &
	$ \unit {t_{i^{-}}} [yr] $ &
	$ \unit {t_{i^{+}}} [yr] $ &
	$ \unit {\Delta t_{r}} [yr] $ \\
	[0.5ex]
	\hline
	0 & $ (0, 1) $ & 2 &
	$ 0.00 $ &
	$ 13.8 \times 10^{3} $ &
	$ 13.8 \times 10^{3} $ &
	4 & $ (29, 37) $ & 9 &
	$ 13.8 \times 10^{6} $ &
	$ 13.8 \times 10^{7} $ &
	$ 13.8 \times 10^{6} $ \\
	1 & $ (2, 10) $ & 9 &
	$ 13.8 \times 10^{3} $ &
	$ 13.8 \times 10^{4} $ &
	$ 13.8 \times 10^{3} $ &
	5 & $ (38, 46) $ & 9 &
	$ 13.8 \times 10^{7} $ &
	$ 13.8 \times 10^{8} $ &
	$ 13.8 \times 10^{7} $ \\
	2 & $ (11, 19) $ & 9 &
	$ 13.8 \times 10^{4} $ &
	$ 13.8 \times 10^{5} $ &
	$ 13.8 \times 10^{4} $ &
	6 & $ (47, 55) $ & 9 &
	$ 13.8 \times 10^{8} $ &
	$ 13.8 \times 10^{9} $ &
	$ 13.8 \times 10^{8} $ \\
	3 & $ (20, 28) $ & 9 &
	$ 13.8 \times 10^{5} $ &
	$ 13.8 \times 10^{6} $ &
	$ 13.8 \times 10^{5} $ &
	7 & $ (56, 64) $ & 9 &
	$ 13.8 \times 10^{9} $ &
	$ 13.8 \times 10^{10} $ &
	$ 1.38 \times 10^{10} $ \\
	\hline
	\end{tabular}
	\caption{
	The $ N_{r} = 7 $ time-runs, denoted by the index $ r = 1, 2, ..., 7 $, into which each realisation is partitioned for a total of $ N_{t} = 64 $ marked by the index $ i = i^{i} = i^{-} = 1, 2, ..., 7 $ and the number of non-overlapping time-slices in each run, as well as the initial time, ending time and time-step factored by the present age of the Universe, $ t_{64} = \unit [13.8 \times 10^{10}] {yr} $, which corresponds to the current age of the Universe $ t_{\textrm{U}} = \unit [(13.800 \pm 0.024) \times 10^{10}] {yr} $ according to \textit{Planck 2018} (TT, TE, EE+lowE, see Ref.~\cite{Aghanim:2018eyx}). The time intervals and time-slices shown in Figure~\ref{fig: Dynamical parameters evolution}, displaying the PBHs positions and velocities at each of these.
	}
	\label{tab: Simulation time-runs}
\end{table*}


Once the simulation starts, $ N_{t}-1 = 64 $ time slices are produced and the masses, positions, velocities and other orbital parameters are recorded for all objects, until simulation time $ t_{64} = \unit [1.38 \times 10^{10}] {yr} $ is reached.

Note that the position and velocity coordinates will henceforth be shown with respect to to the cluster barycentre, in the hereby called Cluster Frame (CF), a non-inertial frame of reference w.r.t the original Simulation Frame (SF), and which is calculated on a snapshot-by-snapshot basis considering only the cluster objects, with positions an velocities transforming as
\begin{align}
	x^{\mathrm{CF}}_{i} &= x^{\mathrm{SF}}_{i}-\langle x^{\mathrm{SF}}_{i,\cpop}\rangle_{\mathrm{CM}}, \\
	v^{\mathrm{CF}}_{i} &= v^{\mathrm{SF}}_{i}-\langle v^{\mathrm{SF}}_{i,\cpop}\rangle_{\mathrm{CM}},
	\label{eq: Cluster Frame}
\end{align}
and where the position and velocity centres of mass are computed, time by time-slice, with
\begin{align}
	\langle x^{\mathrm{SF}}_{i,\cpop}\rangle_{\mathrm{CM}} &= \frac {\sum_{i}^{N_\ocap} \delta_{\mathrm{Z_{i}},\cpop} m_{i} x^{\mathrm{CF}}_{i}} {\sum_{i}^{N_\ocap} \delta_{\mathrm{Z_{i}},\cpop} m_{i}}, \\
	\langle v^{\mathrm{SF}}_{i,\cpop}\rangle_{\mathrm{CM}} &= \frac {\sum_{i}^{N_\ocap} \delta_{\mathrm{Z_{i}},\cpop} m_{i} v^{\mathrm{CF}}_{i}} {\sum_{i}^{N_\ocap} \delta_{\mathrm{Z_{i}},\cpop} m_{i}},
	\label{eq: Cluster barycentre}
\end{align}
where $ \delta_{\mathrm{Z_{i}},\cpop} = 1 $ if object $ i $ in the simulation remains bounded to the cluster or $ \delta_{\mathrm{Z_{i}},\cpop} = 0 $ if said object has been ejected and is no longer bounded to the cluster.

Then, by definition the Cluster Frame origin does not drift away from the cluster over time, unlike the case with the Simulation Frame, because of the ejected PBHs carrying mass away from the cluster in a non-perfectly isotropic manner.

We proceed now to describe the criteria chosen for the output choice of the time slices, the simulation averaging of the different realisations in order to provide our predictions with reliable confidence bands, the distinct phases through which the simulations go, as well as the algorithm design to reconstruct the merger and parent trees of each realisation.



\subsubsection{Simulation period}
\label{subsubsec: Simulation period}

All realisations of the simulated PBH cluster start with the IC at $ t_{0} = \unit [0] {yr} $ and end with the last snapshot exactly at $ t_{64} = \unit [1.38 \times 10^{10}] {yr} $, where 64 is the total number of evolution time-slices, the last of which corresponding to the current age of the Universe $ t_{\textrm{U}} = \unit [(13.800 \pm 0.024) \times 10^{10}] {yr} $ according to \textit{Planck 2018} (CMB TT, TE, EE+lowE spectra, see Ref.~\cite{Aghanim:2018eyx}).

In order to capture the fast dynamics of PBHs at early simulation times and the much slower dynamics at late times, we settle then for a data collection algorithm in which the data collection time is updated in every time-run being multiplied by a factor of ten. Following this scheme, the simulation evolution is arranged in the following manner:
\begin{enumerate}[label=\roman*)]
	\item Each simulation is divided into $ N_{r} = 7 $ consecutive intervals, or runs, the first taking as the starting point the IC, and each of the following runs taking as the starting point the last time slice of the preceding run.
	\item Each successive run lasts for ten times the duration of the preceding one and outputs data every ninth of the run time, so that each run consists of nine non-overlapping time slices linearly binned in time.
	\item It can be shown, then, that per each realisation, there are in total, $ N_{t} = 65 $ time-slices: 64 time-slices corresponding to the time slices at $ t_j \in \unit [13.8 \times [1000, 10^{10}]] {yr} $, the first of which is the effective IC for the first run, and an extra time-slices at the beginning for the global IC at $ t_{0} = \unit [0] {yr} $, as detailed in Table~\ref {tab: Simulation time-runs}.
\end{enumerate}



\subsubsection{Evolution tests}
\label{subsubsec: Evolution tests}

It has been checked for a small number of realisations that both total mass $ \sum_{i} M_{i} (t) $, total energy $ \sum_{i} E_{i} (t) $, linear momentum $ \sum_{i} \vec{P}_{i} (t) = 0 $ and angular momentum $ \sum_{i} \vec{L}_{i} (t) = 0 $ are conserved for all times in the simulation period, in accordance with the ICs, where both quantities were set to zero by choosing an appropriate frame of reference as shown in Eq.~\eqref{eq: Conservation of linear momentum} and Eq.~\eqref{eq: Conservation of angular momentum}. In particular we find that energy conservation is upheld up to $ \order{10^{-6}} $ and stable in time evolution, while total linear and angular momenta are conserved to within $0.01\% $ with respect to their corresponding initial values.



\subsubsection{Realisation averaging}
\label{subsubsec: Realisation averaging}

In order to better quantify and constrain the evolution of the simulated PBH clusters and in particular of their masses $ m $, radii $ r^{\mathrm{S}} $, absolute positions $ \vec{x} = (x_{1}, x_{2}, x_{3}) $ and velocities $ \vec{v} = (v_{1}, v_{2}, v_{3}) $ with respect to the simulation box frame that have been set at $ t_{0} $ in the previous Section~\ref{subsec: Simulation Initial Conditions}, we make exactly $ N_{\rcap} = 5000 $ different realisations, each having for the IC a different random realisation of the same probabilistic distribution of the physical parameters that fully determine the ICs.

As will be later described in Section~\ref{sec: Close encounters}, two distinct phases can be easily differentiated during this time evolution. Borrowing from similarly looking features in Monte Carlo Markov Chain simulations we describe the first phase as a ``burn-in" (BI) period, followed by a ``quasi-static'' (QS) phase. The algorithm by which the time of transition between the burn-in period and the quasi-static phase of the simulation is found will be detailed next.



\subsubsection{Burn-in period}
\label{subsubsec: Burn-in period}


\begin{table}[t!]
	\centering
	\begin{tabular}[c]{| p{10mm} | p{17mm} p{17mm} | p{17mm} p{17mm}|}
	\hline
	\multicolumn{5}{| c |}{Snapshot burn-in times:} \\
	[0.5ex]
	\hline\hline
	$ \unit {t_{\bicap}} [yr] $ &
	$ N_{\rpop} $ &
	$ N_{\bpop} $ &
	$ N_{\pos} $ &
	$ N_{\vel} $ \\
	[0.5ex]
	\hline
	$ N_{\cpop} $ &
	$ 4.92 \times 10^{5} $ &
	$ 3.50 \times 10^{5} $ &
	$ 2.95 \times 10^{5} $ &
	$ 2.95 \times 10^{5} $ \\
	$ N_{\epop} $ &
	$ 3.00 \times 10^{5} $ &
	$ 5.93 \times 10^{5} $ &
	$ 5.72 \times 10^{5} $ &
	$ 5.72 \times 10^{5} $ \\
	$ N_{\tpop} $ &
	$ 0.00 $ &
	$ 3.59 \times 10^{5} $ &
	$ 5.72 \times 10^{5} $ &
	$ 5.72 \times 10^{5} $ \\
	\hline
	\end{tabular}
	\caption{
	Computed burn-in time of a number of populations of the remaining objects, mainly the cluster $ N_{\cpop} $ and ejecta $ N_{\epop} $ populations, the population of objects in bounded $ N_{\bpop} $ systems, that is, those objects part of multiple, usually binary, system, or even binary within a binary system, and combinations of those. These burn-in times are shown in Figure~\ref{fig: Dynamical parameters evolution} when computed from position and velocity evolution, along with Figure~\ref{fig: Remaining population} and \ref{fig: Bounded population} when computed from population evolution.
	}
	\label{tab: Simulation burn-in times}
\end{table}


One can easily observe in Figure~\ref{fig: Dynamical parameters evolution} that the very first few stages of the time evolution, usually covering a fraction no more than the first $ \order {10^{-4}} $ of the simulation time, show an exceedingly fast dynamic with respect to later times, forcefully expelling a large number of PBHs from the cluster to the background.

This feature occurs because, unlike other seemingly similar structures to this PBH clusters, like ordinary stellar globular clusters, the density in the ICs of this PBHs clusters is very large, much more so than in the case of stellar globular clusters.

In particular, in the simple model where there is no realisation variance and the IC random sample perfectly traces the IC distribution moments, at $ t_{0} $ the total simulation mass is
\begin{align}
	m_{\tot} &= N_{\ocap} \bar{m} \nn \\
	&= N_{\ocap} \exp (\mu_{\mass}+\sigma_{\mass}^{2}/2) \nn \\
	&\approx \unit [2.3 \times 10^{4}] {\msun},
\end{align}
mostly contained in a spherical region with
\begin{equation}
	x_{\tot} \approx 2 \bar{x} \approx \unit [3.2] {pc}.
\end{equation}

In such a case, the mean total density of the PBH cluster at the initial time is
\begin{equation}
	\rho = \frac {3 m_{\tot}} {4 \pi x_{\tot}^{3}} \approx \unit [1.7 \times 10^{3}] {\msun \ pc^{-3}},
\end{equation}
in line with the mass density of the central regions of globular clusters, of $ \rho \approx \unit [1000-10^{5}] {\msun \ pc^{-3}} $ (see Ref.~\cite{Chatterjee:2013}), and orders of magnitude above that of the stellar density in the solar neighbourhood, $ \rho \approx \unit [0.040 \pm 0.002] {\msun \ pc^{-3}} $ (see Ref.~\cite{Bovy:2017}).

This results in that while PBHs move through their trajectories in between the first few snapshots at $ \unit [0] {yr} \leq t < \unit [10^{6}] {yr} $ they observe large local gradients in the PBH cluster gravitational potential, often enough to expel these objects with speeds greater than the escape velocity of the cluster within their respective radial shell, and helping them escape as ejecta to the background. For similar reasons, the majority of hyperbolic encounters found within the simulations do occur within this short period, due to the much smaller mean inter-object distance in between PBHs.

This is ultimately due to the fact that long term cluster stability imposes a harsh boundary to the minimum spatial volume of a cluster for a given mass. It is therefore natural, as it is also seen in Section~\ref{sec: Dynamical distributions}, that after this brief period of fast evolution and complex dynamics, the cluster puffs-up and settles in the next stage.

The particular time at which the burn-in period is considered to transition to the quasi-static phase is found as follows. We compute the cluster population $ N_{\cpop} (t) $ and that of ejected objects $ N_{\epop} (t) $ at any given simulation snapshot, and average this quantity over the $ N_{\rcap} = 1000 $ realisations so that the resulting evolution is well behaved and constrained, quasi-monotonously decreasing and increasing respectively for the two populations as it is shown in Figure~\ref{fig: Bounded population}.

Prior to this transition time $ t_{\bicap} $, there is essentially no evolution and their numbers remain static; after this the evolution of the cluster and ejected objects fit well to a power-law
\begin{equation}
	N_{i = \cpop,\epop} (t \geq t_{\bicap}) \sim \beta_{i} t^{\alpha_{i}},
	\label{eq: Power-Law fit}
\end{equation}
with $ \alpha_{\cpop} > 0 $ for a decreasing cluster population, $ \alpha_{\epop} < 0 $ for an increasing ejecta population, and $ \beta_{\cpop} > \beta_{\epop} > 0 $ since the initial ejecta population is for all realisations exceedingly small, sometimes nil.

Due to the discrete nature of the output, the resulting transition time $ t_{\bicap} $ at which the change between these two different regimes cannot be found exactly, but it can be estimated. At each time slice, the evolution of the cluster and ejecta populations is fitted to the logarithmic law of Eq.~\eqref{eq: Power-Law fit}, from that particular time slice time to the end time of the run, of which there are $ N_{r} = 7 $, in which such time slice is found, for all time slices $ N_{t}-1 = 64 $ time-slices.

The motivation to perform this fits in a run-by-run manner is to avoid overfitting some regions to the detriment of others since the separation in between time-slices varies by seven orders of magnitude in between the first and the last fit, and time-slices are globally neither linearly spaced nor logarithmically spaced, but rather a combination between the two. After the fitting, the array of times and determination coefficients, $ (t, r^{2}) $, is interpolated with a cubic spline, again in a run-by-run manner.

The resulting interpolated curve $ r^{2} (t) $ generally has low values for $ t \rightarrow t_{0} $, and unit values for $ t \rightarrow t_{64} $. Inside each run $ i \in [1, 2, ..., N_{\icap}] $, the interpolated curve $ r^{2} (t) $ will tend to have lower values for $ t \rightarrow t_{0,i} $, and higher values for $ t \rightarrow t_{\textrm{64,i}} $. The burn-in time, $ t_{\bicap} $, is then, the time where there is a first run with a maximum value of the $ r^{2} (t) $ curve within 1\% from the maximum of the range in the subsequent runs $ j \geq i+1 $
\begin{align}
	t_{\bicap} & \equiv t : \max [r^{2} (t)_\textrm{i}] \in \left[t_{*,j}, \frac {t_{*,j}-t_{*,j}} {100} \right], \\
	t_{*,j} & = \min (r^{2} (t)_\textrm{j}), j \geq i+1 ,\\
	t_{*,j} & = \max (r^{2} (t)_\textrm{j}), j \geq i+1,
	\label{eq: Burn-in time}
\end{align}
the results of which for the different populations and their combinations are shown in Table~\ref {tab: Simulation burn-in times}. Note that, in all cases, $ t_{\bicap} \leq \unit [5.93 \times 10^{6}] {yr} $.



\subsubsection{Quasi-static phase}
\label{subsubsec: Quasi-static phase}

After the cluster has undergone the puffing up phase and expanded by a factor that varies greatly on the ICs, but usually up to $ \order {10} $ during its very first stages, then the density decreases significantly and becomes comparable to that of stellar globular clusters.

Starting at this point, the evolution is much slower than in the previous phase. The fraction of merged objects per realisation barely amounts to $ 117 $ mergers out of a possible total of $ N_{\rcap} N_{\ocap} = 10^{6} $ so there are barely any mergers in each of the individual clusters in the period that we have considered, so the number of objects states almost constant in our simulations.

Also starting at this point, the core of the cluster attracts the most massive and slower objects, sending the least massive and fastest of this objects to the periphery of the cluster through the well-known processes of mass segregation and dynamical friction, in a manner qualitatively similar to that of stellar globular clusters, but enhanced with respect to these due to the larger masses involved.

In this stage and for the rest of the PBH cluster evolution, the vast majority of would-be sources of gravitational radiation would come from hyperbolic encounters between pairs of objects, often ending in the slingshot and ejection of one of the PBHs at high speeds out of the cluster, resulting in a slow cluster population decrease, while releasing high-frequency GWs. Be reminded that our code does not include GR effects and will source cluster evaporation only as a consequence of close Newtonian 2-body encounters resulting in the ejection of a PBH, again in a very similar manner to the case of stellar globular clusters, with no emission of GWs and the associated velocity dampening. This slow cluster depletion of objects increases the matter density of the background at the expense of the cluster's own, which results in a bimodal distribution in the DM power spectrum.

Also in this stage a number of PBH binary systems are formed inside the cluster, with their survival depending on their possible ejection. Cluster binaries are rapidly disrupted by third-object encounters, usually lasting less that $ \tau_{\mathrm{CB}} \sim \unit [\order {10^{7}}] {yr} $. However, having escaped to the much depopulated background, ejected binaries are stable and long lived since the time to merger is well approximated in Ref.~\cite{Peters:1964zz}. In this scenario, only relatively infrequent massive, close PBHs could slowly inspiral into each other within a Hubble time, while emitting low-frequency gravitational radiation with regular peaks at the periastron crossing.



\subsubsection{Parent trees computation}
\label{subsubsec: Parent trees computation}

In each time-slice we locate the possible bounded object sub-systems and their coordinates in their Orbital Frame (OF) of reference. Note that the orbital coordinate frame of reference is one centred on the barycentre of all the inferior objects of the sub-system, that is, all objects excluding those outside the shell defined by each PBH and centred around the cluster. This method of choice for the frame of reference is best when, as this is the case, no object is significantly more massive than all of the others combined, and the number of ejected objects dripping from the cluster grows overtime.

This way we identify sub-systems of objects within the simulations, corresponding to binary and multiple systems, either within the cluster or the ejecta, and even in a few brief instances sub-sub-systems corresponding to binaries orbiting a binary within the simulations. As for how the parent object is identified, in each time snapshot we find the parent tree of the simulation objects, that is, for all objects we find the most massive object that they are orbiting, in order to identify multiple systems. Depending on the different population to which the object may belong, then
\begin{enumerate}[label=\roman*)]
	\item Cluster isolated PBHs do orbit the cluster barycentre, alone, and not as a part of more compact bounded sub-systems. They will then be assigned as their main-parent the cluster most massive PBH, centrally located, unless they they show up in bounded sub-systems in which a sequence of sub-parents follows this tracing the hierarchical substructure of the subsystem.
	\item Ejected isolated PBHs will always be assigned as their main parent again the cluster most massive, centred object unless they show up in bounded sub-systems in which then a list of sub-parents follows just as it was the case for cluster objects. This is so even if for all other effects ejecta PBHs are not part of the cluster and exert only very weak and ever decreasing gravitational influence over it.
	\item Only the remaining fraction of cluster and ejecta objects, typically of order less than 10\% for each of these respectively, do show up in the form of binary pairs of PBHs, and much more rarely, in the form of binary pairs already within a sub-system, with their sub-parent automatically assigned to the sub-system most massive object, and their parent tree tracing back to the aforementioned main parent of the cluster.
\end{enumerate}

When an object belongs to a sub-system within a sub-system of say, the cluster or ejecta, we label the former as a $ 2^\textrm{nd} $ generation system and the latter we label a $ 1^\textrm{st} $ generation sub-system, while the cluster or ejecta is $ 0^\textrm{th} $ generation. Note that the parent tree registers all the hierarchy up to the most massive, centred object in the cluster, thus the tree always ends up with the same $ 0^\textrm{th} $ generation primogenitor for all objects, tracing generations in between to the last.

We compute orbital the semi-major axis, semi-minor axis and eccentricity from the relative position and velocity vectors in the Orbital Frame for each bounded pair $ i, \ j$, which are computed similarly as those for the Cluster Frame in Eq.~\eqref{eq: Cluster Frame},
\begin{align}
	x^{\mathrm{OF}}_{i,j} &= x^{\mathrm{CF}}_{i}-\langle x^{\mathrm{CF}}_{i,j,\bpop}\rangle_{\mathrm{CM}}^{*}, \\
	v^{\mathrm{OF}}_{i,j} &= v^{\mathrm{CF}}_{i}-\langle v^{\mathrm{CF}}_{i,j,\bpop}\rangle_{\mathrm{CM}}^{*},
	\label{eq: Orbital Frame}
\end{align}
and where again the position and velocity centres of mass are computed, time by time-slice, with
\begin{align}
	\langle x^{\mathrm{CF}}_{i,j,\bpop}\rangle_{\mathrm{CM}} &= \frac {\sum_{i > j}^{N_{\ocap}} \delta_{\mathrm{Z}_{i,j},\bpop} m_{i} x^{\mathrm{CF}}_{i}} {\sum_{i}^{N_\ocap} \delta_{\mathrm{Z}_{i,j},\bpop} m_{i}}, \\
	\langle v^{\mathrm{CF}}_{i,j,\bpop}\rangle_{\mathrm{CM}} &= \frac {\sum_{i > j}^{N_{\ocap}} \delta_{\mathrm{Z}_{i,j},\bpop} m_{i} v^{\mathrm{CF}}_{i}} {\sum_{i}^{N_\ocap} \delta_{\mathrm{Z}_{i,j},\bpop} m_{i}},
	\label{eq: Orbital barycentre}
\end{align}
where $ \delta_{\mathrm{Z}_{i,j},\bpop} = 1 $ if the pair $ i, \ j $ is bounded and $ \delta_{\mathrm{Z}_{i,j},\bpop} = 0 $ if the pair $ i, \ j $ is unbounded within a subsystem.

From Eq.~\eqref{eq: Orbital Frame} the semi-major axis, semi-minor axis and eccentricity can be extracted by first computing the specific angular momentum of the pair with
\begin{equation}
	\vec{h}_{i,j} = \mu_{i,j} \vec{x}_{i,j}^{\mathrm{OF}} \times \vec{v}_{i,j}^{\mathrm{OF}},
\end{equation}
where $ \gamma = x_{i,j}^{\mathrm{OF}} (v_{i,j}^{\mathrm{OF}})^{2} - 2\mu_{i,j} $ and $ \mu_{i,j} = G (m_{i} + m_{j}) $ is the gravitational parameter. Then, the semi-major axis and orbital eccentricity can be computed with
\begin{align}
	a &= \sign ({\gamma}) \gamma^{-1} \mu_{i,j} x^{\mathrm{OF}}_{i,j}, \\
	e^{2} &= 1 + \sign ({\gamma}) \frac {h^{2}_{i,j}} {a \mu_{i,j}}, \\
	b^{2} &= a^{2} \sign (\gamma) (1 - e^{2}),
\end{align}
which, for $ \sign (\gamma) < 0 $ yields an elliptical orbit, for $ \sign (\gamma) > 0 $ yields a hyperbolic orbit, and $ \gamma \rightarrow 0 $ constituting the parabolic limit where $ e = 1 $, $ a \rightarrow \infty $.



\subsubsection{Merger trees computation}
\label{subsubsec: Merger trees computation}

Also in each time-slice we compute the merger history of the bodies, tracking every occurring PBH absorption and in each merger event adding the identity of the least massive PBHs to the history of the more massive PBHs, thus providing a reconstruction of which particular objects have been merged to produce more massive ones.

This is done by means of tracing every mass change in between every conservative pair of the $ N_{t} -1 = 64 $ last ESs, then solving $ N_{t} -1 = 64 $ constrained system of the diophantic equations, one per each pair of ESs, in to find out the identity of the pairs of objects that have merged and store the data
\begin{equation}
	(\Delta M)_{+, i} = C_{ij} (\Delta M)_{-, j},
	\label{eq: Merger diophantic equation}
\end{equation}
where $ (\Delta M)_{+, i} $ is the mass gained by object $ i $, $ (\Delta M)_{-, j} $ is the mass lost by object $ j $, for all $ i = 1, 2, ..., N_{+} $ objects that gain mass and for all $ j = 1, 2, ..., N_{-} $ objects that loose mass in between snapshots, so $ N_{+} \leq N_{-} $. $ C $ is the mixing sparse matrix with elements $ C_{i, j} = 0, 1 $ relating the absorber $ i $ with the absorbed $ j $, of unit value in the case of a merger and zero in the opposite case.

The mixing sparse matrix $ C $ can be very slow to solve in the case where a single PBH, or worse, many PBHs, have absorbed many other PBHs each in the same interval within snapshots. In that case, a number of matrix with elements $ C_{i, j} $ could be found first before solving the full diophantic system of Eq.~\eqref{eq: Merger diophantic equation} and thus easing the problem if any of the easier merger cases or lack of thereof are identified first, mainly those of
\begin{enumerate}[label=\roman*)]
	\item Impossible mergers:
	\begin{align}
		& (\Delta M)_{+, i} < (\Delta M)_{-, j}, \\
		& C_{i, j} = 0.
		\label{eq: Merger impossible case}
	\end{align}
	\item Necessary mergers:
	\begin{align}
		& (\Delta M)_{+, i} > (\Delta M)_{-, j}, \\
		& \sum_{i} C_{i, j} \in (2, 3, ...).
		\label{eq: Merger necessary case}
	\end{align}
	\item Single one-to-one mergers:
	\begin{align}
		(\Delta M)_{+, i} = & (\Delta M)_{-, j}, \\
		& j = j_{1}, \\
		C_{i, j} = & 1, C_{i, j' \neq j} = C_{i' \neq i, j} = 0.
		\label{eq: Merger single case}
	\end{align}
	\item Multiple $ k $ -to-one mergers:
	\begin{align}
		(\Delta M)_{+, i} = & \sum_{j} (\Delta M)_{-, j}, \\
		& j = j_{1}, ..., j_{k}, \\
		C_{i, j} = & 1, C_{i, j' \neq j} = C_{i' \neq i, j} = 0.
		\label{eq: Merger multiple case}
	\end{align}
\end{enumerate}
After solving the aforementioned, one may proceed to solve the simplified diophantic system of Eq.~\eqref{eq: Merger diophantic equation}.



\section{Population statistics}
\label{sec: Population statistics}

\begin{figure*}[t!]
	\centering
	\subfloat[ESs's remaining PBHs position norms in the CF.]{%
	\hspace*{-0.00cm}
	\includegraphics[width = 0.50\textwidth]
	{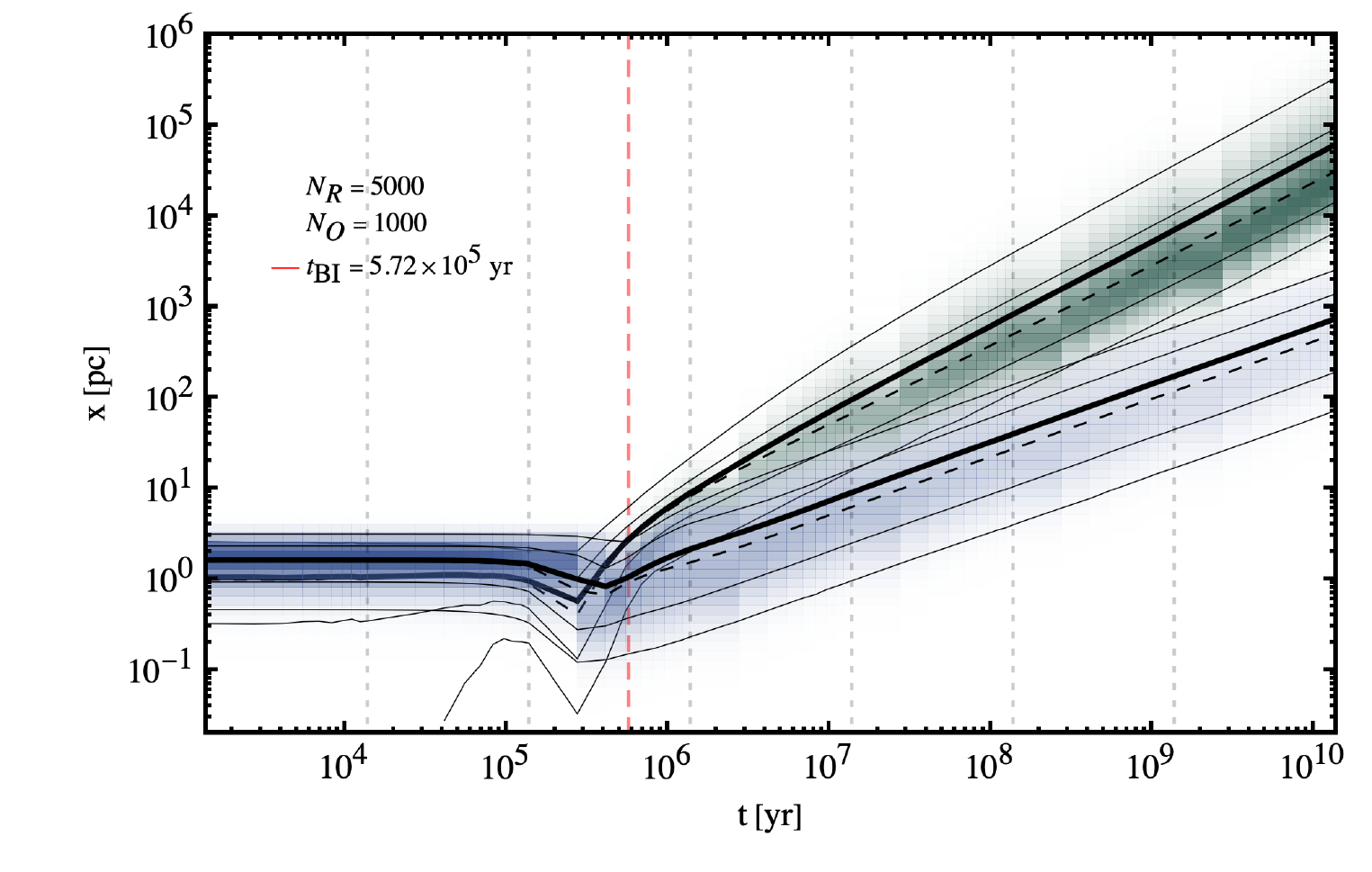}
	\label{subfig: Time positions histogram}}
	\subfloat[ESs's remaining PBHs velocity norms in the CF.]{%
	\hspace*{-0.00cm}
	\includegraphics[width = 0.50\textwidth]
	{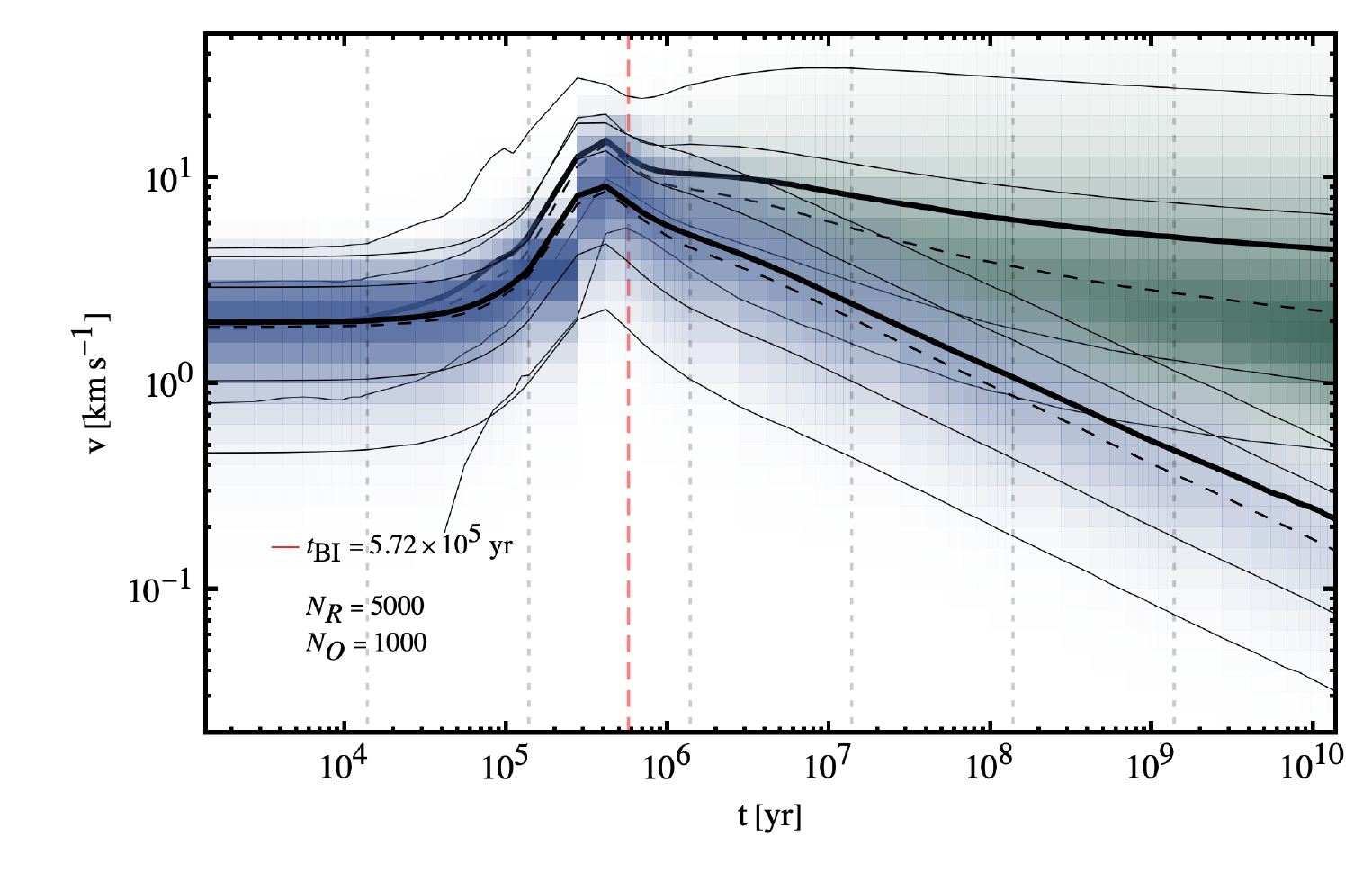}
	\label{subfig: Time velocities histogram}}
	\caption{
	Positions and velocities time evolution of a subset 10\% of objects in one particular realisation. In blue and teal, the position and velocity curves for cluster and ejecta objects respectively. For both positions and velocities, the thick black line mark the mean of the discrete distribution of evolution trajectories per snapshot, while the thin black line in the same panel represents the median, and the thin black lines mark the 68\% and 95\% confidence regions. The mean fits obtained for the position and velocity profiles at each time-slice are shown in Table~\ref{tab: Dynamical parameters evolution}.
	}
	\label{fig: Dynamical parameters evolution}
\end{figure*}

The time evolution of this positions and velocities is shown in Figure~\ref{fig: Dynamical parameters evolution}. Note that the cluster puffs-up to a factor of $ \order {1000} $ to a final size of $ x_{\cpop} (\unit [1.38 \times 10^{10}] {yr}) \approx \unit [100-1000] {pc} $, decreasing its density by a factor of $ \order {10^{9}} $, while the ejecta expands further to distances in the range of $ x_{\epop} (\unit [1.38 \times 10^{10}] {yr}) \approx \unit [10^{5}] {pc} $ during the simulation period.

It is apparent in Figure~\ref{fig: Dynamical parameters evolution} that the cluster time-trajectories exhibit a far wigglier pattern in comparison to the ejecta time-trajectories, as expected since the PBHs in the former ones move in quasi-random trajectories within the cluster while the latter ones very quickly arrive to the asymptotic regime where they move in straight lines.

The ejecta velocity dependence with time, however, is small in most cases, as it corresponds to asymptotically uniform rectilinear time-trajectories, the exception to this rule being the a few instances where wiggles are manifest, corresponding to PBHs expelled from the cluster in binary pairs, so that the velocity in the Cluster Frame frame is composed of two modes: a time-dependent orbital velocity over which the constant velocity of the binary barycentre is summed.

The Power Law (PL) fits of the time evolution of positions and velocities are given in Table~\ref{tab: Dynamical parameters evolution}. Note that no fit is made before the burn-in time for the ejecta positions and velocities of the simulations, given that the data variability is too large for this subset of objects, as very few of these exist in the time snapshots prior to the burn-in time. In particular, a threshold of 20 ejecta objects present collectively in all the $ N_{\rpop} = 5000 $ realisation considered has been required as the minimum over which the fits are computed both for the cluster and ejecta objects, which is only the case starting in the third time-run at the burn-in time $ t_{\bicap} = \unit [5.72 \times 10^{5}] {yr} $ in Figure~\ref{fig: Dynamical parameters evolution}


\begin{table*}[t!]
	\centering
	\begin{tabular}[c]{
	| l l |
	p{6 mm} p{7 mm} l l |
	p{6 mm} p{7 mm} l l |}
	\hline
	\quad & $ t_{\bicap} = \unit [5.72 \times 10^{5}] {yr} $ &
	\multicolumn{4}{c |}{Cluster objects positions:} &
	\multicolumn{4}{c |}{Cluster objects velocities:} \\
	\hline\hline
	$ r $ & $ \unit {(t_{i^{-}}, t_{i^{+}})} [yr] $ &
	Fit: & $ r_{i,\pos,\cpop}^{2} $ &
	$ \unit {\beta_{i,\pos,\cpop}} [\textrm{pc}] $ &
	$ \alpha_{i,\pos,\cpop} $ &
	Fit: & $ r_{i,\vel,\cpop}^{2} $ &
	$ \unit {\beta_{i,\vel,\cpop}} [\mathrm{km / s}] $ &
	$ \alpha_{i,\vel,\cpop} $ \\
	[0.5ex]
	\hline
	$ 3 $ & $ (\unit [t_{\bicap}] {yr^{-1}}, 1.38 \times 10^{6}) $ &
	[PL] & $ 1.000 $ &
	$ (3.13 \pm 0.90) \times 10^{-5} $ &
	$ 0.788 \pm 0.021 $ &
	[PL] & $ 1.000 $ &
	$ (3.8 \pm 1.1) \times 10^{3} $ &
	$ -0.468 \pm 0.021 $ \\
	$ 4 $ & $ 1.38 \times (10^{6}, 10^{7}) $ &
	[PL] & $ 1.000 $ &
	$ (2.34 \pm 0.21) \times 10^{-4} $ &
	$ 0.6407 \pm 0.0055 $ &
	[PL] & $ 1.000 $ &
	$ 577 \pm 49 $ &
	$ -0.3311 \pm 0.0055 $ \\
	$ 5 $ & $ 1.38 \times (10^{7}, 10^{8}) $ &
	[PL] & $ 1.000 $ &
	$ (2.374 \pm 0.081) \times 10^{-4} $ &
	$ 0.6410 \pm 0.0019 $ &
	[PL] & $ 1.000 $ &
	$ 911 \pm 23 $ &
	$ -0.3602 \pm 0.0014 $ \\
	$ 6 $ & $ 1.38 \times (10^{8}, 10^{9}) $ &
	[PL] & $ 1.000 $ &
	$ (2.553 \pm 0.061) \times 10^{-4} $ &
	$ 0.6374 \pm 0.0012 $ &
	[PL] & $ 1.000 $ &
	$ 912 \pm 23 $ &
	$ -0.3599 \pm 0.0013 $ \\
	$ 7 $ & $ 1.38 \times (10^{9}, 10^{10}) $ &
	[PL] & $ 1.000 $ &
	$ (2.525 \pm 0.069) \times 10^{-4} $ &
	$ 0.6374 \pm 0.0012 $ &
	[PL] & $ 1.000 $ &
	$ 499 \pm 31 $ &
	$ -0.3309 \pm 0.0028 $ \\
	\hline
	\multicolumn{10}{c}{\quad} \\
	\hline
	\quad & $ t_{\bicap} = \unit [5.72 \times 10^{5}] {yr} $ &
	\multicolumn{4}{c |}{Ejecta objects positions:} &
	\multicolumn{4}{c |}{Ejecta objects velocities:} \\
	\hline\hline
	$ r $ & $ \unit {(t_{i^{-}}, t_{i^{+}})} [yr] $ &
	Fit: &	$ r_{i,\pos,\epop}^{2} $ &
	$ \unit {\beta_{i,\pos,\epop}} [\textrm{pc}] $ &
	$ \alpha_{i,\pos,\epop} $ &
	Fit: &	$ r_{i,\vel,\epop}^{2} $ &
	$ \unit {\beta_{i,\vel,\epop}} [\mathrm{km / s}] $ &
	$ \alpha_{i,\vel,\epop} $ \\
	[0.5ex]
	\hline
	$ 3 $ & $ (\unit [t_{\bicap}] {yr^{-1}}, 1.38 \times 10^{6}) $ &
	[PL] & $ 1.000 $ &
	$ (2.29 \pm 0.84) \times 10^{-7} $ & $ 1.236 \pm 0.026 $ &
	[PL] & $ 1.000 $ &
	$ 63 \pm 22 $ & $ -0.128 \pm 0.025 $ \\
	$ 4 $ & $ 1.38 \times (10^{6}, 10^{7}) $ &
	[PL] & $ 1.000 $ &
	$ (7.79 \pm 0.83) \times 10^{-6} $ & $ 0.9917 \pm 0.0066 $ &
	[PL] & $ 1.000 $ &
	$ 48.6 \pm 4.3 $ & $ -0.1072 \pm 0.0057 $ \\
	$ 5 $ & $ 1.38 \times (10^{7}, 10^{8}) $ &
	[PL] & $ 1.000 $ &
	$ (1.956 \pm 0.058) \times 10^{-5} $ & $ 0.9362 \pm 0.0016 $ &
	[PL] & $ 1.000 $ &
	$ 59.0 \pm 2.3 $ & $ -0.1201 \pm 0.0022 $ \\
	$ 6 $ & $ 1.38 \times (10^{8}, 10^{9}) $ &
	[PL] & $ 1.000 $ &
	$ (1.896 \pm 0.072) \times 10^{-5} $ & $ 0.9370 \pm 0.0018 $ &
	[PL] & $ 1.000 $ &
	$ 30.0 \pm 1.3 $ & $ -0.0841 \pm 0.0022 $ \\
	$ 7 $ & $ 1.38 \times (10^{9}, 10^{10}) $ &
	[PL] & $ 1.000 $ &
	$ (1.555 \pm 0.018) \times 10^{-5} $ & $ 0.94624 \pm 0.00049 $ &
	[PL] & $ 1.000 $ &
	$ 18.26 \pm 0.39 $ & $ -0.06032 \pm 0.00095 $ \\
	\hline
	\end{tabular}
	\caption{
	Mean cluster and ejecta position and velocity PL fits of Figure~\ref{fig: Dynamical parameters evolution} from Eq.~\eqref{eq: Power-Law fit}: $ z_\textrm{P} (t \in r) = \beta_{i,z,\textrm{P}} t^{\alpha_{i,z,\textrm{P}}} $ where $ z = x, v $ and $ r = 3, ..., 7 $ refers to the separate run where the fit is performed, $ \beta_{i,z,\textrm{P}} $ corresponds to the amplitude of the PL while $ \alpha_{i,z,\textrm{P}} $ is the PL index, and $ \textrm{P} = \cpop, \epop $ denotes the cluster and ejecta populations respectively.
	}
	\label{tab: Dynamical parameters evolution}
\end{table*}


Now we proceed to identify the different populations arising in the simulations and show their evolution, which are the clustered vs. ejected PBH populations of Section~\ref{subsec: Cluster and ejecta populations}, the isolated vs. bounded PBH populations of Section~\ref{subsec: Isolated and Bounded populations}, and primitive vs merged PBH populations of Section~\ref{subsec: Primitive and merged populations}.

The PBHs in our simulations can merge onto each other, so the total remaining PBH population is a monotonically decreasing function, while the absorbed PBH population is, in contrast, monotonically increasing over time. Once made this distinction, PBHs in the simulations can be, then, classified according to three different, independent criteria
\begin{enumerate}[label=\roman*)]
	\item Depending on the eccentricity of the PBHs with respect to the cluster barycentre, one object belongs to the Cluster (C) for less than unity eccentricities:
	\begin{equation}
	N_{i} \in \cpop \leftrightarrow e_{i}^{0} < 1,
	\end{equation}
	corresponding to instantaneously-closed orbits about the cluster barycentre, and it belongs to the Ejecta (E) for greater than unity eccentricities:
	\begin{equation}
	N_{i} \in \epop \leftrightarrow e_{i}^{0} \geq 1,
	\end{equation}
	corresponding to instantaneously-parabolic and hyperbolic orbits, where super-index $ 0^{\textrm{th}} $ denotes that it is the eccentricity with respect to the zero level, that is, the complete system, the one considering, and not with respect to the barycentre of a subsystem should the $ i $ PBH be part of one.
	\item Depending on the dimensionality of the parent tree, that is, on whether there is more than the unique $ 0^{\textrm{th}} $ PBHs common parent all objects share, the object may be classified as being part of the Bounded (B) population or the Isolated (I) population.
	\item Depending on the dimensionality of the merger tree, that is, on whether there is at least a $ 1^{\textrm{st}} $ PBHs merger event, the object may be classified as being part of the Merged (M) population or the Primitive (P) population.
\end{enumerate}

There are a number of caveats to this classification algorithm that need further clarification.

First, and concerning the classification of PBHs into either cluster or ejecta objects, note that by ``instantaneously'' we mean that, at any given time, orbits can be assimilated with their respective Keplerian equivalents from their body to barycentre interaction, even though trough time no orbit will obviously remain a static Keplerian when successive interactions with other PBH takes place, and indeed, a PBH may flip from the cluster to ejecta population a number of times in short intervals, unless it quickly escapes the gravitational potential well of the cluster and joins indeed the ejecta population for the remainder of the simulations.

Second, and concerning the classification of PBHs into either bounded or isolated objects, note that any PBH part of a bounded system will be classified as such independently of whether the PBH is the more massive in the pair, in which case it will be labelled as the parent in the binary, or the least massive, in which case it will reference its parent in the parent tree. Also note that any bounded system save the cluster itself irrespective of the number of PBHs in it will be considered a binary system. However, in the unlikely event that the system is ternary or larger, this will be noted by subsequent parent trees following the hierarchy of masses in such multiple system, that is, by adding further levels in the parent trees.

Third, and concerning the classification of PBHs into either merged or primitive objects, note that, unlike in the other two cases there a PBH could flip-flop from one population to another and vice-versa, in this case the PBH can only transit either from the primitive to the merged or absorbed population it it is the more massive PBH in the merger or from the merged to the absorbed population it it is the less massive PBH in the merger, with no going back to the primitive population in any case, which is a decreasing function.

Note at last that all PBHs necessarily fall into one of the two options in the three separate categories, and therefore all PBHs must have exactly three tags: $ (\cpop $ or $ \epop, \bpop $ or $ \ipop, \mpop $ or $ \ppop) $, Additionally, the tags $ \rpop $ and $ \apop $ stand for the remaining and absorbed objects, and it is clear then that $ N_{\rpop, X} = N_{\cpop} $ and $ N_{\apop} \geq N_{\mpop} $
\begin{align}
	N_{\ocap} & = N_{\rpop} (t)+N_{\apop} (t), \\
	N_{\rpop} (t) & = N_{\cpop} (t)+N_{\epop} (t), \nn \\
	& = N_{\ipop} (t)+N_{\bpop} (t), \nn\\
	& = N_{\ppop} (t)+N_{\mpop} (t),
	\label{eq: Population classification}
\end{align}
for all possible populations $ X = (\cpop, \epop), (\bpop, \ipop), (\mpop, \ppop) $, where the populations in-between parenthesis are mutually exclusive.

We have quantified the statistical behaviour and evolution over time of all these populations by computing, snapshot-by-snapshot, their mean, median and $ 1 \sigma $ and $ 2 \sigma $ C.B., as well as fits to the PL expression of Eq.~\eqref{eq: Power-Law fit}. Note two things, however, in relation with such statistical quantities
\begin{enumerate}[label=\roman*)]
	\item First, that an outlier-exclusion algorithm has been applied to all data removed by more than $ 3 \sigma $ from the best-fit of the calculated empirical distributions, which in effect removes at any given time-snapshot $ \order {10} $ realisations from the computation of such statistical quantities, precisely those that exhibit very rare behaviour.
	\item Second, that a weak softening kernel has been applied to these quantities, in order to erase the short-lived, that is, single-snapshot, features that less rare but still atypical realisations have in the computation of such quantities, mainly in the form of irregularities of small amplitude and short wavelength that may render the corresponding time-evolved lines too wiggly.
	\item Third, that the kernel is weakest when applied to the total populations shown in the larger panels of Figure~\ref{fig: Remaining population}, since prior to softening the curves are already soft enough, and stronger when applied to the differential of populations shown in the smaller panels of same Figure~\ref{fig: Remaining population}, as those curves are, unsurprisingly, more wiggly.
\end{enumerate}

The effects of both the outlier-exclusion algorithm and the softening kernel are, in fact, very minor, to the point of being barely visible in the case of the total population plots. They are, however, more relevant in the case of the differential population plots as they portray better the right levels of these statistical quantities, albeit at the expense of the small cyclic jumps visible in the $ 1 \sigma $ and $ 2 \sigma $ C.B. of Figure~\ref{fig: Remaining population}, a boundary effect due to the ten-fold increase in the time slices time-step $ \Delta t_{r} $ across different runs $ r = 0, 1, 2, ..., 7 $.



\subsection{Cluster \& ejecta populations}
\label{subsec: Cluster and ejecta populations}


\begin{figure*}[t!]
	\centering
	{\subfloat[Remaining cluster population of objects, $ N_{\rpop, \cpop} (t) $.]{
	\hspace*{-0.00cm}
	\includegraphics[width = 0.50\textwidth]
	{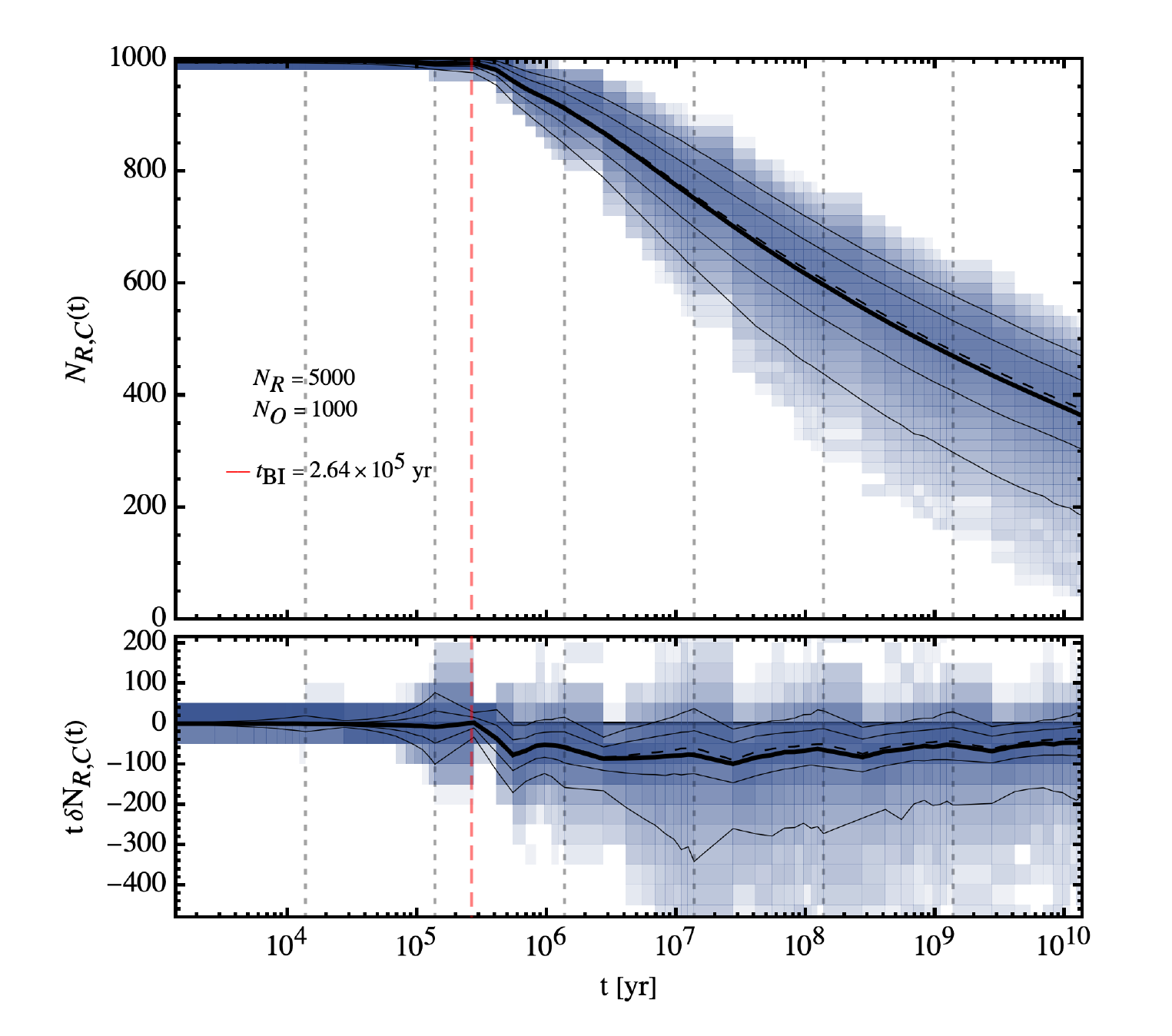}
	\label{subfig: Remaining population cluster}}
	\subfloat[Remaining ejecta population of objects, $ N_{\rpop, \epop} (t) $.]{
	\hspace*{-0.00cm}
	\includegraphics[width = 0.50\textwidth]
	{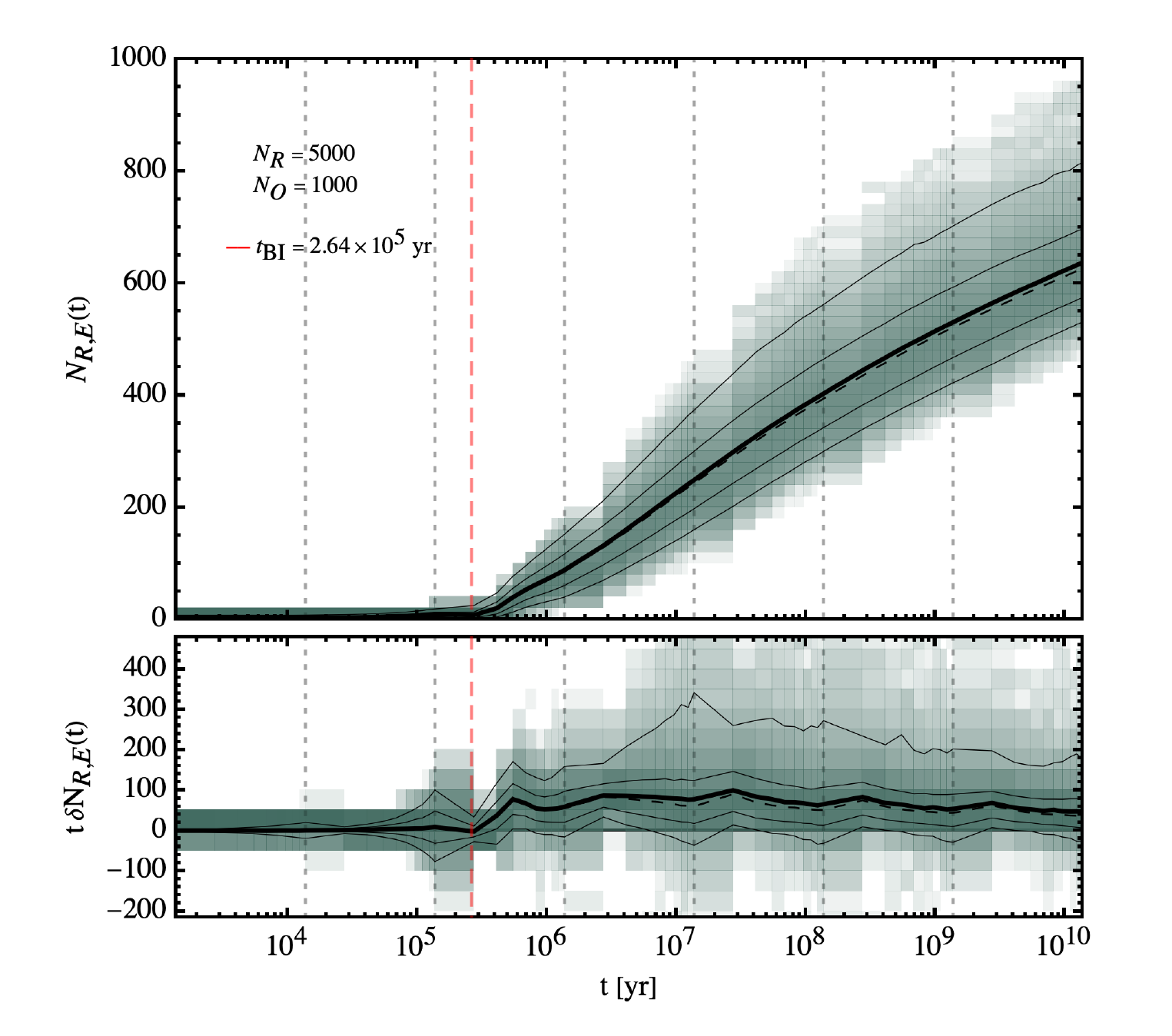}
	\label{subfig: Remaining population ejecta}}}
	\caption{
	Top: Total population of remaining cluster and ejecta objects in blue and teal respectively, $ N_{\rpop, X} (t) $ where $ X = \cpop, \epop $. The thick black line mark the mean of the discrete distribution of evolution trajectories per snapshot, while the thin black line in the same panel represents the median.
	Bottom: Differential of the total population of remaining cluster/ejecta objects, or loss/gain counts in between snapshots, rescaled by the simulation time, $ t \delta N_{\rpop, X_{2}} (t) $ where $ X = \cpop, \epop $. The thick black line in the lower panels mark the mean differential counts, while the thin black line represents the median differential counts and the dotted black line corresponds the $ B_{i} $ coefficient of the logarithmic decrease/increase, or alternatively, the mean differential counts themselves.
	All: The darker and lighter regions represent the 68\% and 95\% C.B. of each individual evolution of the cluster and ejecta populations for all $ N_{\rcap} = 5000 $ realisations. The dotted vertical grey lines separate the $ N_{r} = 7 $ consecutive runs, and the dotted vertical grey line signals the burn-in time.
	}
	\label{fig: Remaining population}
\end{figure*}

\begin{table*}[t!]
	\centering
	\begin{tabular}[c]{
	| l l |
	p{6 mm} p{7 mm} l l |
	p{6 mm} p{7 mm} l l |}
	\hline
	\quad & $ t_{\bicap} = \unit [2.64 \times 10^{5}] {yr} $ &
	\multicolumn{4}{c |}{Cluster population fits:} &
	\multicolumn{4}{c |}{Ejecta population fits:} \\
	[0.5ex]
	\hline\hline
	$ r $ & $ \unit {(t_{i^{-}}, t_{i^{+}})} [yr] $ &
	Fit: & $ r_{i,\rpop,\cpop}^{2} $ &
	$ \beta_{i,\rpop,\cpop} $ &
	$ \alpha_{i,\rpop,\cpop} $ &
	Fit: & $ r_{i,\rpop,\epop}^{2} $ &
	$ \beta_{i,\rpop,\epop} $ &
	$ \alpha_{i,\rpop,\epop} $ \\
	[0.5ex]
	\hline
	$ 3 $ & $ (\unit [t_{\bicap}] {yr^{-1}}, 1.30 \times 10^{6}) $ &
	[PL] & $ 1.000 $ &
	$ (1.982 \pm 0.049) \times 10^{3} $ & $ -0.0549 \pm 0.0018 $ &
	[PL] & $ 0.991 $ &
	$ (5.4 \pm 8.7) \times 10^{-5} $ & $ 1.00 \pm 0.12 $ \\
	$ 4 $ & $ 1.38 \times (10^{6}, 10^{7}) $ &
	[PL] & $ 1.000 $ &
	$ (3.037 \pm 0.085) \times 10^{3} $ & $ -0.0847 \pm 0.0018 $ &
	[PL] & $ 0.999 $ &
	$ 0.320 \pm 0.095 $ & $ 0.0410 \pm 0.018 $ \\
	$ 5 $ & $ 1.38 \times (10^{7}, 10^{8}) $ &
	[PL] & $ 1.000 $ &
	$ (3.8684 \pm 0.0088) \times 10^{3} $ & $ -0.09968 \pm 0.00013 $ &
	[PL] & $ 1.000 $ &
	$ 9.9 \pm 1.4 $ & $ 0.2000 \pm 0.0076 $ \\
	$ 6 $ & $ 1.38 \times (10^{8}, 10^{9}) $ &
	[PL] & $ 1.000 $ &
	$ (4.210 \pm 0.019) \times 10^{3} $ & $ -0.10418 \pm 0.00023 $ &
	[PL] & $ 1.000 $ &
	$ 46.0 \pm 2.9 $ & $ 0.1200 \pm 0.0031 $ \\
	$ 7 $ & $ 1.38 \times (10^{9}, 10^{10}) $ &
	[PL] & $ 1.000 $ &
	$ (4.874 \pm 0.082) \times 10^{3} $ & $ -0.11107 \pm 0.00075 $ &
	[PL] & $ 1.000 $ &
	$ 100.0 \pm 3.1 $ & $ 0.0780 \pm 0.0013 $ \\
	\hline
	\end{tabular}
	\caption{
	Mean remaining in-cluster and in-ejecta object population PL fits of Figure~\ref{fig: Remaining population} from Eq.~\eqref{eq: Power-Law fit}: $ N_\textrm{P} (t \in r) = \beta_{i,\textrm{P}} t^{\alpha_{i,\textrm{P}}} $. $ \alpha_{i,\textrm{P}} $ is the PL index, and corresponds roughly to the constant value of the differential counts, while $ \beta_{i,\textrm{P}} $ in turn, corresponds to the amplitude of the PL, $ r = 3, ..., 7 $ refers to the separate run where the fit is performed and $ \textrm{P} = (\rpop, \cpop), (\rpop, \epop) $ denotes the remaining cluster and ejecta populations respectively.
	}
	\label{tab: Remaining population}
\end{table*}
	
\begin{table*}[t!]
	\centering
	\begin{tabular}[c]{| l l | l l l l | l l | l l l l|}
	\hline
	\multicolumn{12}{| c |}{Remaining cluster and ejecta population fraction and median:} \\
	[0.5ex]
	\hline\hline
	$ i $ & $ \unit [t_{i}] {[yr]} $ &
	$ f_{i,\rpop,\cpop} $ &
	$ \tilde{f}_{i,\rpop,\cpop} $ &
	$ f_{i,\rpop,\epop} $ &
	$ \tilde{f}_{i,\rpop,\epop} $ &
	$ i $ & $ \unit [t_{i}] {[yr]} $ &
	$ f_{i,\rpop,\cpop} $ &
	$ \tilde{f}_{i,\rpop,\cpop} $ &
	$ f_{i,\rpop,\epop} $ &
	$ \tilde{f}_{i,\rpop,\epop} $ \\
	[0.5ex]
	\hline
	$ 0 $ & $ 0 $ &
	$ 0.9975 \substack {+0.0002 \\ -0.0025} $ &
	$ 1.000 $ &
	$ 0.0025 \substack {+0.0025 \\ -0.0002} $ &
	$ 0.000 $ &
	$ 37 $ & $ 1.38 \times 10^{7} $ &
	$ 0.76 \substack {+0.08 \\ -0.13} $ &
	$ 0.763 $ &
	$ 0.25 \substack {+0.12 \\ -0.09} $ &
	$ 0.237 $ \\
	$ 10 $ & $ 1.38 \times 10^{4} $ &
	$ 0.9950 \substack {+0.0015 \\ -0.0015} $ &
	$ 0.999 $ &
	$ 0.0050 \substack {+0.0015 \\ -0.0015} $ &
	$ 0.001 $ &
	$ 46 $ & $ 1.38 \times 10^{8} $ &
	$ 0.59 \substack {+0.11 \\ -0.16} $ &
	$ 0.595 $ &
	$ 0.40 \substack {+0.16 \\ -1.10} $ &
	$ 0.505 $ \\
	$ 19 $ & $ 1.38 \times 10^{5} $ &
	$ 0.990 \substack {+0.002 \\ -0.010} $ &
	$ 0.991 $ &
	$ 0.009 \substack {+0.011 \\ -0.004} $ &
	$ 0.009 $ &
	$ 55 $ & $ 1.38 \times 10^{9} $ &
	$ 0.47 \substack {+0.11 \\ -0.17} $ &
	$ 0.476 $ &
	$ 0.53 \substack {+0.17 \\ -0.11} $ &
	$ 0.524 $ \\
	$ 28 $ & $ 1.38 \times 10^{6} $ &
	$ 0.912 \substack {+0.051 \\ -0.072} $ &
	$ 0.910 $ &
	$ 0.092 \substack {+0.072 \\ -0.050} $ &
	$ 0.090 $ &
	$ 64 $ & $ 1.38 \times 10^{10} $ &
	$ 0.36 \substack {+0.11 \\ -0.18} $ &
	$ 0.368 $ &
	$ 0.63 \substack {+0.17 \\ -0.11} $ &
	$ 0.632 $ \\
	\hline
	\end{tabular}
	\caption{
	Mean remaining in-cluster and in-ejecta object population fractions $ \tilde{f}_{i,\rpop,\mathrm{P}} $ and median $ \tilde{f}_{i,\rpop,\mathrm{P}} $ where $ \mathrm{P} = (\cpop, \epop) $. The fraction is centred around the mean and shown are the 95\% C.B. and is computed along with the median by averaging over the $ N_{\rcap} = 5000 $ realisations at selected times.}
	\label{tab: Remaining population fraction and median}
\end{table*}


The simulated evolution of the cluster $ N_{\cpop} (t) $ and ejecta $ N_{\epop} (t) $ population (see Figure~\ref{subfig: Remaining population cluster} and Figure~\ref{subfig: Remaining population ejecta} respectively) shows, as it is expected, the gradual depopulation of the cluster and its loss of mass to the background, inter-cluster medium.

Note that the means, medians and confidence bands in said Figure~\ref{fig: Remaining population} are cleaned with both the outlier-exclusion and the softening kernel algorithms mentioned previously in Section~\ref{sec: Population statistics}, as otherwise they would show excessive variance in between time-slices due the very rare outlier $ \order {0.001} $ realisation in which the cluster becomes much more rapidly depopulated than usual, generally because of the presence of an initially very massive PBH in the IC that slingshots many of its less massive companions, which is an artefact produced by the relatively large tails of the Log-Normal distribution that is used to sample the initial masses in the IC, and which will occasionally produce this very large PBHs with masses of $ \unit [\order {0.001}] {\msun} $ in a few of the $ N_{\rcap} = 5000 $ realisations.

The population's evolution, showing an increase for the ejecta PBHs or decrease for the cluster objects has been fitted to the PL model of Eq.~\eqref{eq: Power-Law fit} and its results are shown in Table~\ref{tab: Remaining population}. The goodness of the fit is shown to be very good with less that $ \order {0.001} $ variance unexplained by the PL model as shown by the goodness of the fit coefficient.

Note that the fits are done separately in each of the runs, and are computed only for the five lasts runs starting the burn-in time of the simulation $ t_{\bicap} = \unit [2.64 \times 10^{5}] {yr} $ for the first of such runs. This is due to the fact that prior to that time the time-step of the simulation is too slow to capture the evolution within the time-runs as the characteristic time is during that period, as will be later computed and shown in Table~\ref{tab: Characteristic time scales} much greater than the time-step.

We provide as well the actual mean ($ 95\% $ C.L.) and median values of the fraction of cluster and ejecta objects within the simulations in Table~\ref{tab: Remaining population fraction and median}, at the eight time-slices that bound the seven time-runs. Note that the cluster and the ejecta populations are almost, but not quite, complementary to each other. This is because some PBHs merge in the process, and is the case indeed here with $ \dot{N}_{\rpop} (t) < 0 $, as will be seen in Section~\ref{subsubsec: Mass profiles evolution}. It will be seen later in Section~\ref{subsec: Primitive and merged populations} that the merging of PBH is so rare (thus the need for the high number of realisations to meaningfully capture it) that both fractions add up to unity.

In particular, we find in Figure~\ref{fig: Remaining population} a high degree of cluster evaporation as roughly two thirds of the initial cluster objects are dispersed out of the cluster. The actual fraction and median number of PBH that remain in the cluster or join the ejecta population given in Table~\ref{tab: Remaining population fraction and median}, show this trend indeed.

It will be later seen in Figure~\ref{subsubsec: Mass profiles evolution} that the mass profiles of both the cluster and the ejected PBHs do not differ significantly throughout the simulations, meaning that both populations can be described at all times by the initial mass distribution, and so the mass fraction that is dispersed by the cluster is coinciding with that of the number of objects.



\subsubsection{Cluster evaporation \& virialisation}
\label{subsubsec: Cluster evaporation and virialisation}

The phenomenon by which a group of gravitating bodies loses mass to the background, be it in stellar globular clusters or PBH clusters, is known as cluster ``evaporation''. In our particular case, starting at the burn-in time onwards , the evaporation rate is quite stable throughout the simulations. Our simulated PBH clusters are consistent with that fact, and indeed show up an evaporation rate roughly constant throughout the whole quasi-static phase, as it follows by the fact that the estimated confidence intervals of the $ \beta_{i} $ coefficient of the PL expression of Eq.~\eqref{eq: Power-Law fit}, shown in Table~\ref{tab: Remaining population}, are all quite close to each other, even if they quite not overlap due to the narrowness of the interval.

We find that, for the cluster, while the totality of PBHs are collectively bounded to the cluster at the initial times $ t < t_{BI, \rpop} = \unit [4.92 \times 10^{5}] {yr} $, already in the third run the cluster starts to loose mass, doing so at a practically constant rate in proportion to the remaining mass, as it is shown in Figure~\ref{subfig: Remaining population cluster}.

As for the ejecta, we find that the reverse is true: at the initial times $ t < t_{BI, \epop} = \unit [3.00 \times 10^{5}] {yr} $ no object has acquired a speed high enough to escape the gravitational pull of the cluster in order to escape the cluster potential well. Only in the third run the background starts to be populated with ejected PBHs, doing so again at a practically constant rate in proportion to the remaining mass.

By the end of our simulations, the PBH cluster population has been reduced to fraction of objects
\begin{equation}
	f^{\loss}_{\cpop,0} = \frac {| m_{\cpop, 64}-m_{\cpop, 0} |} {m_{\cpop, \loss, 64}} = 0.36 \substack {+0.11 \\ -0.18}.
\end{equation}
Conversely, the PBH ejecta population has gained throughout the simulations a fraction of objects
\begin{equation}
	f^{\gain}_{\epop,64} = \frac {| m_{\epop, 64}-m_{\cpop, 0} |} {m_{\epop, \loss,0}} = 0.36 \ \substack {+0.11 \\ -0.18}.
\end{equation}

Note that both quantities being almost complementary to each other due to the lack of significant merging. Also note that, for the reasons shown in Section~\ref{subsubsec: Mass profiles evolution}, these fractions are good approximations to the fraction of mass, since the cluster and ejecta population don't develop differentiated mass profiles.

Cluster evaporation is apparent in Figure~\ref{fig: Remaining population} starting from the burn-in time of $ t_{\bicap} = \unit [2.64 \times 10^{5}] {yr} $, a time at which the characteristic dynamical time of the cluster is comparable to the time-step in between the snapshots, and when the evolution is fast enough to be captured by the time-step. From that time onwards, then, the cluster proceeds to expel PBHs at a declining rate $ \delta N_{\rpop, \cpop} \approx-N_{\rpop, \epop} < 0 $, roughly given by
\begin{equation}
	t \delta N_{\rpop, \cpop} (t) = -t \delta N_{\rpop, \epop} (t) = -60 \substack {+65 \\ -145}.
\end{equation}

The cluster and ejecta populations as a function of time are then well approximated by PLs, declining in the former case and growing in the latter case, whose fits are shown, time by time interval, in Table~\ref{tab: Remaining population}. It can be checked there that the fits are indeed very accurate with measures of the goodness of these with $ r^2 > 0.9 $ even right after the burn-in time.

Note that these PBH clusters are originated in the radiation era, and while they form this lasts, they do not, then, evaporate at all, their characteristic dynamical time being comparable to the total radiation domination era duration, and only right after cluster evaporation begins to be significant, therefore rendering this process irrelevant to cluster formation at least when the clusters are originated with parsec-sized characteristic length. Later evolution is then consistent with increasingly a majority of the PBH mass contained in diffuse haloes surrounding the proper PBH clusters.

It is shown in Figure~\ref{subfig: Time positions histogram} that ejecta haloes reach at present sizes of
\begin{equation}
	\bar{x}_{\epop,64} \in \unit [(7.0 \times 10^{3}, 3.5 \times 10^{5})] {pc},
\end{equation}
while cluster haloes reach at present sizes of
\begin{equation}
	\bar{x}_{\cpop,64} \in \unit [(80, 2.3 \times 10^{3})] {pc}.
\end{equation}

Also shown in Figure~\ref{subfig: Time positions histogram} is the characteristic PBH ejecta halo typical velocities in the present to be in the range
\begin{equation}
	\bar{v}_{\epop,64} \in \unit [(0.49, 23)] {km / s},
\end{equation}
while the later cluster haloes reach typical velocities of
\begin{equation}
	\bar{v}_{\cpop,64} \in \unit [(0.030, 0.51)] {km / s}.
\end{equation}

These are indeed very interesting results, as one finds that ejecta haloes are comparable in size to typical galactic haloes, while cluster haloes reach sizes are comparable in size with dwarf galaxies and stellar globular cluster sizes, perhaps suggesting a link between the two.



\subsection{Isolated \& Bounded populations}
\label{subsec: Isolated and Bounded populations}


\begin{figure*}[t!]
	\centering
	%
	%
	{\subfloat[Bounded cluster population of objects, $ N_{\bpop, \cpop} (t) $.]{
	\hspace*{-0.00cm}
	\includegraphics[width = 0.50\textwidth]
	{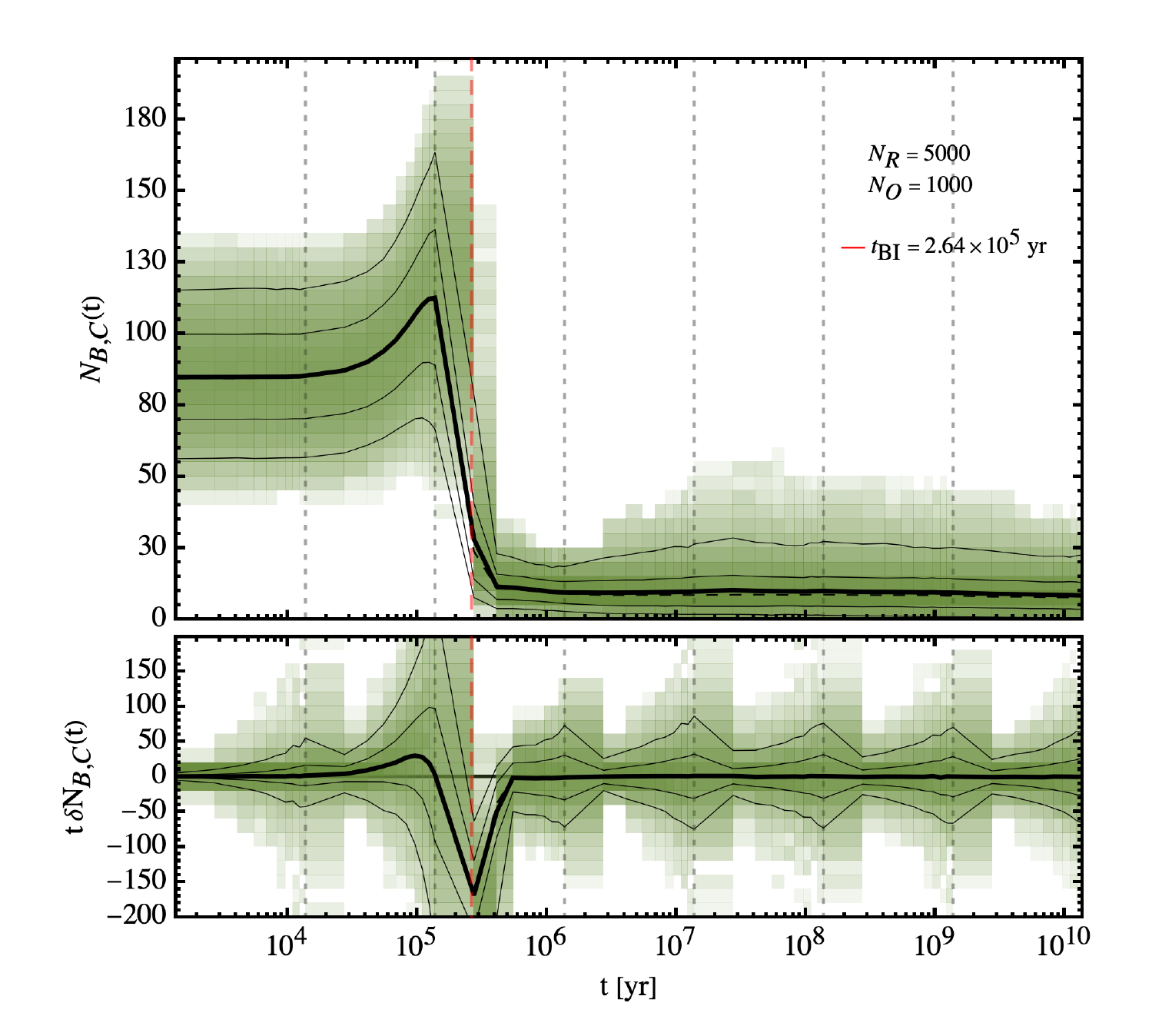}
	\label{subfig: Bounded population cluster}}
	\subfloat[Bounded ejecta population of objects, $ N_{\bpop, \epop} (t) $.]{
	\hspace*{-0.00cm}
	\includegraphics[width = 0.50\textwidth]
	{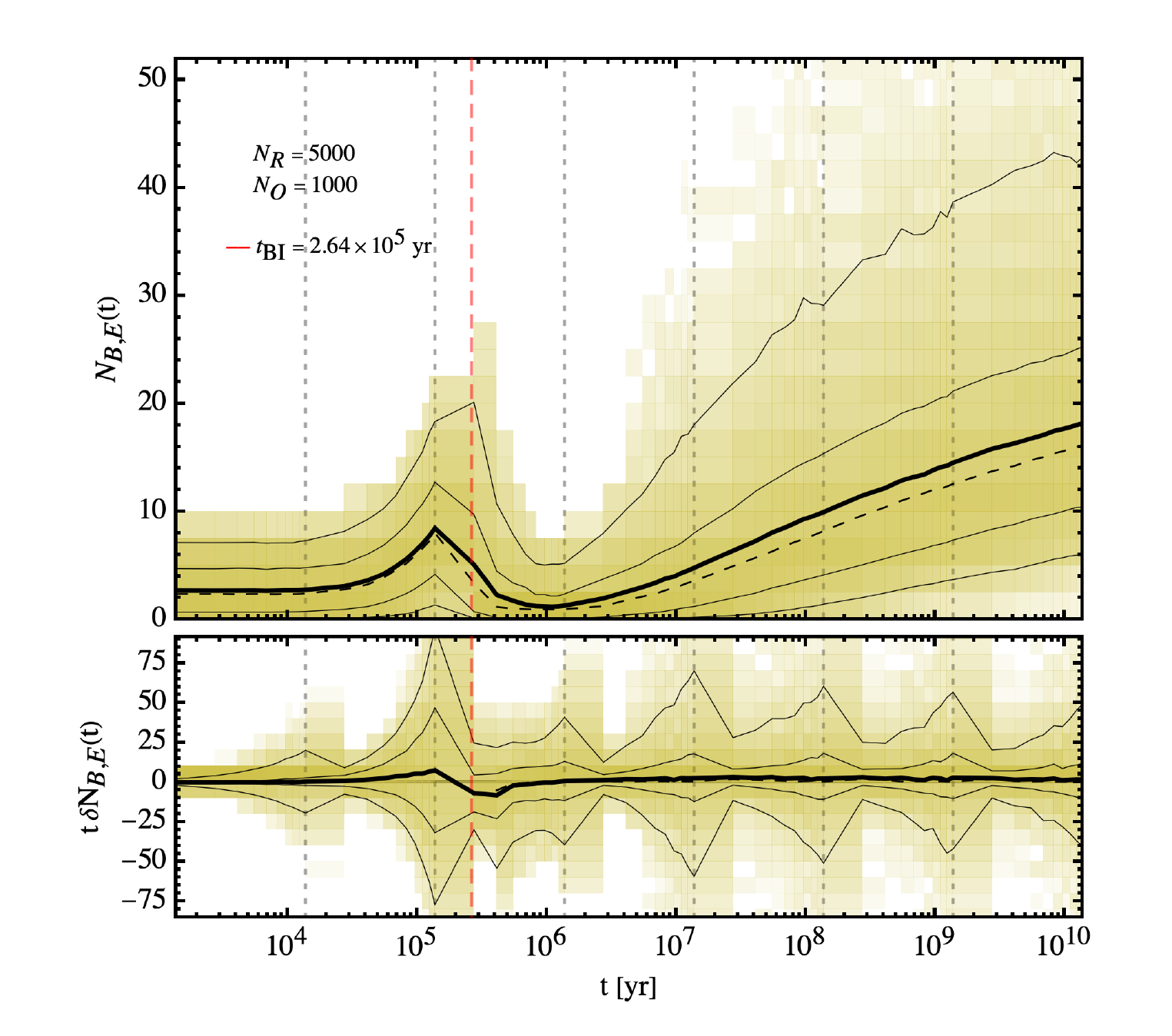}
	\label{subfig: Bounded population ejecta}}}
	\caption{
	Top: Total population of bounded in-cluster and in-ejecta objects in green and yellow, $ N_{\bpop, X} (t) $ where $ X = \cpop, \epop $. The thick black line mark to the mean of the discrete distribution of evolution trajectories per snapshot, while the thin black line represents the median.
	Bottom: Differential of the total population of bounded in-cluster/in-ejecta objects, or loss/gain counts in between snapshots, rescaled by the simulation time, $ t \delta N_{\bpop, X} (t) $ where $ X = \cpop, \epop $. The thick black line marks the mean differential counts, while the thin black line represents the median differential counts. The dotted black line corresponds the $ B_{i} $ coefficient of the logarithmic decrease/increase, or alternatively, the mean differential counts themselves.
	All: The darker and lighter regions represent the 68\% and 95\% C.B. of each individual evolution of the cluster and ejecta populations for all $ N_{\rcap} = 5000 $ realisations. The dotted vertical grey lines separate the $ N_{r} = 7 $ consecutive runs, and the dotted vertical grey line signals the burn-in time. The population's evolution is fitted to PLs in Table~\ref{tab: Bounded population}.
	}
	\label{fig: Bounded population}
\end{figure*}

\begin{table*}[t!]
	\centering
	\begin{tabular}[c]{
	| l l |
	p{6 mm} p{7 mm} l l |
	p{6 mm} p{7 mm} l l |}
	\hline
	\quad & $ t_{\bicap} = \unit [2.64 \times 10^{5}] {yr} $ &
	\multicolumn{4}{c |}{Bounded in-cluster population fits:} &
	\multicolumn{4}{c |}{Bounded in-ejecta population fits:} \\
	[0.5ex]
	\hline\hline
	$ r $ & $ \unit {(t_{i^{-}}, t_{i^{+}})} [yr] $ &
	Fit: & $ r_{i,\bpop,\cpop}^{2} $ &
	$ \beta_{i,\bpop,\cpop} $ &
	$ \alpha_{i,\bpop,\cpop} $ &
	Fit: & $ r_{i,\bpop,\epop}^{2} $ &
	$ \beta_{i,\bpop,\epop} $ &
	$ \alpha_{i,\bpop,\epop} $ \\
	[0.5ex]
	\hline
	$ 3 $ & $ (\unit [t_{\bicap}] {yr^{-1}}, 1.30 \times 10^{6}) $ &
	[PL] & $ 0.953 $ &
	$ (4.4 \pm 9.4) \times 10^{5} $ & $ -0.78 \pm 0.16 $ &
	[PL] & $ 0.969 $ &
	$ (1.6 \pm 3.1) \times 10^{7} $ & $ -1.20 \pm 0.15 $ \\
	$ 4 $ & $ 1.38 \times (10^{6}, 10^{7}) $ &
	[PL] & $ 1.000 $ &
	$ 7.00 \pm 0.47 $ & $ 0.0200 \pm 0.0043 $ &
	[PL] & $ 1.000 $ &
	$ (5.9 \pm 1.5) \times 10^{-4} $ & $ 0.550 \pm 0.16 $ \\
	$ 5 $ & $ 1.38 \times (10^{7}, 10^{8}) $ &
	[PL] & $ 1.000 $ &
	$ 9.80 \pm 0.81 $ & $ (1.0 \pm 4.6) \times 10^{-3} $ &
	[PL] & $ 0.999 $ &
	$ 0.044 \pm 0.012 $ & $ 0.290 \pm 0.015 $ \\
	$ 6 $ & $ 1.38 \times (10^{8}, 10^{9}) $ &
	[PL] & $ 1.000 $ &
	$ 16.00 \pm 0.84 $ & $ -0.0260 \pm 0.0026 $ &
	[PL] & $ 1.000 $ &
	$ 0.520 \pm 0.057 $ & $ 0.1600 \pm 0.0053 $ \\
	$ 7 $ & $ 1.38 \times (10^{9}, 10^{10}) $ &
	[PL] & $ 1.000 $ &
	$ 24.0 \pm 1.4 $ & $ -0.0450 \pm 0.0026 $ &
	[PL] & $ 1.000 $ &
	$ 2.20 \pm 0.11 $ & $ 0.0920 \pm 0.0023 $ \\
	\hline
	\end{tabular}
	\caption{
	Mean bounded in-cluster and in-ejecta object population PL fits of Figure~\ref{fig: Bounded population} from Eq.~\eqref{eq: Power-Law fit}: $ N_\textrm{P} (t \in r) = \beta_{i,\textrm{P}} t^{\alpha_{i,\textrm{P}}} $. $ \alpha_{i,\textrm{P}} $ is the PL index, again corresponding to the constant level of the differential counts, while $ \beta_{i,\textrm{P}} $, corresponds to the amplitude of the PL, $ r = 3, ..., 7 $ refers to the separate run where the fit is performed and $ \textrm{P} = (\rpop, \cpop), (\rpop, \epop) $ denotes the bounded cluster and ejecta populations respectively.
	}
	\label{tab: Bounded population}
\end{table*}

\begin{table*}[t!]
	\centering
	\begin{tabular}[c]{| l l | l l l l | l l | l l l l|}
	\hline
	\multicolumn{12}{| c |}{Bounded in-cluster and in-ejecta population fraction and median:} \\
	[0.5ex]
	\hline\hline
	$ i $ & $ \unit [t_{i}] {[yr]} $ &
	$ f_{i,\rpop,\cpop} $ &
	$ \tilde{f}_{i,\rpop,\cpop} $ &
	$ f_{i,\rpop,\epop} $ &
	$ \tilde{f}_{i,\rpop,\epop} $ &
	$ i $ & $ \unit [t_{i}] {[yr]} $ &
	$ f_{i,\rpop,\cpop} $ &
	$ \tilde{f}_{i,\rpop,\cpop} $ &
	$ f_{i,\rpop,\epop} $ &
	$ \tilde{f}_{i,\rpop,\epop} $ \\
	[0.5ex]
	\hline
	$ 0 $ & $ 0 $ &
	$ 0.086 \substack {+0.032 \\ -0.030} $ &
	$ 0.085 $ &
	$ 0.0026 \substack {+0.0040 \\ -0.0025} $ &
	$ 0.002 $ &
	$ 37 $ & $ 1.38 \times 10^{7} $ &
	$ 0.009 \substack {+0.017 \\ -0.009} $ &
	$ 0.008 $ &
	$ 0.005 \substack {+0.013 \\ -0.005} $ &
	$ 0.004 $ \\
	$ 10 $ & $ 1.38 \times 10^{4} $ &
	$ 0.087 \substack {+0.032 \\ -0.030} $ &
	$ 0.086 $ &
	$ 0.0027 \substack {+0.004 \\ -0.0026} $ &
	$ 0.002 $ &
	$ 46 $ & $ 1.38 \times 10^{8} $ &
	$ 0.010 \substack {+0.017 \\ -0.009} $ &
	$ 0.008 $ &
	$ 0.010 \substack {+0.019 \\ -0.009} $ &
	$ 0.008 $ \\
	$ 19 $ & $ 1.38 \times 10^{5} $ &
	$ 0.123 \substack {+0.051 \\ -0.052} $ &
	$ 0.121 $ &
	$ 0.008 \substack {+0.010 \\ -0.008} $ &
	$ 0.007 $ &
	$ 55 $ & $ 1.38 \times 10^{9} $ &
	$ 0.009 \substack {+0.016 \\ -0.008} $ &
	$ 0.008 $ &
	$ 0.014 \substack {+0.024 \\ -0.011} $ &
	$ 0.0012 $ \\
	$ 28 $ & $ 1.38 \times 10^{6} $ &
	$ 0.008 \substack {+0.010 \\ -0.007} $ &
	$ 0.007 $ &
	$ 0.0012 \substack {+0.0044 \\ -0.0012} $ &
	$ 0.001 $ &
	$ 64 $ & $ 1.38 \times 10^{10} $ &
	$ 0.008 \substack {+0.014 \\ -0.007} $ &
	$ 0.007 $ &
	$ 0.018 \substack {+0.024 \\ -0.012} $ &
	$ 0.0016 $ \\
	\hline
	\end{tabular}
	\caption{
	Mean bounded in-cluster and in-ejecta object population fractions $ \tilde{f}_{i,\bpop,\mathrm{P}} $ and median $ \tilde{f}_{i,\bpop,\mathrm{P}} $ where $ \mathrm{P} = (\cpop, \epop) $. The fraction is centred around the mean and shown are the 95\% C.B. and is computed along with the median by averaging over the $ N_{\rcap} = 5000 $ realisations at selected times.
	}
	\label{tab: Bounded population fraction and median}
\end{table*}


A small portion of the PBHs in the simulations will end up in bounded systems, in which a larger object captures a smaller object and form a binary, as shown in Figure~\ref{subfig: Bounded population cluster} and Figure~\ref{subfig: Bounded population ejecta}. These can be classified into two categories: transient, when the binaries last less that the simulation time, and stable, when they do last for more that the simulation time.

Note again that the means, medians and confidence bands in Figure~\ref{fig: Bounded population} are cleaned with both the outlier-exclusion and the softening kernel algorithms mentioned previously in Section~\ref{sec: Population statistics} in order to remove the noisier data, particularly in this case at the burn-in time crossing where the number of in-cluster binaries spikes to only to later be quickly reduced to very low levels as the IC is erased.

The population's evolution is fitted to PLs of Eq.~\eqref{eq: Power-Law fit} whose results are shown in Table~\ref{tab: Bounded population}. The fit is performed run by run, and for similar reasons as we had with the cluster and ejecta population, no fit is made before the burn-in time of the simulation $ t_{\bicap} = \unit [2.64 \times 10^{5}] {yr} $ as the evolution is too slow to be captured by the time-step in the time-steps prior to it, or, in other words, the characteristic dynamical time the simulation prior to the burn-in time is, as will be later computed and shown in Table~\ref{tab: Characteristic time scales} much greater than the time-step at these time-slices.

Figure~\ref{subfig: Bounded population cluster} shows that while a respectably high fraction of PBHs appear in bounded systems initially in the cluster, the fraction quickly drops and stabilises at a low level of $ \order {0.001} $ relative to the total number of PBHs.

As for the in-ejecta bounded PBHs, Figure~\ref{subfig: Bounded population ejecta} shows as it would be expected that their number consistently grows from almost none at all, as there is no ejecta population whatsoever in the initial stages of the simulations, to $ \order {0.01} $ at present relative to the total number of PBHs. The actual fraction and median number of bounded PBH that remain in the cluster or join the ejecta population are given in Table~\ref{tab: Bounded population fraction and median}.

Initially, the vast majority of the binaries are bound to be contained within the cluster, which contains a fraction of bounded objects of
\begin{equation}
	f_{\cpop, \bpop, 0} = 0.086 \substack {+0.0032 \\ -0.030},
\end{equation}
as there has not yet elapsed enough time to evaporate a significant fraction of objects to the ejecta, which contains a fraction of bounded objects of:
\begin{equation}
	f_{\cpop, \epop, 0} = 0.0026 \substack {+0.0040 \\ -0.0025},
\end{equation}
both at the initial simulation time of $ t_{0} = \unit [0] {yr} $, and with respect to the total population.

Note that the large fraction of bounded cluster objects contained is an artefact of the IC, and quickly disappears after the burn-in time, which is natural considering that around this time the cluster undergoes a transformation, moving from the initial Multi-Normal position distribution of Figure~\ref{fig: IC distributions} to a cuspier one as will later be seen in \ref{subfig: Quasi-static cluster density profiles}.

As a result from this, the density is increased at the PBH cluster core at the same time that the cluster slightly contracts about 5\%-10\% as seen in Figure~\ref{subfig: Time positions histogram}. As the PBHs fall to this dense environment they quickly acquire speeds as seen in Figure~\ref{subfig: Time velocities histogram} that are able to disrupt the majority of bounded systems and transition from a Gaussian to a PL density profile.

After the transition and while in the cluster region, such binaries are typically very short lived, as they interact with neighbours multiple times in each run and so they are still disrupted easily. As a consequence of this, the population of cluster binaries is small overall and remains roughly constant, as the generation rate of binaries is matched by their disruption rate at any time-slice $ t > t_{\bicap} $.

The fraction of cluster binaries with respect to the total population varies then from an absolute minimum with respect to the total population of
\begin{equation}
	f_{\cpop, \bpop,28} = \frac {N_{\cpop, \bpop, 2}} {N_{\rpop, 0}} = 0.008 \substack {+0.010 \\ -0.007},
\end{equation}
at $ t = \unit [1.38 \times 10^{6}] {yr} $ to a present relative maximum with respect to the total population of:
\begin{equation}
	f_{\cpop, \bpop,64} = \frac {N_{\cpop, \bpop, 64}} {N_{\rpop, 0}} = 0.008 \substack {+0.014 \\ -0.007},
\end{equation}
at $ t = \unit [1.38 \times 10^{10}] {yr} $, as seen in Figure~\ref{subfig: Bounded population cluster}. Note that, as the cluster itself looses population, while the cluster bounded population remains nearly constant, then the fraction of PBHs contained in binaries within the cluster grows at a rate inverse to the evaporation rate, more that doubling from the burn-in time to present.

However, when such a binary is ejected from the cluster then it very quickly ceases to be perturbed by nearest neighbour encounters and then, in the code classic treatment, its constituents remain in stable Keplerian orbits indefinitely, as seen in Figure~\ref{subfig: Bounded population ejecta}.

It follows from this fact that the population of binaries in the ejected background naturally increases overtime, eventually overcoming that of the population cluster binaries already at $ t \approx \unit [\order {10^{8}}] {yr} $ and constituting at late times a non negligible fraction of all ejecta objects, which in our simulations is found to be
\begin{equation}
	f_{\epop, \bpop,28} = \frac {N_{\epop, \bpop, 28}} {N_{\rpop, 0}} = 0.0012 \substack {+0.0044 \\ -0.0012},
\end{equation}
at $ t = \unit [1.38 \times 10^{6}] {yr} $ to a present relative maximum of:
\begin{equation}
	f_{\epop, \bpop,64} = \frac {N_{\epop, \bpop, 64}} {N_{\rpop, 0}} = 0.018 \substack {+0.014 \\ -0.012},
\end{equation}
at $ t = \unit [1.38 \times 10^{10}] {yr} $, as seen in Figure~\ref{subfig: Bounded population ejecta}.

Last, a more detailed analysis of binary pairs is given in Section~\ref{subsubsec: hyperbolic encounter rates}, where we extract the mass profile of bounded PBHs and their population characteristics.



\subsubsection{Simulation parent trees}
\label{subsubsec: Simulation parent trees}


\begin{table*}[t!]
	\centering
	\begin{tabular}[c]{| p{6mm} p{18mm} | p{18mm} p{18mm} p{18mm} | p{6mm} p{18mm} | p{18mm} p{18mm} p{18mm} |}
	\hline
	\multicolumn{10}{| c |}{Parent tree hierarchy levels:} \\
	[0.5ex]
	\hline\hline
	$ i $ & $ \unit [t_{i}] {[yr]} $ &
	$ l_{\bpop} = 2 $ &
	$ l_{\bpop} = 3 $ &
	$ l_{\bpop} = 4 $ &
	$ i $ & $ \unit [t_{i}] {[yr]} $ &
	$ l_{\bpop} = 2 $ &
	$ l_{\bpop} = 3 $ &
	$ l_{\bpop} = 4 $ \\
	[0.5ex]
	\hline
	0 & $ 0 $ &
	$ 8.07 \times 0.01 $ &
	$ 9.88 \times 10^{-4} $ &
	$ 2.14 \times 10^{-6} $ &
	37 & $ 1.38 \times 10^{7} $ &
	$ 1.13 \times 0.01 $ &
	$ 4.25 \times 10^{-5} $ &
	$ 0.00 $ \\
	10 & $ 1.38 \times 10^{4} $ &
	$ 8.07 \times 0.01 $ &
	$ 9.88 \times 10^{-4} $ &
	$ 1.98 \times 10^{-6} $ &
	46 & $ 1.38 \times 10^{8} $ &
	$ 1.67 \times 0.01 $ &
	$ 1.49 \times 10^{-4} $ &
	$ 5.63 \times 10^{-7} $ \\
	19 & $ 1.38 \times 10^{5} $ &
	$ 9.45 \times 0.01 $ &
	$ 1.06 \times 10^{-3} $ &
	$ 1.56 \times 10^{-6} $ &
	55 & $ 1.38 \times 10^{9} $ &
	$ 2.29 \times 0.01 $ &
	$ 3.28 \times 10^{-4} $ &
	$ 2.93 \times 10^{-6} $ \\
	28 & $ 1.38 \times 10^{6} $ &
	$ 1.10 \times 0.1 $ &
	$ 3.21 \times 10^{-3} $ &
	$ 2.65 \times 10^{-5} $ &
	64 & $ 1.38 \times 10^{10} $ &
	$ 2.90 \times 0.01 $ &
	$ 6.16 \times 10^{-4} $ &
	$ 6.45 \times 10^{-6} $ \\
	\hline
	\end{tabular}
	\caption{
	Fraction of binaries at each of the eight time-slices dividing the seven runs. Each level $ l_{\bpop} $ indicates the level in the binary hierarchy, with $ l_{\bpop} = 2 $ indicating two bound PBHs in a binary system, $ l_{\bpop} = 3 $ indicating a binary system within another binary system, and $ l_{\bpop} = 4 $ indicating a binary system within another two binary systems. Note that there are $ N_{r} = 7 $ time-runs, for a total of eight bounding time-slices, corresponding to the times shown in Table~\ref{tab: Simulation time-runs}.
	}
	\label{tab: Parent tree hierarchy levels}
\end{table*}


While it is true that the majority of PBHs do not constitute a part of a binary system at any given time, a fraction of PBHs that form part of such systems has been found to lie in the interval
\begin{equation}
	f_{\bpop,28} = \frac {N_{\bpop, 28}} {N_{\rpop, 0}} = 0.010 \substack {+0.010 \\ -0.007},
\end{equation}
after recombination at $ t = \unit [1.38 \times 10^{6}] {yr} $ and
\begin{equation}
	f_{\bpop,64} = \frac {N_{\bpop, 64}} {N_{\rpop, 0}} = 0.025 \substack {+0.028 \\ -0.014},
\end{equation}
at the present time $ t = \unit [1.38 \times 10^{10}] {yr} $, with $ f_{\bpop} = f_{\bpop, \cpop}+f_{\bpop, \epop} $, as previously explained in Section~\ref{subsec: Isolated and Bounded populations}.

However, we have found as well a rich structure of subsequent binary sub-systems within multiple PBH systems. It has been found using the methods from Section~\ref{subsubsec: Parent trees computation} that a hierarchical substructure of binaries arises in the simulation IC and is maintained throughout the whole evolution period, the results of which can be seen in Table~\ref{tab: Parent tree hierarchy levels}.
	
Generally speaking, we find that a majority of PBHs do not participate in any binary system. However, of those which take part in such systems, it has been found that
\begin{enumerate}[label=\roman*)]
	\item A 99\%-90\% majority of those belong to binary ($ 2 $-component) systems in which two PBHs, one larger and one smaller, orbit each other, which we label on the binary hierarchy $ 0^{\textrm{th}} $ level and $ 1^{\textrm{st}} $ level respectively.
	\item Another 10\%-1\% of bounded objects are present in tertiary ($ 3 $-component) systems, in which the least massive PBH ($ 2^{\textrm{nd}} $ level) orbits the intermediate PBH ($ 1^{\textrm{st}} $ level), which itself orbits the most massive PBH ($ 0^{\textrm{th}} $ level), amounting to a second level in the hierarchy.
	\item Another 0.11\%-0.01\% of bounded objects are present in quaternary ($ 4 $-component) systems, adding an extra third level in the hierarchy.
	\item Last, 0.001\%-0.0001\% of bounded objects are part of quinary $ 5 $-component) systems, adding a final forth level in the parent tree.
\end{enumerate}
Thus, it is found that each subsequent level in the hierarchy is populated by roughly a hundredth the number of PBHs than the level immediately above. Note as well that, as shown in Table~\ref{tab: Parent tree hierarchy levels}, the population in each level is less stable over time the deeper the level in the hierarchy.



\subsection{Primitive \& merged populations}
\label{subsec: Primitive and merged populations}

Last, an even smaller portion of the PBHs in the simulations will, by the end of the simulation period, have merged with another and form larger PBHs. We have found in our simulation $ (N_{\rpop})_{\mpop} = 117 $ such mergers in $ N_{\rpop} = 5000 $ realisations, leading to a total fraction of merged objects of $ f (N_{\rpop})_{\mpop} = 2.34 \times 10^{-5} $. Out of these,
\begin{enumerate}[label=\roman*)]
	\item $ (N_{\rpop, \cpop, \ipop})_{\mpop} = 50 $ such mergers occur in cluster, isolated objects, leading to a sub-fraction of merged objects throughout all simulations at the last time-slice of:
	\begin{equation}
		f (N_{\rpop, \cpop, \ipop})_{\mpop} = 1.00 \times 10^{-5}.
	\end{equation}
	\item $ (N_{\rpop, \cpop, \bpop})_{\mpop} = 4 $ such mergers occur in cluster, bounded objects leading to a total sub-fraction of merged objects of:
	\begin{equation}
		f (N_{\rpop, \cpop, \bpop})_{\mpop} = 8.00 \times 10^{-7}.
	\end{equation}
	\item $ (N_{\rpop, \epop, \ipop})_{\mpop} = 1 $ such mergers occur in ejecta, isolated objects leading to a total sub-fraction of merged objects of:
	\begin{equation}
		f (N_{\rpop, \epop, \ipop})_{\mpop} = 2.00 \times 10^{-7}.
	\end{equation}
	\item Last, $ (N_{\rpop, \epop, \bpop})_{\mpop} = 62 $ such mergers occur in ejecta, bounded objects leading to a total sub-fraction of merged objects of:
	\begin{equation}
		f (N_{\rpop, \epop, \bpop})_{\mpop} = 1.24 \times 10^{-5}.
	\end{equation}
\end{enumerate}

These results, summarised in Table~\ref{tab: Merger tree hierarchy levels}, show that the large majority of merged PBHs belong to two populations of objects, mainly isolated, in-cluster PBHs that constitute about 43\% of mergers and bounded, in-ejecta PBHs which constitute about 53\% of mergers.

Indeed, the latter case of bounded, in-ejecta PBHs can be attributed to the fact that the PBHs that partake in the mergers are almost exclusively counted among the most massive in the simulations, as shown in Table~\ref{tab: Merger mass distributions}, and as the PBHs mutually approach each other within the cluster, they acquire large infall velocities as they fall in their absorber's potential well, that can overcome the cluster escape velocity and be expelled to the background. The former case of isolated, in-cluster PBHs, however, has a simpler explanation, and is merely due to the fact that cluster objects, by definition, live in a denser environment where collision is far more likely.

Even though the numbers are small, of the remaining cases, bounded, in-cluster PBHs merely constitute about 4\% of PBH mergers, while isolated, in-ejecta objects which constitute a very minor 1\% of PBH mergers. The cause of this can be attributed to the environment density where the mergers do take place.

In the former case, the merger takes place between bounded objects within the cluster sphere-of-influence, and their low number count can be attributed to the fact that the bounded in-cluster population is already quite small at any given time, even though it can be observed that events of this kind are enhanced with respect to the mergers arising from isolated in-cluster objects, as shown by the quotient of merger counts being larger than the population quotient at the last time-slice at present
\begin{equation}
	\frac {(N_{\rpop, \cpop, \bpop})_{\mpop}} {(N_{\rpop, \cpop, \ipop})_{\mpop}} = 0.080 > 0.025 = \frac {f_{\cpop, \bpop}} {f_{\cpop, \ipop}} \approx \frac {f_{\cpop, \bpop}} {1-f_{\cpop, \bpop}}.
\end{equation}

However, in the latter case, the merger takes place just within the periphery of the cluster with massive, high-velocity PBH, just evaporated and fast enough to overcome the cluster escape velocity so it is considered as part of the ejecta population even if its least massive merger pair still is part of the cluster, and so it cannot be consider a proper ejecta merger, consistent with the fact that, given the radial profile of ejecta velocities and the very low ejecta density, the chances of such mergers from objects incoming from the same single cluster are vanishingly small.

Be reminded that the number of PBH mergers that we find in our simulated cluster underestimates the amount of actual mergers that would be produced due to the non-relativistic behaviour of the code. Given the typically low object velocities shown in Figure~\ref{subfig: Time velocities histogram}, it would expected first that many of the identified hyperbolic encounters and binary captures in our simulations would produce a binary capture or inspiral and merger in a relativistic code, and as such, the numbers provided in this section must be understood as lower bound to the actual merger amount.

A more detailed analysis of these merger events will given in Section~\ref{subsubsec: Merger event rates}, where we compute the mass profile of merging PBHs and their population characteristics.

\begin{table}[t!]
	\centering
	\begin{tabular}[c]{| p{9mm} p{16mm} p{16mm}| p{16mm} p{16mm} |}
	\hline
	\multicolumn{5}{| c |}{Merger tree hierarchy levels:} \\
	[0.5ex]
	\hline\hline
	$ r $ &
	$ \unit [t_{i^{-}}] {[yr]} $ &
	$ \unit [t_{i^{+}}] {[yr]} $ &
	$ l_{\mpop} = 2 $ &
	$ l_{\mpop} = 3 $ \\
	[0.5ex]
	\hline
	$ 1-5 $ & $ 1.38 \times 10^{3} $ & $ 1.38 \times 10^{9} $ &
	$ 0.00 $ &
	$ 0.00 $ \\
	$ 6 $ & $ 1.38 \times 10^{8} $ & $ 1.38 \times 10^{9} $ &
	$ 6.00 \times 10^{-7} $ &
	$ 0.00 $ \\
	$ 7 $ & $ 1.38 \times 10^{9} $ & $ 1.38 \times 10^{10} $ &
	$ 2.24 \times 10^{-5} $ &
	$ 4.00 \times 10^{-7} $ \\
	\hline
	\end{tabular}
	\caption{
	Fraction of mergers grouped from the time-slices dividing the first five time-runs. Each level $ l_{\mpop} $ indicates the level in the merger tree hierarchy, with $ l_{\mpop} = 2 $ indicating the merger two PBHs onto a final one, and $ l_{\mpop} = 3 $ indicating the merger of three PBHs onto single PBH, accumulated in the runs.
	}
	\label{tab: Merger tree hierarchy levels}
\end{table}



\subsubsection{Simulation merger trees}
\label{subsubsec: Simulation merger trees}

The identification of merger pairs in order to construct the merger trees has been done by the procedure described in Section~\ref{subsubsec: Merger trees computation}, looking at mass differences of objects in between snapshots. Typically, the mergers happen at the last simulation time-runs, a time when average velocities are the smallest for all PBH populations at the core of the cluster, and where nearest neighbour interactions are most frequent, often arising from disruption of a transient binary.

Merger identification then straightforward, at least for most cases in our simulations, as there are not more that two mergers in between time snapshots in any realisation, so that, out of the $ (N_{\rpop})_{\mpop} = 117 $ merger events identified in our simulations, we find that
\begin{enumerate}[label=\roman*)]
	\item $ (N_{\rpop})^{1,1}_{\mpop} = 115 $ events are one-to-one mergers in which a single PBH absorbs another less massive PBH in between two particular time-slices belonging to the last simulation time-run.
	\item $ (N_{\rpop})^{2,1}_{\mpop} = 2 $ events are two-to-one mergers, in which a single PBH subsequently absorbs two PBHs in between particular time-slices that may or may not be contiguous, within the last time-run.
\end{enumerate}

Last, it should be noted that mergers are so rare in each of the simulations that, out of the methods devised in Section \ref{subsubsec: Merger trees computation}, only the first three of impossible, necessary and single-to-one mergers are relevant, as there has been no case found in all of the $ N_{\rcap} = 5000 $ realisations where, in between two consecutive time-slices, may have been many-to-one mergers.



\section{Close encounters}
\label{sec: Close encounters}

In this Section we take a closer look to the identified binary and merger pairs, in Sections~\ref{subsec: Binary pairs} and \ref{subsec: Merger pairs} respectively. In particular, we study the collisional and short-distance interaction dynamics of PBHs in these pairs, and extract the mass distributions of the PBHs participating in such systems. Also, we offer a lower bound prediction of the rates of occurrence of PBH mergers and of the PBH hyperbolic encounters per cluster.



\subsection{Binary pairs}
\label{subsec: Binary pairs}


\begin{table*}[t!]
	\centering
	\begin{tabular}[c]{| l l | l l | l l | l l | l l |}
	\hline
	$ \quad $ & $ \quad $ &
	\multicolumn{2}{c |}{Binary populations:} &
	\multicolumn{6}{c |}{Binary distribution parameters:} \\
	[0.5ex]
	\hline
	\multicolumn{2}{| c |}{Pop.} &
	\multicolumn{2}{c |}{Frequency:} &
	\multicolumn{2}{c |}{Total mass:} &
	\multicolumn{2}{c |}{Chirp mass:} &
	\multicolumn{2}{c |}{Mass ratio:} \\
	[0.5ex]
	\hline\hline
	$ P_{i} $ & $ P_{j} $ &
	$ N_{\rpop, \bpop}^{P_{i},P_{j}} $ &
	$ f_{\rpop, \bpop}^{i,j} $ &
	$ \unit {\bar{m}_{T}} [\msun] $ &
	$ \unit {\tilde{m}_{T}} [\msun] $ &
	$ \unit {\bar{m}_{C}} [\msun] $ &
	$ \unit {\tilde{m}_{C}} [\msun] $ &
	$ \bar{q} $ & $ \tilde{q} $ \\
	[0.5ex]
	\hline
	$ \quad $ & $ \quad $ &
	$ 1.27 \times 10^{6} $ & $ 2.55 \times 10^{-3} $ &
	$ 303 $ & $ 181 $ &
	$ 45.0 $ & $ 25.2 $ &
	$ 0.154 $ & $ 0.0552 $ \\
	$ \cpop $ & $ \ppop $ &
	$ 1.05 \times 10^{5} $ & $ 2.10 \times 10^{-3} $ &
	$ 305 $ & $ 185 $ &
	$ 39.9 $ & $ 26.2 $ &
	$ 0.200 $ & $ 0.0446 $ \\
	$ \epop $ & $ \ppop $ &
	$ 2.27 \times 10^{4} $ & $ 4.54 \times 10^{-4} $ &
	$ 371 $ & $ 202 $ &
	$ 61.2 $ & $ 29.0 $ &
	$ 0.404 $ & $ 0.0715 $ \\
	\hline
	\end{tabular}
	\caption{
	Distribution mean and median values of the distributions in Figure~\ref{fig: Binary mass distributions}, representing the binary profiles segregated by the population (cluster and ejecta) type integrated across all time-slices in the simulations starting at the burn-in time $ t_{\bicap} = \unit [2.64 \times 10^{5}] {yr} $ up to the present, for a total of $ \sum_{i = 1}^{N_{\rpop}}(N_{\bpop})_{\rpop} = 2.32 \times 10^{6} $ binaries throughout all realisations. Note that we use our convention that C, E, P, R and B stand for the cluster, ejecta, primitive, remaining and bounded populations respectively. Note as well that we do not show the corresponding mean and median values of merged PBHs in bounded systems in this table since their number is negligible compared to the other cases, but we do show them however on Table \ref{tab: Merger mass distributions}.
	}
	\label{tab: Binary mass distributions}
\end{table*}

\begin{figure*}[t!]
	\centering
	\subfloat[Total mass profile $ f (N_{\bpop})_{\rpop} (m_{T}) $.]{
	\hspace*{-0.00cm}
	\includegraphics[width = 0.50\textwidth]
	{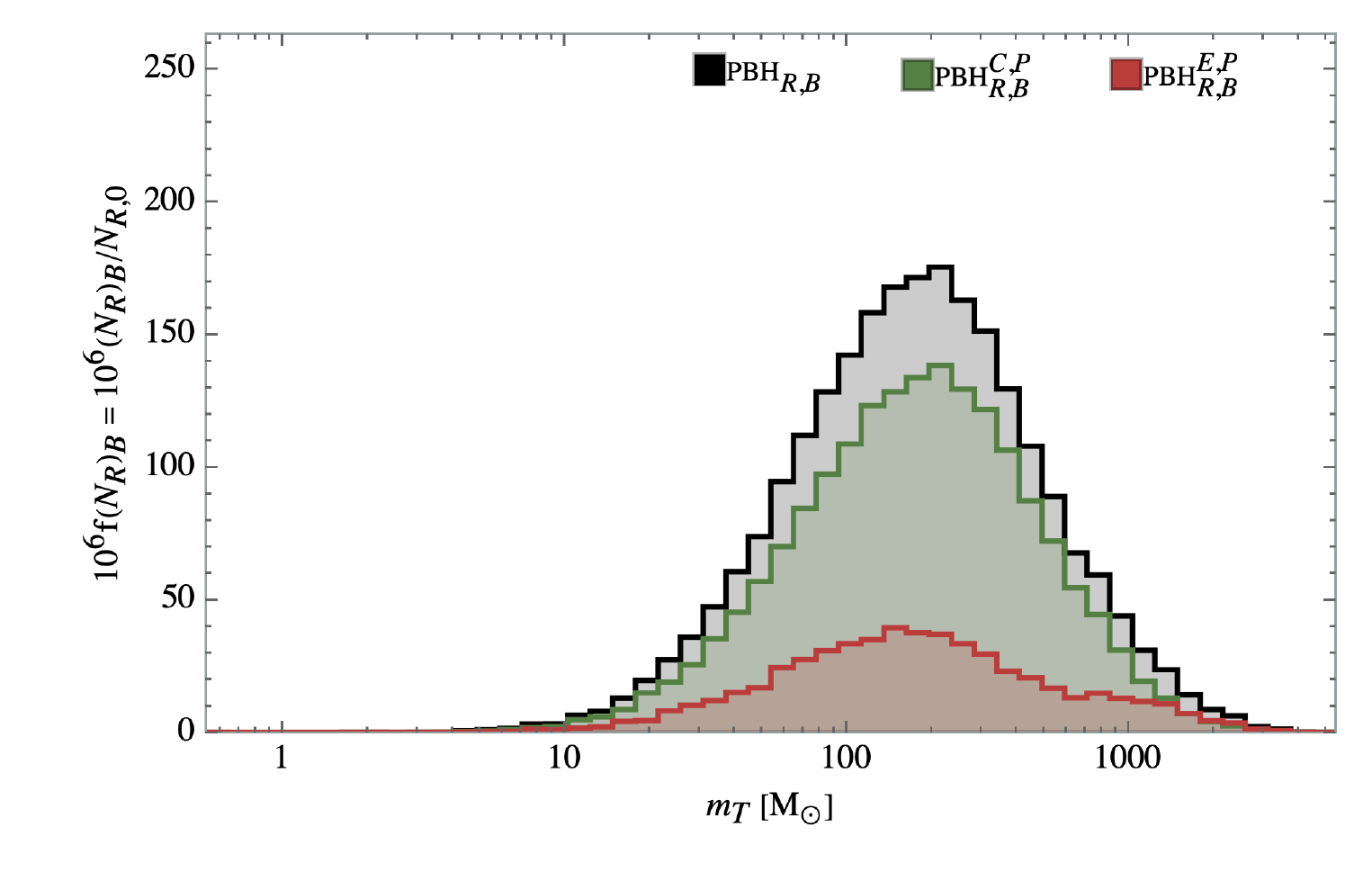}
	\label{subfig: Binaries 1D total mass profile}}
	\subfloat[Distribution of masses $ f (N_{\bpop})_{\rpop} (m_{1}, m_{2}) $.]{
	\hspace*{-0.00cm}
	\includegraphics[width = 0.34\textwidth]
	{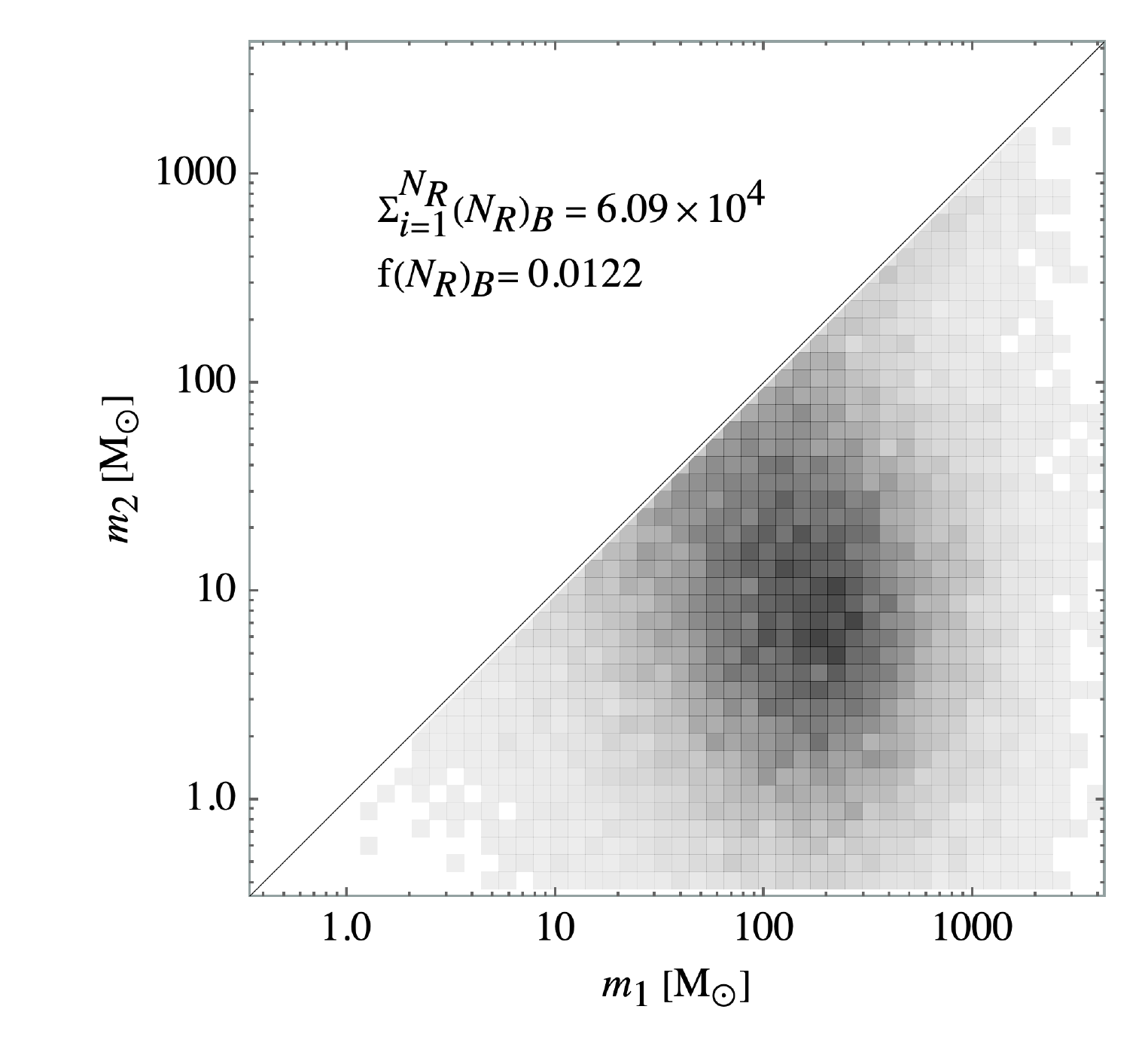}
	\label{subfig: Binaries 2D objects distribution}}
	\newline
	\subfloat[Chirp mass profile $ f (N_{\bpop})_{\rpop} = (m_{C}) $.]{
	\hspace*{-0.00cm}
	\includegraphics[width = 0.50\textwidth]
	{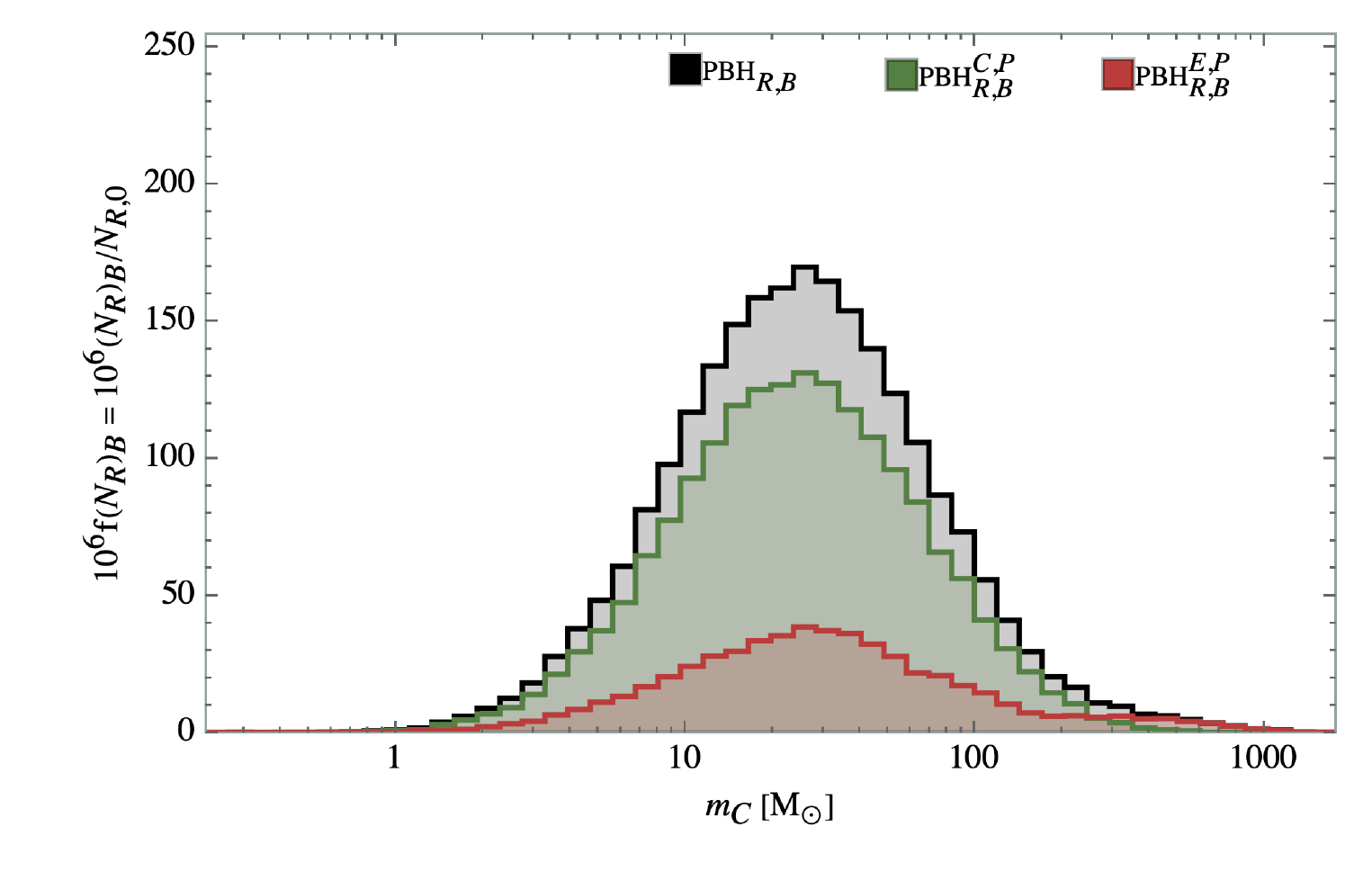}
	\label{subfig: Binaries 1D chirp mass profile}}
	\subfloat[Cluster dist. of masses $ f (N_{\bpop, \cpop})_{\rpop} (m_{1}, m_{2}) $.]{
	\hspace*{-0.00cm}
	\includegraphics[width = 0.34\textwidth]
	{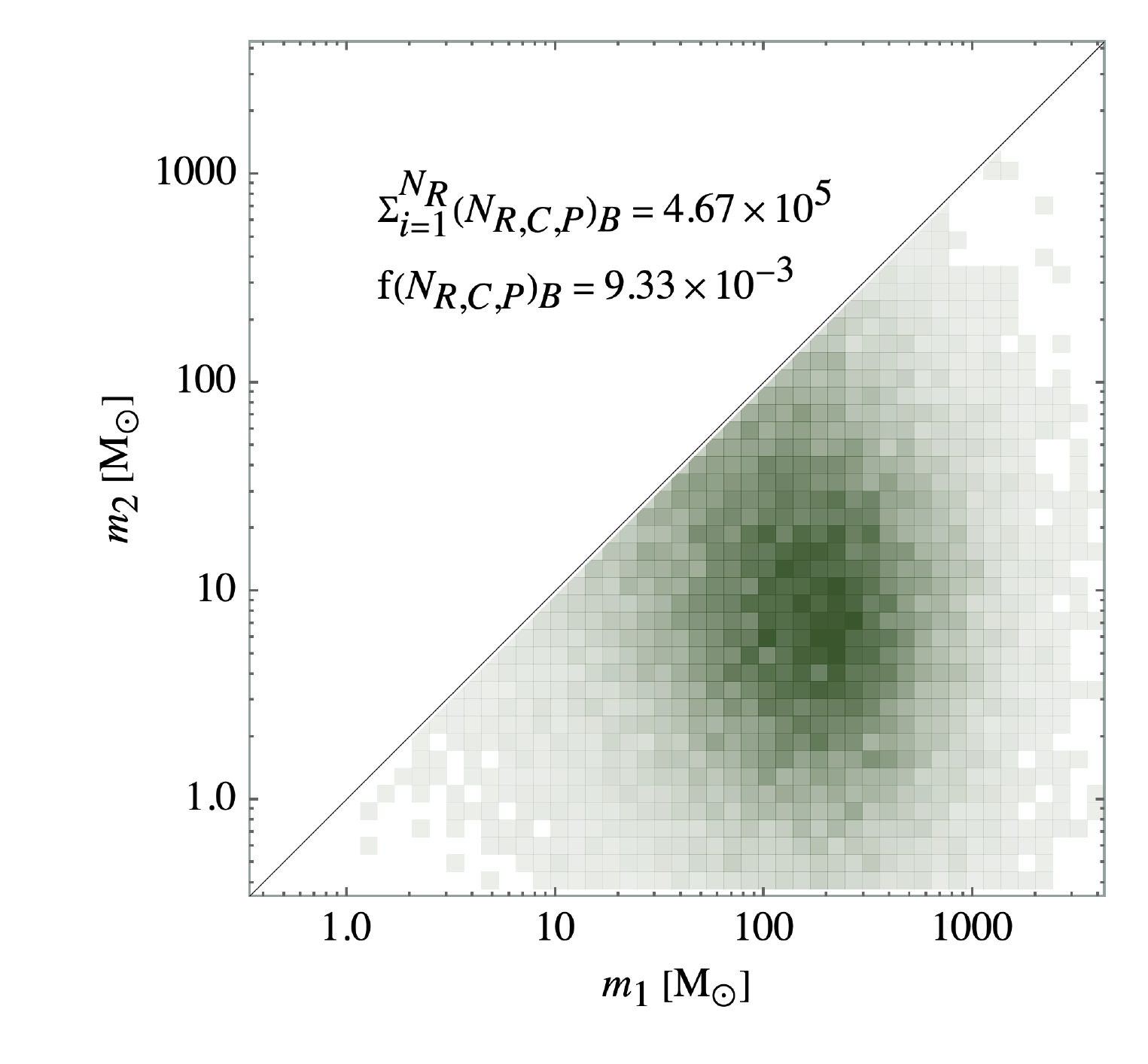}
	\label{subfig: Binaries 2D cluster distribution}}
	\newline
	\subfloat[Mass ratio profile $ f (N_{\bpop})_{\rpop} = (q) $.]{
	\hspace*{-0.75cm}
	\includegraphics[width = 0.50\textwidth]
	{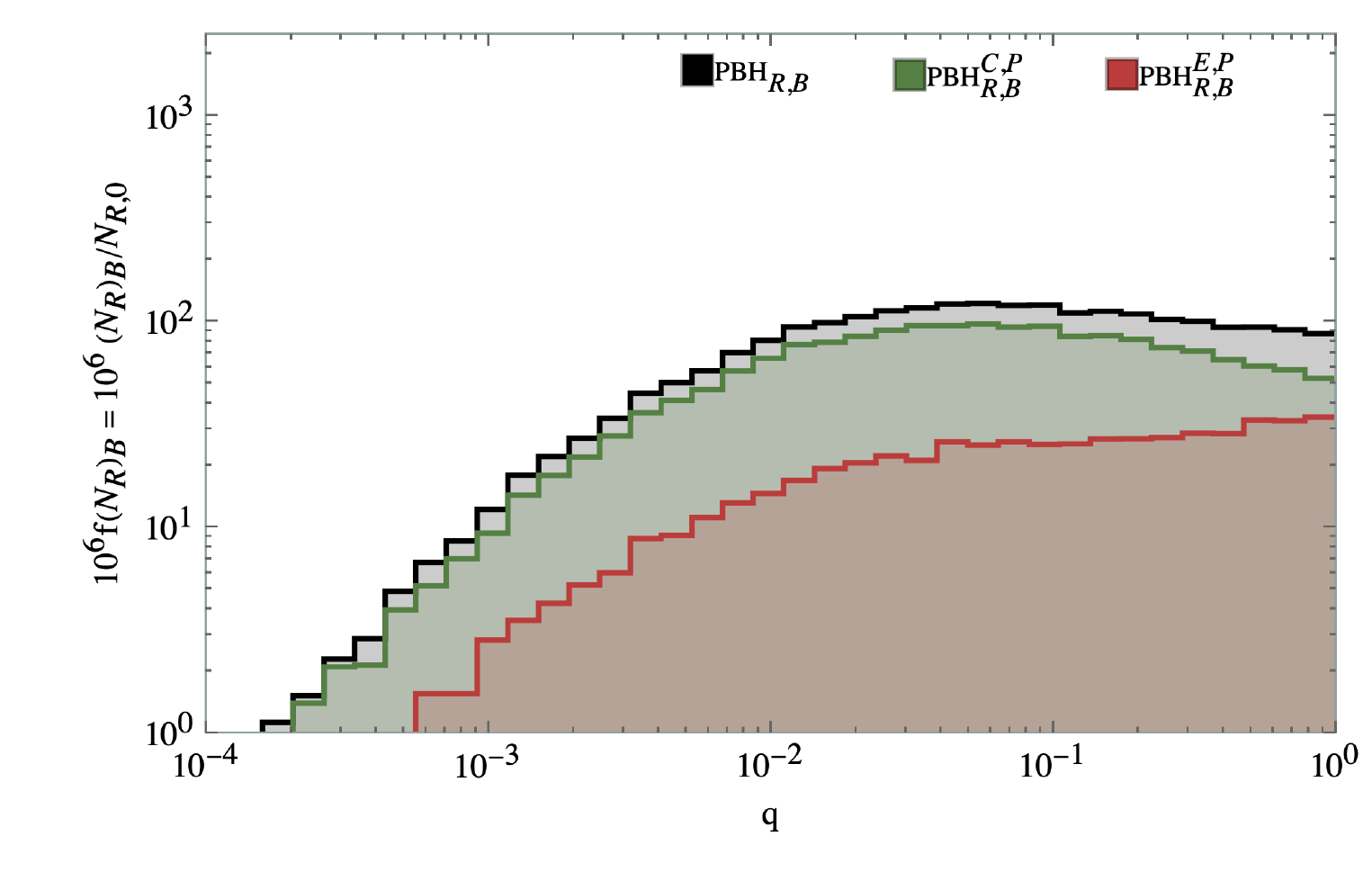}
	\label{subfig: Binaries 1D mass ratio profile}}
	\subfloat[Ejecta dist. of masses $ f (N_{\bpop, \epop})_{\rpop} (m_{1}, m_{2}) $.]{
	\hspace*{-0.00cm}
	\includegraphics[width = 0.34\textwidth]
	{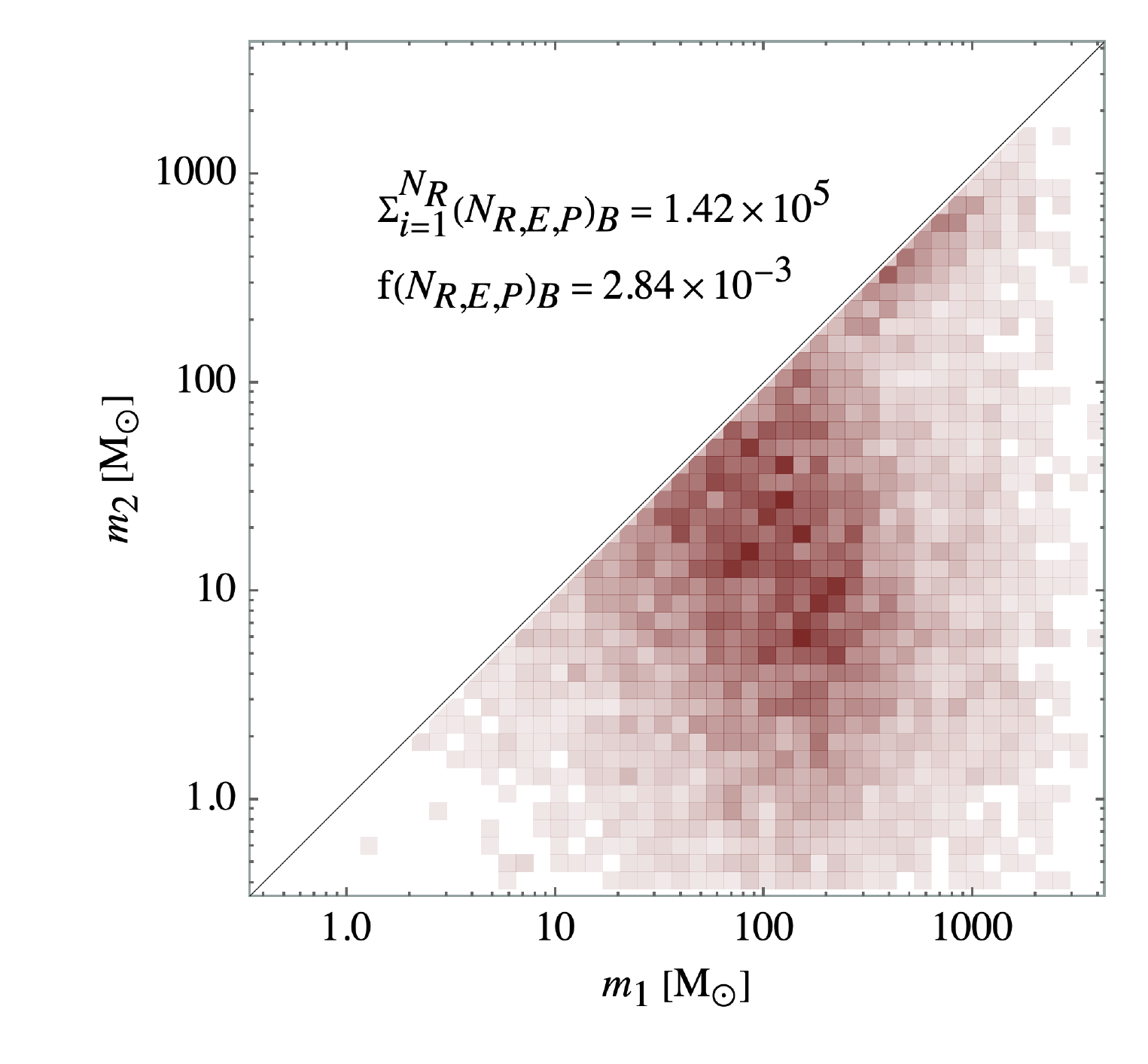}
	\label{subfig: Binaries 2D ejecta distribution}}
	\caption{
	Binary mass-related profiles segregated by their population (cluster and ejecta) type integrated across all time-slices in the simulations from the burn-in time, $ t_{\bicap} = \unit [2.64 \times 10^{5}] {yr} $, to the present. Profiles are normalised by the total number of objects in the simulations at the initial time, $ N_{\tim_{0}} $, and distribution mean and median values are given in Table~\ref{tab: Binary mass distributions}.
	Note that panels (b), (d) and (f) are arranged by choosing $ m_{1} > m_{2} $ to avoid double-plotting and therefore all binary systems are represented by a single point at or below the first quadrant diagonal.
	}
	\label{fig: Binary mass distributions}
\end{figure*}

Shown in Figure~\ref{fig: Binary mass distributions}, we have computed the binary pair number and frequency in the simulations along with the total mass, chirp mass and mass ratio profile of the pair, as well as the component mass distributions. We have as well extracted the mean, median and modal values of the total mass, chirp mass and mass ratio profiles, which are shown in Figure~\ref{tab: Binary mass distributions}.

In particular, we find that, integrated across all the time-slices, binary pairs constitute about $ \order {0.001} $ of all bodies in the simulations. About 85\% of these remain in the cluster while the remaining 15\% of these have been ejected in two-to-one encounters.

Note that this ratio is heavily influenced both by the IC, as the abundance of bounded in-cluster PBHs at early times is overwhelmingly larger than that of bounded in-ejecta PBHs, as well as the later evolution, as we had found in Section~\ref{subsec: Cluster and ejecta populations} that as the simulations approach the present time, the fraction of in-cluster PBHs remains roughly constant despite the gradual evaporation and depopulation of the cluster, while the corresponding number of in-ejecta PBHs grows monotonically, even if at an ever decreasing rate as shown in the PL-fits, also as a consequence of cluster evaporation.

If however, we restrict ourselves to time-slices posterior to the burn-in time, so that the IC has been already erased, then the numbers at present change significantly. About 65\% of bounded PBHs do remain in the cluster while the remaining 35\% of bounded PBHs will have been have been ejected in two-to-one encounters, as shown Section~\ref{subsec: Cluster and ejecta populations}.

In any case, the masses of the bodies participating in binary systems are naturally found to be substantially larger than those of the mean or median PBH in the simulations. As seen in Section~\ref{subsubsec: Mass distribution}, the PBH average and mean masses in the IC are $ \bar{m} = \unit [21.9] {\msun} $ and $ \tilde{m} = \unit [7.55] {\msun} $. Given both the lack of any meaningful number of mergers and that the mass profiles of the cluster and ejecta populations do not evolve differently throughout the simulations, the mean and median masses remain nearly constant for both the in-cluster and in-ejecta bounded populations up until the present time.

However, for binary pairs, the mean and median total masses in the pair are $ \bar{m}_{\tcap} = \unit [303] {\msun} $ and $ \tilde{m}_{\tcap} = \unit [181] {\msun} $, which is far more than twice the simulation mean or modal mass. This indicates that only very massive PBHs with sufficiently deep and far reaching gravitational potential wells are able to overcome the large, $ \unit [\order {1}] {pc}-\unit [\order {1000}] {pc} $-sized and monotonically growing distances between PBHs in between the burn-in time and the present time as the cluster puffs up and become a parent in the binary pair.

Moreover, the binary pair mean and median mass ratios of $ \bar{q}_{\tcap} = \unit [0.154] {} $ and $ \tilde{q}_{\tcap} = \unit [0.0552] {} $ indicate that the less massive partner in the binary is typically between 6 and 20 times less massive than its parent, establishing a clear mass hierarchy between both, as more equally distribute binary pairs may be easier to disrupt in the long term.

In addition to this, and taking into account both the mean and median total mass and mass ratio, we find that just like its parent PBH, the secondary PBH in the pair is also typically larger than the typical PBH in the simulations, although to a lesser extent determined by the mass ratio. In particular, given that $ m_{1}+m_{2} = m_{\tcap} $, $ m_{2} / m_{1} = q $, and the LIGO convention $ m_{1} \geq m_{2} $, two PBHs in a binary pair, according to the previous results would have masses close to
\begin{align}
	(\bar{m}_{1}, \bar{m}_{2})_{\bpop} & = \unit [(261, 40.9)] {\msun}, \\
	(\tilde{m}_{1}, \tilde{m}_{2})_{\bpop} & = \unit [(172, 9.46)] {\msun},
\end{align}
when computed from the mean or the median values of the total mass and mass ratio respectively, showing masses for the smaller partner that are bigger than the simulations' mean and median mass in both cases.

This is maintained in the in-cluster and in-ejecta bounded populations of PBHs as well, although the degree to which the masses are larger than the typical mass or not varies a by a small amount. In particular, we find
\begin{enumerate}[label=\roman*)]
	\item For the bounded, in-cluster PBHs,
	\begin{align}
		(\bar{m}_{1}, \bar{m}_{2})_{\bpop,\cpop} & = \unit [(254, 50.8)] {\msun}, \\
		(\tilde{m}_{1}, \tilde{m}_{2})_{\bpop,\cpop} & = \unit [(177, 7.90)] {\msun}.
	\end{align}
	\item For the bounded in-ejecta PBH,
	\begin{align}
		(\bar{m}_{1}, \bar{m}_{2})_{\bpop,\epop} & = \unit [(264, 107)] {\msun}, \\
		(\tilde{m}_{1}, \tilde{m}_{2})_{\bpop,\epop} & = \unit [(186, 13.5)] {\msun}.
\end{align}
\end{enumerate}

Overall, this indicates that the more one-sided distributions of mass in a binary pair as well as the more massive binary pairs are biased towards the ejecta population, as expected.



\subsubsection{Hyperbolic encounter rates}
\label{subsubsec: hyperbolic encounter rates}


We define a hyperbolic encounter between two bodies in the simulations as one in which the eccentricity of the approaching crosses the thresholds $ e = 1 $ in any direction in between snapshots $ t $ and $ t+\Delta t $ or at any point during the close encounter, irrespective of the impact parameter $ b $ of such encounter or other orbital characteristics, and due to a close one-to-one body interaction. However, there are two issues with this approach that makes the interpretation of this quantity as the close encounter, GW-generating event a bit problematic.

First, that the number of hyperbolic encounters is by these means underestimated as one particular PBH with an orbital eccentricity crossing the parabolic threshold two times in opposite directions in between consecutive time slices will not register as a hyperbolic encounter at all despite that there have been, in fact two total encounters in such time interval.

Second, that such hyperbolic encounters most often happen at distances large enough that the amplitude of the emitted GWs will be too small to be registered on Earth detectors within the foreseeable future, meaning that this rates are not to be confused with the rate of detection of a gravitational wave event.

We can thus only provide a minimum threshold of the number of this encounters, given that an irreducible number of these hyperbolic encounters will be missing in any case from the fact that the time resolution is limited. This effect is not, in fact, very large, since, as seen in Table~\ref{tab: Characteristic time scales}, the time-step in each runs is adapted to capture most of such interactions, being of an order of magnitude as the cluster crossing time, but being of the same order of magnitude the effect cannot be neglected either.

We now proceed to compute the minimum hyperbolic encounter rate per time-run and realisation as
\begin{equation}
	\Gamma_{r}^{\mathrm{S},\mathrm{P}} = \frac {N_{r}^{\mathrm{S}}-N_{r-1}^{\mathrm{S}}} {\Delta t_{r}},
\end{equation}
where $ N_{r}^{\mathrm{S}} $ is the cumulative number of slingshots identified in the simulations with the aforementioned procedure at the end of time-run $ r $ and $ \Delta t_{r} = t_{i^{+}}-t_{i^{-}} $ the time-run total period, leading to a minimum hyperbolic encounter rate of:
\begin{align}
	\Gamma_{1}^{\mathrm{S}} (0 \leq t_{i} \leq 10) & = \unit [(2.0 \substack {+1.6 \\ -1.4}) \times 10^{6}] {Gyr^{-1}}, \\
	\Gamma_{2}^{\mathrm{S}} (11 \leq t_{i} \leq 19) & = \unit [(2.2 \substack {+1.2 \\ -1.8}) \times 10^{6}] {Gyr^{-1}}, \\
	\Gamma_{3}^{\mathrm{S}} (20 \leq t_{i} \leq 28) & = \unit [(2.2 \substack {+2.3 \\ -1.7}) \times 10^{5}] {Gyr^{-1}}, \\
	\Gamma_{4}^{\mathrm{S}} (29 \leq t_{i} \leq 37) & = \unit [(0.9 \substack {+2.9 \\ -0.6}) \times 10^{4}] {Gyr^{-1}}, \\
	\Gamma_{5}^{\mathrm{S}} (38 \leq t_{i} \leq 46) & = \unit [(1.0 \substack {+3.5 \\ -0.6}) \times 10^{4}] {Gyr^{-1}}, \\
	\Gamma_{6}^{\mathrm{S}} (47 \leq t_{i} \leq 55) & = \unit [98 \substack {+400 \\ -63}] {Gyr^{-1}}, \\
	\Gamma_{7}^{\mathrm{S}} (56 \leq t_{i} \leq 64) & = \unit [9 \substack {+95 \\ -6}] {Gyr^{-1}}.
\end{align}

Note that these slingshot rates have been extracted for non-relativistic simulations, adding up to the underestimation the underestimation of the rates as the emission of gravitational radiation results over time in a more compact clusters of PBHs, were one-to-one PBH encounters are more frequent.

However, given the large scale of the cluster, of $ \unit [\order {1}] {pc} $ already at the burn-in time, the effect is nearly negligible. Also, note that these rates have been computed for a single cluster. Latter, in Section~\ref{sec: Background DM implications} we will compute this quantity for a comoving cosmological volume of $ \unit [1] {Gpc^{3}} $, and thus provide a more meaningful value for the slingshot rate of PBHs.



\subsection{Merger pairs}
\label{subsec: Merger pairs}


\begin{table*}[t!]
	\centering
	\begin{tabular}[c]{| l l | l l | l l | l l | l l |}
	\hline
	$ \quad $ & $ \quad $ &
	\multicolumn{2}{c |}{Merger populations:} &
	\multicolumn{6}{c |}{Merger distribution parameters:} \\
	[0.5ex]
	\hline
	\multicolumn{2}{| c |}{Pop.} &
	\multicolumn{2}{c |}{Frequency:} &
	\multicolumn{2}{c |}{Total mass:} &
	\multicolumn{2}{c |}{Chirp mass:} &
	\multicolumn{2}{c |}{Mass ratio:} \\
	[0.5ex]
	\hline\hline
	$ P_{i} $ & $ P_{j} $ &
	$ N_{\rpop, \mpop}^{P_{i},P_{j}} $ &
	$ f_{\rpop, \mpop}^{i,j} $ &
	$ \unit {\bar{m}_{T}} [\msun] $ &
	$ \unit {\tilde{m}_{T}} [\msun] $ &
	$ \unit {\bar{m}_{C}} [\msun] $ &
	$ \unit {\tilde{m}_{C}} [\msun] $ &
	$ \bar{q} $ & $ \tilde{q} $ \\
	[0.5ex]
	\hline
	$ \quad $ & $ \quad $ &
	$ 117 $ & $ 2.34 \times 10^{-5} $ &
	$ 1670 $ & $ 1510 $ &
	$ 642 $ & $ 567 $ &
	$ 0.965 $ & $ 0.545 $ \\
	$ \cpop $ & $ \ipop $ &
	$ 50 $ & $ 1.00 \times 10^{-5} $ &
	$ 2120 $ & $ 1690 $ &
	$ 793 $ & $ 667 $ &
	$ 0.517 $ & $ 0.517 $ \\
	$ \cpop $ & $ \bpop $ &
	$ 4 $ & $ 8.00 \times 10^{-7} $ &
	$ 967 $ & $ 925 $ &
	$ 368 $ & $ 368 $ &
	$ 0.634 $ & $ 0.641 $ \\
	$ \epop $ & $ \ipop $ &
	$ 1 $ & $ 2.00 \times 10^{-7} $ &
	$ 1590 $ & $ 1590 $ &
	$ 689 $ & $ 689 $ &
	$ 0.867 $ & $ 0.867 $ \\
	$ \epop $ & $ \bpop $ &
	$ 62 $ & $ 1.24 \times 10^{-5} $ &
	$ 1350 $ & $ 1230 $ &
	$ 538 $ & $ 477 $ &
	$ 0.594 $ & $ 0.561 $ \\
	\hline
	\end{tabular}
	\caption{
	Distribution mean and median values of the distributions in Figure~\ref{fig: Merger mass distributions}, representing the merger profiles segregated by the population (cluster and ejecta) type integrated across all time-slices in the simulations starting at the time of first-merger (1M) $ t_{\fmcap} = \unit [1.38 \times 10^{9}] {yr} $ up to the present, for a total of $ \sum_{i = 1}^{N_{\rpop}}(N_{\mpop})_{\rpop} = 117 $ merger events throughout all realisations. Note that, as in the rest of the paper, we use our convention that C, E, I, B, R M stand for the cluster, ejecta, isolated, bounded, remaining and merged populations respectively.
	}
	\label{tab: Merger mass distributions}
\end{table*}

\begin{figure*}[t!]
	\centering
	\subfloat[Total mass profile $ f (N_{\mpop})_{\rpop} (m_{T}) $.]{
	\hspace*{-0.6cm}
	\includegraphics[width = 0.50\textwidth]
	{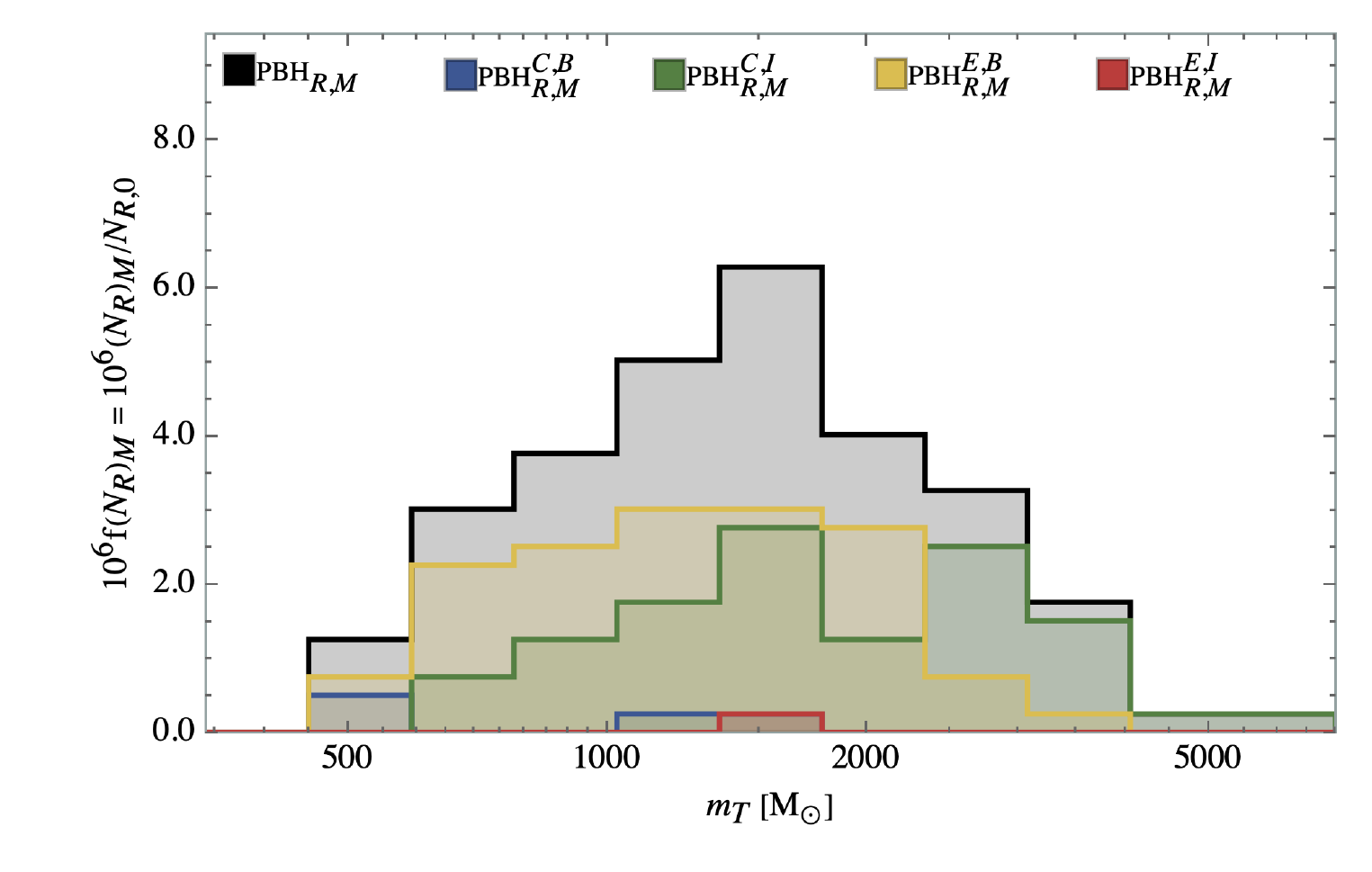}
	\label{subfig: Mergers 1D total mass profile}}
	\subfloat[Distribution of masses $ f (N_{\mpop})_{\rpop} (m_{1}, m_{2}) $.]{
	\hspace*{-0.5cm}
	\includegraphics[width = 0.34\textwidth]
	{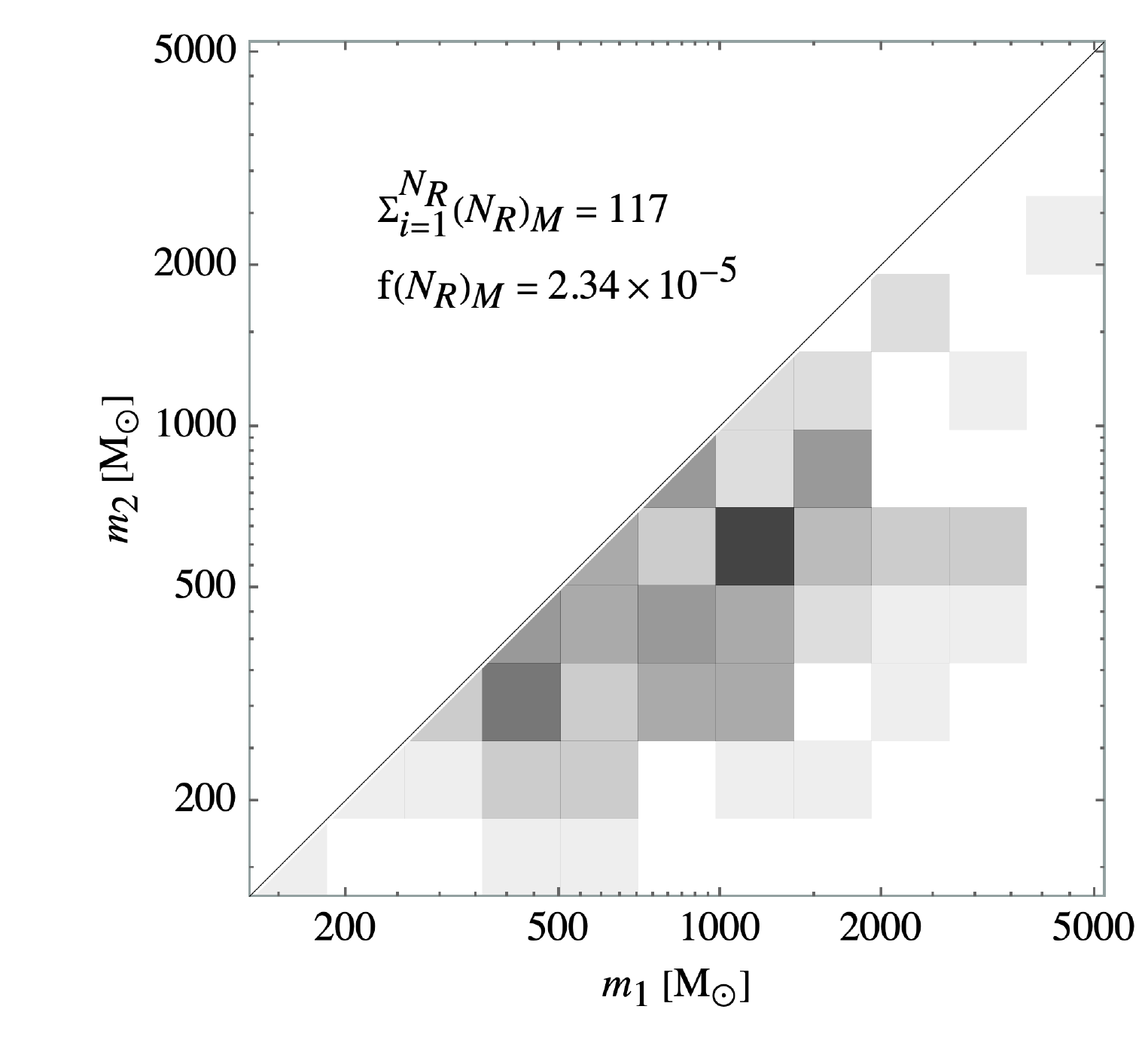}
	\label{subfig: Mergers 2D objects distribution}}
	\newline
	\subfloat[Chirp mass profile $ f (N_{\mpop})_{\rpop} = (m_{C}) $.]{
	\hspace*{-0.6cm}
	\includegraphics[width = 0.50\textwidth]
	{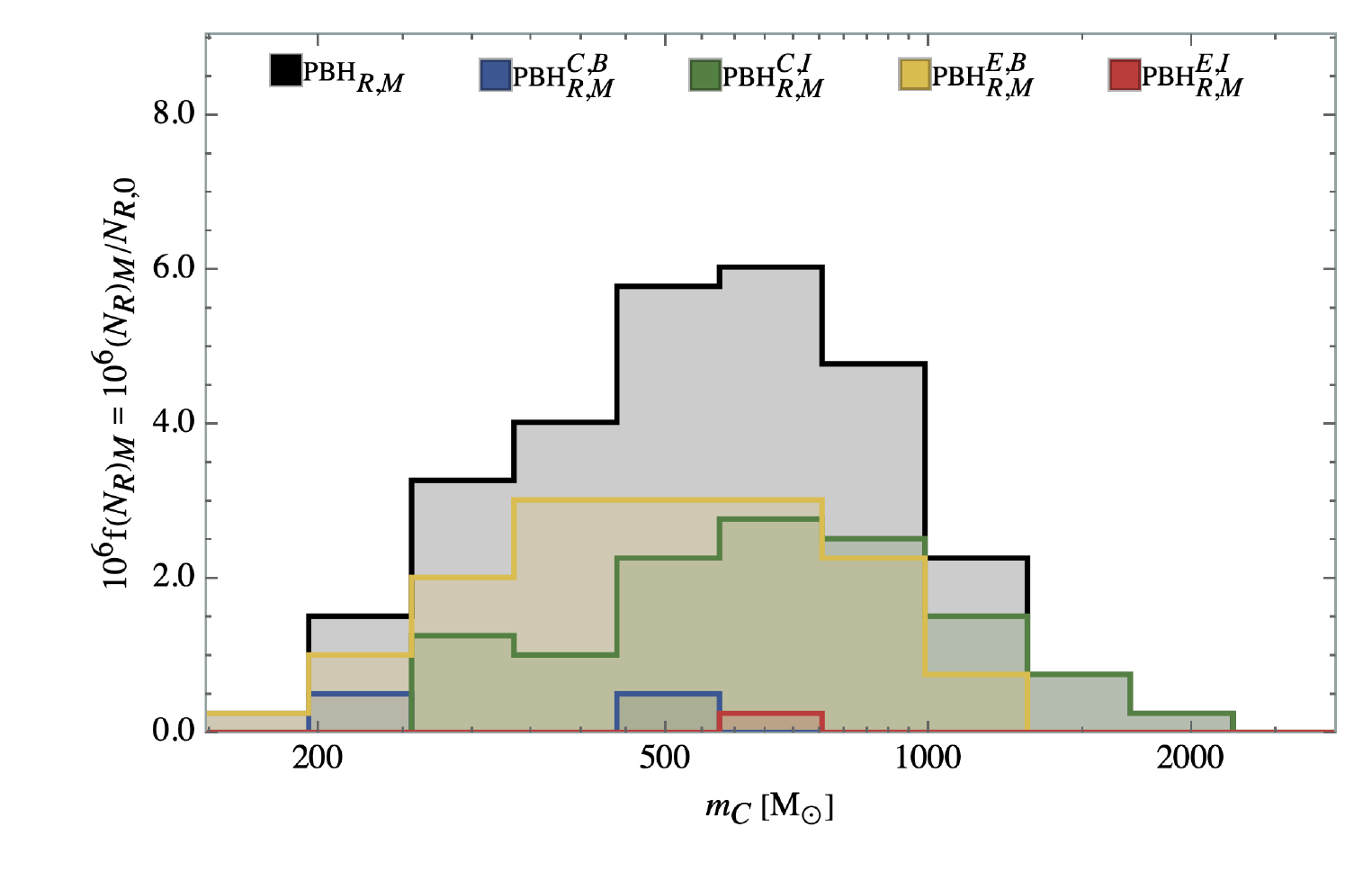}
	\label{subfig: Mergers 1D chirp mass profile}}
	\subfloat[Dist. of masses $ f (N_{\mpop, \cpop})_{\rpop} (m_{1}, m_{2}) $.]{
	\hspace*{-0.5cm}
	\includegraphics[width = 0.34\textwidth]
	{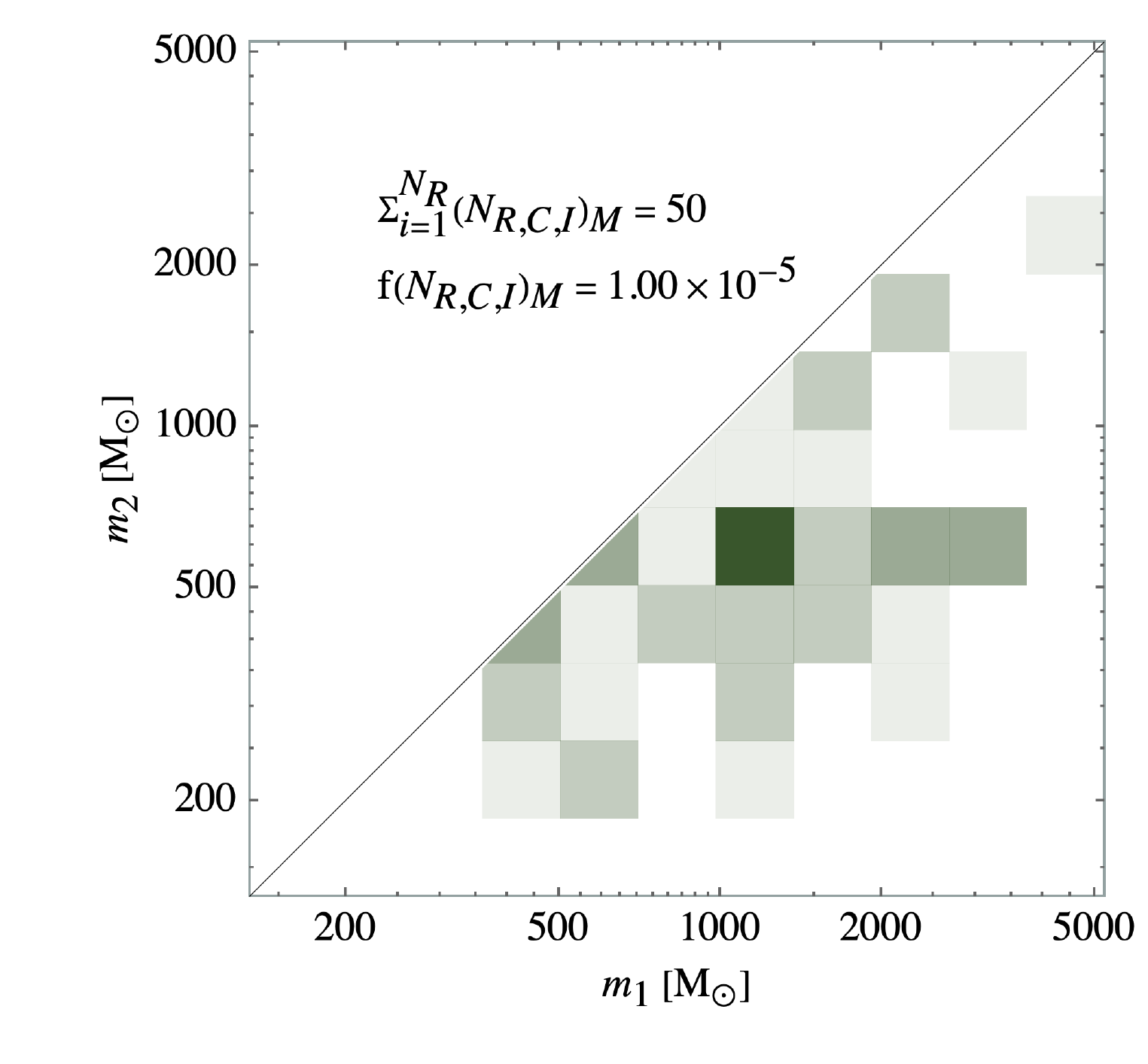}
	\label{subfig: Mergers 2D cluster, unbounded distribution}}
	\newline
	\subfloat[Mass ratio profile $ f (N_{\mpop})_{\rpop} = (q) $.]{
	\hspace*{-1.40cm}
	\includegraphics[width = 0.50\textwidth]
	{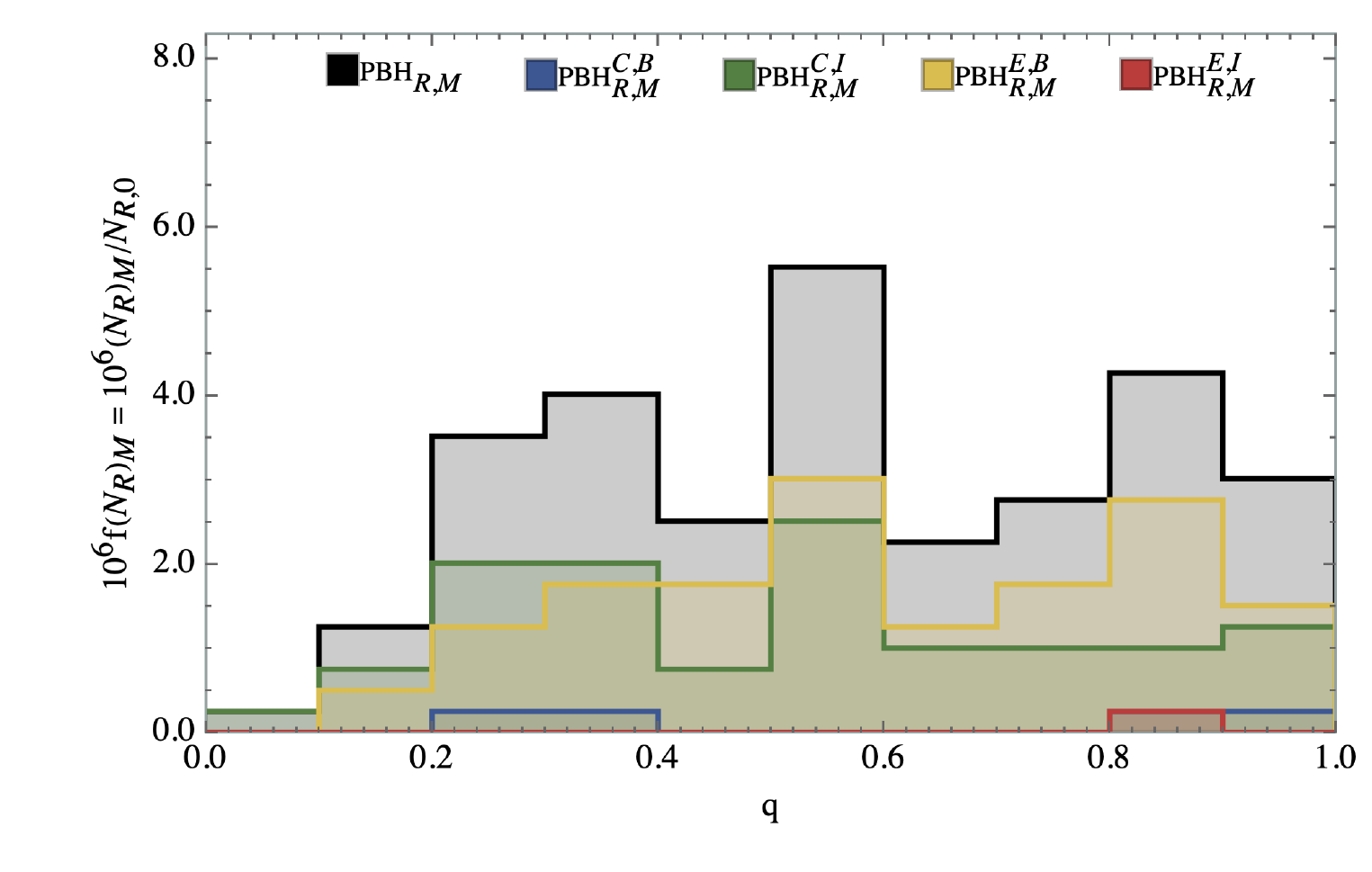}
	\label{subfig: Mergers 1D mass ratio profile}}
	\subfloat[Dist. of masses $ f (N_{\mpop, \epop})_{\rpop} (m_{1}, m_{2}) $.]{
	\hspace*{-0.50cm}
	\includegraphics[width = 0.34\textwidth]
	{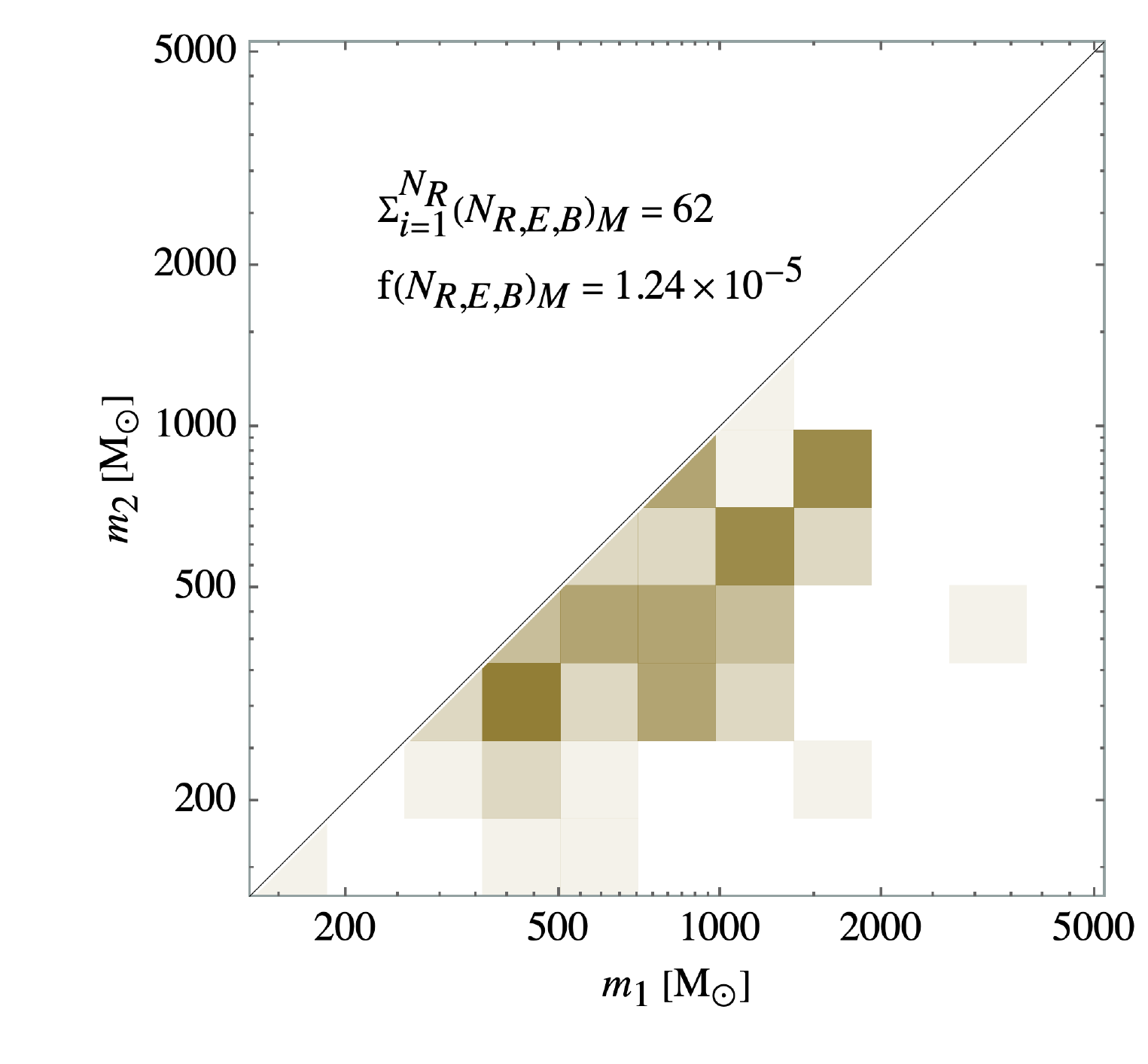}
	\label{subfig: Mergers 2D ejecta, bounded distribution}}
	\caption{
	Merger mass-related profiles segregated by the population (cluster and ejecta) type integrated across all time-slices in the simulations staring at the time of first-merger $ t_{\fmcap} = \unit [1.38 \times 10^{9}] {yr} $ up to the present. Profiles are normalised by the total number of objects in the simulations at the initial time, $ N_{\tim_{0}} $, and distribution mean and median values are given in Table~\ref{tab: Merger mass distributions}.
	Note that panels (b), (d) and (f) are arranged again as in Figure~\ref{fig: Binary mass distributions} by choosing $ m_{1} > m_{2} $ to avoid double-plotting and so again all binary systems are represented by a single point at or below the first quadrant diagonal.
	}
	\label{fig: Merger mass distributions}
\end{figure*}

Similarly to what we have done for the binary pairs of Section~\ref{subsec: Binary pairs}, we have computed the merger pair number and frequency in the simulations along with the total mass, chirp mass and mass ratio profile of the pair, as well as the component mass distributions, shown in Figure~\ref{fig: Merger mass distributions}. The mean, median and modal values of these distributions are shown in Figure~\ref{tab: Merger mass distributions}.

In particular, we find that, integrated across all the time-slices, merged pairs constitute $ \order {10^{-5}} $ of all bodies in the simulations, though note that, in contrast to what was the case with binary pairs, mergers occur late in the simulations, with most of them occurring in the last time-run, and none occurring during the first five runs.

Of these, about 46\% remain in the cluster while the remaining 54\% have been ejected in two-to-one encounters, occasionally because of the merger itself, as we had found in Section~\ref{subsec: Primitive and merged populations}. Unlike what happened with binary pairs, however, this ratio is not influenced at all by the IC as there are no mergers during that stage, but it may still be influenced by the later evolution as the cluster evaporates. Regrettably, with only 2.6\% of mergers in the sixth run and the remaining 97.4\% of mergers in the seventh run, we lack data to quantify if later evolution does indeed alter the balance between the in-cluster and in-ejecta merged population.

In any case, and for similar reasons as what was the case with binary pairs, the masses of the bodies participating in mergers are naturally found to be substantially larger than those of the mean or median PBH in the simulations, only even more so. We had extracted in Section~\ref{subsubsec: Mass distribution} the PBH average and mean masses in the IC are $ \bar{m} = \unit [21.9] {\msun} $ and $ \tilde{m} = \unit [7.55] {\msun} $ and found those to be practically invariant for the cluster and ejecta populations throughout time-slices up to the present.

For merged pairs, the mean and median total masses in the pair are indeed very large, far more so than for binary pairs, with $ \bar{m}_{\tcap} = \unit [1.67 \times 10^{3}] {\msun} $ and $ \tilde{m}_{\tcap} = \unit [1.51 \times 10^{3}] {\msun} $, which is $ 55-200 $ times more than the simulation mean or modal mass respectively. This is because only the truly massive PBHs, and most often the most massive PBH in the simulations, are sufficiently dominant over the other less massive PBHs that they may be able to overcome the order kpc-sized distances between PBHs in the later stages of evolution and merge still with a small number of them.

This typically happens for a merger pair of mean and median mass ratios of $ \bar{q}_{\tcap} = \unit [0.965] {} $ and $ \tilde{q}_{\tcap} = \unit [0.545] {} $, indicating that the less massive partner in the merger is typically bigger than half the mass of the more massive partner, with no pronounced mass hierarchy between both, as it would be expected from the fact that in this classical computation the merger cross section is quadratic with the Schwarzschild radius and therefore quadratic with the mass, so the merger of two large PBHs is dramatically more likely to occur than the merger of a more uneven pair, especially so as in our non-relativistic simulations the bodies neither inspiral nor emit GWs.

Therefore, and taking into account the mean and median total mass and mass ratio, we find that, given that $ m_{1}+m_{2} = m_{\tcap} $, $ m_{2} / m_{1} = q $, and the LIGO convention $ m_{1} \geq m_{2} $, two PBHs in a merged pair, according to the previous results would have masses close to
\begin{align}
	(\bar{m}_{1}, \bar{m}_{2})_{\mpop} & = \unit [(850, 820)] {\msun}, \\
	(\tilde{m}_{1}, \tilde{m}_{2})_{\mpop} & = \unit [(977, 533)] {\msun},
\end{align}
when computed from the mean or the median values of the total mass and mass ratio respectively, showing masses for the smaller partner that are bigger than the simulation mean and median mass in both cases.

We find as well that this feature is upheld in the in-cluster and in-ejecta bounded populations of PBHs as well, although the degree to which the masses are larger than the typical mass or not varies a by a small amount. In particular, we find that
\begin{enumerate}[label=\roman*)]
	\item For the merged, isolated and in-cluster PBHs,
	\begin{align}
		(\bar{m}_{1}, \bar{m}_{2})_{\mpop,\ipop,\cpop} & = \unit [(1400, 712)] {\msun}, \\
		(\tilde{m}_{1}, \tilde{m}_{2})_{\mpop,\ipop,\cpop} & = \unit [(1110, 575)] {\msun}.
	\end{align}
	\item For the merged, bounded and in-ejecta PBH,
	\begin{align}
		(\bar{m}_{1}, \bar{m}_{2})_{\mpop,\bpop,\epop} & = \unit [(847, 503)] {\msun}, \\
		(\tilde{m}_{1}, \tilde{m}_{2})_{\mpop,\bpop,\epop} & = \unit [(788, 442)] {\msun}.
	\end{align}
\end{enumerate}

This shows again that less one-sided distributions of mass than in the binary pairs as well as the fact that more massive binary pairs are biased towards the cluster population, which is to be expected from dynamical friction, significant in this mass range as will be shown in Section~\ref{subsec: Mass segregation and dynamical friction}. Also, we do not compute the cases of the merged, bounded, in-cluster PBHs and merged, isolated, in-ejecta PBHs as with only 4 and 1 cases respectively we are far from having enough sampling to give meaningful results.



\subsubsection{Merger event rates}
\label{subsubsec: Merger event rates}


We have defined a merger between two bodies in the simulations in Section~\ref{subsubsec: Simulation merger trees}, and Tables~\ref{tab: Merger tree hierarchy levels} and \ref{tab: Merger mass distributions} show the frequency and population distribution of such events.

Note that while we on the one hand underestimate the number of PBH mergers in the simulations because of the fact that the $ N $-body solver computes the evolution of the system in a non-relativistic manner, and as such there are no inspiral events that enhance the number of mergers from close binary systems, on the other hand we may be overestimating merging at later times since a rich merger history at early times will concentrate more mass at the core of the cluster and depopulate faster its outer layers, resulting in a diminished likelihood for merging at later times.

In this case, and considering the 117 merger events identified in total in the $ N_{\rcap} = 5000 $ realisations during the last two runs, we can roughly estimate the merger event rate per time-run and realisation with
\begin{equation}
	\Gamma_{r}^{\mathrm{M},\mathrm{P}} = (l_{\mpop}-1) \frac {N_{r}^{\mathrm{M}, \mathrm{P}}-N_{r-1}^{\mathrm{M}, \mathrm{P}}} {\Delta t_{r}},
\end{equation}
where $ \mathrm{P} = (\cpop, \epop), (\ipop, \bpop) $ is the population of interest and $ \Delta t_{r} = t_{i^{+}}-t_{i^{-}} $ the time-run total period. Then, replacing with the results of Table~\ref{tab: Merger tree hierarchy levels} we find that,
\begin{align}
	\Gamma_{6}^{\mathrm{M}} (46 \leq t_{i} \leq 55) & = \unit [6.44 \times 10^{-4}] {Gyr^{-1}}, \\
	\Gamma_{7}^{\mathrm{M}} (56 \leq t_{i} \leq 64) & = \unit [1.82 \times 10^{-3}] {Gyr^{-1}}.
\end{align}

We can also compute these results segregating by their origin, which can be of great importance when identifying one such merger as mergers arising from a hyperbolic encounter between two isolated PBHs or an inspiral encounter between two bounded PBHs are easy to discern as they leave very different signals in detectors. Then
\begin{enumerate}[label=\roman*)]
	\item For all 51 mergers occurring from previously isolated PBHs, out of which 1 occurs in the sixth time-run and 50 do occur in the seventh time-run, we find that,
	\begin{align}
		\Gamma_{6}^{\mathrm{M},\cpop,\ipop} (46 \leq t_{i} \leq 55) & = \unit [1.61 \times 10^{-4}] {Gyr^{-1}}, \\
		\Gamma_{7}^{\mathrm{M},\cpop,\ipop} (56 \leq t_{i} \leq 64) & = \unit [8.05 \times 10^{-4}] {Gyr^{-1}}.
	\end{align}
	\item For all 66 mergers occurring originating in previously bounded systems, and out of which 3 occur in the sixth time-run and 63 do occur in the seventh time-run, we find that,
	\begin{align}
		\Gamma_{6}^{\mathrm{M},\epop,\bpop} (46 \leq t_{i} \leq 55) & = \unit [4.83 \times 10^{-4}] {Gyr^{-1}}, \\
		\Gamma_{7}^{\mathrm{M},\epop,\bpop} (56 \leq t_{i} \leq 64) & = \unit [1.01 \times 10^{-3}] {Gyr^{-1}}.
\end{align}
\end{enumerate}

Note that, in this last step we have taken into account the fact that most PBH mergers that are accounted in the ejecta do in fact occur within the cluster, only to be soon expelled to the ejecta as the simulations progress. This typically happens because of the large velocity kicks associated with the merger process, and so we have segregated the merged pair by whether they originate from single PBHs or bounded PBHs.

Also, these are the merger event rates for a non-relativistic evolution for a single cluster. In Section~\ref{sec: Background DM implications} we will compute this quantity for a comoving cosmological volume of $ \unit [1] {Gpc^{3}} $, and thus offer a more meaningful value for the merger event rate of PBHs, and one that can be interpreted as a minimum for the actual GR evolution.



\section{Dynamical distributions}
\label{sec: Dynamical distributions}

Now we proceed to extract in Section~\ref{subsec: Dynamical profiles} the density, mass, position and velocity distributions of PBHs evolution, slice by slice in time and segregated by the sub-population type. We do as well quantify the cluster evaporation rate in Section~\ref{subsubsec: Cluster mass segregation} and the degree of both mass segregation and dynamical friction within it in Section~\ref{subsubsec: Cluster dynamical friction}.



\subsection{Dynamical profiles}
\label{subsec: Dynamical profiles}

The PBH density profile, position profile and velocity profiles are shown in Figure~\ref{fig: Quasi-static profiles} segregated by cluster and ejecta populations for selected times binding the runs, while the fits to the data are shown in Table~\ref{tab: Quasi-static profiles}.



\subsubsection{Mass profile evolution}
\label{subsubsec: Mass profiles evolution}

The evolved mass distribution of PBHs, $ \rho_{\mass, i} \vert_{t} $, for $ i = \cpop, \epop $ the cluster and ejecta populations respectively, has not been found to do develop differentiated characteristics or significant evolution throughout the whole simulation period and so no Figure is given showing it. Shown, however, in Table~\ref{tab: Quasi-static profiles} are both the cluster and ejecta mass distribution parameters, which remain purely LN, and virtually indistinguishable from the masses IC $ (\mu_\mass, \sigma_\mass) = \unit [(2.0, 1.5)] {\msun} $.

This is a somewhat surprising result, as it would be naively expected from dynamical arguments that the ejected object's mass profile would noticeably skew towards the lower end of the initial mass range, while the cluster object mass profile would in turn skew towards the higher end of the mass range, given the large mass differences in between PBHs, spanning $ \order {4} $ orders of magnitude as seen in Figure~\ref{fig: Initial mass distribution}.

 The underlying reasoning behind mass segregation is as follows. As more massive PBHs do concentrate in the inner core of the cluster, as shown in Figure~\ref{subfig: Cluster mass segregation mass weighted}, and acquire in the process large kinetic energies, as shown in Figure~\ref{subfig: Cluster dynamical friction mass weighted}. Then, due to the large number of encounters with massive PBHs in the core, the least massive objects are in turn expelled with increased likelihood from the core. In the end this results in an emerging mass-depleted population of PBHs that would be segregated from the mass-enhanced population of PBHs in the cluster, and particularly, from the cluster core.

Indeed, this is in a very restricted sense what happens. Despite the fact that the ejecta population comprises about 66\% of PBHs at the last time-slice in the simulations, cases where the largest PBH ends up in the ejecta background are very rare, exceedingly so with respect to what would be expected if no correlation between mass and the likelihood of a PBH being slingshot away to the background is assumed.

In practice, this means that phenomena such as mass segregation and dynamical friction will be significant for about the top $ \order {10} $ most massive objects in the simulations, which are well into the tail of the mass distribution. For the remaining $ \order {1000} $ of objects, however, there is little evidence of significant mass segregation and dynamical friction, either in the cluster or in the ejecta populations. The cluster and ejecta-specific mass distributions do in fact skew towards larger and smaller masses respectively during the simulated period of time, but only so slightly, with their respective Log-Normal distribution parameters $ \mu^{\mass}_{C} $ and $ \sigma^{\mass}_{C} $ increasing only by
\begin{align}
	\Delta \mu^{\mass}_{C} &= \frac {\mu^{\mass}_{C,64}-\mu^{\mass}_{C,0}} {\mu^{\mass}_{C,0}} = 2.8\% ,\\
	\Delta \sigma^{\mass}_{C} &= \frac {\mu^{\mass}_{C,64}-\mu^{\mass}_{C,0}} {\mu^{\mass}_{C,0}} = 1.0\%,
\end{align}
in the case of the in-cluster PBHs between the initial $ t_{0} = \unit [0] {yr} $ and final simulation times $ t_{64} = \unit [1.38 \times 10^{10}] {yr} $, and $ \mu^{\mass}_{E} $ and $ \sigma^{\mass}_{E} $ conversely decreasing just by
\begin{align}
	\Delta \mu^{\mass}_{C} &= \frac {\mu^{\mass}_{C,64}-\mu^{\mass}_{C,46}} {\mu^{\mass}_{C,46}} = -0.20\%, \\
	\Delta \sigma^{\mass}_{C} &= \frac {\mu^{\mass}_{C,64}-\mu^{\mass}_{C,46}} {\mu^{\mass}_{C,46}} = -0.13\%,
\end{align}
in the case of the in-ejecta PBHs between the beginning of the $ t_{46} = \unit [1.38 \times 10^{8}] {yr} $ and the final simulation times $ t_{64} = \unit [1.38 \times 10^{10}] {yr} $. Note that, the difference is very small for both parameters and populations, of $ \order {0.01} $ in the case of mass expected values, and of $ \order {0.001} $ in the case of the standard deviation of the logarithmic mass at the last time-slice, $ t_{64} = \unit [1.38 \times 10^{10}] {yr} $, consistent with a small degree of mass segregation between the cluster and ejecta populations.



\subsubsection{Density profile evolution}
\label{subsubsec: Density profile evolution}

Given in Figure~\ref{subfig: Quasi-static cluster density profiles} and Figure~\ref{subfig: Quasi-static ejecta density profiles} are the PBH volume-factored density profiles $ x^{3} \rho_{i}(x) \vert_{t} $, with $ i = \cpop, \epop $ being the cluster and ejecta populations, computed at the five different time-slices that bound the last four time-runs plus the initial time.

The PL fits of the density profiles are shown in Table~\ref{tab: Quasi-static profiles}, once removed the outer 20\% of the data points at which the cluster and ejecta densities suddenly fall off as there is less and less sampling of the density distribution as one transitions from the cluster sphere-of-influence to the ejecta sphere-of-influence in the first case, and from the ejecta sphere-of-influence to the void in the second place.

Note that we have tried pseudo-isothermal, Navarro-Frenk-White, and Einasto profiles for the fits, but it has been found that the both the cluster and ejecta density profiles were best described by simple PLs, with the PL index as a free parameter. Two different observations are noteworthy.

The first is that, for time-slices posterior to the reconfiguration of bodies that follows the burn-in time $ t_{\text{BI}} $ when the IC is erased, the cluster volume-factored density profile exhibits a plateau up to the time-dependent cut-off radius $ x_{\cpop}^{\max} (t) $ where the cluster is sparse enough, that effectively can be considered as the cluster radius.

The second concerns the ejecta objects. The volume-factored distribution density profile of these, unlike in the previous case, does not exhibit any plateau, but rather a peak at a time-increasing length scale corresponding to the median distance travelled from the cluster barycenter. Before the peak, the actual ejecta density of objects gradually increases achieving a maximum at the cluster boundary. After the peak, the density PBHs falls abruptly, sampling only the more infrequent high-velocity PBHs. Last, the density profile arrives as well at each time-slice to a time-increasing cut-off length scale $ x_{\epop}^{\max} (t) $ that corresponds to the maximum reach of PBHs expelled from the cluster.

Taking into account these two then we find that:
\begin{enumerate}[label=\roman*)]
	\item The proper cluster density profile up to the cut-off distance roughly behaves as as inverse nearly-cubic PL:
	\begin{equation}
		\rho_{\cpop}(x) \vert_{t} \propto x^{-2.8}, \ t > t_{\text{BI}}, \ x < x_{\cpop}^{\max} (t).
		\label{eq: Characteristic density cluster}	
	\end{equation}
	\item The actual ejecta density profile up to the cut-off distance roughly behaves again as an inverse PL, with a slightly smaller power than in the cluster case:
	\begin{equation}
		\rho_{\epop} (x) \vert_{t} \propto x^{-2.6}, \ t > t_{\text{BI}}, \ x < x_{\cpop}^{\max} (t).
		\label{eq: Characteristic density ejecta}	
	\end{equation}
\end{enumerate}

Note that the volume-factor $ x^{3} $ that multiplies the actual density profiles $ \rho (x) \vert_{t} $ masks the large range covered by the density within the cluster and ejecta spheres of influence, as distances extend to $ x_{\cpop}^{\max} \rightarrow \unit [10^{4}] {pc} $ and $ x_{\epop}^{\max} \rightarrow \unit [10^{4}] {pc} $ respectively. In practice, we find in our simulations, after averaging all realisations, that densities are in the range of
\begin{enumerate}[label=\roman*)]
	\item In the case of the initial time-slice of cluster objects:
	\begin{align}
		& \rho_{\cpop} (x) \vert_{t_{0}} \in \unit [(10^{6}, 10^{-6})] {\msun \mbox pc^{-3}}, \\
		& \qquad x \in \unit [(0.001, 10)] {pc}, \\
		& \qquad t_{0} = \unit [0] {yr}.
	\end{align}
	\item In the case of the last time-slice of cluster objects:
	\begin{align}
		& \rho_{\cpop} (x) \vert_{t_{64}} \in \unit [(10^{-6}, 10^{-22})] {\msun \mbox pc^{-3}}, \\
		& \qquad x \in \unit [(10, 10^{4})] {pc}, \\
		& \qquad t_{64} = \unit [1.38 \times 10^{10}] {yr}.
	\end{align}
	\item In the case of the first sufficiently populated time-slice of ejecta objects, with $ N_{\epop} (t) > 100 $ :
	\begin{align}
		& \rho_{\epop} (x) \vert_{t_{28}} \in \unit [(10^{5}, 10^{-7})] {\msun \mbox pc^{-3}} ,\\
		& \qquad x \in \unit [(0.001, 10)] {pc}, \\
		& \qquad t_{28} = \unit [1.38 \times 10^{6}] {yr}.
	\end{align}
	\item In the case of the final time-slice of ejecta objects:
	\begin{align}
		& \rho_{\epop} (x) \vert_{t_{64}} \in \unit [(10^{-8}, 10^{-32})] {\msun \mbox pc^{-3}}, \\
		& \qquad x \in \unit [(10, 10^{7})] {pc}, \\
		& \qquad t_{64} = \unit [1.38 \times 10^{10}] {yr}.
	\end{align}
\end{enumerate}

It is worth noting, however, that cluster layers with densities smaller than $ \rho = \unit [0.08] {\msun \mbox pc^{-3}} $ (see Ref.~\cite{Bovy:2017}) will already be disrupted at solar neighbourhood-like environments, and stripped of PBHs. Therefore, considering the more realistic case of clusters not in isolation but on top of a background mass density and in interaction with other clusters will put more stringent bounds both to the cluster and ejecta radial reach, coinciding with the local background matter density at their outer layers.



\subsubsection{Position profile evolution}
\label{subsubsec: Position profile evolution}

The PBH evolved position distributions, $ \rho_{\pos, i} \vert_{t} $, with $ i = \cpop, \epop $ being the cluster and ejecta populations, are given in Figure~\ref{subfig: Quasi-static cluster position profiles} and Figure~\ref{subfig: Quasi-static ejecta density profiles} respectively at the five different time-slices that bound the last four time-runs and the IC. Log-Normal fits to these position profiles are shown in Table~\ref{tab: Quasi-static profiles}, when such fit is possible. There, a number of features become apparent.

The first is that right after the ICs, the cluster PBHs compress around $ \unit [1.38 \times 10^{6}] {yr} $, only to later rebound and subsequently puffs up. At this time the PBHs do, on average, falling closest than they will ever do during the whole simulation period afterwards towards the cluster barycenter, approaching it to distances up to $ \unit [\order {0.1}] {pc} $. This short compression-expansion phase is due to the choice of Maxwell-Boltzmann ICs in the process described in the previous section, which do not survive after this time.

Secondly, that as the cluster puffs-up, its characteristic length scale increases greatly as described in Eq.~\eqref{eq: Cluster mean total radius}, with its radius expanding by three orders of magnitude, as seen for the last four time-slices in Figure~\ref{subfig: Quasi-static cluster position profiles}, after a brief phase of slight contraction on the second time-run after the IC. This transition happens right around the burn-in time $ t_{\bicap} = \unit [2.64 \times 10^{5}] {yr} $, and marks the transition from the Maxwell-Boltzmann IC that best describes the IC to the Log-Normal time slices later on, in the last four time-runs.

This late-time cluster positions have been fitted to a variety of functions, including Maxwell-Boltzmann and Log-Normal distributions, finding out that it is the latter that best describes the data, in particular to accommodate the large tails in the later position distribution that develops as the cluster grows a subpopulation of far-off components with very elongated orbits with respect to the cluster barycenter, with periods comparable to the entire simulation time $ \delta t_{\cpop} \vert_{x \gg x_{\cpop}^{\max}} \approx t_{64} = \unit [1.38 \times 10^{10}] {yr} $ and nearly unit eccentricities $ e_{\cpop} \vert_{x \gg x_{\cpop}^{\max}} \rightarrow 1^{-} $.

Cluster positions are constrained to be in the range
\begin{align}
	& \unit [0.02] {pc} \leq x_{\cpop} (t) \leq \unit [2 \times 10^{3}] {pc} \ \forall \ t > t_{0}, \\
	& \qquad x_{\cpop} (\unit [0] {yr}) \in \unit [(0.02, 20)] {pc}, \\
	& \qquad x_{\cpop} (\unit [1.38 \times 10^{10}] {yr}) \in \unit [(10, 2 \times 10^{3})] {pc},
\end{align}
while the width of the constraint at a particular time is found to always be roughly of two orders of magnitude between the minimum and maximum values that enclose 99\% of the variable's distribution.

The cut-off distance, akin to the cluster radius, roughly behaves as
\begin{align}
	& x_{\cpop}^{\max} (t)
	 \approx \unit [\beta_{\pos, \cpop} t^{\alpha_{\pos, \cpop}}] {pc}, \quad t > t_{\text{BI}}, \\
	& \qquad \beta_{\pos ,\cpop} = \unit [(2.96 \pm 0.56) \times 10^{-4}] {pc}, \\
	& \qquad \alpha_{\pos, \cpop} = 0.639 \pm 0.010.
	\label{eq: Characteristic position cluster}
\end{align}

Thirdly, that the stability of these PBHs in orbits bound to the cluster is almost guaranteed in our simulations, since the cluster lives in isolation, and the likelihood of an encounter that may transfer enough kinetic energy to overcome the cluster escape velocity is very low, as the ejected objects travel the cluster periphery where this far-off objects live very fast.

However, it will be seen in Section~\ref{subsubsec: Cluster DM energy density} that at late time-slices, in the very last run, the stability of cluster objects and particularly those on the periphery, far off the cluster barycenter, is far from assured in realistic scenarios where the cluster does not live in isolation, and inter-cluster interactions become likely.

A last interesting feature is that the ejecta reach, or alternatively, the ejecta sphere-of-influence, grows linearly in time, as it is natural, since the ejecta sphere-of-influence is bounded by the fastest bodies that escape the cluster with isotropic and asymptotically constant velocities. As in the cluster case, the characteristic length scale increases linearly in time in the manner described in Eq.~\eqref{eq: Ejecta mean total radius}.

Ejecta positions are constrained to be in the range
\begin{align}
	& \unit [0.4] {pc} \leq x_{\epop} (t) \leq \unit [10^{5}] {pc} \ \forall \ t > t_{0}, \\
	& \qquad x_{\epop} (\unit [1.38 \times 10^{6}] {yr}) \in \unit [(0.8, 10)] {pc}, \\
	& \qquad x_{\epop} (\unit [1.38 \times 10^{10}] {yr}) \in \unit [(2, 100) \times 10^{3}] {pc},
\end{align}
while the width of the previous constraint at any given time in the quasi-static phase is found to always be of one order of magnitude between the minimum and maximum bound that enclose 99\% of the individual positions.

The cut-off distance, which can be understood as the reach of the ejecta sphere-of-influence, scales as
\begin{align}
	& x_{\epop}^{\max} (t)
	 \approx \unit [\beta_{\pos, \epop} t^{\alpha_{\pos, \epop}}] {pc}, \quad t > t_{\text{BI}}, \\
	& \qquad \beta_{\pos ,\epop} = \unit [(3.70 \pm 0.80) \times 10^{-5}] {pc}, \\
	& \qquad \alpha_{\pos ,\epop} = 0.917 \pm 0.012.
	\label{eq: Characteristic position ejecta}
\end{align}

Late time ejecta positions have been as well been fitted to a large collection of distributions, including the MB, and Multi-Normal distributions.

However, and unlike the cluster case, no good single parametric fit has been found to accurately describe the at all post burn-in times simultaneously. The reason why the fits fail, particularly at late times, is that as PBHs are constantly emitted from the cluster at all times, the low end of the ejecta position distribution is constantly repopulated, which does not behave well with the exponentially-suppressed Log-Normal distribution in the logarithmic variable at that scale, while the long tails of the ejecta position distribution due to high speed objects are ill-described by the exponential suppression on the variable's higher end of the Maxwell-Boltzmann distribution, itself only good to describe the IC and the immediately following time slices.

Note that these ejecta PBHs remain in rectilinear asymptotically-uniform motion trajectories, gradually approaching their end-velocity in our simulations at infinity, and no longer taking part in the dynamics of the simulations, as seen in Figure~\ref{subfig: Time velocities histogram}. This greatly increases the computational speed of the simulations at late times, since the characteristic discrete dynamical time-step in the $ N $-body solver algorithm increases accordingly.

However, it will be seen in Section~\ref{subsubsec: Ejecta DM energy density} that again in the last runs, the stability of such ejecta objects is no longer assured in realistic scenarios where the cluster does not live in isolation, as interactions between the ejecta objects and other clusters become likely at even earlier times.



\subsubsection{Velocity profile evolution}
\label{subsubsec: Velocity profile evolution}

\begin{figure*}[t!]
	\centering
	%
	%
	%
	\subfloat[Cluster density profile, by a volume factor, $ x^{3} \rho (x)_{\cpop} $.]{
	\hspace*{-1.00cm}
	\includegraphics[width = 0.50\textwidth]{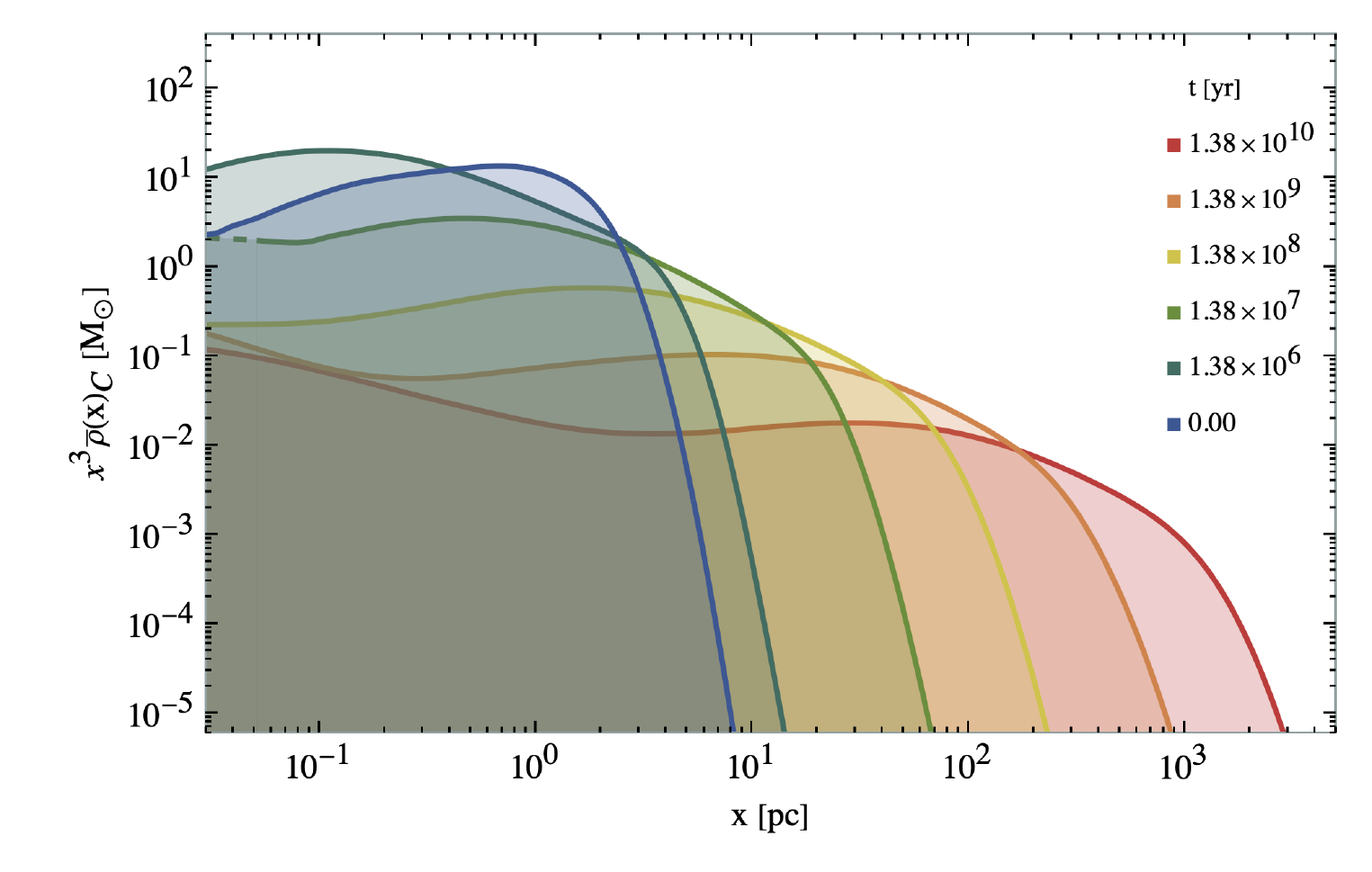}
	\label{subfig: Quasi-static cluster density profiles}}
	\subfloat[Ejecta density profile, by a volume factor, $ x^{3} \rho (x)_{\epop} $.]{
	\hspace*{-0.00cm}
	\includegraphics[width = 0.50\textwidth]{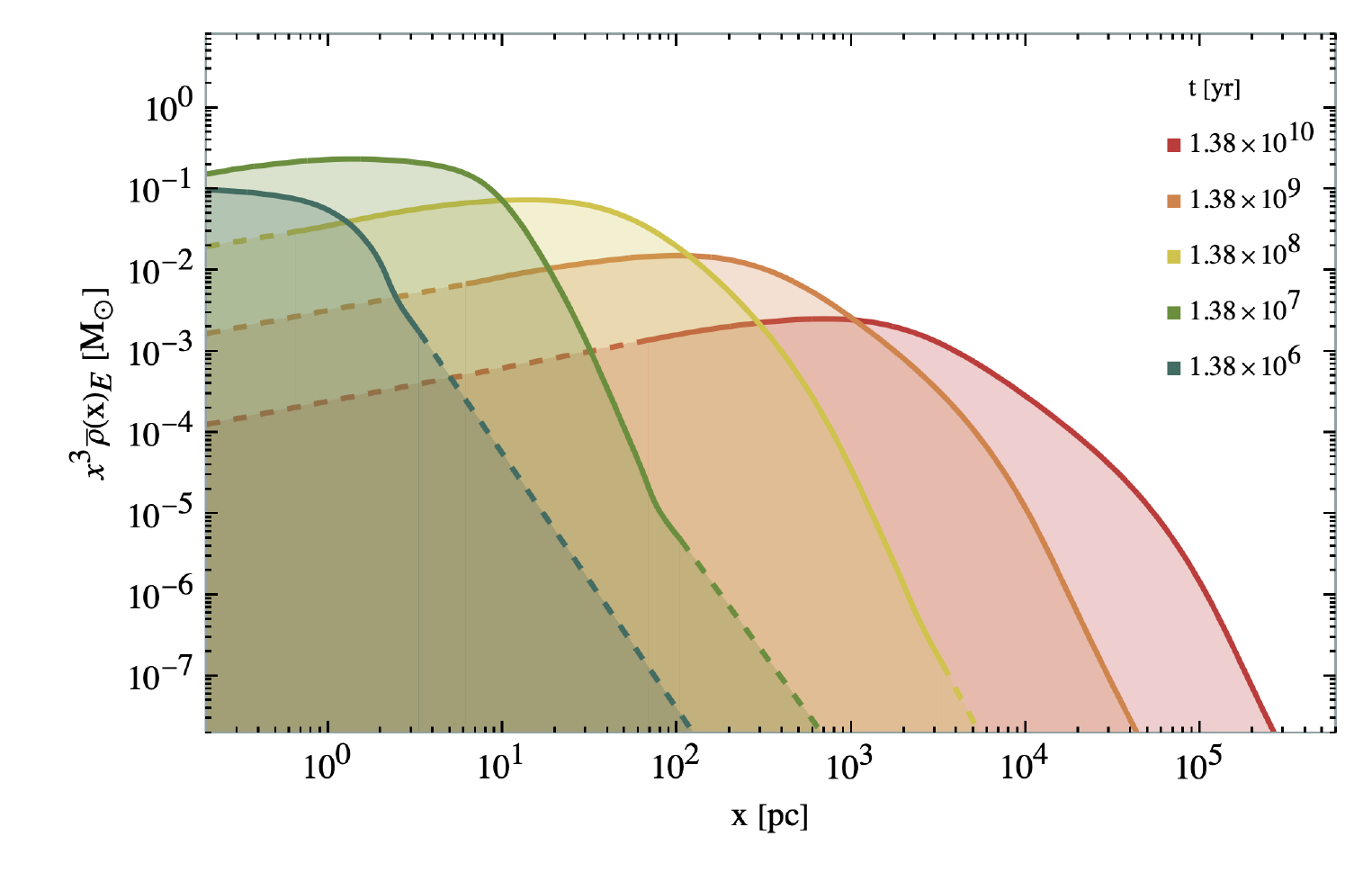}
	\label{subfig: Quasi-static ejecta density profiles}}
	\newline
	\subfloat[Cluster position profile, $ x_{\cpop} $, rescaled.]{
	\hspace*{-1.00cm}
	\includegraphics[width = 0.50\textwidth]{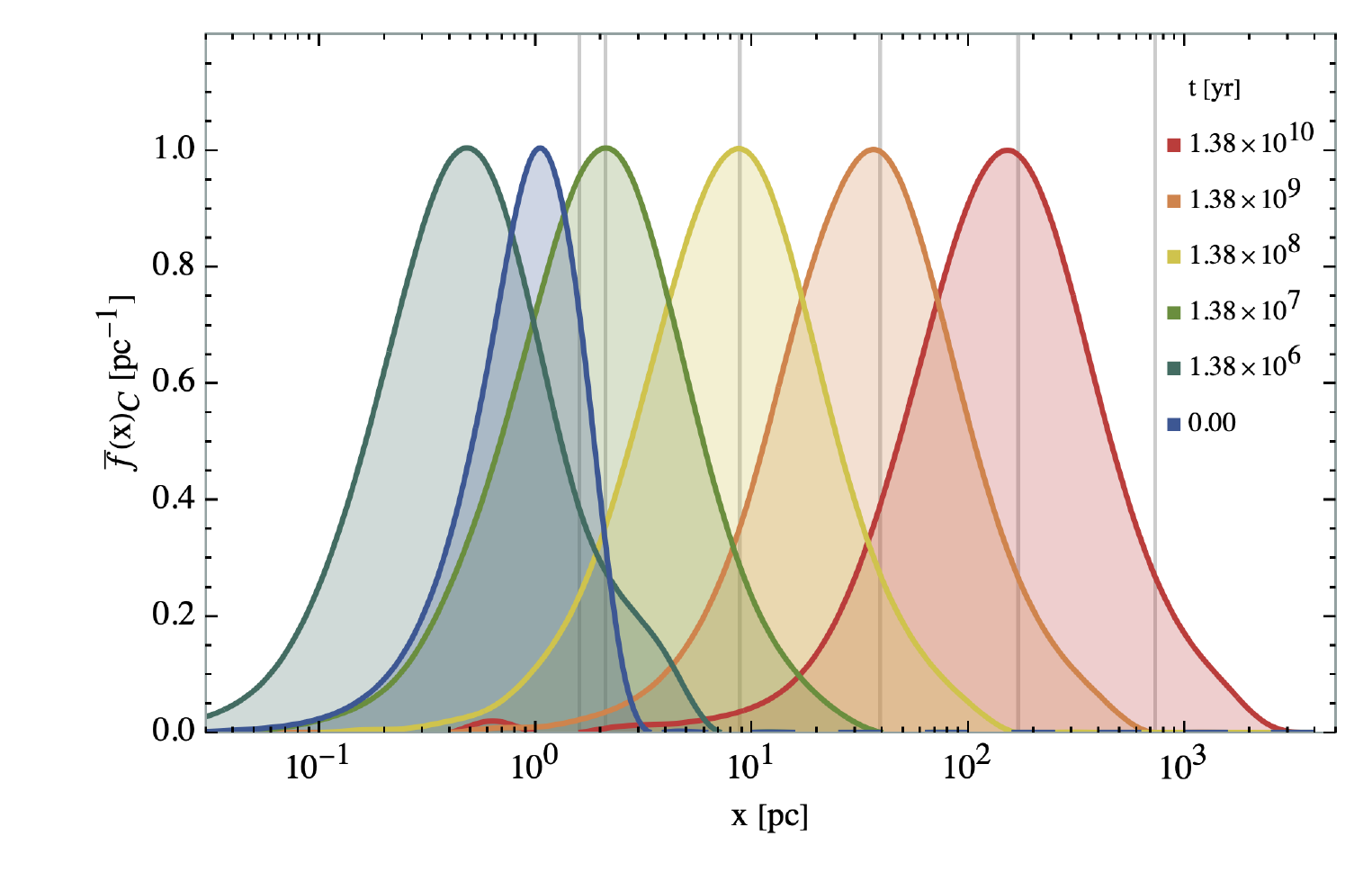}
	\label{subfig: Quasi-static cluster position profiles}}
	\subfloat[Ejecta density profile, $ x_{\epop} $, rescaled.]{
	\hspace*{-0.00cm}
	\includegraphics[width = 0.50\textwidth]{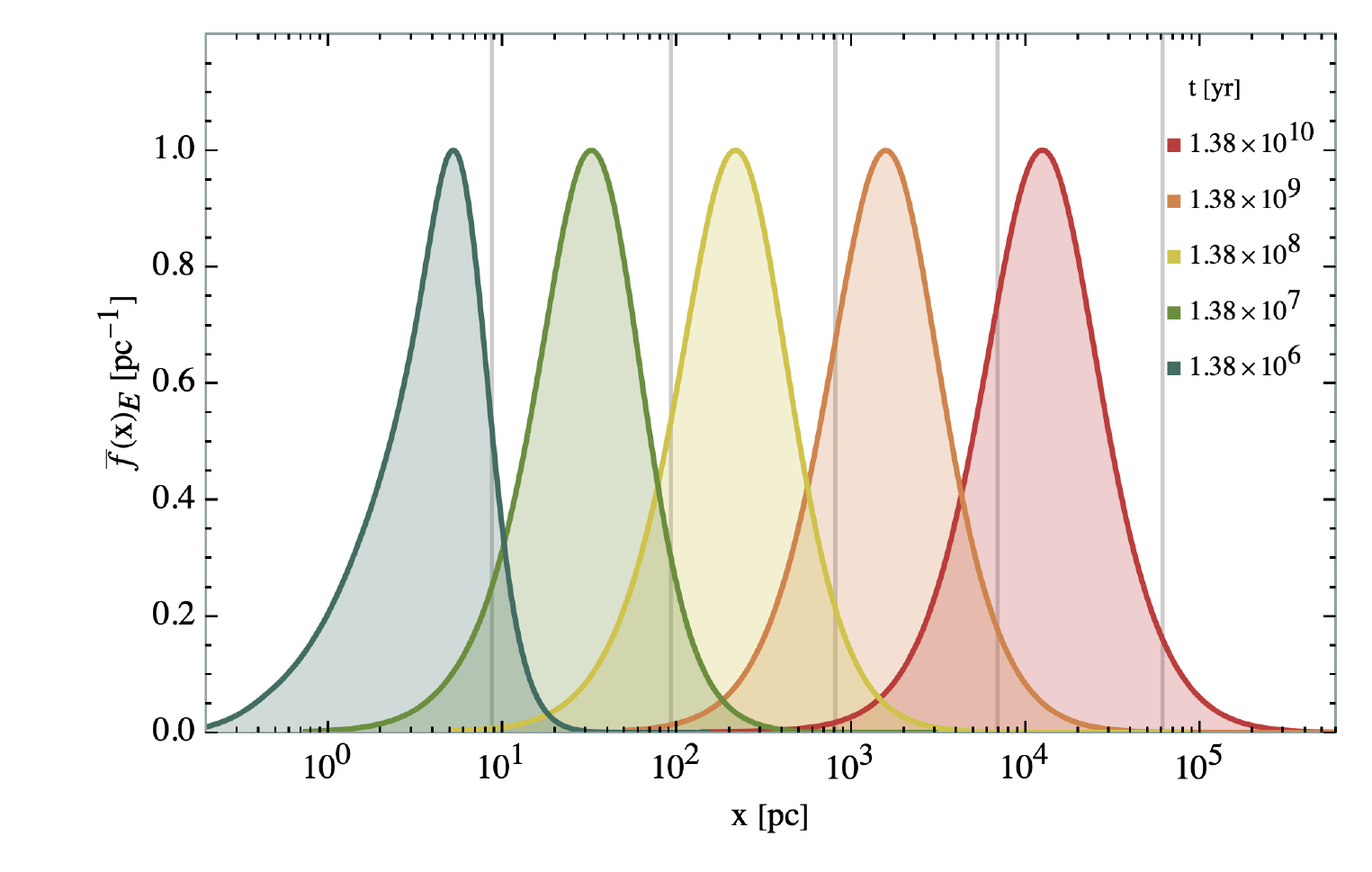}
	\label{subfig: Quasi-static ejecta position profiles}}
	\newline
	\subfloat[Cluster velocity profile, $ v_{\cpop} $, rescaled.]{
	\hspace*{-1.00cm}
	\includegraphics[width = 0.50\textwidth]{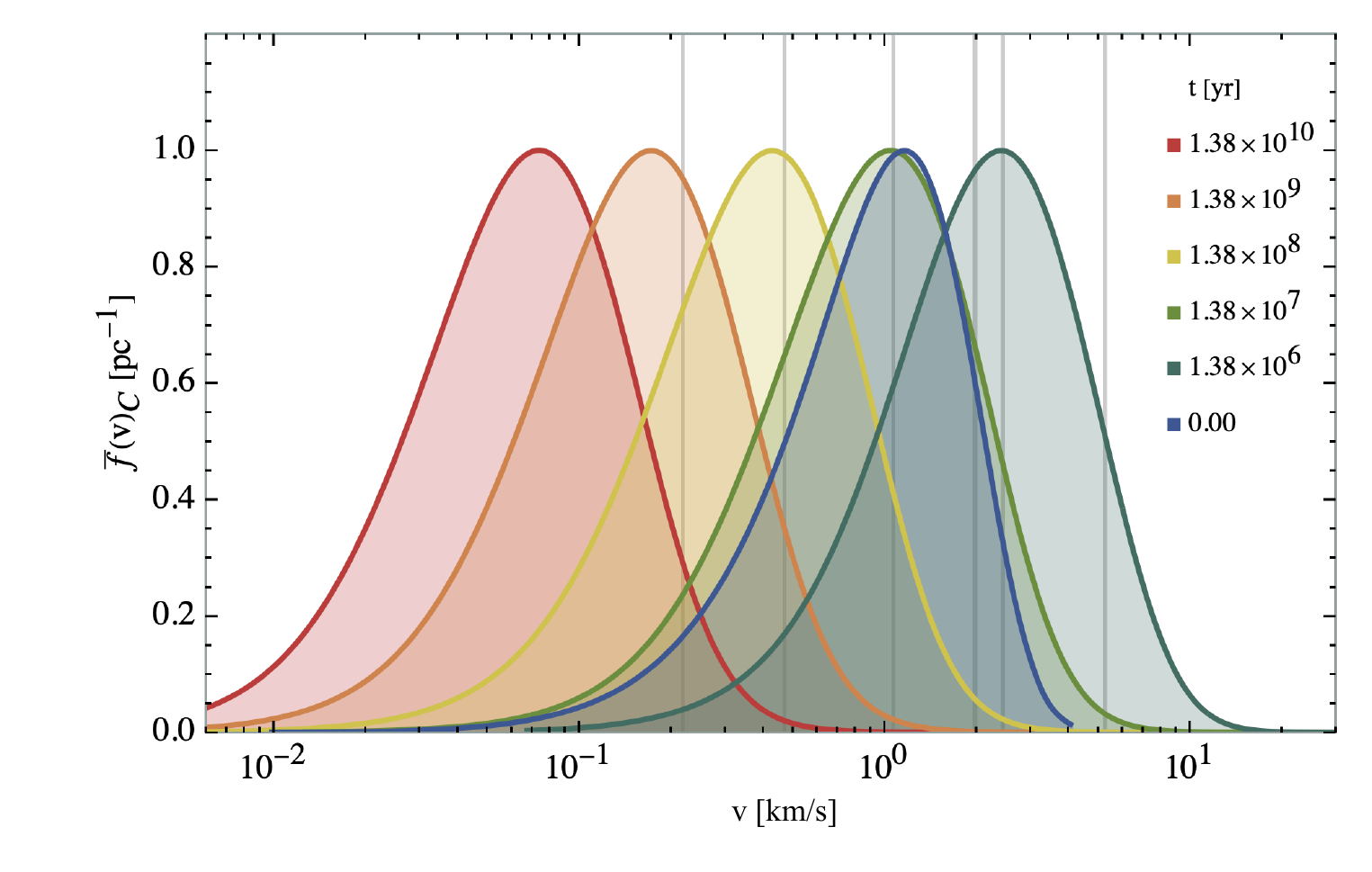}
	\label{subfig: Quasi-static cluster velocity profiles}}
	\subfloat[Ejecta density profile, $ v_{\epop} $, rescaled.]{
	\hspace*{-0.00cm}
	\includegraphics[width = 0.50\textwidth]{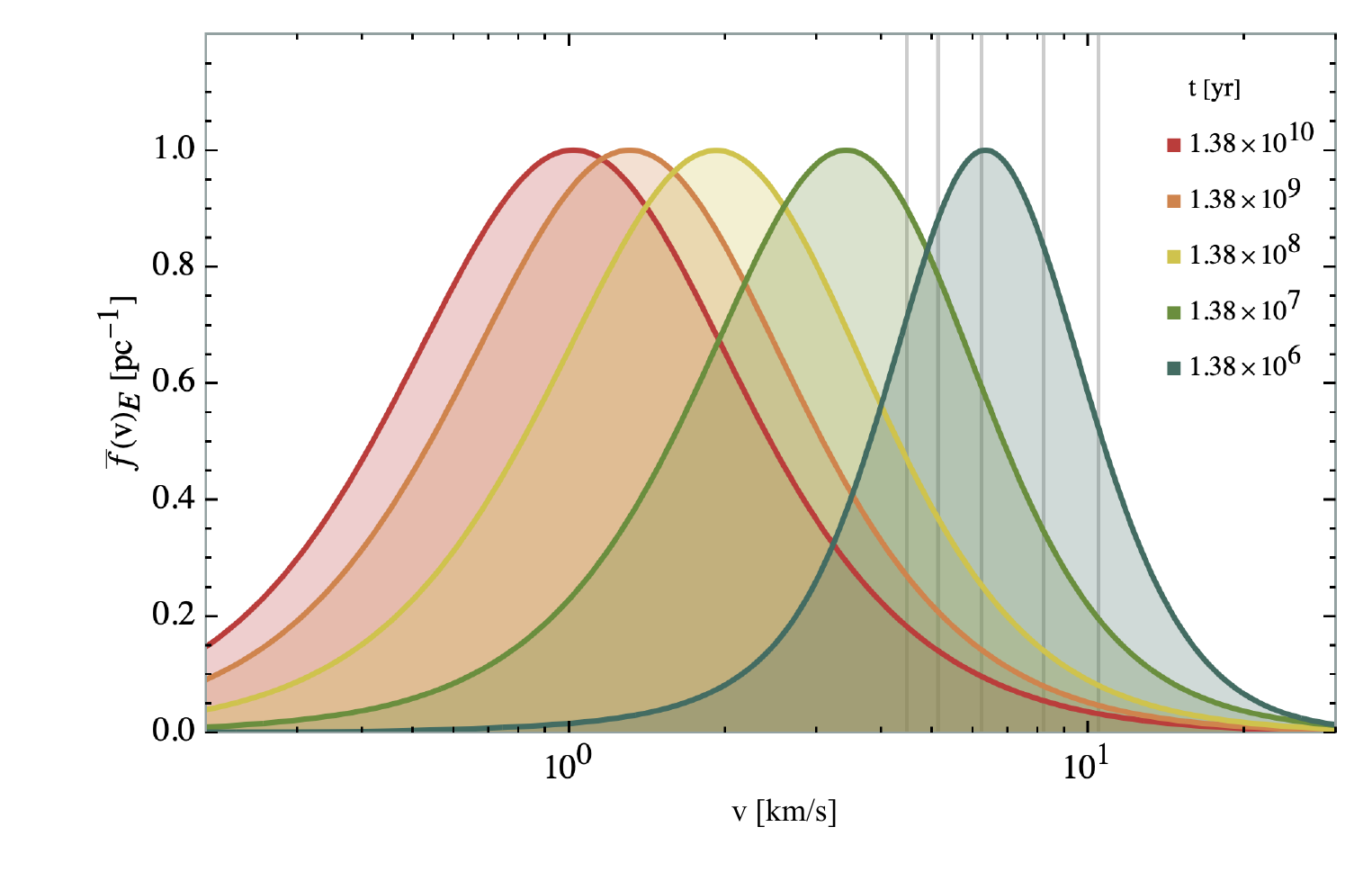}
	\label{subfig: Quasi-static ejecta velocity profiles}}
	\caption{
	Density, position and velocity quasi-static profiles segregated by population (cluster and ejecta) type at selected times.
	Fits for the mass, position, velocity and density probability distributions are given in Table~\ref{tab: Quasi-static profiles}. Note the mass distribution is not displayed since, first, the absence of a significant number of mergers implies that the evolution of the total mass profile of objects has an almost negligible evolution, and second, that as seen in said Table~\ref{tab: Quasi-static profiles}, the cluster and ejecta mass profiles do not segregate significantly over time.
	}
	\label{fig: Quasi-static profiles}
\end{figure*}

The PBH evolved velocity distribution, $ \rho_{\vel, i} \vert_{t} $, for $ i = \cpop, \epop $ being the cluster and ejecta populations, is given in Figure~\ref{subfig: Quasi-static cluster velocity profiles} and Figure~\ref{subfig: Quasi-static ejecta velocity profiles} respectively at the five different time-slices that bound the last four time-runs and the IC. Log-Normal fits to these velocity profiles are shown in Table~\ref{tab: Quasi-static profiles}, when such fits are appropriate, although, like it was the case with the position profiles computation of Section~\ref{subsubsec: Position profile evolution}, other fits to Maxwell-Boltzmann and Rayleigh distributions have been tried and proved to worse describe the data.

From Figure~\ref{subfig: Quasi-static cluster velocity profiles} and \ref{subfig: Quasi-static ejecta velocity profiles} it is apparent that the range covered by velocities for both the cluster and ejecta PBHs is much narrower than it was for masses or positions, as it would be expected from equipartition of energy arguments. In particular, cluster velocities are constrained to a narrow range of
\begin{align}
	& \unit [0.01] {km / s} \leq v_{\epop} (t) \leq \unit [10] {km / s} \ \forall \ t > t_{0}, \\
	& \qquad v_{\cpop} (\unit [0] {yr}) \in \unit [(0.2, 3)] {km / s}, \\
	& \qquad v_{\cpop} (\unit [1.38 \times 10^{10}] {yr}) \in \unit [(0.01, 0.3)] {km / s},
\end{align}
and the width of the constraint at any given time is estimated to be approximately of one order of magnitude between the minimum and maximum bound that enclose 99\% of the individual velocities.

A feature in Figure~\ref{subfig: Quasi-static cluster velocity profiles} stands out: right after the simulations starts, the cluster PBHs accelerate and reach the largest velocities they achieve on average during the whole simulated period of even $ \unit [\order {10}] {km /s} $ as the cluster contracts and then later rebounds due to the inward fall process described in Section~\ref{subsubsec: Position profile evolution}. After this phase, nonetheless, cluster velocities continue to decrease indefinitely as PBHs continue to expand and be diluted.

The time-dependent characteristic velocity scale is then
\begin{align}
	& v_{\cpop}^{\max} (t)
	 \approx \unit [\beta_{\vel, \cpop} t^{\alpha_{\vel, \cpop}}] {pc}, \quad t > t_{\text{BI}}, \\
	& \qquad \beta_{\vel, \cpop} = \unit [(383 \pm 259) \times 10^{-4}] {km / s}, \\
	& \qquad \alpha_{\vel, \cpop} = -0.276 \pm 0.041,
	\label{eq: Characteristic velocity cluster}
\end{align}
Ejecta velocities are even more constrained than cluster velocities to a range of
\begin{align}
	& \unit [0.08] {km / s} \leq v_{\epop} (t) \leq \unit [20] {km / s} \ \forall \ t > t_{0}, \\
	& \qquad v_{\epop} (\unit [1.38 \times 10^{6}] {yr}) \in \unit [(2, 20)] {pc}, \\
	& \qquad v_{\epop} (\unit [1.38 \times 10^{10}] {yr}) \in \unit [(0.08, 8) \times 10^{3}] {pc},
\end{align}
while the constraint width between the minimum and maximum bounds that enclose 99\% of the distribution velocities is estimated to be approximately of half an order of magnitude at any given time after the burn-in period ends.

The characteristic velocity scale is even less time-dependent, and is given by
\begin{align}
	& v_{\epop}^{\max} (t)
	 \approx \unit [\beta_{\vel, \epop} t^{\alpha_{\vel, \epop}}] {pc}, \quad t > t_{\text{BI}}, \\
	& \qquad \beta_{\vel, \epop} = \unit [(29.6 \pm 4.8) \times 10^{-4}] {km / s}, \\
	& \qquad \alpha_{\vel, \epop} = -0.0826 \pm -0.0094.
	\label{eq: Characteristic velocity ejecta}
\end{align}

The reason for this lack of time dependence in the ejecta velocity distribution is due to the balance between two competition effects. On the one hand, the dynamical relaxation of the cluster as it expands produces an ever increasingly slower motion of PBHs within the cluster. On the other hand, the cluster escape velocity, which is well approximated by
\begin{equation}
	V^\mathrm{esc}_{\cpop} ( X^{*}_{\cpop}, t) = \left( 2 G_{\mathrm{N}} \frac {M (x < X^{*}_{\cpop})} {X^{*}_{\cpop}} \right)^{1/2},
\end{equation}
given the symmetries of out clusters, and where $ X^{*}_{\cpop} (t) $ is the time-dependent cluster radius and
\begin{equation}
	M (x < X^{*}_{\cpop}) = \sum_{i = 1}^{N_\ocap} m_{i} \theta (x < X^{*}),
\end{equation}
is the total mass contained in the sphere bounded by it, decreases over time. In the end, the difference between the decreasing kinetic energy kick to escape the cluster and the also decreasing typical cluster PBH kinetic energies is roughly constant and naturally produces in turn roughly constant ejecta PBH velocities at infinity.


\begin{table*}[t!]
	\centering
	\begin{tabular}[c]{| l l | l l l l | l l l l |}
	\hline
	\quad & \quad &
	\multicolumn{4}{c |}{Cluster density distributions:} &
	\multicolumn{4}{c |}{Ejecta density distributions:} \\
	[0.5ex]
	\hline\hline
	$ i $ & $ \unit [t_{i}] {[yr]} $ &
	Fit: & $ r^{2}_{i,\cpop,\rho} $ &
	$ \unit {\beta^{(\rho)}_{i,\cpop}} [pc] $ & $ \alpha^{(\rho)}_{i,\cpop} $ &
	Fit: & $ r^{2}_{i,\epop,\rho} $ &
	$ \unit {\beta^{(\rho)}_{i,\epop}} [pc] $ & $ \alpha^{(\rho)}_{i,\epop} $ \\
	[0.5ex]
	\hline
	$ 0 $ & $ 0 $ &
	[PL] & $ 0.998 $ &
	$ 3.84 \pm 0.75 $ & $ -2.879 \pm 0.050 $ &
	$ - $ & $ - $ &
	$ - $ & $ - $ \\
	$ 28 $ & $ 13.8 \times 10^{6} $ &
	[PL] & $ 1.000 $ &
	$ 5.85 \pm 0.90 $ & $ -3.023 \pm 0.031 $ &
	[PL] & $ 0.999 $ &
	$ 0.183 \pm 0.027 $ & $ -2.898 \pm 0.027 $ \\
	$ 37 $ & $ 13.8 \times 10^{7} $ &
	[PL] & $ 1.000 $ &
	$ 1.93 \pm 0.17 $ & $ -3.005 \pm 0.030 $ &
	[PL] & $ 1.000 $ &
	$ 0.2487 \pm 0.0019 $ & $ -2.7023 \pm 0.0046 $ \\
	$ 46 $ & $ 13.8 \times 10^{8} $ &
	[PL] & $ 1.000 $ &
	$ 0.113 \pm 0.016 $ & $ -3.257 \pm 0.022 $ &
	[PL] & $ 1.000 $ &
	$ 0.035018 \pm 0.000062 $ & $ -2.6077 \pm 0.0047 $ \\
	$ 55 $ & $ 13.8 \times 10^{9} $ &
	[PL] & $ 1.000 $ &
	$ 0.1057 \pm 0.0097 $ & $ -3.246 \pm 0.019 $ &
	[PL] & $ 1.000 $ &
	$ (3.166 \pm 0.013)0.001 $ & $ -2.5875 \pm 0.0022 $ \\
	$ 64 $ & $ 13.8 \times 10^{10} $ &
	[PL] & $ 1.000 $ &
	$ 0.0588 \pm 0.0020 $ & $ -3.2288 \pm 0.0074 $ &
	[PL] & $ 1.000 $ &
	$ (2.592 \pm 0.046)10^{-4} $ & $ -2.6072 \pm 0.0041 $ \\
	\hline
	\multicolumn{10}{c}{ $ \quad $ } \\
	\hline
	\quad & \quad &
	\multicolumn{4}{c |}{Cluster mass distributions:} &
	\multicolumn{4}{c |}{Ejecta mass distributions:} \\
	[0.5ex]
	\hline\hline
	$ i $ & $ \unit [t_{i}] {[yr]} $ &
	Fit: & $ r^{2}_{i,\cpop,\mass} $ &
	$ \unit [\mu^{(\mass)}_{i,\cpop}] {[pc]} $ &
	$ \sigma^{(\mass)}_{i,\cpop} $ &
	Fit: & $ r^{2}_{i,\epop,\mass} $ &
	$ \unit [\mu^{(\mass)}_{i,\epop}] {[pc]} $ &
	$ \sigma^{(\mass)}_{i,\epop} $ \\
	[0.5ex]
	\hline
	$ 0 $ & $ 0 $ &
	[LN] & $ 0.997 $ &
	$ 2.115 \pm 0.019 $ & $ 1.602 \pm 0.010 $ &
	$ - $ & $ - $ &
	$ - $ & $ - $ \\
	$ 28 $ & $ 13.8 \times 10^{6} $ &
	[LN] & $ 0.997 $ &
	$ 2.126 \pm 0.019 $ & $ 1.604 \pm 0.011 $ &
	[LN] & $ 0.998 $ &
	$ 2.039 \pm 0.020 $ & $ 1.584 \pm 0.011 $ \\
	$ 37 $ & $ 13.8 \times 10^{7} $ &
	[LN] & $ 0.997 $ &
	$ 2.128 \pm 0.022 $ & $ 1.605 \pm 0.012 $ &
	[LN] & $ 0.997 $ &
	$ 2.093 \pm 0.020 $ & $ 1.598 \pm 0.011 $ \\
	$ 46 $ & $ 13.8 \times 10^{8} $ &
	[LN] & $ 0.996 $ &
	$ 2.134 \pm 0.023 $ & $ 1.605 \pm 0.013 $ &
	[LN] & $ 0.998 $ &
	$ 2.097 \pm 0.019 $ & $ 1.600 \pm 0.011 $ \\
	$ 55 $ & $ 13.8 \times 10^{9} $ &
	[LN] & $ 0.995 $ &
	$ 2.152 \pm 0.025 $ & $ 1.611 \pm 0.014 $ &
	[LN] & $ 0.997 $ &
	$ 2.094 \pm 0.019 $ & $ 1.598 \pm 0.011 $ \\
	$ 64 $ & $ 13.8 \times 10^{10} $ &
	[LN] & $ 0.994 $ &
	$ 2.175 \pm 0.027 $ & $ 1.617 \pm 0.015 $ &
	[LN] & $ 0.997 $ &
	$ 2.093 \pm 0.020 $ & $ 1.598 \pm 0.011 $ \\
	\hline
	\multicolumn{10}{c}{ $ \quad $ } \\
	\hline
	\quad & \quad &
	\multicolumn{4}{c |}{Cluster position distributions:} &
	\multicolumn{4}{c |}{Ejecta position distributions:} \\
	[0.5ex]
	\hline\hline
	$ i $ & $ \unit [t_{i}] {[yr]} $ &
	Fit: & $ r^{2}_{i,\cpop,\pos} $ &
	$ \unit [a^{(\pos)}_{i,\cpop}] {[pc]} $ &
	$ - $ &
	$ - $ & $ - $ &
	$ - $ &
	$ - $ \\
	[0.5ex]
	\hline
	$ 0 $ & $ 0 $ &
	[MB] & $ 0.988 $ &
	$ 0.813 \pm 0.021 $ & $ - $ &
	$ - $ & $ - $ &
	$ - $ & $ - $ \\
	\hline
	$ i $ & $ \unit [t_{i}] {[yr]} $ &
	Fit: & $ r^{2}_{i,\cpop,\pos} $ &
	$ \unit [\mu^{(\pos)}_{i,\cpop}] {[pc]} $ &
	$ \sigma^{(\pos)}_{i,\cpop} $ &
	Fit: & $ r^{2}_{i,\epop,\pos} $ &
	$ \unit [\mu^{(\pos)}_{i,\epop}] {[pc]} $ &
	$ \sigma^{(\pos)}_{i,\epop} $ \\
	[0.5ex]
	\hline
	$ 28 $ & $ 13.8 \times 10^{6} $ &
	[LN] & $ 0.994 $ &
	$ 0.448 \pm 0.078 $ & $ 1.043 \pm 0.036 $ &
	[LN] & $ 0.994 $ &
	$ 2.368 \pm 0.085 $ & $ 0.943 \pm 0.044 $ \\
	$ 37 $ & $ 13.8 \times 10^{7} $ &
	[LN] & $ 0.997 $ &
	$ 1.791 \pm 0.086 $ & $ 1.014 \pm 0.044 $ &
	[LN] & $ 0.996 $ &
	$ 4.37 \pm 0.12 $ & $ 0.909 \pm 0.062 $ \\
	$ 46 $ & $ 13.8 \times 10^{8} $ &
	[LN] & $ 0.999 $ &
	$ 3.306 \pm 0.088 $ & $ 1.052 \pm 0.040 $ &
	[LN] & $ 0.998 $ &
	$ 6.50 \pm 0.12 $ & $ 1.003 \pm 0.063 $ \\
	$ 55 $ & $ 13.8 \times 10^{9} $ &
	[LN] & $ 0.999 $ &
	$ 4.781 \pm 0.087 $ & $ 1.066 \pm 0.038 $ &
	[LN] & $ 0.999 $ &
	$ 8.54 \pm 0.12 $ & $ 1.009 \pm 0.058 $ \\
	$ 64 $ & $ 13.8 \times 10^{10} $ &
	[LN] & $ 1.000 $ &
	$ 6.260 \pm 0.095 $ & $ 1.083 \pm 0.049 $ &
	[LN] & $ 0.999 $ &
	$ 10.75 \pm 0.14 $ & $ 1.071 \pm 0.076 $ \\
	\hline
	\multicolumn{10}{c}{ $ \quad $ } \\
	\hline
	\quad & \quad &
	\multicolumn{4}{c |}{Cluster velocity distributions:} &
	\multicolumn{4}{c |}{Ejecta velocity distributions:} \\
	[0.5ex]
	\hline\hline
	$ i $ & $ \unit [t_{i}] {[yr]} $ &
	Fit: & $ r^{2}_{i,\cpop,\vel} $ &
	$ \unit [a^{(\vel)}_{i,\cpop}] {[pc]} $ &
	$ - $ &
	$ - $ & $ - $ &
	$ - $ &
	$ - $ \\
	[0.5ex]
	\hline
	$ 0 $ & $ 0 $ &
	[MB] & $ 0.968 $ &
	$ 0.913 \pm 0.023 $ & $ - $ &
	$ - $ & $ - $ &
	$ - $ & $ - $ \\
	\hline
	$ i $ & $ \unit [t_{i}] {[yr]} $ &
	Fit: & $ r^{2}_{i,\cpop,\vel} $ &
	$ \unit [\mu^{(\vel)}_{i,\cpop}] {[pc]} $ &
	$ \sigma^{(\vel)}_{i,\cpop} $ &
	Fit: & $ r^{2}_{i,\epop,\vel} $ &
	$ \unit [\mu^{(\vel)}_{i,\epop}] {[pc]} $ &
	$ \sigma^{(\vel)}_{i,\epop} $ \\
	[0.5ex]
	\hline
	$ 28 $ & $ 13.8 \times 10^{6} $ &
	[LN] & $ 0.989 $ &
	$ 1.638 \pm 0.096 $ & $ 0.883 \pm 0.047 $ &
	[LN] & $ 0.989 $ &
	$ 2.137 \pm 0.063 $ & $ 0.478 \pm 0.038 $ \\
	$ 37 $ & $ 13.8 \times 10^{7} $ &
	[LN] & $ 0.985 $ &
	$ 1.00 \pm 0.10 $ & $ 0.986 \pm 0.051 $ &
	[LN] & $ 0.990 $ &
	$ 1.715 \pm 0.079 $ & $ 0.693 \pm 0.042 $ \\
	$ 46 $ & $ 13.8 \times 10^{8} $ &
	[LN] & $ 0.965 $ &
	$ 0.17 \pm 0.11 $ & $ 1.008 \pm 0.054 $ &
	[LN] & $ 0.990 $ &
	$ 1.51 \pm 0.10 $ & $ 0.881 \pm 0.055 $ \\
	$ 55 $ & $ 13.8 \times 10^{9} $ &
	[LN] & $ 0.825 $ &
	$ 0.00 \substack {+0.25 \\ -0.00} $ & $ 1.34 \pm 0.13 $ &
	[LN] & $ 0.989 $ &
	$ 1.21 \pm 0.10 $ & $ 0.912 \pm 0.051 $ \\
	$ 64 $ & $ 13.8 \times 10^{10} $ &
	[LN] & $ 0.753 $ &
	$ 0.00 \substack {+0.63 \\ -0.00} $ & $ 1.89 \pm 0.44 $ &
	[LN] & $ 0.985 $ &
	$ 1.05 \pm 0.12 $ & $ 0.949 \pm 0.062 $ \\
	\hline
	\end{tabular}
	\caption{
	Density, mass, position, and velocity quasi-static profile fits segregated by population (cluster and ejecta) type of Figure~\ref{fig: Quasi-static profiles}.
	The density profile has been fitted to a PL [PL] in the range delimited by the $ 0^{\textrm{th}} $ and $ 80^{\textrm{th}} $ percentiles in position so as to ensure that the density is computed within the cluster and ejecta spheres. The fit is, following Eq.~\eqref{eq: Power-Law fit}, $ \rho_z (t_i) = \beta_{i,\textrm{P}} t_{i}^{\alpha_{i,\textrm{P}}} $, where $ \alpha_{i,\textrm{P}} $ is the PL index, and corresponds roughly to the constant value of the differential counts, while $ \beta_{i,\textrm{P}} $ in turn, corresponds to the amplitude of the PL.
	The mass, position and velocity profiles has been fitted to either a Log-Normal or a Maxwell-Boltzmann (MB) distribution. In the case of the Log-Normal fit, then $ \rho_{z,i,\textrm{P}} = (z \sigma_{z,i,\textrm{P}} \sqrt{2 \pi})^{-1} \exp (- (\textrm{Log-Normal}~ z-\mu_{z,i,\textrm{P}})^{2} / 2 \sigma_{z,i,\textrm{P}}^{2}) $ where $ \mu_{i,\textrm{P}} $ is the expected value and $ \sigma_{i,\textrm{P}} $ corresponds to the standard deviation, both for the $ z $ variable's natural logarithm. In the case of a Maxwell-Boltzmann fit, however, $ \chi^{2}_{3} = \sqrt {2 / \pi} z^{2} a^{-3}_{z,i,\textrm{P}} \exp (- z^{2} / 2a^{2}_{z,i,\textrm{P}} ) $. In any case, $ z = x, v $ is the fitted variable, and $ \textrm{P} = \cpop, \epop $ stands for the cluster and ejecta populations.
	The present table then shows the fit that better adjusts to the data at selected times. Overall, we have found that the Log-Normal fit beats the other choices, except for the initial time-slices before the burn-in time in position and velocities.
	}
	\label{tab: Quasi-static profiles}
\end{table*}



\subsection{Mass segregation \& dynamical friction}
\label{subsec: Mass segregation and dynamical friction}

In this section we comment on the degree to which mass segregation and dynamical friction are observed within the cluster and characterise both phenomena.



\subsubsection{Cluster mass segregation}
\label{subsubsec: Cluster mass segregation}


\begin{table*}[t!]
	\centering
	\begin{tabular}[c]{| l l | l l | l l|}
	\hline
	\quad & \quad &
	\multicolumn{2}{c |}{Cluster time scales:} &
	\multicolumn{2}{c |}{Ejecta time scales:} \\
	[0.5ex]
	\hline
	$ i $ &	$ \unit [t_{i}] {[yr]}	$ &
	$ \unit [T_{\cpop}^{\mathrm{cross}}] {[yr]} $ &
	$ \unit [T_{\cpop}^{\rel}] {[yr]} $ &
	$ \unit [T_{\epop}^{\mathrm{cross}}] {[yr]} $ &
	$ \unit [T_{\epop}^{\rel}] {[yr]} $ \\
	[0.5ex]
	\hline\hline
	$ 0 $ & $ 0 $ &
	$ (4.2 \pm 0.7) \times 10^{5} $ &	
	$ (2.3 \substack {+0.7 \\ -2.3}) \times 10^{5} $ &	
	$ - $ &	
	$ - $ \\
	$ 28 $ & $ 1.38 \times 10^{6} $ &
	$ (3.5 \pm 0.5) \times 10^{5} $ &	
	$ (8.0 \substack {+5.1 \\ -7.2}) \times 10^{5} $ &
	$ (1.72 \pm 0.20) \times 10^{6} $ &	
	$ (2.93 \substack {+0.41 \\ -0.43}) \times 10^{7} $ \\
	$ 37 $ & $ 1.38 \times 10^{7} $ &
	$ (2.22 \pm 0.33) \times 10^{6} $ &	
	$ (1.21 \substack {+0.52 \\ -0.73}) \times 10^{7} $ &	
	$ (1.65 \pm 0.26) \times 10^{7} $ &	
	$ (2.42 \substack {+0.52 \\ -0.53}) \times 10^{8} $ \\
	$ 46 $ & $ 1.38 \times 10^{8} $ &
	$ (2.4 \pm 0.5) \times 10^{7} $ &	
	$ (2.05 \substack {+0.63 \\ -0.86}) \times 10^{8} $ &	
	$ (1.61 \pm 0.28) \times 10^{8} $ &	
	$ (1.89 \substack {+0.51 \\ -0.59}) \times 10^{9} $ \\
	$ 55 $ & $ 1.38 \times 10^{9} $ &
	$ (8.4 \pm 2.7) \times 10^{7} $ &	
	$ (8.9 \substack {+3.4 \\ -4.0}) \times 10^{8} $ &	
	$ (1.64 \pm 0.28) \times 10^{9} $ &	
	$ (1.57 \substack {+0.50 \\ -0.56}) \times 10^{10} $ \\
	$ 64 $ & $ 1.38 \times 10^{10} $ &
	$ (1.5 \substack {+1.6 \\ -1.5}) \times 10^{8} $ &	
	$ (1.9 \substack {+2.0 \\ -1.9}) \times 10^{9} $ &	
	$ (1.80 \pm 0.40) \times 10^{10} $ &	
	$ (1.39 \substack {+0.49 \\ -0.68}) \times 10^{11} $ \\
	\hline
	\end{tabular}
	\caption{
	Shown are the crossing time, $ T_{\textrm{P}}^{\mathrm{cross}} $, and relaxation time, $ T_{\textrm{P}}^{\rel} , $ of the cluster and ejecta bubbles that populate the galactic halo at selected times, and where $ \textrm{P} = \cpop, \epop $ stands for the cluster and ejecta populations.
	}
	\label{tab: Characteristic time scales}
\end{table*}

\begin{table*}[t!]
	\centering
	\begin{tabular}[c]{| l l | l l l | l l l|}
	\hline
	\quad & \quad &
	\multicolumn{3}{c |}{Cluster characteristic profiles:} &
	\multicolumn{3}{c |}{Ejecta characteristic profiles:} \\
	[0.5ex]
	\hline
	$ i $ &	$ \unit [t_{i}] {[yr]}	$ &
	$ \unit [\bar{m}_{i,\cpop}] {[\msun]} $ &
	$ \unit [\bar{x}_{i,\cpop}] {[pc]} $ &
	$ \unit [\bar{v}_{i,\cpop}] {[km / s]} $ &
	$ \unit [\bar{m}_{i,\epop}] {[\msun]} $ &
	$ \unit [\bar{x}_{i,\epop}] {[pc]} $ &
	$ \unit [\bar{v}_{i,\epop}] {[km / s]} $ \\
	[0.5ex]
	\hline\hline
	$ 0 $ & $ 0 $ &
	$ 29.92 \pm 0.75 $ &	
	$ 1.39 \pm 0.16 $ &	
	$ 1.462 \pm 0.040 $ &	
	$ - $ & 	
	$ - $ &	
	$ - $ \\
	$ 28 $ & $ 1.38 \times 10^{6} $ &
	$ 30.30 \pm 0.84 $ &	
	$ 2.70 \pm 0.23 $ &	
	$ 3.85 \pm 0.13 $ &	
	$ 26.90 \pm 0.83 $ &	
	$ 16.7 \pm 1.6 $ &	
	$ 9.50 \pm 0.56 $ \\
	$ 37 $ & $ 1.38 \times 10^{7} $ &
	$ 30.45 \pm 0.94 $ &	
	$ 10.0 \pm 1.0 $ &	
	$ 4.42 \pm 0.38 $ &	
	$ 29.14 \pm 0.79 $ &	
	$ 119 \pm 16 $ &	
	$ 7.12 \pm 0.59 $ \\
	$ 46 $ & $ 1.38 \times 10^{8} $ &
	$ 30.6 \pm 1.0 $ &	
	$ 47.0 \pm 5.1 $ &	
	$ 1.97 \pm 0.24 $ &	
	$ 29.33 \pm 0.81 $ &	
	$ (1.100 \pm 0.150) \times 10^{3} $ &	
	$ 6.66 \pm 0.75 $ \\
	$ 55 $ & $ 1.38 \times 10^{9} $ &
	$ 31.5 \pm 1.1 $ &	
	$ 210 \pm 20 $ &	
	$ 2.51 \pm 0.68 $ &	
	$ 29.14 \pm 0.84 $ &	
	$ (8.5 \pm 1.1) \times 10^{3} $ &	
	$ 4.15 \pm 0.58 $\\
	$ 64 $ & $ 1.38 \times 10^{10} $ &
	$ 32.5 \pm 1.2 $ &	
	$ 940 \pm 100 $ &	
	$ 6.4 \pm 5.5 $ &	
	$ 29.10 \pm 0.80 $ &	
	$ (8.3 \pm 1.3) \times 10^{4} $ &	
	$ 4.49 \pm 0.58 $ \\
	\hline
	\end{tabular}
	\caption{
	Cluster and ejecta mean PBHs mean mass, position, and velocity, at selected time-slices. These are extracted from the fitted Log-Normal and Maxwell-Boltzmann distribution parameters of Table~\ref{tab: Quasi-static profiles} and by making use of Eqs.~\eqref{eq: Log-Normal mean} and \eqref{eq: Maxwell-Boltzmann mean} of Section~\ref{subsec: Simulation Initial Conditions}.
	}
	\label{tab: Characteristic profile properties}
\end{table*}


Mass segregation is the dynamical process by which the heavier bodies in a gravitationally bound system tend to fall to the central area of the system while the lighter bodies move farther away from the centre. It can be observed at a large range of scales is astrophysical systems, from star clusters to galaxy clusters and results in the negative cross-correlation of the mass and position distribution with heavier masses skewing towards smaller radial distances from the cluster barycentre.

The particular way this phenomenon works is as follows. During a close encounter of two bodies such as PBHs, the bodies exchange both kinetic energy and momentum. After a large number of encounters there is a tendency towards the bodies having similar energies from equipartition of energy arguments. Since energy is quadratic in velocity and linear in mass, the most massive objects move at slower speeds than the least massive objects in order to equipartition of energy be sustained.

This effect works for any range of body masses and radial positions, but is stronger the denser the medium and the more compressed the bodies' motions are, as shown by the time taken by the cluster bodies to achieve equipartition, called the relaxation time, which has been shown in Ref.~\cite{Binney:2008} to approximately be
\begin{equation}
	t^{\rel} (t) = \frac{N_{\rpop, \cpop} (t) t^{\mathrm{cross}} (t)}{8 \log N_{\rpop, \cpop} (t)},
	\label{eq: Relaxation time}
\end{equation}
where $ N_{\rpop, \ocap} (t) $ is our time-dependent number of cluster PBHs, and $ t^{\mathrm{cross}} (t) $ is the time taken by PBHs to cross the cluster. In our simulations, Eq.~\eqref{eq: Relaxation time} yields for the IC a relaxation time of
\begin{equation}
	t^{\rel}_{\cpop, 0} = \unit [(1.323 \pm 0.071) \times 10^{7}] {yr},
	\label{eq: Relaxation time initial}
\end{equation}
since at $ t_0 = \unit [0] {yr} $ we have $ N_{\rpop,\cpop} (t_0) = N_{\ocap} = 1000 $ ; and $ X_{\cpop} (t_0) = \unit [1.710 \pm 0.060] {pc} $ and $ V_{\cpop} (t_0) = \unit [2.301 \pm 0.091] {pc} $ from Table~\ref{tab: Background properties} yields an initial crossing time of
\begin{equation}
	t^{\mathrm{cross}}_{\cpop, 0} = \frac {X_{\cpop,0}} {V_{\cpop,0}} = \unit [(7.31 \pm 0.41) \times 10^{5}] {yr},
	\label{eq: Crossing time initial}
\end{equation}
which is of the same order of magnitude than the burn-in time of $ t_{\bicap} = \unit [2.64 \times 10^{5}] {yr} $, an issue of great importance since, as Table~\ref{tab: Characteristic profile properties} shows, shortly after the burn-in time, the cluster starts to expand, which forces the crossing time to increase dramatically as cluster PBHs bulk speeds do not catch up with the increase in the characteristic length scale
\begin{equation}
	t^{\mathrm{cross}}_{\cpop, 64} = \frac {X_{\cpop, 64}} {V_{\cpop, 64}} = \unit [(1.72 \pm 0.92) \times 10^{8}] {yr}.
	\label{eq: Crossing time final}
\end{equation}

This increase, on top of the population loss from cluster evaporation, eventually increases the relaxation time by two orders of magnitude by the time of the last time-slice
\begin{equation}
	t^{\rel}_{\cpop} (t_{64}) = \unit [(1.32 \substack {+ 2.00 \\ -2.10}) \times 10^{9}] {yr}.
	\label{eq: Relaxation time final}
\end{equation}

We illustrate cluster mass segregation in our simulations by extracting the $ \rho (m_{\cpop}, x_{\cpop}, t) $ distribution at the time-slices that divide the simulation time-runs. We show our results in Figure~\ref{subfig: Cluster mass segregation mass weighted} and give the best-fit values in Table~\ref{tab: Dynamical evolution}. There, a number of features are apparent
\begin{enumerate}[label=\roman*)]
	\item First, that the phase space $ (m_{\cpop}, x_{\cpop}) \vert_{t} $ occupied by the cluster PBHs undergoes an expansion of the position profile from a narrow Maxwell-Boltzmann distribution in the IC to a broader nearly Log-Normal distribution after the burn-in time at $ t_{\bicap} = \unit [2.64 \times 10^{5}] {yr} $, consistently with other observations.
	\item Second, that around the burn-in time, again the aforementioned phase in which the cluster contracts is apparent, in that the best-fit value of the distribution at $ t_{28} = \unit [1.34 \times 10^{6}] {yr} $ has moved towards the lower end of the position range, the only time to do so throughout the simulations, again consistently with other observations throughout the simulations.
	\item Third, that for later times $ t > t_{37} = \unit [1.34 \times 10^{7}] {yr} $ the cluster expands as the distribution's best-fit is displaced towards the higher end of the position range, while not changing significantly its mass value, in agreement with the finding in Section~\ref{subsubsec: Mass profiles evolution} that neither the cluster nor the ejecta distributions develop a mass profile that is substantially different from the IC throughout the entire simulation period.
	\item Last, that mass segregation is clearly apparent in Figure~\ref{subfig: Cluster mass segregation mass weighted} in that both the $ 1 \sigma $ and particularly the $ 2 \sigma $ contours are increasingly tilted leftwards as the cluster evolves, suggesting the presence of a distinct population of massive PBHs in the cluster core.
\end{enumerate}

This last point is evidenced too in that we have plotted on top of the distributions a random sample of 1000 PBHs, that is, 0.02\% of available objects in each time-slice, with a PBH selection likelihood proportional to its mass to better capture the most massive of these. The largest of these PBHs are visible then in Figure~\ref{subfig: Cluster mass segregation mass weighted} as black dots moving rightwards over time as the cluster puffs-up, but always to the left of the bulk of PBHs at their corresponding time-slice.



\subsubsection{Cluster dynamical friction}
\label{subsubsec: Cluster dynamical friction}

Dynamical friction is the loss of momentum and kinetic energy of bodies by the gravitational interaction with their surrounding neighbours in space. Also called gravitational drag, this effect result on the negative cross-correlation of the mass and velocity distribution, with heavier masses skewing towards smaller cluster velocities. It has been observed in a variety of astrophysical systems, from galaxies and galaxy clusters at large scales to proto-planetary disks at smaller scales.

In our case, where the PBHs cover a wide range of masses of $ \order {0.1} \msun-\order {1000} \msun $, the lighter PBHs have an effect on the large PBHs being surrounded by them: the lighter, more abundant of these accelerate during slingshot encounters with their more massive partners gaining kinetic energy and momentum, slowing the more massive PBH of the pair in the process as energy is conserved.

Another way this mechanism is enhanced is conversely by the effect a large PBH has on the lighter surrounding PBHs on the denser cluster core: the larger of these PBHs wake up and accelerate the lighter partners towards himself, but by the time the lighter PBH arrives at its position the larger PBH has moved on already. This results in the tendency of large PBHs to be followed by an anisotropic trail of lighter PBHs that exert a gravitational pull opposite to the motion of the large PBH, slowing it down.

Generally speaking, both effects work the same for any range of masses and relative velocities between objects, but the effects are stronger the denser the medium and the slower the bodies' motions, since the gravitational pull is proportional to the square of the masses of the object and to the inverse square of the velocity. Moreover, dynamical friction is heavily suppressed at high velocities, since the larger the velocity, the less time available for the least massive objects to wake up and accelerate towards the most massive ones.

\begin{figure*}[t!]
	\centering
	\subfloat[Mass-weighted cluster mass segregation, $ \rho^{\textrm{w}} (x_{\cpop}, m_{\cpop}, t) $.]{
	\hspace*{-1.00cm}
	\includegraphics[width = 0.50\textwidth]
	{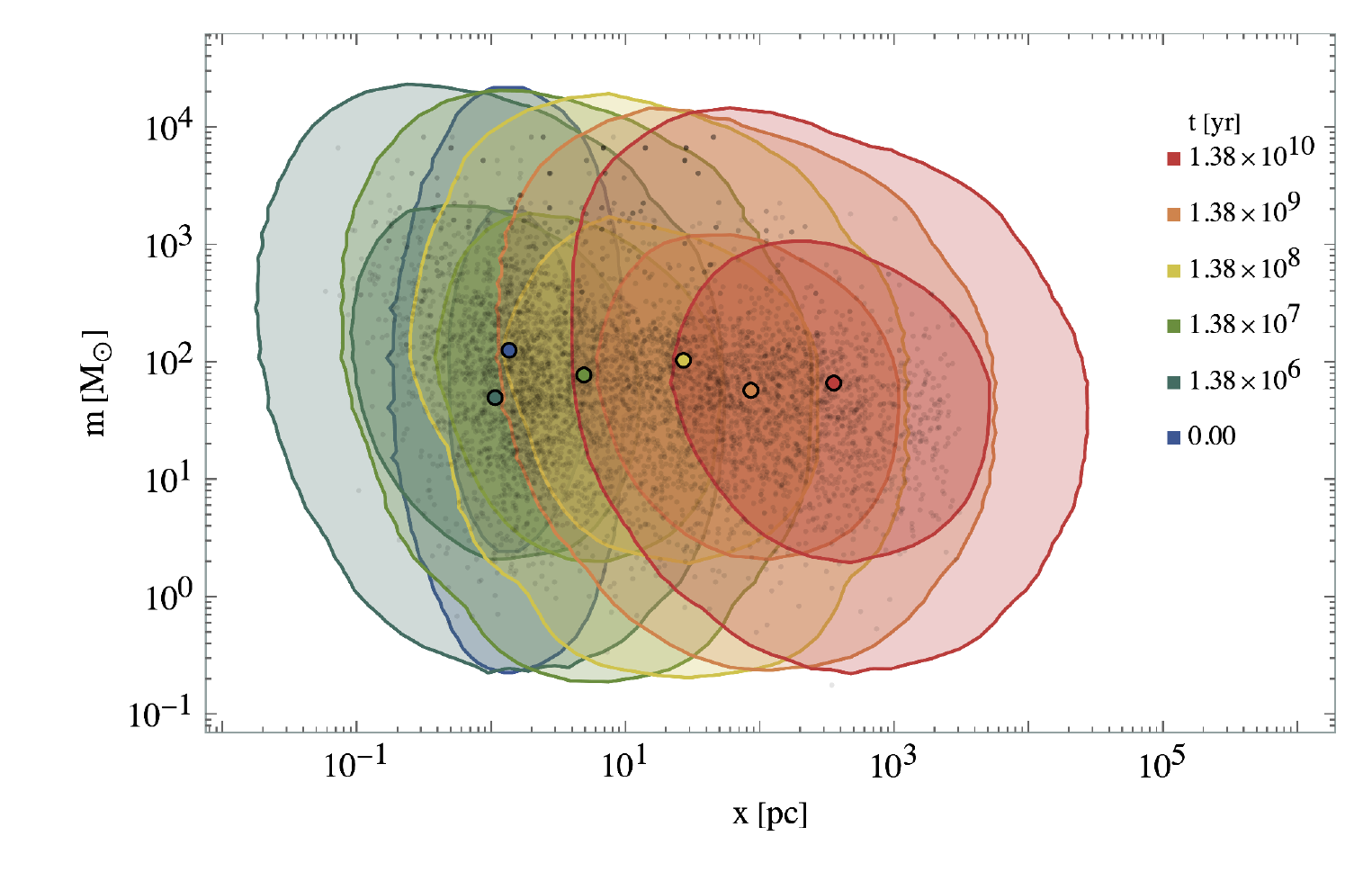}
	\label{subfig: Cluster mass segregation mass weighted}}
	\subfloat[Mass-weighted cluster dynamical friction $ \rho^{\textrm{w}} (v_{\cpop}, m_{\cpop}, t) $.]{
	\hspace*{-0.00cm}
	\includegraphics[width = 0.50\textwidth]
	{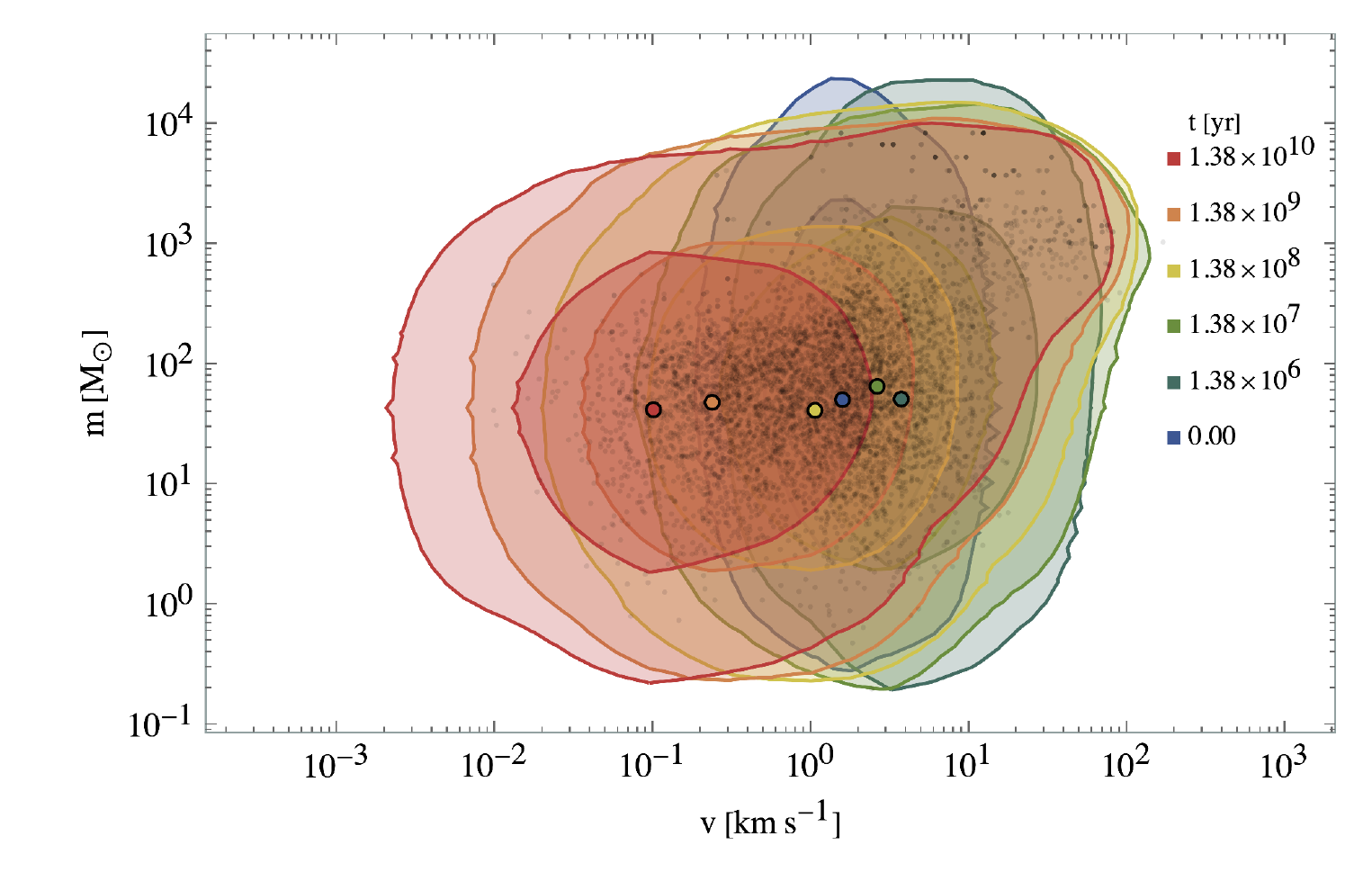}
	\label{subfig: Cluster dynamical friction mass weighted}}
	\newline
	\subfloat[Bare (unweighted) cluster mass segregation $ \rho^{\textrm{b}} (x_{\cpop}, m_{\cpop}, t) $.]{
	\hspace*{-1.00cm}
	\includegraphics[width = 0.50\textwidth]
	{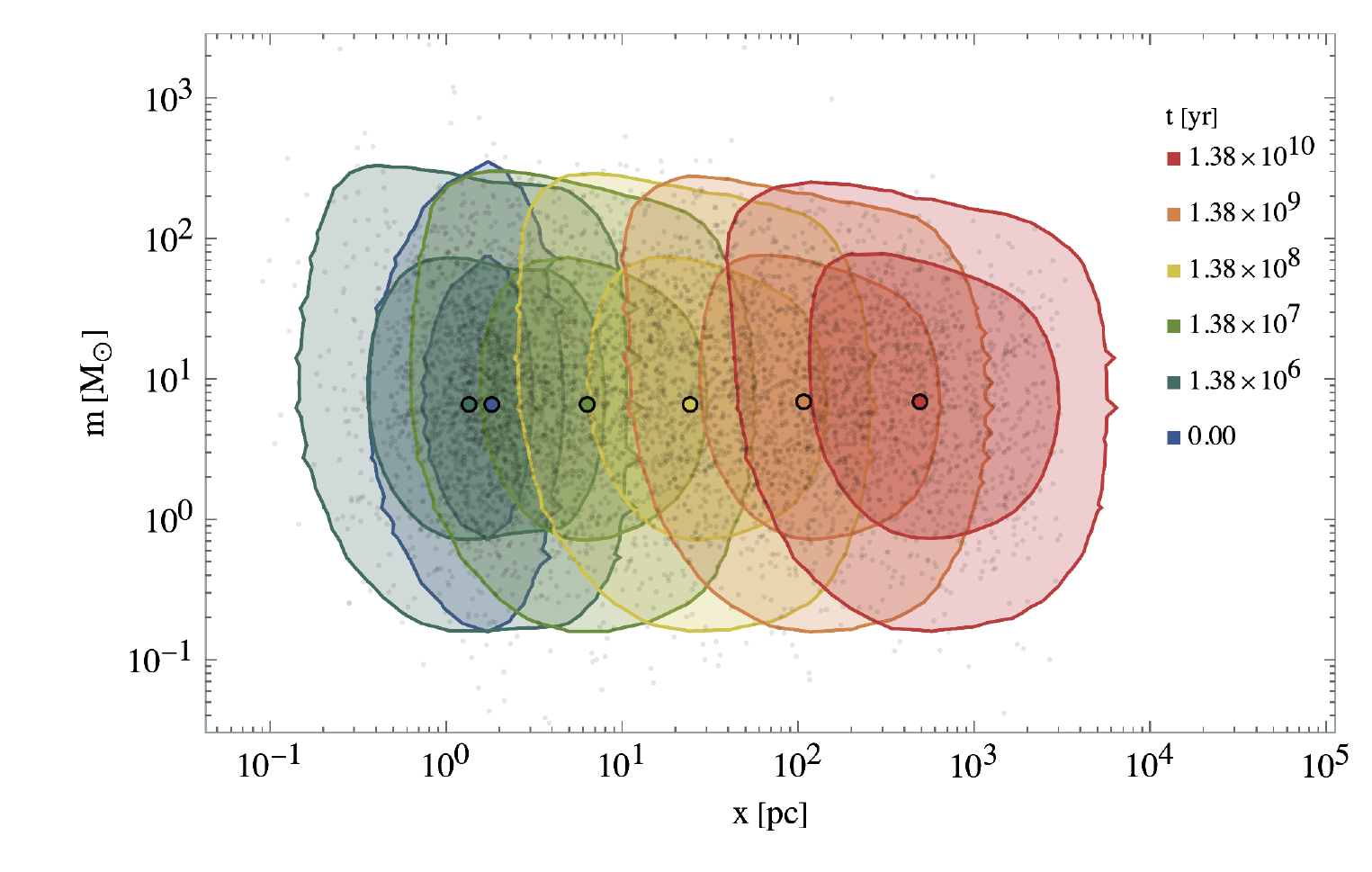}
	\label{subfig: Cluster mass segregation not weighted}}
	\subfloat[Bare (unweighted) cluster dynamical friction $ \rho^{\textrm{b}} (v_{\cpop}, m_{\cpop}, t) $.]{
	\hspace*{-0.00cm}
	\includegraphics[width = 0.50\textwidth]
	{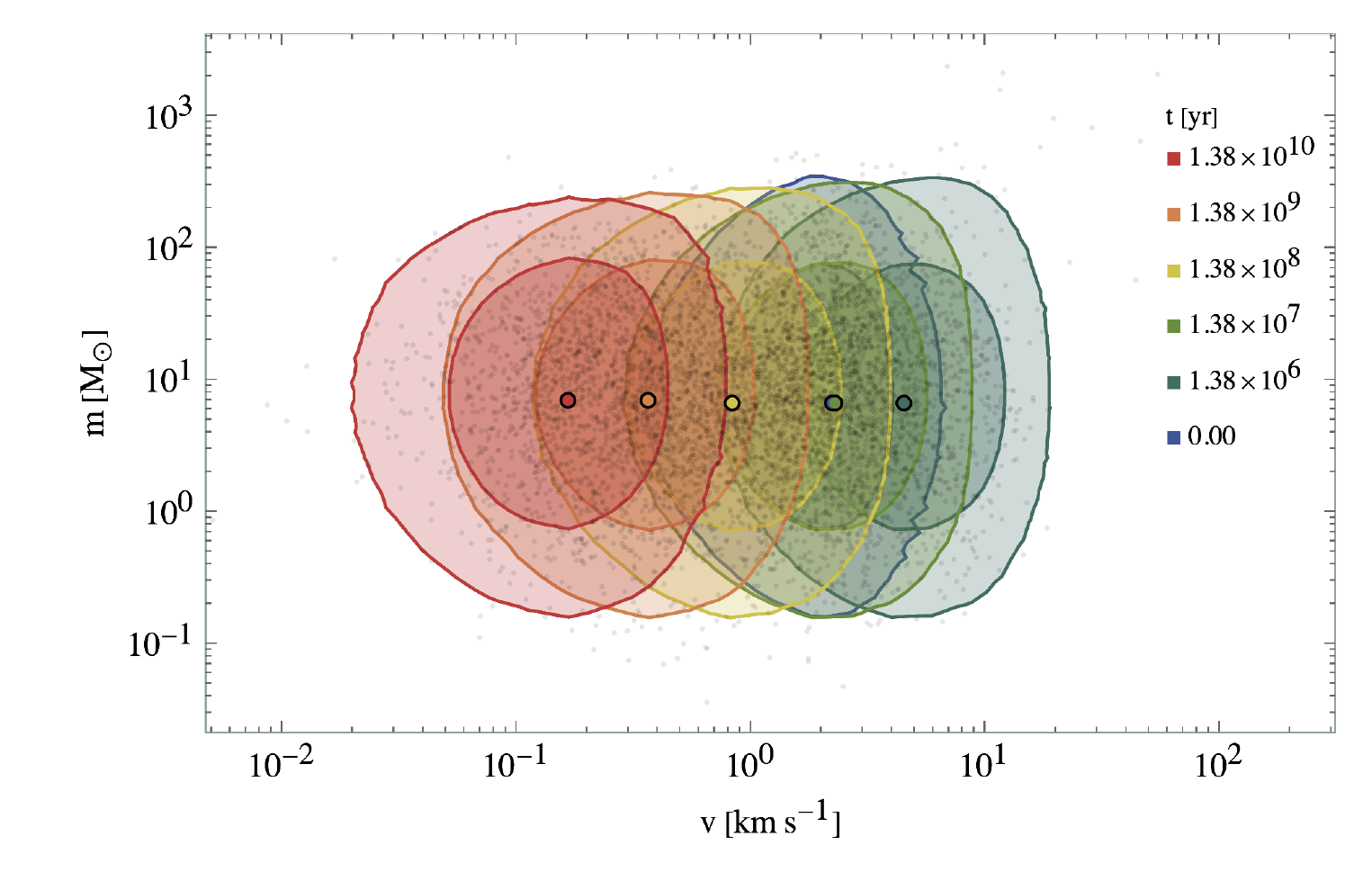}
	\label{subfig: Cluster dynamical friction not weighted}}
	\caption{
	Shown are the $ 1\sigma $ and $ 2\sigma $ contours for the cluster PBHs mass vs. position and velocity distributions at selected times.
	Fits for the mass, position, velocity and density probability distributions are given in Table~\ref{tab: Dynamical evolution}. Black dots show the scatter a 1000 sample of individual PBHs in each time-slice, randomly selected in the case of the bare distribution and selected with a probability proportional to their mass in the case of the mass weighted in order to better illustrate mass segregation and dynamical friction, since otherwise such very massive PBHs are easy to miss in the sample given that there are not more than $ \order {1} $ $ m_{i} > \unit [1000] {\msun} $ massive objects per cluster. The Best-Fit (BF) values of the distributions, marked by the large coloured dots, are given in Table~\ref{tab: Dynamical evolution}.
	}
	\label{fig: Dynamical evolution}
\end{figure*}

In our simulations, cluster dynamical friction is shown by extracting the $ \rho (m_{\cpop}, v_{\cpop}, t) $ distribution at time-slices that divide the simulation time-runs, shown in Figure~\ref{subfig: Cluster dynamical friction mass weighted} and whose best-fit values we give in Table~\ref{tab: Dynamical evolution}, in which a number of features similar to those in the case for mass segregation of Section~\ref{subsubsec: Cluster mass segregation} are apparent
\begin{enumerate}[label=\roman*)]
	\item First is that the phase space $ (m_{\cpop}, v_{\cpop}) \vert_{t} $ occupied by the cluster undergoes an expansion of the velocity profile from the IC's narrow Maxwell-Boltzmann distribution before the $ t_{\bicap} = \unit [2.64 \times 10^{5}] {yr} $ to the later broader nearly Log-Normal distribution.
	\item Second, that around the burn-in time, just as the cluster contracts to later rebound, the velocities best-fit value of the distribution in the interval $ (t_{28}, t_{37}) = \unit [1.34 \times (10^{6}, 10^{7})] {yr} $ moves towards the higher end of the available range, in agreement with the accelerating infall velocities, the only time to do so throughout the simulations.
	\item Third, that for later times $ t > t_{46} = \unit [1.34 \times 10^{8}] {yr} $ the distribution's velocity best-fit is displaced towards the higher end of the available range as the cluster expands, again while not changing significantly its mass value, consistently with the finding in Section~\ref{subsubsec: Mass profiles evolution}.
	\item Last, that Figure~\ref{subfig: Cluster dynamical friction mass weighted} shows velocities that are in fact, for the high mass objects, larger than the bulk velocities of the cluster as seen in that both the $ 1 \sigma $ and particularly the $ 2 \sigma $ contours are increasingly tilted rightwards as the cluster evolves to the later time intervals. This indicates that the population of core massive PBHs move at high speeds relative to each other.
\end{enumerate}

Dynamical friction is then evidenced by the gradual kinetic energy loss of bulk cluster PBHs and the leftward-moving velocity profile of Figure~\ref{subfig: Cluster mass segregation mass weighted}, particularly so for the $ 1 \sigma $ area, less so for the $ 2 \sigma $, but still clear if excluding the largest PBHs.

Note, however that, in contrast to the previous argument that gravitational drag is strongest for massive objects, Figure~\ref{subfig: Cluster mass segregation mass weighted} shows that mass positively correlates with velocity. This is apparent too in that we have plotted on top of the distributions a random sample of $ 1000 $ PBHs or 0.02\% of total objects in each time-slice, with a PBH selection likelihood proportional to its mass to better capture the most massive of these, showing that the most massive objects, with masses of $ \unit [\order {1000}] {\msun} $ are slingshot to high velocities up to $ \unit [\order {100}] {km / s} $.

The reason to this positive correlation stems from the fact that the cluster is not very dense, with a population of $ N_{\ocap, 0} = 1000 $ PBHs at most at the initial time-slice. Its quick expansion makes that the condition for gravitational drag from the trailing PBHs following massive PBHs is not realised in practise, since these large objects do not in fact move in a medium of lighter objects but in one of objects with comparable masses, falling into each other large potential wells and being accelerated to large velocities.


\begin{table*}[t!]
	\centering
	\begin{tabular}[c]{| l l | l l l l l | l l l l l |}
	\hline
	\quad & \quad &
	\multicolumn{5}{c |}{Cluster mass segregation:} &
	\multicolumn{5}{c |}{Cluster dynamical friction:} \\
	[0.5ex]
	\hline\hline
	$ i $ & $ \unit [t_{i}] {[yr]} $ &
	$ \quad $ &
	$ \unit {x_{i,\cpop}^{\textrm{(w)}}} [\textrm{pc}] $ &
	$ \unit{m_{i,\cpop}^{\textrm{(w)}}} [\msun] $ &
	$ \unit {x_{i,\cpop}^{\textrm{(b)}}} [\textrm{pc}] $ &
	$ \unit{m_{i,\cpop}^{\textrm{(b)}}} [\msun] $ &
	$ \quad $ &
	$ \unit {v_{i,\cpop}^{\textrm{(w)}}} [\mathrm{km / s}] $ &
	$ \unit{m_{i,\cpop}^{\textrm{(w)}}} [\msun] $ &
	$ \unit {v_{i,\cpop}^{\textrm{(b)}}} [\mathrm{km / s}] $ &
	$ \unit{m_{i,\cpop}^{\textrm{(b)}}} [\msun] $ \\
	[0.5ex]
	\hline
	$ 0 $ & $ 0 $ &
	[BF] & $ 1.35 $ & $ 125 $ & $ 1.82 $ & $ 6.62 $ &
	[BF] & $ 1.60 $ & $ 49.9 $ & $ 2.24 $ & $ 6.62 $ \\
	$ 28 $ & $ 13.8 \times 10^{6} $ &
	[BF] & $ 1.07 $ & $ 49.3 $ & $ 1.35 $ & $ 6.62 $ &
	[BF] & $ 3.76 $ & $ 50.5 $ & $ 4.54 $ & $ 6.62 $ \\
	$ 37 $ & $ 13.8 \times 10^{7} $ &
	[BF] & $ 4.93 $ & $ 77.3 $ & $ 6.33 $ & $ 6.62 $ &
	[BF] & $ 2.64 $ & $ 64.1 $ & $ 2.30 $ & $ 6.62 $ \\
	$ 46 $ & $ 13.8 \times 10^{8} $ &
	[BF] & $ 27.2 $ & $ 103 $ & $ 24.4 $ & $ 6.62 $ &
	[BF] & $ 1.07 $ & $ 40.8 $ & $ 0.839 $ & $ 6.62 $ \\
	$ 55 $ & $ 13.8 \times 10^{9} $ &
	[BF] & $ 86.1 $ & $ 57.2 $ & $ 107 $ & $ 6.87 $ &
	[BF] & $ 0.240 $ & $ 47.6 $ & $ 0.367 $ & $ 6.87 $ \\
	$ 64 $ & $ 13.8 \times 10^{10} $ &
	[BF] & $ 354 $ & $ 65.9 $ & $ 493 $ & $ 6.87 $ &
	[BF] & $ 0.101 $ & $ 41.3 $ & $ 0.166 $ & $ 6.87 $ \\
	\hline
	\end{tabular}
	\caption{
	Cluster mass segregation and dynamical friction from Figure~\ref{fig: Dynamical evolution}.
	BF values for the mass segregation $ \rho^{j}_{i} (x_{\cpop}, m_{\cpop}, t) $ and dynamical friction $ \rho^{j}_{i} (x_{\cpop}, m_{\cpop}, t) $ plots, where $ j = \textrm{w}, \textrm{b} $ for the mass-weighed and bare (homogeneously weighted) distributions respectively, from which we extract the BF values at selected times.
	}
	\label{tab: Dynamical evolution}
\end{table*}



\section{Orbital distributions}
\label{sec: Orbital distributions}

Here in Sections~\ref{subsec: Custer orbital properties} and \ref{subsec: Ejecta orbital properties} we find constraints to orbital parameter space by computing the $ 1 \sigma $ and $ 2 \sigma $ contours of the $ (a_{i}, e_{i}) \vert_{t} $ distributions at the $ t $ logarithmically spaced time-slices that divide the simulation time-runs, for the $ i = \cpop, \epop $ cluster and ejecta populations respectively, where $ e (t) $ and $ a (t) $ is the eccentricity and semi-major axis of PBHs. Also, in Section~\ref{subsubsec: hyperbolic encounter rates} we find the periods $ \Delta t_{i} $ at time $ t_{i} $, with the cluster and ejecta objects as well as the abundance of ejecta binaries and their orbital parameter distributions with respect to the binary barycentre.

We make a distinction between bounded and unbounded systems, splitting both the cluster and ejecta populations into isolated and bounded systems. as it has been found that they exhibit different properties from the earliest times. Note that eccentricities, semi-major axis and periods are computed with respect to the cluster barycenter in the case of cluster and ejecta isolated objects.

In the case of bounded cluster and ejecta objects, they are computed instead with respect to the bounded system barycenter, regardless of whether they may be part of binary systems, ternary, quaternary and quinary systems. Note yet that the orbital parameters of these bounded systems' barycenters are still included in the isolated population.

\begin{figure*}[t!]
	\centering
	\subfloat[Cluster, isolated CF orbital parameters, $ \rho (e_{\cpop}, a_{\ipop}, t) $.]{
	\hspace*{-1.00cm}
	\includegraphics[width = 0.50\textwidth]{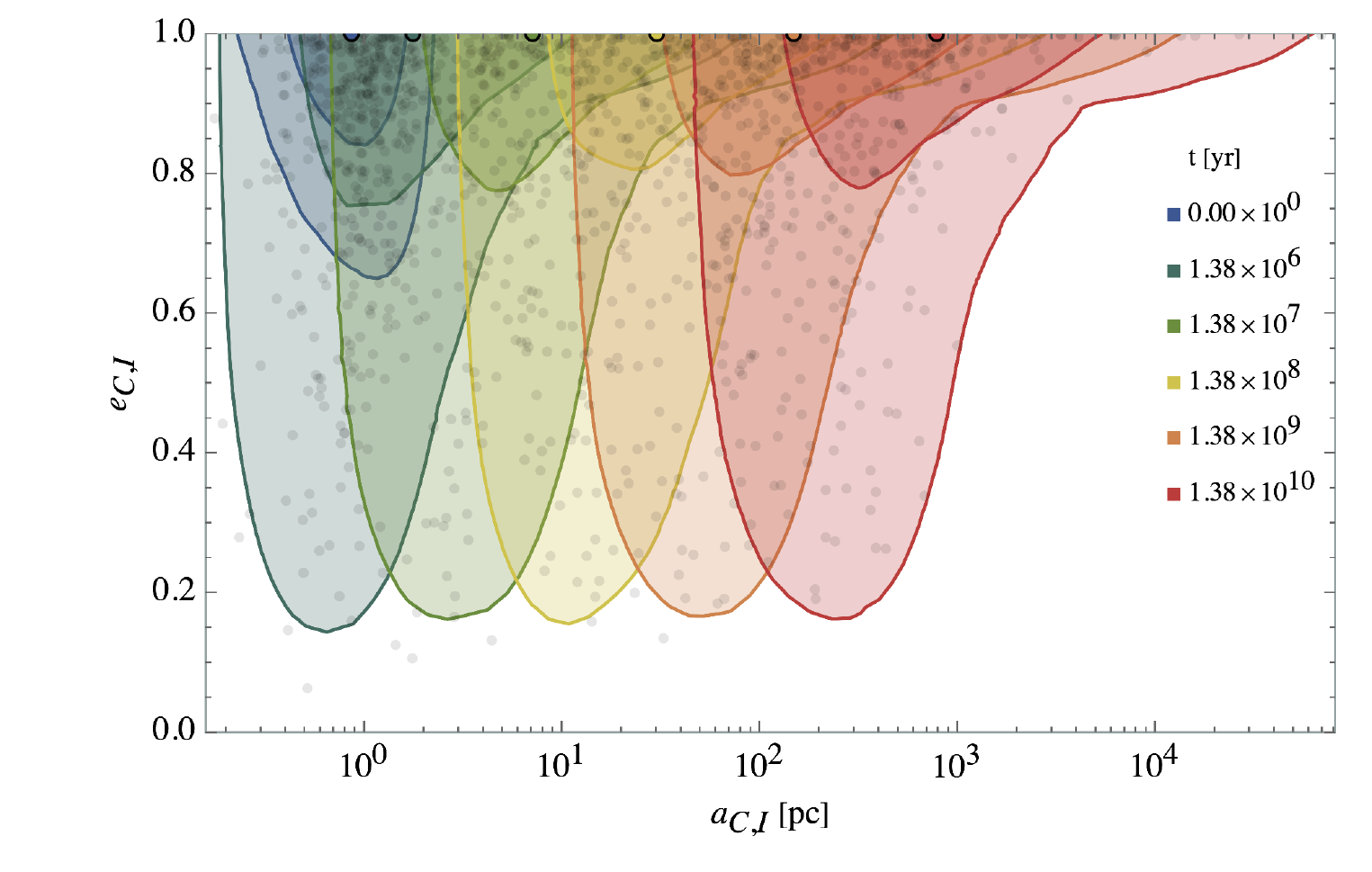}
	\label{subfig: Orbital properties cluster}}
	\subfloat[Ejecta, isolated CF orbital parameters, $ \rho (e_{\cpop}, a_{\ipop}, t) $.]{
	\hspace*{-0.00cm}
	\includegraphics[width = 0.50\textwidth]{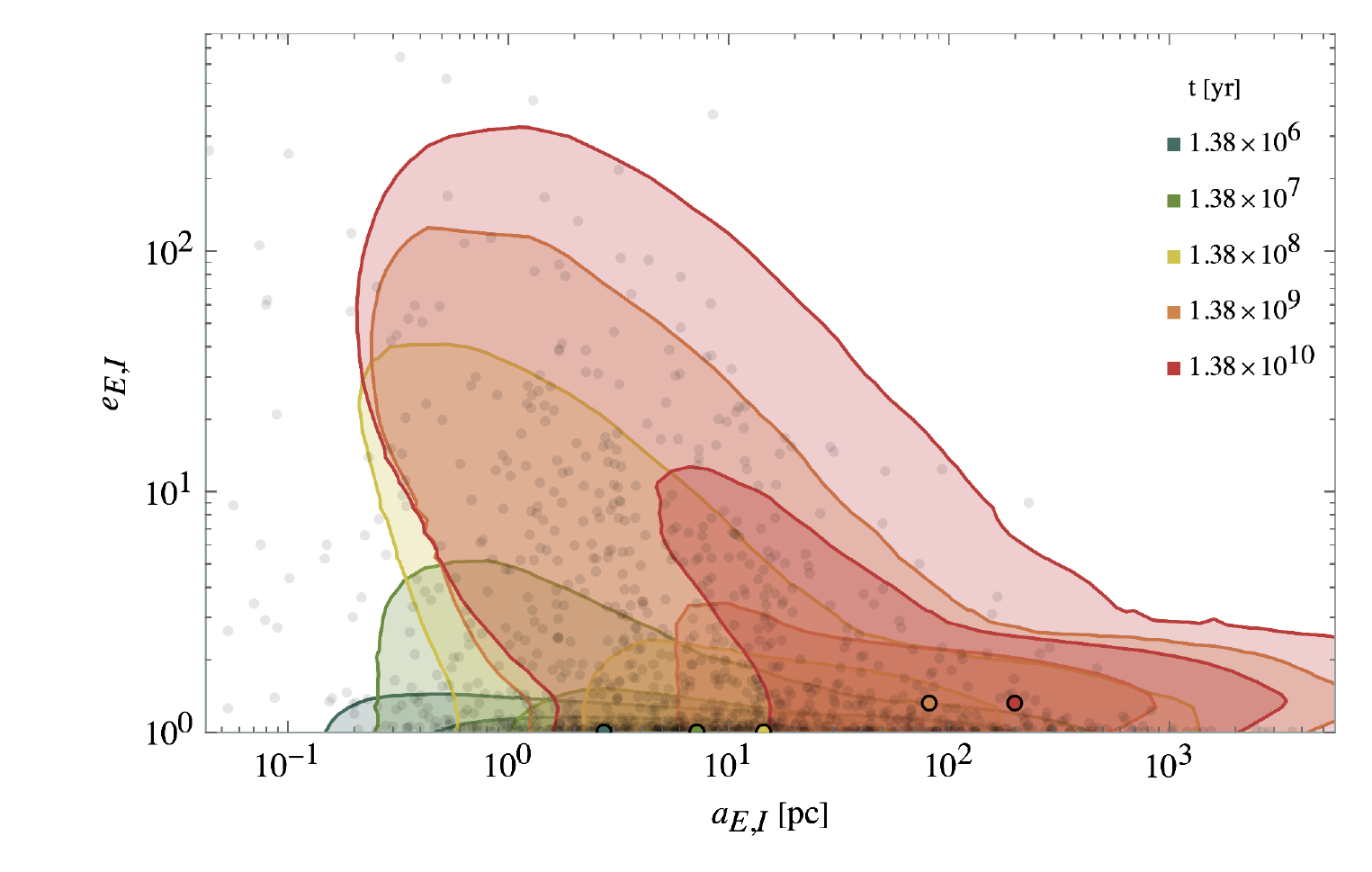}
	\label{subfig: Orbital properties ejecta}}
	\newline
	\subfloat[Cluster, bounded CF orbital parameters, $ \rho (e_{\cpop}, a_{\bpop}, t) $.]{
	\hspace*{-1.00cm}
	\includegraphics[width = 0.50\textwidth]{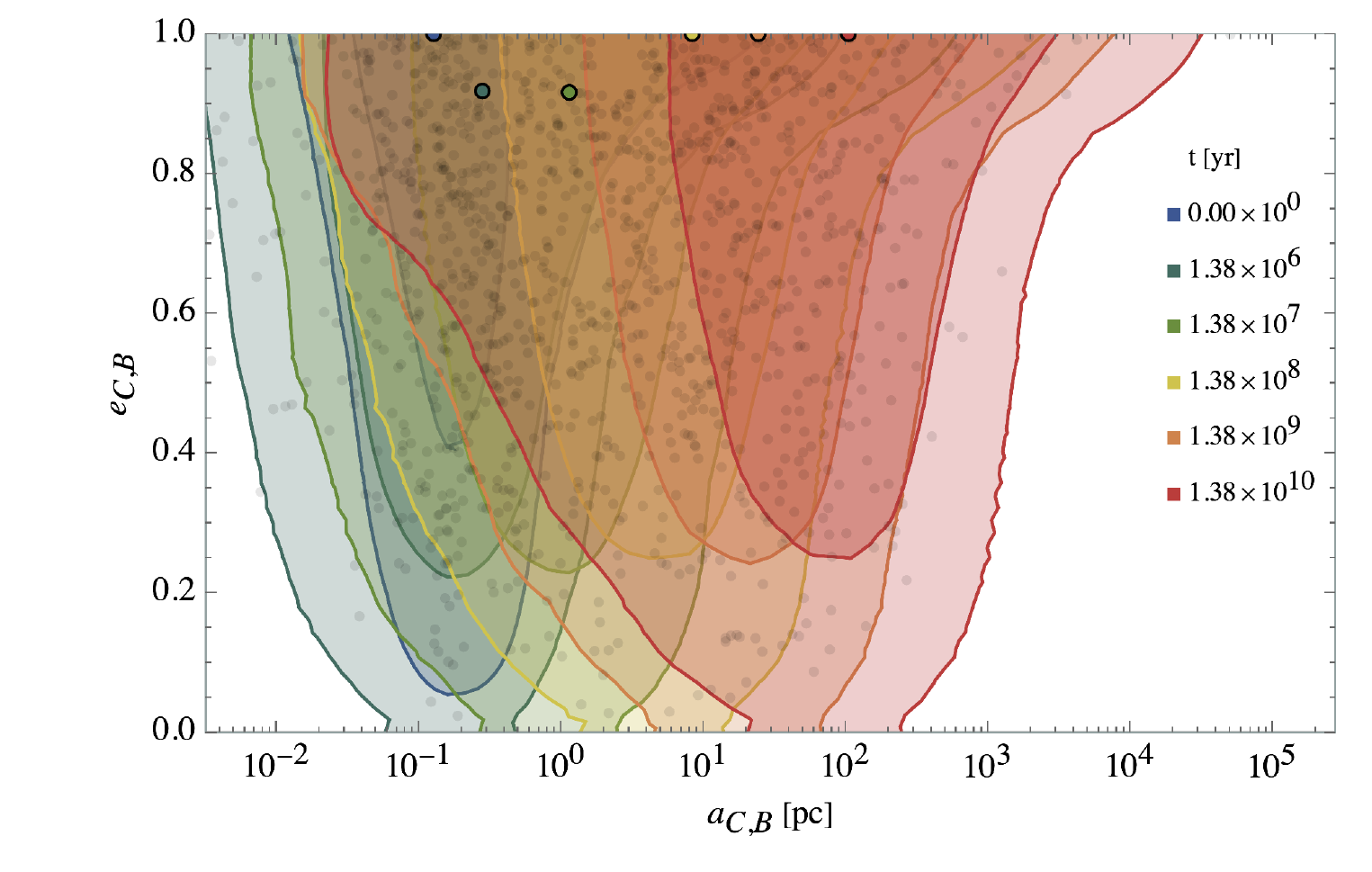}
	\label{subfig: Orbital properties cluster binary}}
	\subfloat[Ejecta, bounded CF orbital parameters, $ \rho (e_{\epop}, a_{\bpop}, t) $.]{
	\hspace*{-0.00cm}
	\includegraphics[width = 0.50\textwidth]{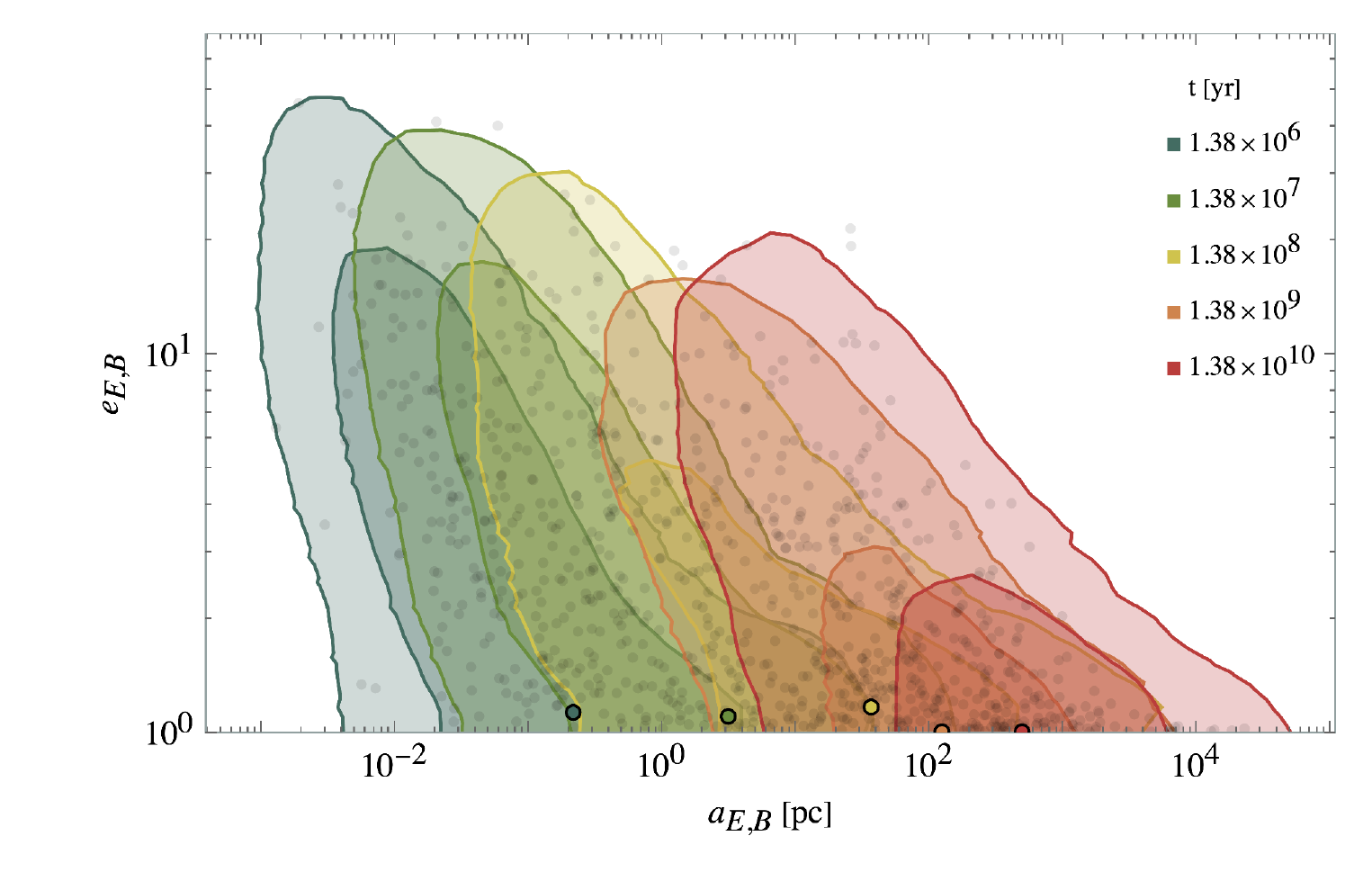}
	\label{subfig: Orbital properties ejecta binary}}
	\caption{
	Shown are the $ 1\sigma $ and $ 2\sigma $ contours PBHs orbital eccentricity vs. semi-major axis segregated by population (cluster and ejecta) type at selected times, $ t_{\bicap} = \unit [2.64 \times 10^{5}] {yr} $ plus the initial time-slice $ t_{0} $.
	All panels: Black dots show the scatter a 100 sample of individual PBHs in each time-slice, randomly selected in the case of the bare distribution and selected with a uniform probability distribution. The Best-Fit (BF) values of the distributions, marked by the large coloured dots in the plots, are given in Table~\ref{tab: Orbital properties evolution}.
	}
	\label{fig: Orbital properties evolution}
\end{figure*}

\begin{table*}[t!]
	\centering
	\begin{tabular}[c]{| l l | l l l l l | l l l l l |}
	\hline
	\quad & \quad &
	\multicolumn{5}{c |}{Cluster objects distributions:} &
	\multicolumn{5}{c |}{Ejecta objects distributions:} \\
	[0.5ex]
	\hline\hline
	$ i $ & $ \unit [t_{i}] {[yr]} $ &
	$ \quad $ &
	$ \unit {a_{i,\cpop,\ipop}} [\textrm{pc}] $ &
	$ \unit{e_{i,\cpop,\ipop}} $ &
	$ \unit {a_{i,\cpop,\bpop}} [\textrm{pc}] $ &
	$ \unit{e_{i,\cpop,\bpop}} $ &
	$ \quad $ &
	$ \unit {a_{i,\epop,\ipop}} [\textrm{pc}] $ &
	$ \unit{e_{i,\epop,\ipop}} $ &
	$ \unit {a_{i,\epop,\bpop}} [\textrm{pc}] $ &
	$ \unit{e_{i,\epop,\bpop}} $ \\
	[0.5ex]
	\hline
	$ 0 $ & $ 0 $ &
	[BF] & $ 0.865 $ & $ 1.000 $ & $ 0.128 $ & $ 1.000 $ &
	[BF] & $ - $ & $ - $ & $ 0.226 $ & $ 1.49 $ \\
	$ 28 $ & $ 13.8 \times 10^{6} $ &
	[BF] & $ 1.77 $ & $ 1.000 $ & $ 0.283 $ & $ 0.918 $ &
	[BF] & $ 2.73 $ & $ 1.00 $ & $ 0.0546 $ & $ 2.82 $ \\
	$ 37 $ & $ 13.8 \times 10^{7} $ &
	[BF] & $ 7.14 $ & $ 1.000 $ & $ 1.16 $ & $ 0.916 $ &
	[BF] & $ 2.73 $ & $ 1.00 $ & $ 2.26 $ & $ 1.04 $ \\
	$ 46 $ & $ 13.8 \times 10^{8} $ &
	[BF] & $ 30.3 $ & $ 1.000 $ & $ 8.43 $ & $ 1.000 $ &
	[BF] & $ 14.4 $ & $ 1.00 $ & $ 35.6 $ & $ 1.06 $ \\
	$ 55 $ & $ 13.8 \times 10^{9} $ &
	[BF] & $ 150 $ & $ 1.000 $ & $ 24.5 $ & $ 1.000 $ &
	[BF] & $ 81.8 $ & $ 1.32 $ & $ 150 $ & $ 1.00 $ \\
	$ 64 $ & $ 13.8 \times 10^{10} $ &
	[BF] & $ 786 $ & $ 1.000 $ & $ 105 $ & $ 1.000 $ &
	[BF] & $ 200 $ & $ 1.32 $ & $ 744 $ & $ 1.07 $ \\
	\hline
	\end{tabular}
	\caption{
	Shown are the cluster and ejecta BF values for the orbital eccentricity and semi-major axis distribution, $ \rho (e_{i, j}, a_{i, j}, t) $, from Figure~\ref{fig: Orbital properties evolution}, where $ i = \cpop, \epop $ stand for the separate cluster and ejecta populations, and $ j = \ipop, \bpop $ stands for the separate isolated and bounded population distributions at selected times.
	}
	\label{tab: Orbital properties evolution}
\end{table*}



\subsection{Cluster orbital properties}
\label{subsec: Custer orbital properties}

In Figure~\ref{subfig: Orbital properties cluster} we show the PBHs occupied orbital parameter space $ \rho (a_{\cpop, \ipop}, e_{\cpop,\ipop}, t) $, for the isolated cluster population, that, is, the PBH cluster population excluding binary, ternary, quaternary and quintic bounded systems. Also, the best-fit values of these distributions are given in Table~\ref{tab: Orbital properties evolution}. As this is the population of PBHs within the cluster, then it is clear that eccentricities lie in the elliptical range $ 0 \leq e_{\cpop,\ipop}, t) < 1 $ and semi-major axis track the typical cluster length scale at all times, $ a_{\cpop, \ipop, t} \approx X_{\cpop} $.

A number of features do stand out. The distribution shape is largely uniform across all time-slices but the IC. It shows for $ t \geq t_{28} = \unit [1.38 \times 10^{6}] {yr} $ a peak at $ (a_{\cpop, \ipop}, e_{\cpop,\ipop}) \vert_{t} \approx (1, X_{\cpop}) \vert_{t} $, characterised by the distribution's best-fit value at $ (a_{\cpop, \ipop}, e_{\cpop,\ipop}) \vert^{\bfcap}_{t} $, and two tails
\begin{enumerate}[label=\roman*)]
	\item The first tail is vertical in $ (a_{\cpop, \ipop}, e_{\cpop,\ipop}) $ orbital phase space, of roughly normally distributed semi-major axes around the time-dependent semi-major axes best-fit value $ \unit [0.865] {pc} \leq a_{\cpop, \ipop} \vert^{\bfcap}_{t} \leq \unit [786] {pc} $ at each time-slice.
	\item The second tail is horizontal, roughly constant in eccentricities, truncated at the best fit value of $ e_{\cpop, \ipop} \vert^{\bfcap}_{t} = 1 $ and saturated at it. The distribution of eccentricities is then heavily skewed to the higher end of the available range at all times, with a distribution best-fit value at $ e_{\cpop,\ipop} \vert_{t} = 1.00 $ and PBHs consequently describing highly elliptical orbits around the cluster barycentre, and very rare circular orbits, with less than 5\% of isolated PBHs with $ 0.00 \leq e_{\cpop,\ipop}, \vert_{t} \leq 0.15 $.
\end{enumerate}



\subsubsection{Custer binaries orbital properties}
\label{subsubsec: Custer binaries orbital properties}

The complementary population of bounded cluster PBHs orbital parameter distributions is given in Figure~\ref{subfig: Orbital properties cluster binary} along with their best-fit values in Table~\ref{tab: Orbital properties evolution}, for $ t \geq t_{28} = \unit [1.38 \times 10^{6}] {yr} $ as there are not enough ejecta PBHs to extract reliable confidence regions before this time.

As it it shown there, the cluster, bounded PBHs distribution shape and position in $ (a_{\cpop, \bpop}, e_{\cpop,\bpop}) $ of orbital parameter space does not differ substantially from that of cluster, unbounded PBHs, again exhibiting the aforementioned two tails, one horizontal of constant semi-major axis in the range of
\begin{equation}
	\unit [0.128] {pc} \leq a_{\cpop,\bpop} \vert_{t} \leq \unit [105] {pc},
\end{equation}
and one vertical of constant eccentricity with
\begin{equation}
	0.916 \leq e_{\cpop, \bpop} \vert^{\bfcap}_{t} \leq 1.000 ,
\end{equation}
asymptotically approaching the parabolic value over time, extending from the best-fit values $ a_{\cpop, \ipop} \vert^{\bfcap}_{t} $.

There is, however, a difference in one particular respect. As seen in Figure~\ref{subfig: Orbital properties cluster binary} and compared to Figure~\ref{subfig: Orbital properties cluster}, binary PBH orbits tend to be more circular than those of cluster PBH orbits. This can be easily seen in the fact that, as opposed to the cluster, isolated case, 95\% C.R. in orbital parameter space do reach indeed the $ e = 0 $ limit of circular orbits in each of the time-slices considered but that of the initial time, while the previous case eccentricities were, for all considered time-slices, above the threshold of $ e = 0.15 $ for 95\% of these PBHs.



\subsubsection{Cluster transient binaries}
\label{subsubsec: Cluster transient binaries}

Figure~\ref{subfig: Binaries lifetime cluster} shows the lifetimes of cluster bounded sub-systems, the majority of which coming as binary pairs as detailed in Section~\ref{subsec: Isolated and Bounded populations}.

We have located in total $ N_{\cpop,\bpop} = 9.61 \times 10^{5} $ of such binary pairs by having, at each time-slice, identified all bounded in-cluster PBH sub-systems parent object and binary, ternary, quaternary and quintic pairs. Note that a large fraction of these sub-systems do exhibit continuity in the following time-slices, preserving the same primary and secondary objects identities, even if most do eventually disrupt, hence the label transient applies to them.

However, a non-negligible fraction of these do form by one time-slice only to disrupt by the following time-slice. This suggests again that the total number of binary pairs is underestimated by a non-negligible fraction of the aforementioned $ N_{\cpop,\bpop} $ total since some binaries will form and disrupt between consecutive time-slices and will not show at all in our simulations.

We find that, prior to the burn-in time $ t_{\bicap} = \unit [2.64 \times 10^{5}] {yr} $, transients have lifetimes never exceeding the $ \delta t_{\cpop,\bpop} = \unit [3.0 \times 10^{7}] {yr} $, indicating that all of the sub-systems present in the IC are disrupted during the first four time-runs, none surviving to the present epoch.

Only later, for $ t > t_{\bicap} $ do cluster binaries start forming with lifetimes typically ranging from a time-varying minimum value, coincident with the time-step at which the binary is formed $ \delta t_{\cpop,\bpop} (t) = \Delta t_{\cpop,\bpop} (t) $, which is natural since, as explained, shorter-lived binaries cannot be captured in our simulations due to the discrete data extraction, up to timer greater than the simulation time of $ \delta t_{\cpop,\bpop} = \unit [1.38 \times 10^{10}] {yr} $.

Even after accounting for these missed binaries, it is clear from Figure~\ref{subfig: Binaries lifetime cluster} that these transient binary pairs do form with a lifetime distribution centred at a time value coinciding with the time-increasing characteristic time of evolution, which roughly equal to the simulation time. The tail of the lifetime distribution is large with lifetime values larger than the simulation time, as indeed some binaries formed as early as $ t = \unit [9.66 \times 10^{4}] {yr} $, shortly-before the burn-in time, do indeed survive the whole simulation period as binary pairs in wide orbits around the PBH cluster, avoiding disruption by other cluster objects thanks to the low PBH density in such region.



\subsection{Ejecta orbital properties}
\label{subsec: Ejecta orbital properties}

In Figure~\ref{subfig: Orbital properties ejecta} we show the PBHs occupied orbital parameter space $ \rho (a_{\epop, \ipop}, e_{e,\ipop}, t) $ for the isolated ejecta population, considering only the PBH cluster population and excluding the binary, ternary, and quaternary bounded systems. Moreover, in Table~\ref{tab: Orbital properties evolution} we give the best-fit values of these distributions. Since this is the population of ejecta PBHs, eccentricities are constrained to the hyperbolic range of $ 1 \leq e_{\cpop,\ipop}, t) \leq \infty $ and semi-major axis are inevitably larger than those of cluster PBHs, $ a_{\cpop, \ipop, t} \approx X_{\epop} \geq X_{\cpop} $.

A number of issues are worth discussing. The distribution shape is again largely uniform across all time-slices, this time with less variation between the IC, and later evolution than it was the case for cluster PBHs. The shape is, naturally, very different in any case from that of cluster PBHs.

It does, however, exhibit some common characteristics. Just like in the cluster case, starting at $ t \geq t_{28} = \unit [1.38 \times 10^{6}] {yr} $ the distribution exhibits a peak at the best-fit value at $ (a_{\epop, \ipop}, e_{\epop,\ipop}) \vert^{\bfcap}_{t} $ that asymptotically approaches the parabolic case $ e = 1 $, deviating from it by less that 5\% by the time the simulations have entered the fourth run at $ t_{37} = \unit [1.38 \times 10^{7}] {yr} $.

Moreover, the distribution also exhibits two tails:

\begin{enumerate}[label=\roman*)]
	\item A first slanted tail in $ (a_{\epop, \ipop}, e_{\epop,\ipop}) $ orbital phase space towards smaller semi-major axis and higher eccentricities, starting at the best-fit value and including PBHs closer to the cluster with lower semi-major-axes. In this case, it is found that eccentricity negatively correlate with semi-major axis and the best-fit value bound to the range $ \unit [2.73] {pc} \leq a_{\epop,\ipop} \vert_{t} \leq \unit [200] {pc} $, with the tails easily extending by a factor of $ \order {100} $ of this value.
	\item A second tail that is horizontal in orbital phase space, roughly constant in eccentricities, truncated at the best fit value of $ e_{\cpop, \ipop} \vert^{\bfcap}_{t} = 1 $ and saturated at it just like it was the case for cluster objects. The distribution of eccentricities is this time heavily skewed to the lower end of the available range of eccentricities at all times, with a distribution best-fit value in the interval $ 1.00 \leq e_{\cpop,\ipop} \vert_{t} \leq 1.32 $ and PBHs consequently describing hyperbolic orbits, asymptotically approaching the inferior bound as time progresses.
\end{enumerate}



\subsubsection{Ejecta binaries orbital properties}
\label{subsubsec: Ejecta binaries orbital properties}

The complementary orbital parameter distributions of for the population of bounded ejecta PBHs is given are Figure~\ref{subfig: Orbital properties ejecta binary} along with their best-fit values in Table~\ref{tab: Orbital properties evolution}.

In this case, the ejecta, bounded PBHs distribution shape and position in $ (a_{\epop, \bpop}, e_{\epop,\bpop}) $ of orbital parameter space does differ substantially from that of ejecta, unbounded PBHs, in contrast with cluster PBHs, where the distribution was roughly universal for both isolated and bounded objects. Now the shape does not exhibit the aforementioned two tails, but only the first slanted one, extending from the best-fit values $ a_{\cpop, \ipop} \vert^{\bfcap}_{t} $. Orbital semi-major axis best-fits are then constrained to
\begin{equation}
	\unit [0.217] {pc} \leq a_{\epop,\bpop} \vert_{t} \leq \unit [498] {pc},
\end{equation}
while orbital eccentricities best fits are constrained to
\begin{equation}
	1.00 \leq e_{\epop, \bpop} \vert^{\bfcap}_{t} \leq 1.16,
\end{equation}
again asymptotically approaching the parabolic threshold as time progresses.

It is again the case that the orbital semi-major axis correlates negatively with eccentricities. Moreover, as seen in Figure~\ref{subfig: Orbital properties ejecta binary}, and compared to Figure~\ref{subfig: Orbital properties ejecta}, binary PBH orbits tend to approach the parabolic limit of hyperbolic orbits faster than those of cluster PBH orbits.

This can be easily seen in the fact that, as opposed to the ejecta, isolated case, the 95\% C.R. in orbital parameter space do never exceed the $ e = 47 $ threshold for the first considered time-slice, decaying to the $ e = 20 $ level in the case of the last time-slice; while in the previous case eccentricities were below the threshold of $ e = 1.1 $ for the first considered time-slice, increasing to the $ e = 330 $ level in the case of the last time-slice, for 95\% of these PBHs.



\subsubsection{Ejecta stable binaries}
\label{subsubsec: Ejecta stable binaries}

\begin{figure*}[t!]
	\centering
	\subfloat[Transient in-cluster binary lifetimes, $ \Delta t_{\cpop, \bpop} (t) $.]{
	\hspace*{-1.00cm}
	\includegraphics[width = 0.50\textwidth]{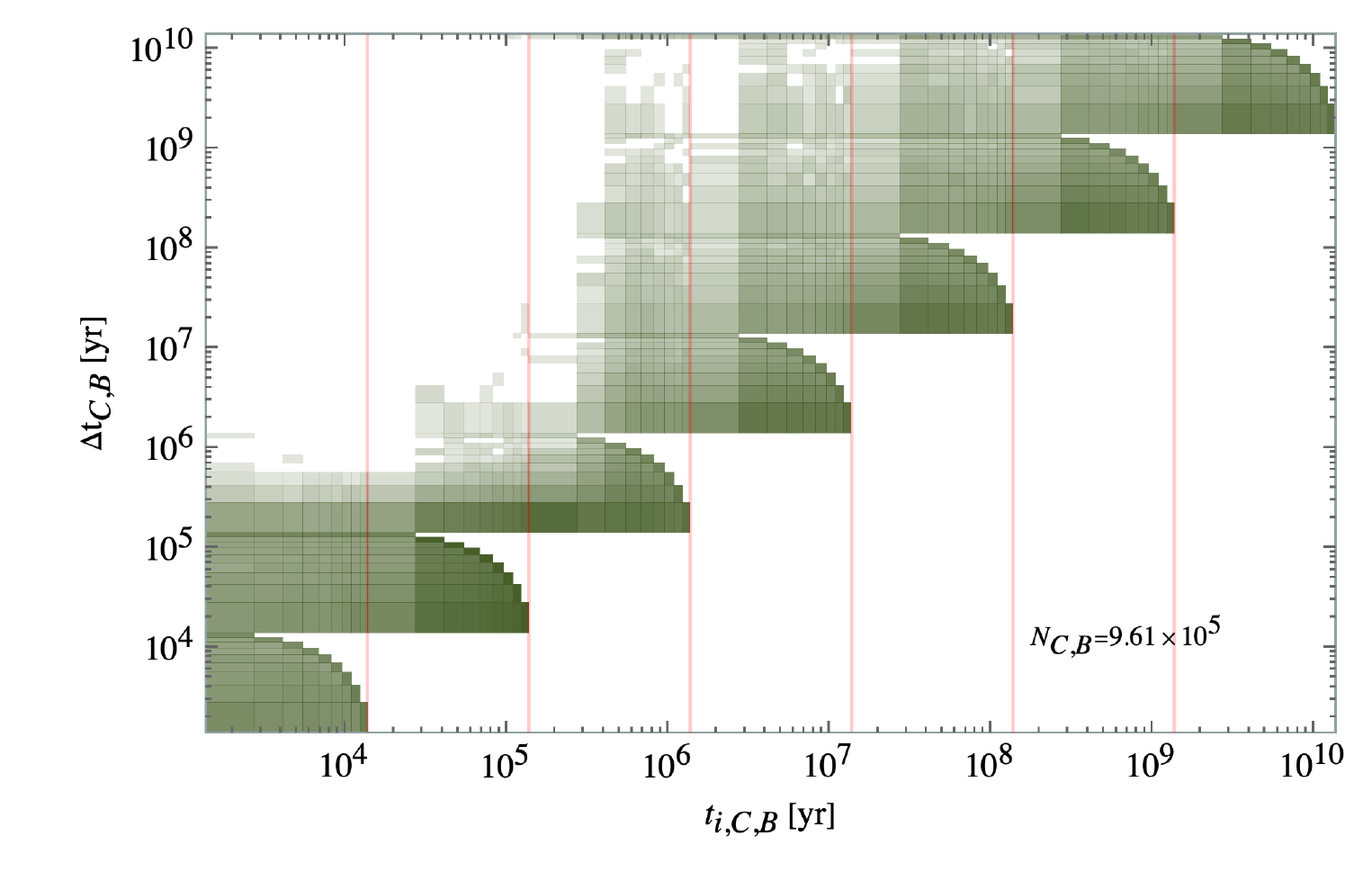}
	\label{subfig: Binaries lifetime cluster}}
	\subfloat[Stable in-ejecta binary lifetimes, $ \Delta t_{\epop, \bpop} (t) $.]{
	\hspace*{-0.00cm}
	\includegraphics[width = 0.50\textwidth]{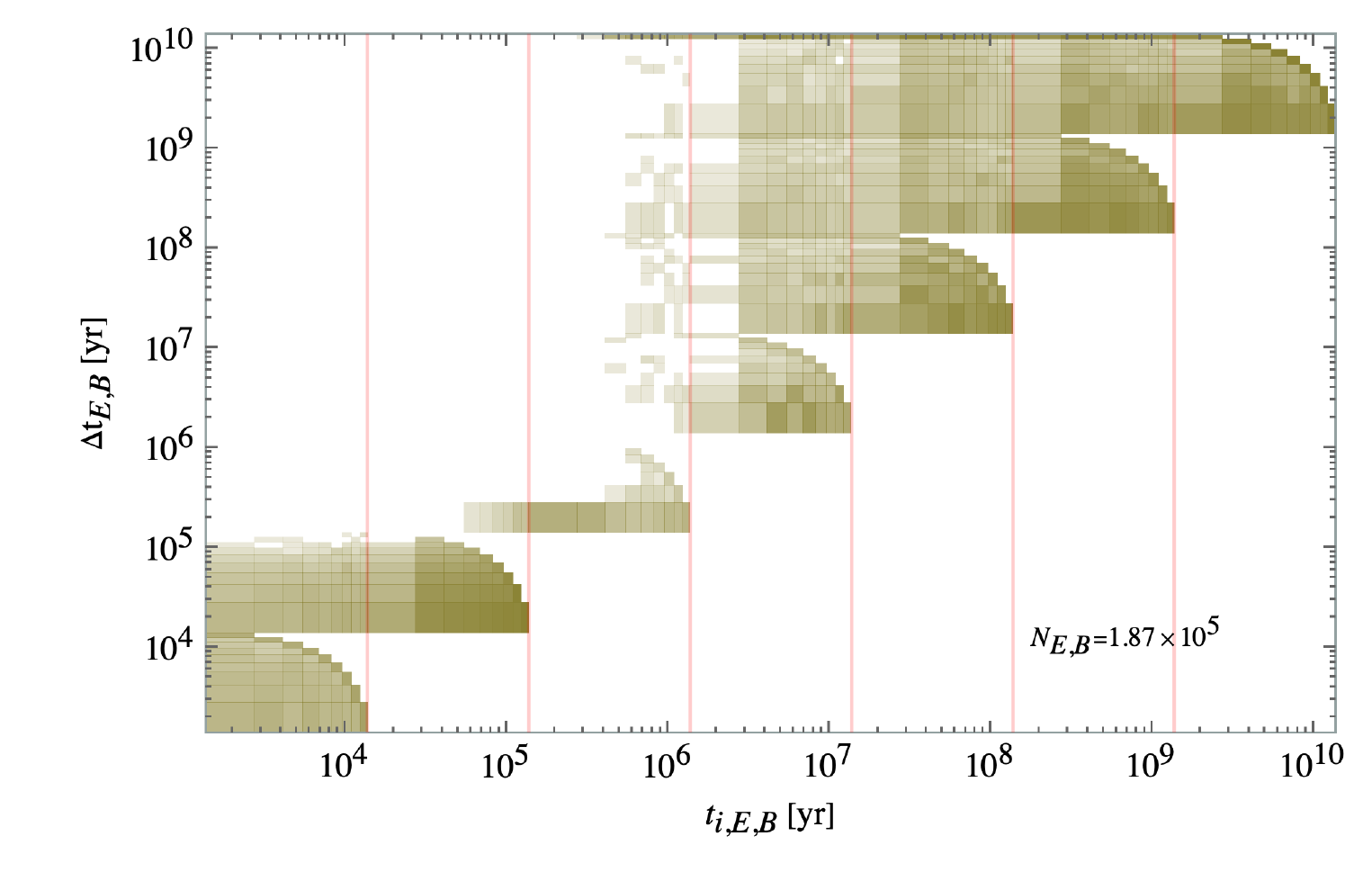}
	\label{subfig: Binaries lifetime ejecta}}
	\caption{
	All panels: Binary lifetime histogram for in-cluster and in-ejecta objects, for each time-slice in the simulations.
	The total number of in-cluster binaries amounts to $ N_{\cpop, \bpop} = 3.85 \times 10^{5} $ objects, while the total number of in-ejecta binaries amounts to $ N_{\epop, \bpop} = 7.46 \times 10^{4} $ objects, throughout all realisations and time-slices.
	}
	\label{fig: Transient and stable binary periods}
\end{figure*}

Figure~\ref{subfig: Binaries lifetime ejecta} shows the lifetimes of ejecta bounded sub-systems, the majority of which as simple binary pairs as detailed in Section~\ref{subsec: Isolated and Bounded populations}.

We have located in total $ N_{\cpop,\bpop} = 1.87 \times 10^{5} $ of such binary pairs by having, at each time-slice, identified all bounded in-ejecta PBH sub-systems parent object and binary, ternary, quaternary and quintic pairs, in a similar way that we did with cluster objects. Note that, in this case, the vast majority of these sub-systems exhibit continuity in the following time-slices, preserving the same primary and secondary objects identities, given that only a negligible fraction of these do form by one time-slice only to be disrupted right after.

This is due to the fact that once in the ejecta population, the chance of a PBH-PBH encounter that may disrupt the binary pair is extremely small and increasingly suppressed over time as our simulated cluster live in isolation and the PBH density quickly and monotonically decays as objects move away from the cluster core. Because of this, and unlike what was the case for cluster transients, this ejecta binary pairs are labelled stable, and are not underestimated by the discrete time-step of collecting information in the simulations.

We find that, prior the burn-in time $ t_{\bicap} = \unit [2.64 \times 10^{5}] {yr} $, transients have lifetimes never exceeding the $ \delta t_{\cpop,\bpop} = \unit [2.0 \times 10^{5}] {yr} $, indicating that all of the sub-systems present in the IC are disrupted during the first three time-runs, none surviving even to the burn-in time.

Only later, for $ t > t_{\bicap} $ do ejecta binaries start forming with lifetimes typically ranging again, as it was the case for cluster binaries, from a time-varying minimum value coincident with the time-step at which the binary is formed $ \delta t_{\cpop,\bpop} (t) = \Delta t_{\cpop,\bpop} (t) $, up to timer greater than the simulation time of $ \delta t_{\cpop,\bpop} = \unit [1.38 \times 10^{10}] {yr} $.

Note that, as Figure~\ref{subfig: Binaries lifetime cluster} suggests, this stable binary pairs do form with a lifetime distribution centred at a time coinciding with the monotonically-increasing characteristic time of evolution, which is roughly equal to the simulation time. The tail of the lifetime distribution is large with lifetime values larger than the simulation time, as indeed some binaries form as early as $ t = \unit [4.14 \times 10^{5}] {yr} $.


\subsection{Hyperbolic encounters power spectra}
\label{subsec: hyperbolic encounters power spectra}

In this section we aim to extract the spectrum of hyperbolic encounters in our simulations. Ideally, we would extract the distribution by identifying the hyperbolic encounters themselves at the moment of the closest approach between the bodies, and computing the distribution of the relative orbital and dynamical parameters from there.

However, we had seen in Section~\ref{subsubsec: hyperbolic encounter rates} that a large majority of these hyperbolic encounters cannot be identified by this method as it is found that the time interval at which the simulation data is extracted is not short enough to capture accurately the trajectories of the participating PBHs. Two PBHs may be closest neighbours at one time-slice, yet at the next time-slice there may be tens of other PBHs closer to them even if they were close enough originally to have undergone a hyperbolic encounter themselves.

Instead, we aim to extract the spectrum of potential hyperbolic encounters in our simulations. We do this by computing the relative eccentricities and impact parameters of all cluster PBHs with respect to their closest $ 100 $ neighbours at selected time-slices, since those are assumed to be produce the majority of the actual encounters in the simulations. For each of the pairs, then we extract the masses of both PBHs in the pair $ m_{i},\ m_{j}) $, as well as the relative positions $ \vec{x}_{i,j}^{\rel} = \vec{x}_{j}-\vec{x}_{i} $ and relative velocities $ \vec{v}_{i,j}^{\rel} = \vec{v}_{j}-\vec{v}_{i} $ where $ i,\ j $ denote the $ j^{\mathrm{the}} $ closest PBH to the $ i^{\mathrm{the}} $ PBH in the cluster.

We compute the relative eccentricity of the pair then by making use of the expression
\begin{equation}
	e_{i,j}^{\rel} = \left | \frac {\vec{v}_{i,j}^{\rel} \times \vec{h}_{i,j}^{\rel}} {G_{N} (m_{i}+m_{j})}-\frac {\vec{x}_{i,j}^{\rel}} {|\vec{x}_{i,j}^{\rel}|} \right |,
	\label{eq: Eccentricity expression}
\end{equation}
where $ \vec{h}_{i,j}^{\rel} = \vec{x}_{i,j}^{\rel} \times \vec{v}_{i,j}^{\rel} $ and $ G_{N} $ is the gravitational constant in the appropriate units.

Then, we compute the pair semi-minor axis by making use of the expression
\begin{equation}
	e_{i,j}^{\rel} = \left | \frac {|\vec{h}_{i,j}^{\rel}|^{2}} {G_{N} (m_{i}+m_{j}) |1-e_{i,j}^{\rel}|^{2}} \right |,
	\label{eq: Impact parameter expression}
\end{equation}
which corresponds to the impact parameter of an hyperbolic encounter when the relative eccentricity exceeds the parabolic limit, for $ e_{i,j}^{\rel} > 1 $.

To ensure that our results are robust, we have repeated the power spectrum extraction algorithm for eccentricities and semi-minor axis where we had considered the 100 closest neighbours for the first, middle and last $ 1/3 $ of objects spanning the $ 1^{\mathrm{st}}$-to-$ 33^{\mathrm{rd}}$, $ 34^{\mathrm{th}}$-to-$ 67^{\mathrm{th}}$ and $ 68^{\mathrm{th}}$-to-$ 99^{\mathrm{th}}$ closest neighbours and found that there power spectra does not change significantly in any of these cases, the only noticeable effect being a widening of the power spectrum to larger impact parameters if more than 100 neighbours are considered as it is be natural to expect in any case.

\begin{figure*}[t!]
	\centering
	\subfloat[Relative eccentricities distribution, rescaled, $ \bar{f}_{1, 100} (e_{i, j}) $.]{
	\hspace*{-1.00cm}
	\includegraphics[width = 0.50\textwidth]{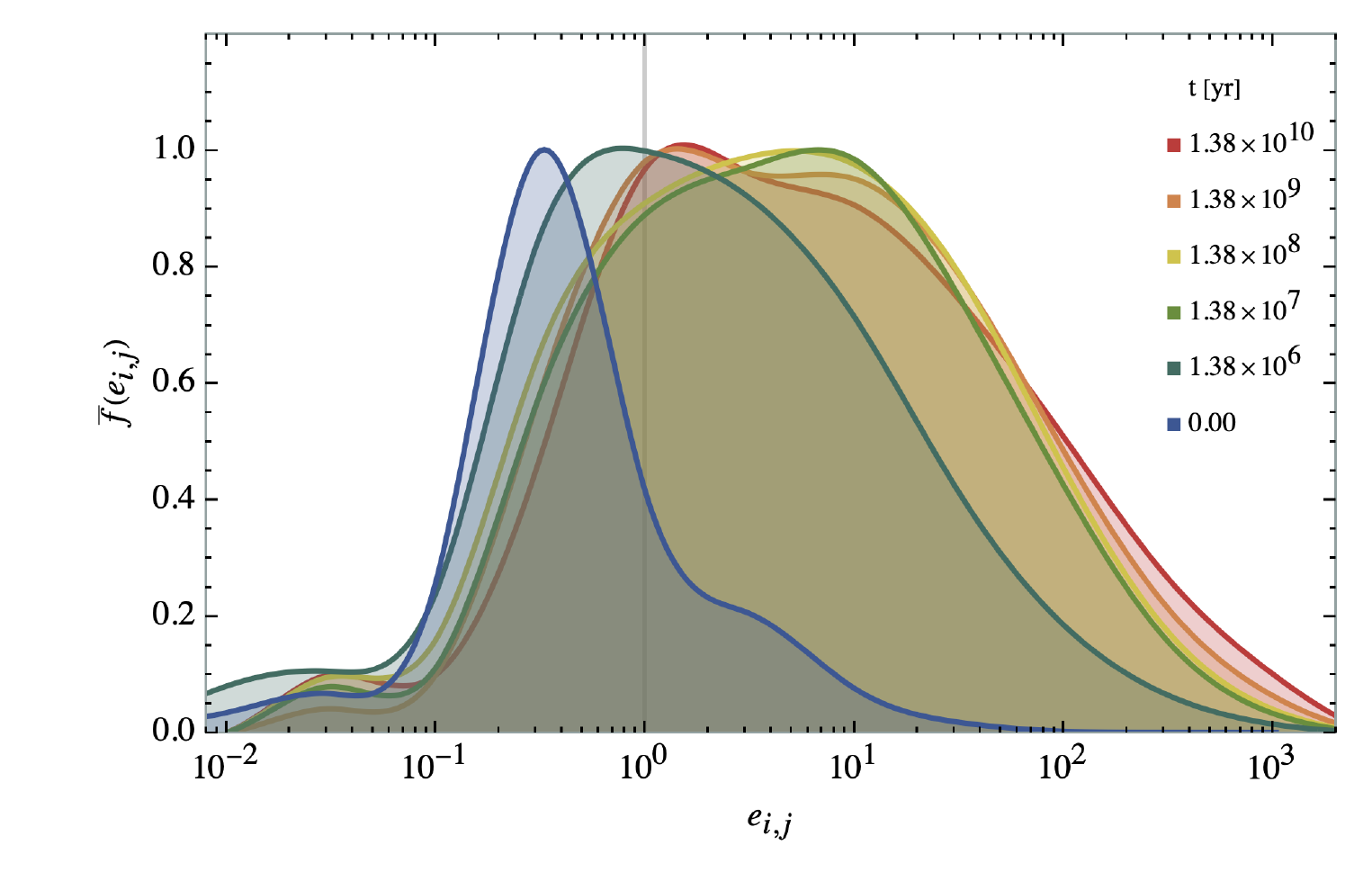}
	\label{subfig: Relative eccentricities profile}}
	\subfloat[Relative impact parameters distribution, rescaled, $ \bar{f}_{1, 100} (b_{i, j}) $.]{
	\hspace*{-0.00cm}
	\includegraphics[width = 0.50\textwidth]{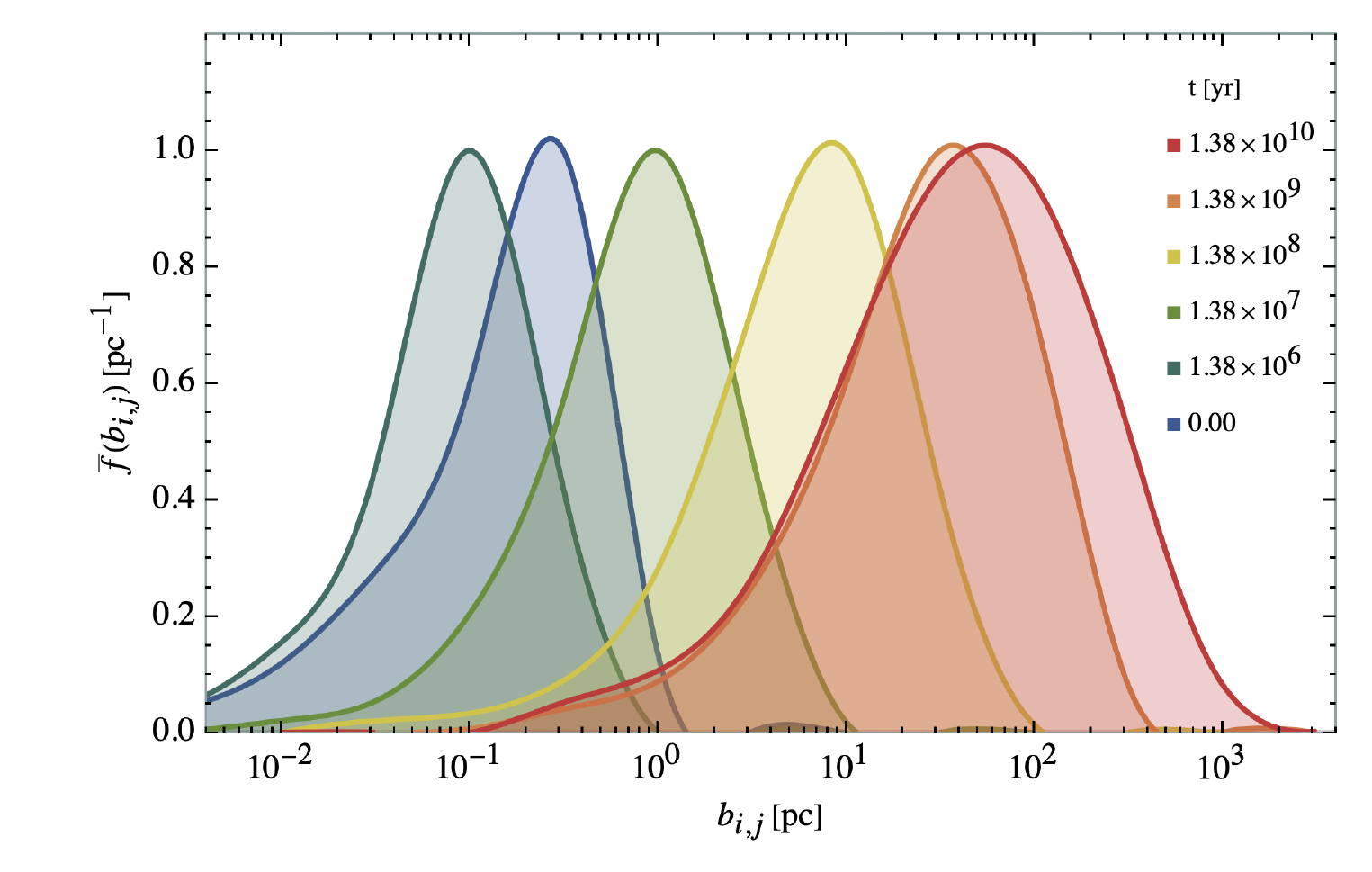}
	\label{subfig: Relative semi-minor axis profile}}
	\caption{
	Shown are the relative orbital eccentricity and impact distributions for the identified two-body encounters at selected times.
	These are computed by taking, for each realisation, the 100 PBHs closest to the cluster barycenter and calculating the array of $ 100 \times 100 $ relative eccentricities among them, for the 6 time-slices that split the last four runs and the IC. Fits for the distributions are given in Table~\ref{tab: Hyperbolic encounters parameters}.
	}
	\label{fig: Hyperbolic encounters parameters}
\end{figure*}

The result of this computation is presented next. We show the results of this computation, for a total of $ N_{i,j}^{\rel} = 100 N_{\rcap} N_{\ocap} = 10^{8} $ pairs of PBHs per time-slice in all considered time-slices but the IC, in Figure~\ref{subfig: Relative eccentricities profile} for the relative eccentricities and Figure~\ref{subfig: Relative semi-minor axis profile} for the relative impact parameters. We do as well fits the resulting curves to Log-Normal distributions in Table~\ref{tab: Hyperbolic encounters parameters}, having tested prior to that other distributions and found that a Log-Normal distribution describes best the data.

We find in Figure~\ref{subfig: Relative eccentricities profile} that, for all the considered time-slices, there is a significant and roughly time-independent fraction of $ i, \ j $ pairs that may potentially produce hyperbolic encounters with $ e_{i,j}^{\rel} > 1 $. Moreover, this encounters may be produced with eccentricities as large as large as $ e_{i,j}^{\rel} = \order {1000} $.

Note that the fraction of hyperbolic encounters is small with a value of 34\% at the IC at the initial time as one would expect from the smaller velocity dispersion of the Maxwell-Boltzmann distribution as compared to the velocity distribution at later stages of the evolution.

However, the fraction of hyperbolic encounters with respect to the total number of encounters does grow monotonically to a value of 64\% at the beginning of the fourth time-run ($ t_{28} = \unit [1.38 \times 10^{6}] {yr} $) and reaching a final value of 71\% at the present ($ t_{64} = \unit [1.38 \times 10^{10}] {yr} $).

Conversely, the fraction of potentially binding encounters decreases from 66\% in the IC at the initial time to 36\% at the beginning of the fourth time-run ($ t_{28} = \unit [1.38 \times 10^{6}] {yr} $) and finally arriving to 29\% at the present ($ t_{64} = \unit [1.38 \times 10^{10}] {yr} $), in line with the overabundance of binaries in the IC and the monotonically decreasing rate of transients throughout as the cluster evolves. The bump in the power spectrum noticeable at eccentricities in the eccentricity range $ e_{i,j}^{\rel} \leq 0.1 $ is produced by real, and not potential, in-cluster binary pairs, and has, as expected, much less power than that the part of the power spectrum dominated by potential encounters in the eccentricity range of $ e_{i,j}^{\rel} \leq \order {0.01} $.

We find as well in Figure~\ref{subfig: Relative semi-minor axis profile} that, again for all the considered time-slices $ t_{i}, i = 0, 28, 37, 46, 55, 64 $ that divide the last four runs and the IC, the range covered by the impact parameter for all potential encounters is constrained to the interval $ \unit[0.001] {pc} \leq b_{i,j}^{\rel} \leq \unit [1000] {pc} $.

However, and unlike what was the case with eccentricities, the range varies greatly in time from the IC at $ t_{0} = \unit [0] {yr} $ where
\begin{equation}
	\unit[0.001] {pc} \leq b_{i,j}^{\rel} (t_{0}) \leq \unit [1] {pc},
\end{equation}
to the present time where
\begin{equation}
	\unit[0.1] {pc} \leq b_{i,j}^{\rel} (t_{0}) \leq \unit [1000] {pc},
\end{equation}
tracking the characteristic length scale of the cluster in time.

It is clear, however, that the vast majority of all pairs here considered will not produce close encounters. Only a very small fraction of the total number of pairs will indeed be at distances of the order of $ \unit [10^{-5}] {pc} \approx \unit [2] {AU} $ as for later times the power spectrum is noticeably suppressed at lengths scales of $ b_{i,j}^{\rel} \leq \unit [\order {0.001}] {pc} $ at the initial time, and of $ b_{i,j}^{\rel} \leq \unit [\order {0.1}] {pc} $ at the present time.

Considering, however, that there are in total $ N_{i,j}^{\rel} = 10^{8} $ pairs potentially producing the encounters, then the likelihood of at least a few of these occurring during the whole evolution time remains large.

Note that the cluster infall and rebound stage described in Sections~\ref{subsubsec: Density profile evolution}-\ref{subsubsec: Velocity profile evolution} at around the burn-in time is apparent as well in the impact parameter profile evolution, which exhibits the same behaviour. The impact parameter profile at the IC is skewed towards values larger than the impact parameter distribution at the beginning of the fourth run ($ t_{28} = \unit [1.38 \times 10^{6}] {yr} $), indicative of the contraction of the cluster up to that time, and more importantly, of the concentration of PBHs in a cuspier core after the IC is erased after the burn-in time. Only after the beginning of the fifth run ($ t_{37} = \unit [1.38 \times 10^{7}] {yr} $) does the IC begin to skew towards values smaller than the impact parameter distribution at that time, increasingly so as the cluster puffs-up and the typical impact parameter grows accordingly.

Last, it is worth mentioning that no significant correlation has been found between eccentricity and semi-minor axis values for this number of pairs. Somewhat contrary to expectation, smaller impact parameters do not correlate with smaller eccentricities, as one would expect from dynamical arguments. The reason for this is that the number of neighbours considered for each PBH has been chosen, as explained in the beginning of the present section, to be small enough that the distributions are consistent with those extracted taking in only the nearest neighbours among those. The shell centred on each of the PBHs typically extends to a distance smaller than the cluster size by $ \order {10} $, and is approximately homogeneous so it does not contain wildly different velocities or masses on average.


\begin{table*}[t!]
	\centering
	\begin{tabular}[c]{| l l | l l l l | l l l l |}
	\hline
	\quad & \quad &
	\multicolumn{4}{c |}{Relative eccentricities fits:} &
	\multicolumn{4}{c |}{Relative impact parameters fits:} \\
	[0.5ex]
	\hline\hline
	$ i $ & $ \unit [t_{i}] {[yr]} $ &
	Fit: & $ r^{2}_{i,e} $ &
	$ \mu_{i,e} $ & $ \sigma_{i,e} $ &
	Fit: & $ r^{2}_{i,b} $ &
	$ \unit {\mu_{i,b}} [\textrm{pc}] $ & $ \unit {\sigma_{i,b}} [\textrm{pc}] $ \\
	[0.5ex]
	\hline
	$ 0 $ & $ 0 $ &
	[LN] & $ 0.992 $ &
	$ 1.81 \pm 0.14 $ & $ 1.559 \pm 0.040 $ &
	[LN] & $ 0.856 $ &
	$ 0.0 \pm 5.7 $ & $ 1.45 \pm 0.14 $ \\
	$ 28 $ & $ 13.8 \times 10^{6} $ &
	[LN] & $ 0.998 $ &
	$ 6.27 \pm 0.31 $ & $ 2.342 \pm 0.088 $ &
	[LN] & $ 0.509 $ &
	$ 0.0 \substack {+3.0 \\ -0.0} $ & $ 1.83 \pm 0.77 $ \\
	$ 37 $ & $ 13.8 \times 10^{7} $ &
	[LN] & $ 0.998 $ &
	$ 6.57 \pm 0.29 $ & $ 2.208 \pm 0.083 $ &
	[LN] & $ 0.771 $ &
	$ 9.8 \pm 3.8 $ & $ 4.0 \pm 1.0 $ \\
	$ 46 $ & $ 13.8 \times 10^{8} $ &
	[LN] & $ 0.999 $ &
	$ 6.74 \pm 0.23 $ & $ 2.268 \pm 0.068 $ &
	[LN] & $ 0.789 $ &
	$ 0.00 \substack {+0.94 \\ -0.00} $ & $ 1.71 \pm 0.24 $ \\
	$ 55 $ & $ 13.8 \times 10^{9} $ &
	[LN] & $ 0.999 $ &
	$ 6.72 \pm 0.23 $ & $ 2.166 \pm 0.064 $ &
	[LN] & $ 0.975 $ &
	$ 3.43 \pm 0.21 $ & $ 1.084 \pm 0.054 $ \\
	$ 64 $ & $ 13.8 \times 10^{10} $ &
	[LN] & $ 0.999 $ &
	$ 7.33 \pm 0.25 $ & $ 2.330 \pm 0.072 $ &
	[LN] & $ 0.987 $ &
	$ 5.55 \pm 0.51 $ & $ 1.45 \pm 0.15 $ \\
	\hline
	\end{tabular}
	\caption{
	Nearby objects eccentricity and impact parameter distributions of Figure~\ref{fig: Hyperbolic encounters parameters} distribution Log-Normal fits: $ \rho_{z,i,\textrm{P}} = (z \sigma_{z,i,\textrm{P}} \sqrt{2 \pi})^{-1} \exp (- (\textrm{Log-Normal} z-\mu_{z,i,\textrm{P}})^{2} / 2 \sigma_{z,i,\textrm{P}}^{2}) $ where $ \mu_{i,\textrm{P}} $ is the expected value and $ \sigma_{i,\textrm{P}} $ corresponds to the standard deviation of the $ z $ variable's natural logarithm at selected times, and $ \textrm{P} = \cpop, \epop $ stands for the cluster and ejecta populations respectively and $ z = e, b $ for the mass, position and velocity functions.	Note that, fits to Rayleigh $ \chi^{2}_{2} $ and Maxwell-Boltzmann $ \chi^{2}_{3} $ distributions have also been tried for the position and velocity profiles, having found that, except for those time-slices before the burn-in time, a Log-Normal fit beats the former trials. Whenever none of the above fits meaningfully describe the data, $ r^{2} < 0.8 $, no fit is provided.
	}
	\label{tab: Hyperbolic encounters parameters}
\end{table*}



\section{Background DM implications}
\label{sec: Background DM implications}

In this section we aim to quantify the implications of the proposed PBHs clusters in the global DM distribution. In particular, Section~\ref{subsec: DM energy density} will deal with predictions for the DM background of cluster and ejecta objects separately for a typically sized galactic halo, while in Section~\ref{subsec: Gravitational event rates} we will compute the distinct hyperbolic encounter and merger event rates of PBHs per comoving volume.



\subsection{DM energy density}
\label{subsec: DM energy density}

It was mentioned in Section~\ref{subsubsec: Position profile evolution} that the trajectory stability of distant objects bound to the cluster or belonging to the ejecta sphere-of-influence is guaranteed in our simulations, since both the cluster and ejecta sphere-of-influence lives in isolation and the likelihood of an encounter that may transfer enough kinetic energy to overcome the cluster escape velocity is very low, as the ejected objects travel the cluster periphery where this far-off objects live very fast, and also given that since the periods of these objects are very high.

However, in a realistic scenario this is not the case anymore. A single galactic halo may contain, as we will soon calculate, $ \order {10^{7}} $ of these PBH clusters with their corresponding cluster and ejecta spheres-of-influence extending up to order kpc in the former case and reaching up to hundreds of kpc in the latter case. These numbers are large enough that eventually the volumes occupied by both the cluster and ejecta spheres-of-influence overlap, and so interaction between PBHs is no longer a remote possibility but a reality.

In the following, we aim to compute precisely the number of these PBH bubbles and the time at which they do interact or overlap within a typical galactic halo. For this computation, we take as our baseline a Milky Way-like halo, whose total mass Ref.~\cite{Eadie:2019} estimates to be
\begin{align}
	M^{*}_{\textrm{MW}} &= M_{\textrm{MW}} (X < X^{*}_{\mw} = \unit [100] {kpc}) \nn\\
	&= \unit [(5.3 \substack {+2.1 \\ -1.2}) \times 10^{11}] {\msun},
\end{align}
at the 50\% C.L. Moreover, we will assume a local DM fraction of $ f_{\dm} = \Omega_{\textrm{DM}} / \Omega_{0} \vert_{\textrm{MW}} $, and second, that these clusters constitute a particular DM fraction $ f_{\pbh} = \Omega_{\pbh} / \Omega_{\textrm{DM}} $.



\subsubsection{Cluster DM energy density}
\label{subsubsec: Cluster DM energy density}

Depending on the total mass profile of the PBH clusters, one can constrain their abundance and compute the mean inter-cluster distance. In our case, considering our clusters have, typically, a mean total mass of
\begin{align}
	M_{\cpop}^{\tot} (t) & = \sum_{i = 1}^{N_{\ocap}} \delta_{i, \cpop} (t) m_{i} (t) \nn \\
	& \approx N_{\ocap} \bar{m}_{\cpop} (t) f_{\cpop} (t) \nn\\
	& \approx N_{\ocap} \exp \left(\mu_{\mass}+\sigma_{\mass}^2 / 2 \right) (1-f_{\cpop}^{\loss} (t)),
	\label{eq: Cluster mean total mass}
\end{align}
given that for our Log-Normal mass distribution the actual mass mean is $ \bar{m} = \exp \left(\mu_{\mass}+\sigma_{\mass}^2 / 2 \right) $. The mass is itself contained in a sphere of a mean effective radius given by
\begin{equation}
	X_{\cpop}^{\tot} (t) = \bar{x}_{\cpop} = \exp \left(\mu_{\pos}+\sigma_{\pos}^2 /2 \right),
	\label{eq: Cluster mean total radius}
\end{equation}
where the input logarithmic mean position, $ \mu_{\pos} $ and and deviation $ \sigma_{\pos} $ can be used again to get the actual position mean, $ \bar{x} = \exp \left(\mu_{\pos}+\sigma_{\pos}^2 / 2 \right) $, as the cluster profile was shown to be best described by a Log-Normal distribution in all time-slices but the very first ones before the IC is erased after the burn-in time.

The cluster mass, position and velocity distribution parameters $ \mu_{i} $ and $ \sigma_{i} $ for $ i = \mass, \pos, \vel $ are given in Table~\ref{tab: Quasi-static profiles}, while the cluster characteristic scales $ \bar{m} $, $ \bar{x} $ and $ \bar{v} $ are shown in Table~\ref{tab: Characteristic profile properties}.

We now aim to estimate the number of simulated PBHs cluster bubbles in a typical galactic halo in order to infer what would be the population os these objects within a typical galaxy, their mass, mean inter-cluster distance and mean inter-PBH distance at the point where the cluster bubbles have expanded to the point of overlapping with each other. In order to do this, we will take in the following our own Milky Way halo as the standard of a typical galactic halo.

The number of these PBH clusters within the $ X < X^{*}_{\mw} $ sphere centred at the Milky Way core is, considering the halo mass and effective radius from Ref.~\cite{Eadie:2019}, found to be
\begin{equation}
	N_{\cpop}^{\tot} = \frac {f_{\dm} f_{\pbh} M^{*}_{\textrm{MW}}} {M_{\cpop}^{\tot} (t_{0})},
	\label{eq: Cluster mean total number}
\end{equation}
where we have not corrected for the possible loss of ejecta PBHs out of the sphere since the fraction of $ M_{\cpop}^{\tot} (t) $ travelling such long distance is negligible, with about $ \order {0.001} $ of PBH being slingshot away to distances larger than $ X < \unit [100] {kpc} $, being among the lightest of simulated PBHs on top of it. Correspondingly, the number of these in-cluster PBHs within the $ X < X^{*}_{\mw} $ sphere centred at the Milky Way core is then
\begin{equation}
	N_{\cpop}^{\ind} (t) = \frac {N_{\ocap} f_{\dm} f_{\pbh} M^{*}_{\textrm{MW}}} {M_{\cpop}^{\ind} (t)}.
	\label{eq: Cluster mean individual number}
\end{equation}

Assuming now for simplicity a homogeneous distribution of PBH clusters within the halo sphere, then the mean inter-cluster distance within said sphere can be well approximated by $ X_{\cpop}^{\tot} (t) \simeq (V_{\mw} / N_{\cpop}^{\tot} (t))^{1/3} $, where $ V_{\mw} $ stands for the Milky Way volume contained in the $ X < \unit [100] {kpc} $ sphere, or after some substitutions
\begin{equation}
	X_{\cpop}^{\tot} \simeq \left(\frac {X^{*3}_{\mw}} {N_{\cpop}^{\tot}} \right)^{1/3},
	\label{eq: Cluster mean total distance}
\end{equation}
and correspondingly, the mean inter-PBH distance within said sphere can be well approximated by
\begin{equation}
	X_{\cpop}^{\ind} (t) \simeq \left(\frac {X^{*3}_{\mw}} {N_{\cpop}^{\ind} (t)} \right)^{1/3},
	\label{eq: Cluster mean individual distance}
\end{equation}
assuming that the cluster have puffed-up enough to start overlapping with each other. Note that the mean inter-PBH distance $ X_{\cpop}^{\ind} (t) $ will be only of significance one clusters begin to overlap with each other, as prior to that point the PBH two-point correlation function will be bimodal function, and only when the clusters sufficiently overlap with each other will the PBH two-point correlation function be a unimodal function as PBH-to-PBH distances are all roughly the same.

Now, each of these PBH cluster bubbles will contain a monotonically decreasing mass over given by the $ 	M_{\cpop}^{\tot} (t) $ of Eq.~\eqref{eq: Cluster mean total mass}, while all the PBH clusters in the galactic halo will total an also monotonically decreasing joint mass of
\begin{equation}
	M_{\cpop}^{\ind} (t) \simeq N_{\cpop}^{\tot} M_{\cpop}^{\tot} (t) .
	\label{eq: Cluster mean individual mass}
\end{equation}

Last, the mean occupied PBH cluster volume fraction of said sphere is then $ F_{\cpop}^{\tot} (t) = N_{\cpop}^{\tot} (t) V_{\cpop}^{\tot} (t) / V_{\mw} $, where $ V_{\cpop}^{\tot} (t) $ stands for the cluster volume contained in the cluster effective cluster radius $ X^{*}_{\cpop} = \unit [100] {kpc} $, which leads to
\begin{equation}
	F_{\cpop}^{\tot} (t) \simeq N_{\cpop}^{\tot} \left(\frac {X^{*}_{\cpop} (t)} {X^{*}_{\mw}} \right)^{3}.
	\label{eq: Cluster mean fraction volume}
\end{equation}

Assuming a DM fraction within the Milky Way halo sphere of $ f_{\dm} = 0.8 $ entirely due to PBHs, $ f_{\pbh} = 1 $ then Eq.~\eqref{eq: Cluster mean total number}, Eq.~\eqref{eq: Cluster mean total distance} and Eq.~\eqref{eq: Cluster mean fraction volume} yield a total cluster number per sphere, $ N_{\cpop}^{\tot} $ of
\begin{align}
	& N_{\cpop}^{\tot}
	 = \unit [(2.7 \substack {+1.7 \\ -0.9}) \times 10^{7}] {},
	\label{eq: Cluster mean total number per shpere}
\end{align}
separated on average by a distance, $ X_{\cpop}^{\tot} $ of
\begin{align}
	& X_{\cpop}^{\tot}
	 = \unit [800 \substack {+190 \\ -100}] {pc},
	\label{eq: Cluster mean total distance homogeneously}
\end{align}
that is, we have in the different realisations $ \order {10^{7}} $ cluster bubbles with their own spheres of influence, separated by distances of $ \unit [\order {100}] {pc} $, which evaporate and expand as the simulations evolve.

The estimated values of the mean individual PBH number $ N_{\cpop}^{\ind} (t) $ and mean individual PBH distance $ X_{\cpop}^{\ind} (t) $ are given in Table~\ref{tab: Background properties} for the cluster population, following the aforementioned procedure along with the mean mass per cluster bubble $ M_{\cpop}^{\tot} (t) $, mean in-cluster mass per halo $ M_{\cpop}^{\ind} (t) $, and cluster occupied halo volume fraction $ F_{\cpop}^{\tot} (t) $, at the te-slices corresponding to those that divide the last four time-runs plus the IC.

An interesting phenomenology arises from this computation. Table~\ref{tab: Background properties} shows that indeed the cluster objects are perfectly isolated initially, even though, at late times and soon after entering in the last time-run when $t^{*} \in ( 1.38,\,10)$
Gyr, that is not longer the case, since $ F_{\cpop, 64}^{\tot} (t > t^{*}) \approx 1 $ and these puffed-up clusters overlap.

Then the typical inter-cluster PBHs and intra-cluster PBHs distances converge to a single value and the cluster distribution within the galactic halo approaches that of a homogeneous distribution where the outer layers of the clusters overlapping with each other. The cuspy cores of the original clusters remain distinct albeit less so as the simulations evolve approaches the present time. In particular, we find that the typical PBH-to-PBH distance regardless of whether the two originate in the same or separate cluster bubbles approximately is $ X_{\cpop}^{\ind} = \unit [112 \substack {+20 \\ -23}] {pc} $ and of the same order of magnitude than the cluster-to-cluster distance.



\subsubsection{Ejecta DM energy density}
\label{subsubsec: Ejecta DM energy density}

We now aim to repeat the same analysis of Section~\ref{subsubsec: Cluster DM energy density} for ejecta objects, and constrain the abundance and compute the mean inter-ejecta spheres-of-influence distance, taking again into account that these PBH ejecta bubbles do not live in isolation.

Given the very low level of merging in our simulations, we can assume that the cluster and ejecta populations are complementary to each other and adding up to the initial cluster population since the number of ejecta objects is negligible at the time of the IC. Then, the ejecta will typically have a mean total mass of
\begin{align}
	M_{\epop}^{\tot} (t) & = \sum_{i = 1}^{N_{\ocap}} \delta_{i, \epop} (t) m_{i} (t) \nn\\
	& \approx N_{\ocap} \bar{m}_{\epop} (t) f_{\epop} (t) \nn\\
	& \approx N_{\ocap} (\exp \left(\mu_{\mass}+\sigma_{\mass}^2 / 2 \right)) f_{\cpop}^{\gain} (t),
	\label{eq: Ejecta mean total mass}
\end{align}
given that, as we had seen in the previous section, the PBH mean mass is $ \bar{m} = \exp\left(\mu_{\mass}+\sigma_{\mass}^2 / 2 \right) $. The bulk of the mass is then contained in a sphere of a mean effective radius given by
\begin{equation}
	X_{\epop}^{\tot} (t) = \bar{x}_{\epop} = \exp \left(\mu_{\pos}+\sigma_{\pos}^2 /2 \right),
	\label{eq: Ejecta mean total radius}
\end{equation}
similarly to what was the case of the cluster bubble, and centred in it, but larger, by a growing factor of roughly $ \order {100} $ at present times. Note that the mean position is $ \bar{x} = \exp \left(\mu_{\pos}+\sigma_{\pos}^2 / 2 \right) $, as the ejecta position distribution is also best described by a Log-Normal distribution in all time-slices.

The ejecta mass, position and velocity distribution parameters $ \mu_{i} $ and $ \sigma_{i} $ for $ i = \mass, \pos, \vel $ are given in Table~\ref{tab: Quasi-static profiles}, while the ejecta characteristic scales $ \bar{m} $, $ \bar{x} $ and $ \bar{v} $ and are shown in Table~\ref{tab: Characteristic profile properties}. Since each ejecta bubble originates by the PBH expelled from the cluster at its centre, the number of these PBH ejecta bubbles within the $ X < X^{*}_{\mw} $ sphere centred at the Milky Way core is,
\begin{equation}
	N_{\epop}^{\tot} = N_{\cpop}^{\tot}.
	\label{eq: Ejecta mean total number}
\end{equation}
Correspondingly, the number of these in-ejecta PBHs within the $ X < X^{*}_{\mw} $ sphere centred at the Milky Way core is then
\begin{equation}
	N_{\epop}^{\ind} (t) = \frac {N_{\ocap} f_{\dm} f_{\pbh} M^{*}_{\textrm{MW}}} {M_{\epop}^{\ind} (t)}.
	\label{eq: Ejecta mean individual number}
\end{equation}

For a homogeneous distribution of PBH clusters within the halo sphere, then the mean inter-ejecta distance within said sphere need to track the mean inter-cluster distance, and so
\begin{equation}
	X_{\epop}^{\tot} = X_{\cpop}^{\tot},
	\label{eq: Ejecta mean total distance}
\end{equation}
and correspondingly, the mean inter-PBH distance within said sphere can be well approximated by
\begin{equation}
	X_{\epop}^{\ind} (t) \simeq \left(\frac {X^{*3}_{\mw}} {N_{\epop}^{\ind} (t)} \right)^{1/3},
	\label{eq: Ejecta mean individual distance}
\end{equation}
assuming that the ejecta have puffed-up enough to start overlapping with each other, a safe assumption given that it was shown that this was indeed the case for the cluster PBHs in the last run, and the fact that the ejecta bubbles reach to far larger distances from early on.

Note that every PBH ejecta bubble will have a monotonically increasing mass over given by the $ 	 M_{\epop}^{\tot} (t) $ of Eq.~\eqref{eq: Ejecta mean total mass}, while all the PBH ejecta in the galactic halo will total as well an also monotonically increasing joint mass of
\begin{equation}
	M_{\epop}^{\ind} (t) \simeq N_{\epop}^{\tot} M_{\epop}^{\tot} (t) .
	\label{eq: Ejecta mean individual mass}
\end{equation}

The mean occupied PBH ejecta volume fraction of the ejecta sphere-of-influence will be then
\begin{equation}
	F_{\epop}^{\tot} (t) \simeq N_{\epop}^{\tot} \left(\frac {X^{*}_{\epop} (t)} {X^{*}_{\mw}} \right)^{3}.
	\label{eq: Ejecta mean fraction volume}
\end{equation}

Last, assuming again a DM fraction within the Milky Way halo sphere of $ f_{\dm} = 0.8 $ entirely due to PBHs, $ f_{\pbh} = 1 $ then Eq.~\eqref{eq: Ejecta mean total number}, Eq.~\eqref{eq: Ejecta mean total distance} and Eq.~\eqref{eq: Ejecta mean fraction volume} a total ejecta number per sphere, $ N_{\cpop}^{\tot} $ of
\begin{align}
	& N_{\epop}^{\tot} = N_{\cpop}^{\tot}
	 = \unit [(2.7 \substack {+1.7 \\ -0.9}) \times 10^{7}] {},
	\label{eq: Ejecta mean total number per sphere}
\end{align}
separated on average by a distance, $ X_{\cpop}^{\tot} $ of
\begin{align}
	& X_{\epop}^{\tot} = X_{\cpop}^{\tot}
	 = \unit [800 \substack {+190 \\ -100}] {pc}.
	\label{eq: Ejecta mean total distance homogeneously}
\end{align}

The estimated values of the mean individual PBH number $ N_{\epop}^{\ind} (t) $ and mean individual PBH distance $ X_{\epop}^{\ind} (t) $ are given in Table~\ref{tab: Background properties} for the ejecta population, along with the mean mass per ejecta bubble $ M_{\epop}^{\tot} (t) $, mean in-ejecta mass per halo $ M_{\epop}^{\ind} (t) $, and ejecta occupied halo volume fraction $ F_{\epop}^{\tot} (t) $, at the time-slices corresponding to those that divide the last four time-runs plus the IC.

Many interesting features regarding the behaviour of the ejecta bubbles are already apparent in Table~\ref{tab: Background properties}, such that, while the ejecta bubbles sphere-of-influence are isolated initially, already by the fifth run they have started to overlap. By the present time the mixing between ejecta bubbles is complete as $ F_{\epop, 64}^{\tot} \gg 1 $ and it makes no longer sense to treat them as anything other than a uniform background, unlike the case of the cluster bubbles which still preserve a distinct, cuspy core.

In particular, we find that the typical PBH-to-PBH distance regardless of whether the two originate in the same or separate ejecta bubbles is approximately of $ \unit [\order {100}] {pc} $, slowly decreasing as the now galactic halo-sized joint ejecta sphere-of-influence slow expansion is overcome by the gradual increase in population of the ejecta bubbles from continuing evaporation in the cluster, from
 \begin{equation}
 	X_{\epop}^{\ind} = \unit [108 \substack {+29 \\ -17}] {pc},
 \end{equation}
 $ $
 at $ t_{46} = \unit [1.38 \times 10^{8}] {yr} $ to a present value of
 \begin{equation}
 	X_{\epop}^{\ind} = \unit [93 \substack {+24 \\ -13}] {pc}.
 \end{equation}

Note, however, that in this computation we have neglected the fact that while our simulations have both the cluster an ejecta bubbles in isolation, at this scales that is clearly not the case anymore. Ejecta objects in particular will by now feel the potential wells of their neighbour bubbles and will at a point cease to expand when they finally populate the entire galactic DM halo. Individual ejected PBHs will not move in the asymptotically uniform rectilinear motion and their trajectories will bend to orbit the galaxy, most of them in the very little dense environment of the galactic halo outskirts, behaving as a virtually collisionless, cold DM component, given the extreme unlikelihood of a close encounter out if the cores of the cluster bubbles, let alone in the far less dense environment of the ejecta background.


\begin{table*}[t!]
	\centering
	\begin{tabular}[c]{| l l | l l | l l l|}
	\hline
	\multicolumn{2}{| c |}{} &
	\multicolumn{2}{c |}{Cluster background number:} &
	\multicolumn{3}{c |}{Cluster background properties:} \\
	[0.5ex]
	\hline
	$ i $ &	$ \unit [t_{i}] {[yr]}	$ &
	$ N_{\cpop}^{\tot} $ &
	$ F_{\cpop}^{\tot} $ &
	$ \unit [M_{\cpop}^{\ind}] {[\msun]} $ &
	$ \unit [M_{\cpop}^{\tot}] {[\msun]} $ &
	$ \unit [X_{\cpop}^{\ind}] {[pc]} $ \\
	[0.5ex]
	\hline\hline
	$ 0 $ & $ 0 $ &
	$ (2.7 \substack {+1.7 \\ -0.9}) \times 10^{10} $ &	
	$ (5.3 \substack {+4.0 \\ -2.7}) \times 10^{-9} $ &	
	$ (3.0 \substack {+2.6 \\ -1.4}) \times 10^{4} $ &	
	$ (8.0 \substack {+5.0 \\ -2.6}) \times 10^{11} $ &	
	$ - $ \\
	$ 28 $ & $ 1.38 \times 10^{6} $ &
	$ (2.4 \substack {+1.5 \\ -0.8}) \times 10^{10} $ &	
	$ (3.8 \substack {+2.9 \\ -1.8}) \times 10^{-8} $ &	
	$ (2.8 \substack {+2.4 \\ -1.3}) \times 10^{4} $ &	
	$ (7.4 \substack {+5.0 \\ -2.5}) \times 10^{11} $ &	
	$ - $ \\
	$ 37 $ & $ 1.38 \times 10^{7} $ &
	$ (2.0 \substack {+1.3 \\ -0.7}) \times 10^{10} $ &	
	$ (2.0 \substack {+1.5 \\ -1.0}) \times 10^{-6} $ &	
	$ (2.3 \substack {+2.0 \\ -1.1}) \times 10^{4} $ &	
	$ (6.1 \substack {+4.0 \\ -2.3}) \times 10^{11} $ &	
	$ - $ \\
	$ 46 $ & $ 1.38 \times 10^{8} $ &
	$ (1.6 \substack {+1.0 \\ -0.7}) \times [10^{10} $ &	
	$ (2.1 \substack {+1.6 \\ -1.0}) \times 10^{-4} $ &	
	$ (1.8 \substack {+1.6 \\ -1.0}) \times 10^{4} $ &	
	$ (4.9 \substack {+3.2 \\ -2.0}) \times 10^{11} $ &	
	$ - $ \\
	$ 55 $ & $ 1.38 \times 10^{9} $ &
	$ (1.3 \substack {+0.8 \\ -0.6}) \times 10^{10} $ &	
	$ 0.018 \substack {+0.14 \\ -0.009} $ &	
	$ (1.5 \substack {+1.3 \\ -0.9}) \times 10^{4} $ &	
	$ (4.0 \substack {+2.6 \\ -1.9}) \times 10^{11} $ &	
	$ - $ \\
	$ 64 $ & $ 1.38 \times 10^{10} $ &
	$ (9.7 \substack {+7.2 \\ -6.0}) \times 10^{9} $ &	
	$ 1.6 \substack {+1.3 \\ -0.8} $ &	
	$ (1.2 \substack {+1.1 \\ -0.8}) \times 10^{4} $ &	
	$ (3.2 \substack {+2.2 \\ -1.8}) \times 10^{11} $ &	
	$ 112 \substack {+29 \\ -23} $ \\
	\hline
	\multicolumn{2}{c}{} &
	\multicolumn{3}{c}{} &
	\multicolumn{2}{c}{} \\
	\hline
	\multicolumn{2}{| c |}{} &
	\multicolumn{2}{c |}{Ejecta background number:} &
	\multicolumn{3}{c |}{Ejecta background properties:} \\
	[0.5ex]
	\hline
	$ i $ &	$ \unit [t_{i}] {[yr]}	$ &
	$ N_{\epop}^{\tot} $ &
	$ F_{\epop}^{\tot} $ &
	$ \unit [M_{\epop}^{\ind}] {[\msun]} $ &
	$ \unit [M_{\epop}^{\tot}] {[\msun]} $ &
	$ \unit [X_{\epop}^{\ind}] {[pc]} $ \\
	[0.5ex]
	\hline\hline
	$ 28 $ & $ 1.38 \times 10^{6} $ &
	$ (6.7 \substack {+8.1 \\ -2.6}) \times 10^{7} $ &	
	$ (9.0 \substack {+7.1 \\ -3.8}) \times 10^{-6} $ &	
	$ 67 \substack {+89 \\ -34} $ &	
	$ (1.8 \substack {+2.1 \\ -0.7}) \times 10^{9} $ &	
	$ - $ \\
	$ 37 $ & $ 1.38 \times 10^{7} $ &
	$ (6.7 \substack {+5.2 \\ -3.2}) \times 10^{9} $ &	
	$ (3.3 \substack {+2.7 \\ -1.8}) \times 10^{-3} $ &	
	$ (7.0 \substack {+7.1 \\ -4.2}) \times 10^{3} $ &	
	$ (1.9 \substack {+1.5 \\ -0.9}) \times 10^{11} $ &	
	$ - $ \\
	$ 46 $ & $ 1.38 \times 10^{8} $ &
	$ (1.12 \substack {+0.81 \\ -0.39}) \times 10^{10} $ &	
	$ 2.6 \substack {+2.1 \\ -1.5} $ &	
	$ (1.2 \substack {+1.1 \\ -0.6}) \times 10^{4} $ &	
	$ (3.2 \substack {+2.3 \\ -1.3}) \times 10^{11} $ &	
	$ 108 \substack {+29 \\ -17} $ \\
	$ 55 $ & $ 1.38 \times 10^{9} $ &
	$ (1.4 \substack {+1.0 \\ -0.5}) \times 10^{10} $ &	
	$ (1.21 \substack {+0.99 \\ -0.67}) \times 10^{3} $ &	
	$ (1.5 \substack {+1.4 \\ -0.8}) \times 10^{4} $ &	
	$ (4.1 \substack {+2.9 \\ -1.6}) \times 10^{11} $ &	
	$ 99 \substack {+26 \\ -14} $ \\
	$ 64 $ & $ 1.38 \times 10^{10} $ &
	$ (1.7 \substack {+1.2 \\ -0.6}) \times 10^{10} $ &	
	$ (1.1 \substack {+1.0 \\ -0.7}) \times 10^{6} $ &	
	$ (1.8 \substack {+1.7 \\ -0.9}) \times 10^{4} $ &	
	$ (4.9 \substack {+3.4 \\ -1.8}) \times 10^{11} $ &	
	$ 93 \substack {+24 \\ -13} $ \\
	\hline
	\end{tabular}
	\caption{
	Shown are the number of bubbles, $ N_{\textrm{P}} $, segregated by the population (cluster and ejecta) type, in a Milky Way-like galactic halo if all DM is made of PBHs, as well as the volume fraction occupied by the spheres of influence of these bubbles in such galaxy, $ F_{\textrm{P}} $at selected times, and where $ \textrm{P} = \cpop, \epop $ stands for the cluster and ejecta populations.
	Shown as well are the individual masses of such bubbles, $ \unit [M_{\textrm{P}}^{\ind}] {[\msun]} $, and the total mass of all such bubbles in a milky Way-like galactic halo, 	$ \unit [M_{\textrm{P}}^{\tot}] {[\msun]} $.
	Last, the average distance between individual PBHs in such bubbles, $ \unit [X_{\textrm{P}}^{\ind}] {[pc]} $, are shown for the time-slices in which the bubble's spheres on influence have undergone sufficient expansion that they start to merge with each other, a point where the volume fraction holds $ F_{\textrm{P}} \approx 1 $.	
	}
	\label{tab: Background properties}
\end{table*}



\subsection{Gravitational event rates}
\label{subsec: Gravitational event rates}

We aim now to compute a minimum threshold for the average hyperbolic encounter rate and merger event rate in time per comoving volume, starting from the computed rates for both per cluster in Sections~\ref{subsubsec: hyperbolic encounter rates} and \ref{subsubsec: Merger event rates}.

In order to do so, we begin by computing the total DM mass contained in a comoving box of volume $ \unit [1] {Gpc^{3}} $. The total DM energy density of the Universe is $ \rho_{\dm} = \unit [(3.965 \pm 0.073) \times 10^{19}] {\msun Gpc^{-3}} $, according to \textit{Planck 2018} (TT, TE, EE+lowE, see Ref.~\cite{Aghanim:2018eyx}).

The initial cluster total mass is given by Eq~\eqref{eq: Cluster mean total mass}, which reduces to $ M_{\cpop,0}^{\tot} = N_{\ocap} \bar{m}_C (t_0) $ thus obtaining
\begin{equation}
	M_{\cpop,0}^{\tot} = \unit [(2.991 \pm 0.074) \times 10^{4}] {\msun},
	\label{eq: Cluster mean total mass computed}
\end{equation}
for a total number of clusters per comoving volume of $ N_{\cpop}^{\tot} = \rho_{\dm} / M_{\cpop,0}^{\tot} $, which yields
\begin{equation}
	N_{\cpop}^{\to,t} = \unit [(1.325 \pm 0.040) \times 10^{15}] {Gpc^{-3}},
	\label{eq: Cluster mean total number computed}
\end{equation}
which is a large number indeed, albeit an expected one since it is known that a $ \unit [1] {Gpc^{3}} $ comoving box contains about $ \order {10^{7}} $ Milky Way-sized galactic haloes.



\subsubsection{Comoving hyperbolic encounter rates}
\label{subsubsec: Comoving hyperbolic encounter rates}

We do now take the hyperbolic encounter rates of Section~\ref{subsubsec: hyperbolic encounter rates} and scale them to a $ \unit [1] {Gpc^{3}} $-sized box with the transformation $ \Gamma_{r}^{\mathrm{S}} \rightarrow N_{\cpop}^{\tot} \Gamma_{r}^{\mathrm{S}} $. The final comoving slingshot rates are then
\begin{align}
	\Gamma_{1}^{\mathrm{S}} (0 \leq t_{i} \leq 10) & = \unit [(2.7 \substack {+7.6 \\ -2.1}) \times 10^{12}] {yr^{-1} Gpc^{-3}}, \\
	\Gamma_{2}^{\mathrm{S}} (11 \leq t_{i} \leq 19) & = \unit [(3.0 \substack {+2.5 \\ -1.6}) \times 10^{12}] {yr^{-1} Gpc^{-3}}, \\
	\Gamma_{3}^{\mathrm{S}} (20 \leq t_{i} \leq 28) & = \unit [(2.9 \substack {+3.9 \\ -2.6}) \times 10^{11}] {yr^{-1} Gpc^{-3}}, \\
	\Gamma_{4}^{\mathrm{S}} (29 \leq t_{i} \leq 37) & = \unit [(1.2 \substack {+4.8 \\ -0.9}) \times 10^{10}] {yr^{-1} Gpc^{-3}}, \\
	\Gamma_{5}^{\mathrm{S}} (38 \leq t_{i} \leq 46) & = \unit [(1.4 \substack {+5.2 \\ -1.1}) \times 10^{9}] {yr^{-1} Gpc^{-3}}, \\
	\Gamma_{6}^{\mathrm{S}} (47 \leq t_{i} \leq 55) & = \unit [(1.3 \substack {+5.5 \\ -1.0}) \times 10^{8}] {yr^{-1} Gpc^{-3}}, \\
	\Gamma_{7}^{\mathrm{S}} (56 \leq t_{i} \leq 64) & = \unit [(1.2 \substack {+5.9 \\ -0.9}) \times 10^{7}] {yr^{-1} Gpc^{-3}}.
\end{align}

These are indeed very large rates, however the same caveats that applied to the slingshot rates of Section~\ref{subsubsec: hyperbolic encounter rates} still apply here; mainly that while the number of hyperbolic encounters is underestimated because of the discrete time-step in each of the time-runs and the non-relativistic nature of the N-body code, we also find too many such events to be understood as gravitational wave generating events. Indeed most encounters most often happen at PBH-to-PBH distances way too large to detect the corresponding emission of GWs with current instruments.



\subsubsection{Comoving merger event rates}
\label{subsubsec: Comoving merger event rates}

We do at last take the merger event rates of Section~\ref{subsubsec: Merger event rates} and scale them again to a $ \unit [1] {Gpc^{3}} $-sized box with the transformation $ \Gamma_{r}^{\mathrm{M}} \rightarrow N_{\cpop}^{\tot} \Gamma_{r}^{\mathrm{M}} $. The final comoving merger event rates are then
\begin{align}
	\Gamma_{6}^{\mathrm{M}} (46 \leq t_{i} \leq 55) & = \unit [852 \pm 26] {yr^{-1} Gpc^{-3}}, \\
	\Gamma_{7}^{\mathrm{M}} (56 \leq t_{i} \leq 64) & = \unit [2409 \pm 74] {yr^{-1} Gpc^{-3}}.
\end{align}

The merger event rates segregated by their origin, would be, for all 51 mergers occurring from previously isolated PBHs
\begin{align}
	\Gamma_{6}^{\mathrm{M},\cpop,\ipop} (46 \leq t_{i} \leq 55) & = \unit [213 \pm 11] {yr^{-1} Gpc^{-3}}, \\
	\Gamma_{7}^{\mathrm{M},\cpop,\ipop} (56 \leq t_{i} \leq 64) & = \unit [1066 \pm 33] {yr^{-1} Gpc^{-3}},
\end{align}
while for all 66 mergers occurring originating in previously bounded systems, we find that
\begin{align}
	\Gamma_{6}^{\mathrm{M},\epop,\bpop} (46 \leq t_{i} \leq 55) & = \unit [639 \pm 20] {yr^{-1} Gpc^{-3}}, \\
	\Gamma_{7}^{\mathrm{M},\epop,\bpop} (56 \leq t_{i} \leq 64) & = \unit [1337 \pm 41] {yr^{-1} Gpc^{-3}}.
\end{align}

Note however, that the same caveats that applied to the merger event rates of Section~\ref{subsubsec: Merger event rates} still apply here; mainly that we do underestimate mergers to some extent since the evolution is computed in a non-relativistic manner, and thus this rates may have to be interpreted in practice as lower bounds to the real PBH merger rates.

Indeed the lack of inspiral may be particularly relevant at the beginning of the simulations, as we have explained how an abundance of merging at early times has the effect of both increasing the mass density at the core of the cluster and decreasing the cluster total mass at later times by stripping the cluster of its outer layer due to in-falling PBHs acquiring large infall velocities only to be dispersed later on on a one-to-one encounter with another PBH at high relative speeds.

Note that, despite the large number of mergers produced per comoving volume, there is no tension between the total comoving merger rate from all population sources of PBHs found at present, $ \Gamma_{7}^{\mathrm{M}} (56 \leq t_{i} \leq 64) = \unit [2409 \pm 74] {yr^{-1} Gpc^{-3}} $ and the typically $ \order {100}-\order {1000} $ values estimated from LIGO's runs O1 and O2 (see Ref.~\cite{Gow:2019pok}), since LIGO is not in any case sensitive merger events in which the participating BHs have total masses as large as those found in our simulation, typically of order $1000\ \msun $, though future experiments such as LISA will (see Ref.~\cite{Bartolo:2016ami}).



\section{Conclusions}
\label{sec: Conclusions}

Understanding the nature of Dark Matter remains one of the key questions in theoretical physics and cosmology and has resulted in a plethora of models ranging from weakly interacting massive particles (WIMPs), sterile neutrinos, axions, modifications of gravity like MOND but also PBHs, see Ref.~\cite{Arun:2017uaw,Li:2017kst} for recent reviews. Also, an interesting possibility is to have a mixed $f(R)$ gravity model with axion, which could make it easier to be detected as it affects both the inflation and DE era \cite{Odintsov:2019mlf,Odintsov:2020nwm}.

Here we focus on the PBHs as it also provides a model that connects the early time and late time physics and remains a plausible candidate for DM. In this case, PBHs not only play the role of DM, but they may also influence DE at late times, as by evaporating they inject energy and behave as a time-dependent DE equation of state, see Ref.~\cite{Nesseris:2019fwr}. PBH may also induce non-adiabatic perturbations on DE, see
Ref.~\cite{Arjona:2020yum}.

In this work though, we focus on the dynamics of PBH in clusters, as it is one of the cornerstones in our understanding of structure formation at small scales, as well the way classical PBH constraints can be resolved. Furthermore, PBHs have unique signatures that could be detected by current and future GW detectors, such as AdvLIGO and LISA. In particular, close hyperbolic PBH encounters in dense clusters, such as the ones we simulated here, emit bursts with millisecond durations which may be detected depending on the duration and peak frequency of the emission, but also the rates and distributions of the PBHs.

We found that PBHs can constitute a viable DM candidate, whose clustering offers a rich phenomenology. We have quantified the stability and rate of evaporation of PBH clusters, obtained the number of PBHs that remain in the cluster and those that are ejected from it to the background, also the fraction of these that show up in gravitationally-bound subsystems or merge with each other, as well as computing their parent and merger tree histories.

Furthermore, these PBH clusters are indeed stable and survive until present times retaining about a third of their total initial mass, while the remaining two thirds is transferred to the background ejecta, providing a more or less uniform spatial distribution of both free PBH and the cores of the original PBH clusters with a wide range of masses. In particular, we find that cluster puffing up and evaporation leads to PBH subhaloes of $ \unit [\order {1}] {kpc} $ in radius containing at present times about 36\% of objects and mass, with $ \unit [\order {100}] {pc} $ cores. We also find that these PBH sub-haloes are distributed within larger PBH haloes of $ \unit [\order {100}] {kpc} $, containing about 63\% of objects and mass, coinciding with the sizes of galactic halos.

We have also extracted the component mass profiles and mass spectrum of binary and merger pairs in these clusters. In particular, we find that binary systems do appear frequently, constituting about 9.5\% of all PBHs at present, with mean and median mass ratios of $ \bar{q}_{\bpop} = 0.154 $ and $ \tilde{q}_{\bpop} = 0.0552 $, and total masses of $ \bar{m}_{\tcap,\bpop} = \unit[303]{\msun} $ and $ \tilde{m}_{\tcap,\bpop} = \unit[183]{\msun} $, and larger, less skewed binary systems for those PBHs in the ejecta rather than those in the cluster.

Alternatively, we find that mergers are very infrequent, with barely 0.0023\% of all PBHs merged at present, having mean and median mass ratios of $ \bar{q}_{\mpop} = 0.965 $ and $ \tilde{q}_{\mpop} = 0.545 $. Merger total masses masses found are $ \bar{m}_{\tcap,\mpop} = \unit[1670]{\msun} $ and $ \tilde{m}_{\tcap,\mpop} = \unit[1510]{\msun} $, while found chirp masses are $ \bar{m}_{c,\mpop} = \unit[642]{\msun} $ and $ \tilde{m}_{c,\mpop} = \unit[567]{\msun} $. Also, we find that the PBHs resulting from the largest mergers tend to segregate themselves to the cluster cores, and that merged PBHs escaping to the ejecta are often less massive and more skewed.

We have looked at the one-to-one close interactions of these PBHs and extracted lower bounds to the hyperbolic encounter and merger event rates produced. The simulations have a number of shortcomings that make a proper comparison with observations somewhat difficult, particularly so for the hyperbolic encounter rates. As for the merger event rates, we have found rates for massive black holes beyond the range of sensitivity of LIGO-VIRGO observations. We find close encounter rates to be $ \Gamma^{\mathrm{S}} = \unit [(1.2 \substack {+5.9 \\ -0.9}) \times 10^{7}] {yr^{-1} Gpc^{-3}}$ and merger rates of $\Gamma^{\mathrm{M}} = \unit [1337 \pm 41] {yr^{-1} Gpc^{-3}} $. These values are thus far compatible with observations, and might be probed by future space-based observatories like LISA.

We have also computed the cluster and ejecta dynamical parameter space distributions, and found that the mass, position, velocity and density profiles and found them to be consistent with that of a cold DM component. Interestingly, at present times the clusters overlap just enough that their outer layers are indeed in contact with each other while retaining their separate cores, having length scales comparable to those of dwarf galaxies and globular clusters. Also interesting is the fact that the ejecta background once the many different ejecta bubbles merge has length scales comparable to medium sized galactic haloes. Moreover, we have studied the degree of cluster mass segregation and dynamical friction, and found those to be of no great significance.

We then determined the cluster and ejecta orbital parameter-space distributions of PBH clusters at selected times and the segregation by bounded and unbounded PBHs and studied their evolution. We have found that the profiles for both the orbital eccentricity and semi-major axis are compatible with current observations, that there is a large number of cluster transient binaries, disrupted by third encounters, and stable ejecta binaries that may be responsible for the events detected by LIGO. Last, we have computed the power spectrum of potential hyperbolic encounters between close cluster PBH pairs.

Overall, we have found that inhomogeneously distributed PBH clusters are a good candidate for a cold, nearly collisionless DM, consistent with astrophysical observations of the DM distribution and replicating the features that a good DM candidate must uphold, throughout the entire history of the Universe starting in the matter era.

It remains an open question how to extrapolate these results to a more general set of initial conditions in which the mass, position and velocity distribution of the PBHs may be different and the PBH cluster initially more concentrated. We leave the study of a more varied set of PBH clusters for future work \footnote{After our paper was submitted to the arXiv we received a message informing us that a related paper had appeared a few days before~\cite{Jedamzik:2020ypm}, addressing the problem of PBH clustering dynamics is a similar way.}.

Our work may also be extended in many other ways. For example, an important limitation in our current analysis is that we did not include GW emission from the PBHs as the AstroGrav code is purely Newtonian. This maybe resolved in the future by extending the equations of motion to higher order in the post-Newtonian expansion. Another possibility is to do simulations with a varying and higher number of PBHs in the simulation, so as to study more realistic clusters of different sizes and more varying in mass.


\begin{acknowledgments}

The authors acknowledge support from the Research Project PGC2018-094773-B-C32 [MINECO-FEDER], the Centro de Excelencia Severo Ochoa Program SEV-2016-0597 and use of the Hydra HPC cluster at the Instituto de F\'isica Te\'orica (UAM/CSIC). S.N. acknowledges support from the Ram\'{o}n y Cajal program through Grant No. RYC-2014-15843. M.T. acknowledges support from the Grant No. BES-2016-077817 (MINECO-FPI).

\end{acknowledgments}



\bibliography{Main}
\bibliographystyle{ieeetr}

\end{document}